# Review of magnetic- and antimagnetic-rotational structures in nuclei

Sushil Kumar[1*], Sukhjeet Singh[1], Balraj Singh[2], Amita[3], Ashok Kumar Jain[1,4,5]

1. Department of Physics, Akal University, Talwandi Sabo, Bathinda-151302, India
2. Department of Physics and Astronomy, McMaster University Hamilton, Ontario L8S 4M1, Canada
3. Department of Physics, R.B. S. College, Agra, UP-282002, India
4. Amity Institute of Nuclear Science and Technology, Amity University UP, Noida-201313, India
5. Department of Physics, Indian Institute of Technology, Roorkee- 247667, India.

**Abstract:** This work is an update of the 2000 publication of magnetic-rotational bands by Amita et al. [1], followed by an unpublished update of 2006 [2], and reviews detailed experimental data extracted from original publications for 228 magnetic-rotational (MR or Shears) structures spread over 117 nuclides, and 40 antimagnetic-rotational (AMR) structures in 28 nuclei, with a brief commentary about each band. Many of these nuclei are located at or near the semi-magic nucleon numbers, mostly for protons. For example, 88 MR bands are currently known for the Pb (Z=82) nuclei, and 29 AMR band in Pd, Cd and In nuclei. It is interesting that the proton magic numbers appear to play a major role in the MR phenomenon, which seems less well understood. A brief discussion of the salient features of the MR and AMR bands and their theoretical interpretation has been presented in the present review. The tables contain gamma-ray energies, associated level energies with spins and parities, level lifetimes, B(M1), B(E2), and B(M1)/B(E2) ratios and probable spherical quasiparticle configurations. We find that many bands claimed in the literature as MR and AMR bands still have tentative assignments, as level lifetimes, thus B(M1) and B(E2) values, for a large number of MR and AMR bands, which can potentially provide critical criteria for firm identification of such structures, are lacking. Additionally, theoretical model calculations for many of these bands, which could provide insight for a better description of nuclear structure, are also lacking in literature. While this review is mainly based on original research articles, nuclear structure databases ENSDF [3], XUNDL [4], and NSR [5] have been consulted for completeness. The literature cut-off date March 31, 2025.

**In Memorium:** We dedicate this work to our collaborator and mentor Dr. Balraj Singh.

[*]Corresponding author: sushil.rathi179@gmail.com

## 1. Introduction

Nuclei display a variety of excitations which can often be grouped into a set of levels called bands, the excitation energies of the levels, in some cases, related to a simple formula. Rotational and vibrational bands are the most common examples of such excitations. Appearance of rotational motion is intimately linked to the non-sphericity of the nuclear mean field and the presence of anisotropic shapes in nuclei. This breaking of the spherical symmetry in nuclei normally happens in mid-shell nuclei, having nucleon numbers intermediate to the magic numbers [6]. However, appearance of well-defined rotational bands at higher excitations have also been noticed in nuclei close to the magic numbers, which is known to give rise to the phenomenon of super-deformed bands [7,8]. These normal deformed (ND) and super-deformed (SD) rotational bands represent the manifestation of the rotation of charged non-spherical nuclei, the levels of the band strongly connected by electric quadrupole (E2) transitions; we may, therefore, also call this rotation as an electric rotation. Last 50 years of nuclear structure physics have been a witness to the observation of a variety of new exotic excitations, which are related to rotational motion in nuclei [9].

Until early 1990's, it was a general belief that only deformed nuclei can exhibit rotational bands. It was therefore a big puzzle for the nuclear physicists when several groups [10,11,12] working independently, observed regular patterns of dipole (M1) gamma rays in the near spherical

Pb isotopes. A detailed study revealed that the intraband transitions in these group of levels were magnetic dipole (M1) in nature in contrast to normal deformed and super-deformed nuclei where the intraband transitions were electric quadrupole (E2). These bands were therefore explained and characterized as Magnetic Rotational (MR) bands by Frauendorf [13], where spherical symmetry is broken by the magnetic moments of nuclei, who later interpreted [14] these excitations in terms of a Tilted Axis Cranking (TAC) model, and a review article highlighting the symmetry breaking in such systems and TAC, and a more complete review of the unified model and beyond, both dealing with the MR bands was published by Frauendorf [15]. A systematic of various properties exhibited by majority of MR bands as compared to normal deformed and super-deformed bands is presented in Table 1. A more extensive list of the theory papers related to the MR bands is included in this article along with a list of the experimental papers.

The present work contains a detailed Table of all the well-established, likely, and tentative candidates of magnetic rotational (MR) bands for 228 bands in 117 nuclides, and 40 antimagnetic rotational (AMR) bands in 28 nuclei. Most of the MR-bands, have been observed in either even–even or odd–*A* nuclei; the largest number discovered in the Pb region, where 88 bands have been observed in 33 nuclides, with the highest number (nine bands) seen in $^{196}$Pb, which has nine bands, and eight in $^{194}$Pb. Nuclei with neutron numbers 112-117 and Z=82 alone have 37 MR bands. The longest sequences of single MR bands have been seen in $^{194}$Pb (spin $15^+$ to $33^+$), $^{196}$Pb (spin $17^+$ to $36^+$, and $16^-$ to $36^-$), and $^{197}$Pb ($27/2^-$ to $65/2^-$), thus showing that Pb region is the most fertile ground for observing MR bands. Presence of high-j orbitals ($i_{13/2}$ and $h_{9/2}$ protons and $i_{13/2}$ neutrons) along with the presence of many nucleons in the core appear to play an important role. It may be noted that this is also a fertile region for observing the super-deformed rotational bands. Most of the antimagnetic rotational (AMR) bands have been reported for the Pd, Cd and In nuclei, for example, eight bands in Pd, twelve in Cd, and nine in In isotopes. During the course of this work, a systematic study on magnetic and antimagnetic rotational bands has been published by J. X. Teng and K. Y. Ma [16]; however, the thematic focus of their study differs from the present research.

Nearly all the data have been searched and taken from the original references listed at the end of the Table. The unique nuclear structure databases ENSDF [3], and XUNDL [4] have been extensively used. Literature search was considerably facilitated by NSR database [5].

## 2. General Characteristics of the MR bands

The general characteristics exhibited by majority of the MR bands may be summarized as follows:

1. The levels in the bands (away from band crossings) have energies that follow a pattern close to the $I(I+1)$ behaviour. Here, $I$ denotes the total angular momentum of a level.
2. The bands consist of strong M1, ΔI=1 intraband transitions with only weak E2 crossovers resulting in large B(M1)/B(E2) ratios, ≥20 $\mu_N^2/(eb)^2$, Weak or absent E2 transitions also imply a small deformation.
3. The M1 transitions have a reduced transition probability B(M1) ~ 2-10 $\mu_N^2$, and B(E2) values may lie in the range of 1-0.01 $(eb)^2$ or less.

4. The ratio of the dynamic moment of inertia ($\Im^{(2)}$) to the B(E2) is large, $\Im^{(2)}$/B(E2)>150 $MeV^{-1}(eb)^{-2}$ when compared with the normal-deformed ~10 $MeV^{-1}(eb)^{-2}$ or super-deformed ~5 $MeV^{-1}(eb)^{-2}$ structures.
5. The B(M1) values decrease with increasing angular momentum in an MR band.

Table 1. Comparison of various properties exhibited by majority of MR bands with the ND and SD bands.

| | MR bands | ND bands | SD bands |
|---|---|---|---|
| E(K) (MeV) | >2 | ~ 0 – 2 | >2 |
| $I^{\pi}_{min.}$ | 6 – 15 ħ | ~ 0 – 10 ħ | >10 ħ |
| $\varepsilon_2$ | ≤ 0.15 | 0.2-0.35 | ~ 0.5 |
| B(M1) $\mu_N^2$ | ≥ 1 | < 0.1 | ~ 0.0 |
| B(M1)/B(E2) $(\mu_N/eb)^2$ | ~10-100 | ~1 | Very Small |
| $\Im^{(2)}$ ($\hbar^2$ $MeV^{-1}$) | ~10-25 | ~50 | ~100 |
| $\Im^{(2)}$/ B(E2) ($\hbar^2$ $MeV^{-1}$ $(eb)^{-2}$) | >150 | ~10 | ~5 |

**3. Physical mechanism of magnetic rotation: shears mechanism**

As a major fraction of the well-defined MR bands has been observed in the Pb region, we discuss basic features of the MR bands as well as an underlying physical mechanism by using specific examples from the Pb isotopes. The ΔI=1 bands observed in the lead isotopes have been mostly assigned weakly oblate configurations consisting of the $K^{\pi}=11^{-}$ excitation arising from two-proton particles in $\{h_{9/2}\otimes i_{13/2}\}$ orbital coupled to low-Ω one/three quasi-neutrons (in odd-A nuclides) or two/four quasi-neutrons (in even-A nuclides) dominated by $i_{13/2}$ neutron holes. At the bandhead, the proton angular momentum vector $\vec{j_{\pi}}$ (which arises from particles in $h_{9/2}$ and $i_{13/2}$ orbitals) is nearly parallel to the symmetry axis due to torus-like density distribution, while the neutron angular momentum $\vec{j_{\nu}}$ (from holes in high-j $i_{13/2}$ subshell) is nearly perpendicular to it due to dumb-bell-like density distribution. It may be noted that the role of high-j orbitals is crucial in these structures as it is in many other rotational phenomena.

The total angular momentum vector $\vec{I}$ then lies along a tilted axis at an angle θ with respect to the symmetry axis. Higher angular momentum states are generated by aligning $\vec{j_{\pi}}$ and $\vec{j_{\nu}}$ in a way similar to the closing of the blades of a pair of shears. This coupling of the proton particles and neutron holes results in a large magnetic dipole moment ($\mu$), which precesses around $\vec{I}$. The perpendicular component of $\mu$ relative to $\vec{I}$, denoted $\mu_{\perp}$, breaks the spherical symmetry. It also decreases in a characteristic manner as $\overrightarrow{j_{\pi}}$ and $\overrightarrow{j_{\nu}}$ align (and the total angular momentum increases) as shown in the Figure 1. This alignment costs energy and the excitation energy of the MR band increases as the angular momentum increases.

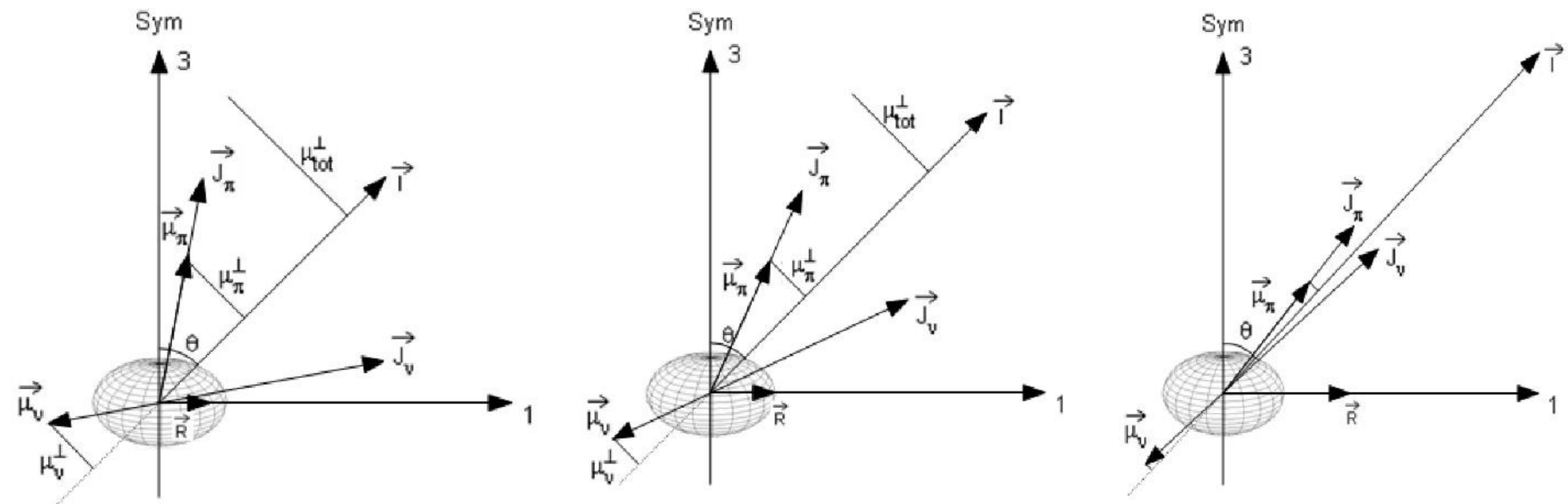

Figure 1: Angular momentum coupling scheme for shears mechanism in a ΔJ=1 MR band in a weakly oblate deformed nucleus, at low spin, i.e. at the bandhead (left), at intermediate spin (middle), and at maximum spin (right). The perpendicular component of magnetic moment shows a decline as the spin rises.

The behaviour of the B(M1) values is the most crucial experimental signature of the shears mechanism. If an MR band arises from the collective motion of a non-spherical shape, then the B(M1) values are expected to be rather constant. If an MR bands involves shears mechanism, then this should lead to a characteristic decline of the B(M1) values with increasing angular momentum. This is because B(M1) is proportional to $\mu_{\perp}^{2}$, and $\mu_{\perp}^{2}$ decreases with rising angular momentum. The maximum angular momentum for a configuration is reached when the shears close fully, terminating the MR band, as shown in Fig. 1 (right panel). Another important quantity which is used for knowing the strength of the M1 transitions is the ratio B(M1)/B(E2). The ratio of B(M1) and B(E2) can be obtained from the experimental data by using the relation

$$\frac{B(M1)}{B(E2)} = \frac{0.6975 E_r^5(E2) I(M1)}{E_r^3(M1) I(E2)} \quad (1)$$

$E_r$ is the gamma energy in keV, *B(E1)* and *B(E2)* are in $e^2b$ and $e^2b^2$ respectively. *B(M1)* is in $\mu_N^2$. This B(M1)/B(E2) ratio for a large number of cases has been tabulated in the data table. We summarize below the conditions required for the MR bands:

1. A particle-hole coupling so that the angular momentum at the bandhead is large and maximum overlap of the proton and neutron wavefunctions occur as the shears close.
2. The active orbitals involved must have large j values. This can give rise to a good tilt angle and a long sequence of levels can be formed.
3. The deformation of the core of the nucleus must be small enough so that the rotational motion does not dominate over the shears mechanism.

Above conditions suggest that regions around the spherical magic numbers are probably the most suitable for the occurrence of MR bands. However, as the data presented in this paper show, MR bands are spread out over many regions of nuclear chart. The common thread possibly is the presence of high-j proton/neutron orbitals, an essential requirement to generate sufficient magnetic moment and angular momentum. Moreover, the proton magic numbers are more important, which is not surprising, as the high-j protons will contribute the most to the magnetic moment.

We briefly discuss some of the observed experimental features of these exotic structures by selecting an example from the Pb region where the MR bands are prominent and well established. We show in Fig. 2, a plot of angular momentum *I* vs the gamma ray energy for an MR band in $^{196}$Pb. As can be seen it is a monotonically rising plot, the $E_\gamma$ increasing with the spin *I*, a behaviour similar to the rotational bands, hence the name Magnetic Rotational band. It is the magnetic dipole that breaks the isotropy in MR bands, and the shears mechanism requires more and more energy as the blades of shears close increasing the spin *I*.

This smoothly rising energy, however, sometimes exhibits a backbending, a feature signifying the crossing of two bands, as shown in the right panel of Fig. 2. It is rather interesting that the second band which crosses the first band is also an MR band. The higher lying MR band has a configuration with an additional pair of neutron hole alignment. The configurations assigned are $\pi(i_{13/2}h_{9/2})_{K=11^-} \otimes \nu(i_{13/2}^{-2})$ before, and $\pi(i_{13/2}h_{9/2})_{K=11^-} \otimes \nu(i_{13/2}^{-4})$ after the band crossing. The two blades of shears, which were beginning to close as the spin was rising in the lower MR band, again open up in the higher lying MR band, and the shears mechanism becomes active once again.

A second order polynomial fit to the level energy E vs. *I* data is shown on the right-hand side of Fig. 2. By polynomial fits of these bands, and from the minimum of the curve, the bandhead spin and energy were deduced to be 21ħ, and 7465.5 keV, respectively. From the crossing point of the two bands, the band crossing energy and spin could also be calculated as 8264.2 keV and 27ħ, respectively.

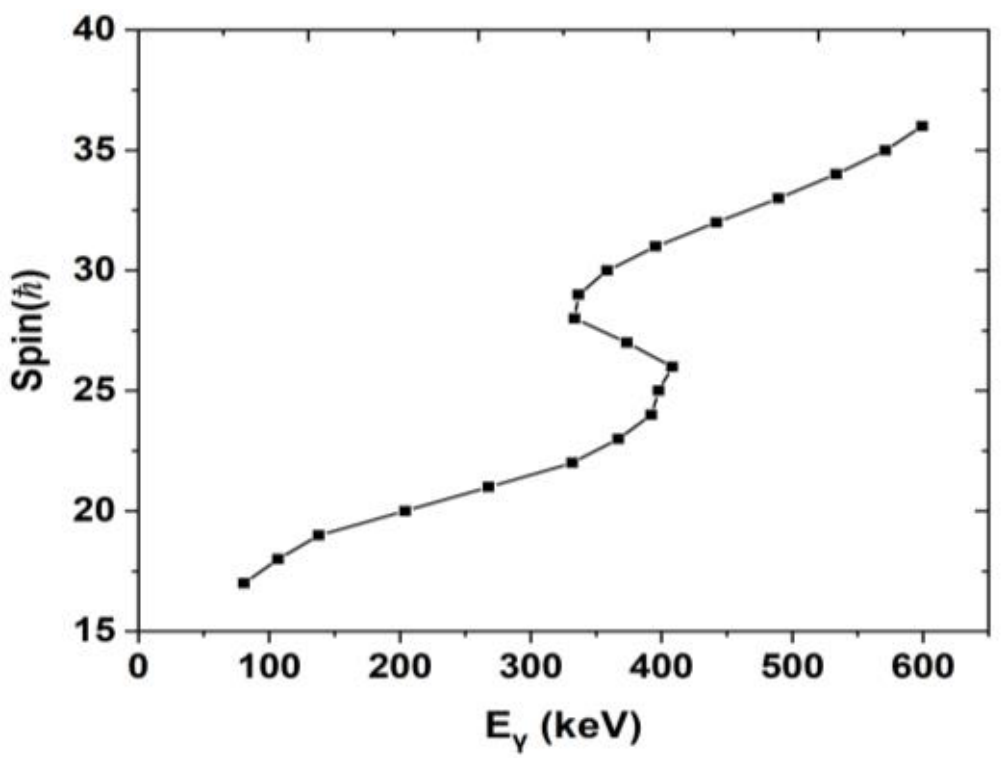

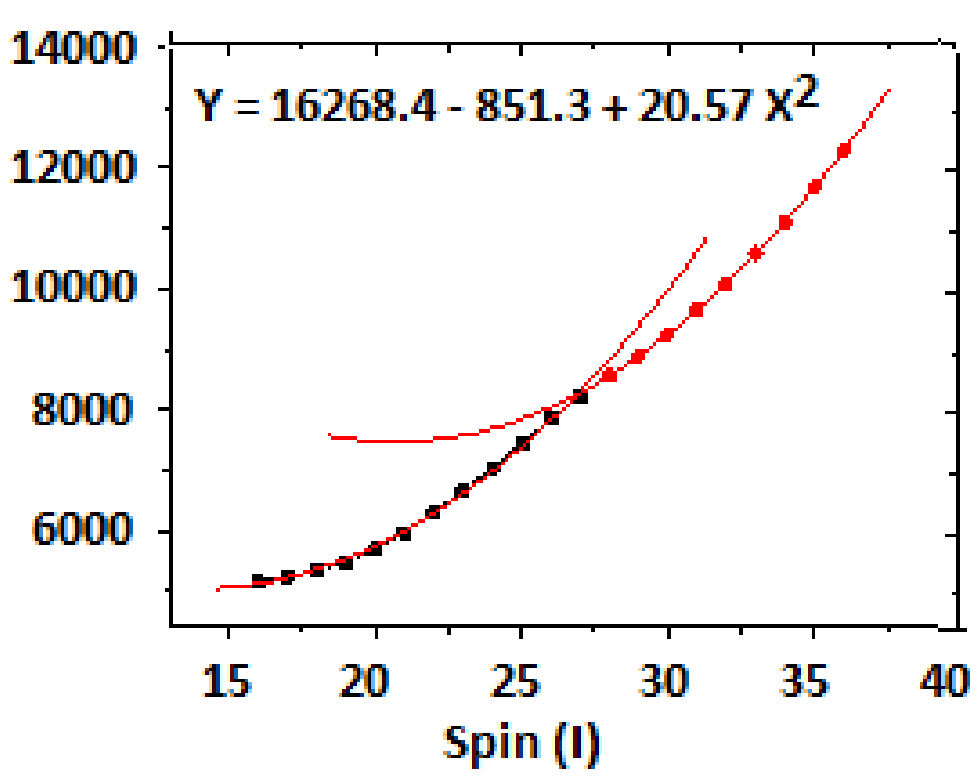

Figure 2: Plots of the angular momentum I vs. gamma ray transition energies Eγ, and level energy E vs. I for the MR band 4 of $^{196}$Pb.

Using these values, we could calculate the shears angle for each spin for the band after the band crossing by using the semi-classical model of Macchiavelli et al. [17,18,19]. At the bandhead, we assume that there is a perpendicular coupling of shears and thereafter with increase in spin the shears close. A value of effective g-factor 0.92 was assumed. We have calculated the variation in the B(M1) values with the change in the shears angle and the results are shown in Fig. 3. It has been observed that the B(M1) values increase near the band crossing. The excited band lies lower in energy than the lower MR band after the band crossing. The shears reorient by increasing the shears angle between the proton and neutron angular momentum vectors, which open up again, to reduce the energy at the band crossing.

Such a backbending and band crossing has been seen in many nuclei in the Pb region because of the observation of long sequences of levels in the MR bands. A crossing of two MR bands with largest number of quasi-particles was seen in the $^{198}$Bi by Pai *et al.* [20], where a crossing of eight quasi-particle (qp) MR band with a six-quasi-particle band was observed. Crossing of a 6-qp and a 4-qp band was also seen in the same nucleus.

It is, however, important to point out that these calculations are kind of idealized and are good for a physical understanding. The most successful microscopic nuclear structure model applicable to the MR bands is the Tilted Axis Cranking model due to Frauendorf [13]. A detailed application of this model and their results may be found in the references [13-15, 21]. For a successful application of the hybrid version of TAC to Rb isotopes, see Amita *et al.* [22]. Many MR bands also display a significant signature splitting or odd-even staggering in energy. This phenomenon generally requires the presence of a small deformation, high-j configuration, and high spin. All the conditions are present in the MR structures. The particle rotor model could suitably be applied to explain the odd-even staggering in the MR bands [23].

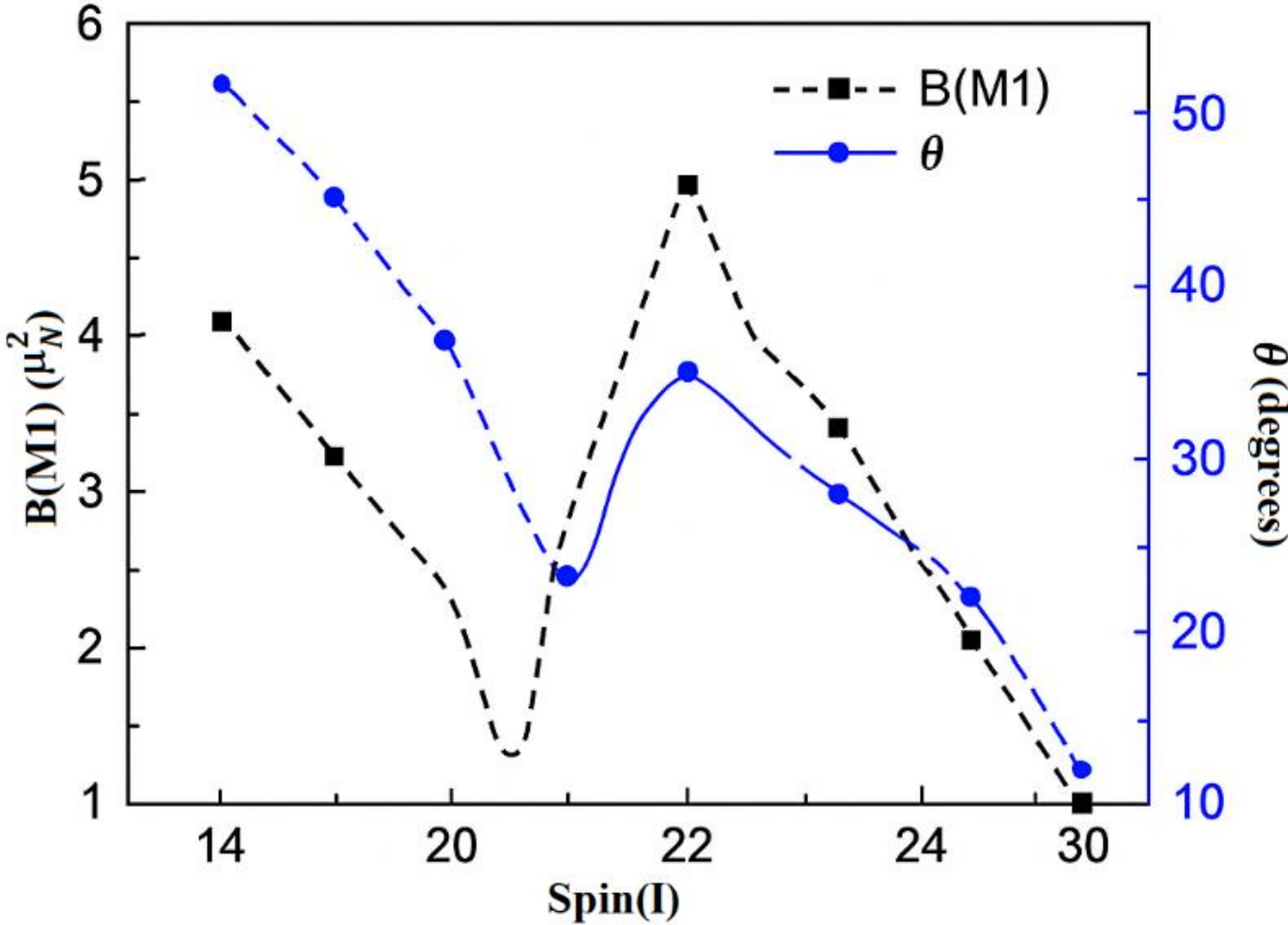

Fig. 3: Plot of the B(M1) vs. I for the MR band 4 of $^{196}$Pb. Also shown is the variation of the tilt angle θ with the spin I, which closely follows the variation in BM1 values.

## 4. Antimagnetic Rotation

A phenomenon closely related to the magnetic rotation bands was also proposed by Frauendrof [24], in an analogy to antiferromagnetism in solids. It involves an alternative arrangement of the proton and neutron angular momentum vectors ($\boldsymbol{j}_\pi$ and $\boldsymbol{j}_\nu$), which also breaks the symmetry about the total angular momentum vector. This leads to a new kind of phenomenon termed as 'Antimagnetic Rotation', where the neutrons are aligned along an axis perpendicular to the symmetry axis and the protons are placed in a stretched position along the symmetry axis as shown in Fig. 4. As the two proton blades close, higher spin states are generated, and the contribution of the single particle mode to total angular momentum rises. The magnetic moments of the two proton spins cancel with each other, and therefore, only the E2 transitions are possible for AMR bands.

The AMR bands are expected to occur in weakly deformed nuclei in the same mass regions as the MR bands. As the deformation is very small, the E2 are weak, which further decline with increase in spin as the blades close, generating angular momentum along the rotation axis. The angular momentum is generated by the gradual alignment of the two proton spin vectors, in contrary to collective angular momentum of a well-deformed nucleus where it is generated from the contributions of many particles. In the case of even-even nuclei, a pair of proton angular momentum vectors (arising due to a pair of proton holes in high-Ω orbitals) are in a stretched mode with the neutron spin vector (arising due to high-*j*, low-neutron particle configuration) in the middle and nearly perpendicular to both of them. It is therefore like having two shears like subsystems operating together. The total angular momentum *I* points along $j_\nu$. Generation of higher angular momenta is by the simultaneous closing of the two proton blades towards the direction of *I* with a small contribution from the core rotation *R*.

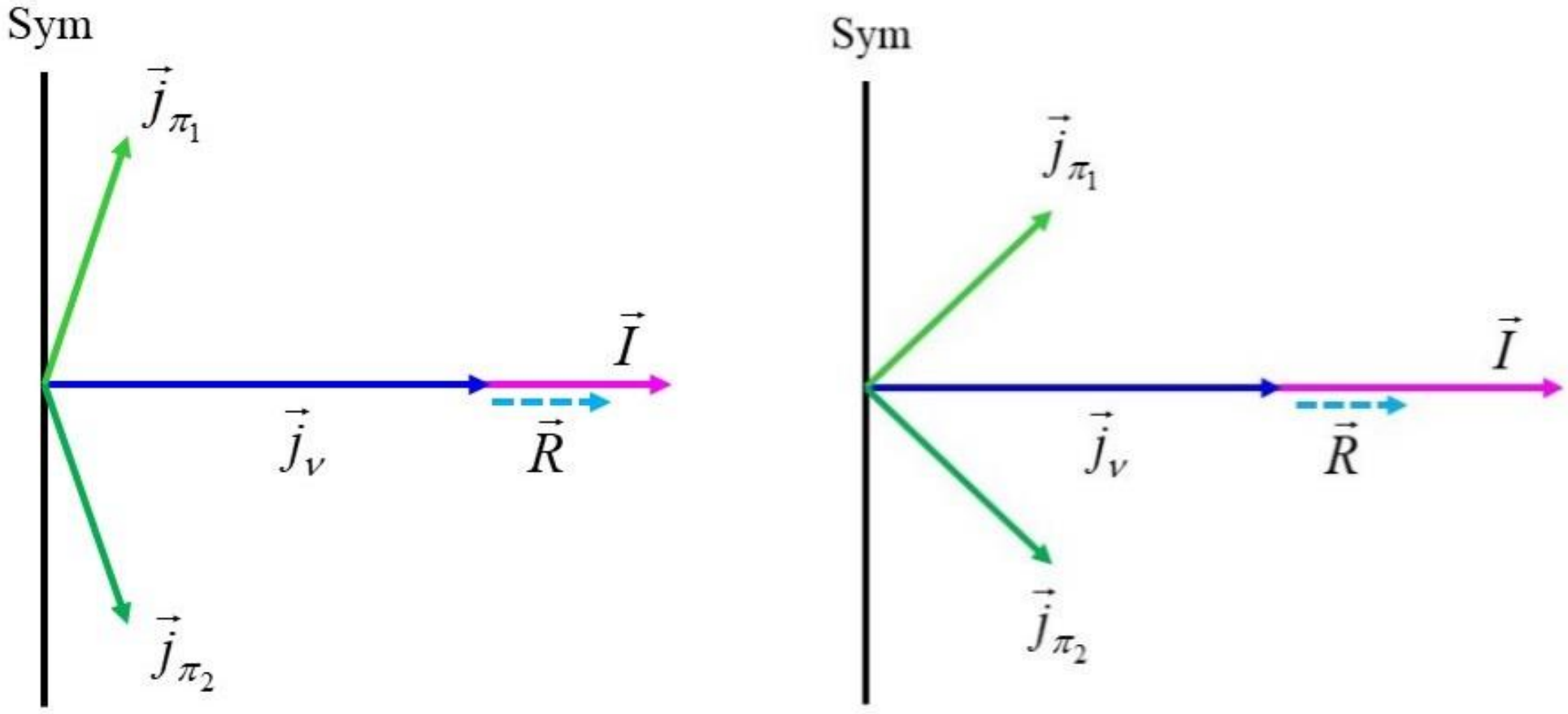

Figure 4: Schematic diagram of the twin-shears mechanism (AMR) showing the coupling of angular momentum of the proton blades ($j_{\pi 1}$ and $j_{\pi 2}$) and the neutron blade $j_\nu$ at lower spin (upper part) and at higher spin (lower part). The total angular momentum $I = I_{sh} + R$, where $R$ is the contribution from the core and $I_{sh}$ is the sum of $j_{\pi 1}$, $j_{\pi 2}$ and $j_\nu$. The core appears to play a prominent role in AMR.

Since the perpendicular components of $\mu$ of the two subsystems are antialigned and cancel each other, there is no net magnetic dipole moment. The magnetic moment of one of the two shears specifies the orientation. As the system is symmetric with respect to a rotation of π about the axis of $j_\nu$, or $I$, *i.e.* the mean field has $R_z(\pi)$ symmetry, the AMR bands consist of regular sequences of energy levels differing in spin by 2, and are connected by weak electric quadrupole ($E2$) transitions (i.e. small $B(E2)$ values) reflecting a small deformation of the nucleus [13].

To distinguish this phenomenon from collective rotation, it is reasonable to say that a band is an AMR band if the ratio $\Im^{(2)}/$ B(E2) ≥ 100 (ħ$^2$ MeV$^{-1}$ (eb)$^{-2}$) is as large as for magnetic rotation, i.e. it should exceed 100 MeV$^{-1}$(eb)$^{-2}$. Some characteristics of AMR bands, based on observed experimental features in the Cd and Pd isotopes, which so far, seem the most well-established examples of such structures, are:

(1) Nearly constant = $\Im^{(2)}$ [ ~ 25 ħ$^2$MeV$^{-1}$]
(2) Small $B(E2)$ values ~ 0.05 - 0.42 (eb)$^2$, which decrease with ascending spins for higher levels.

Even though the phenomenon of AMR mode of nuclear excitations was proposed 21 years ago [14], there has been limited success in observing such structures experimentally, with firm evidence even rarer. First observation of an antimagnetic rotational band, with lifetime measurements for four levels in the band, was suggested in $^{109}$Cd nuclide by Chiara *et al.* [25], where two MR bands were also reported, later followed by an observation of two AMR bands as signature partner bands in $^{108}$In and $^{110}$In [26] nuclei, but with limited measurements of level lifetimes. A contemporary work by Zhu *et al.* [27] reported an AMR band in $^{100}$Pd, but with lifetimes for only two levels in the band, the assignment remains tentative. Based on level lifetime measurements, and from decreasing trend of B(E2) values with increasing spin Datta *et al*. [28] identified an AMR structure in $^{108}$Cd, while, Simons *et al*. [29] reported AMR bands in $^{106}$Cd and $^{108}$Cd, although the assignment of AMR band to the latter nuclide was less certain. In a series of experimental investigations from 2010 to 2015, Choudhury *et al.* [30-32], discovered an AMR band based on level lifetime measurements for five levels in a band in $^{105}$Cd [30], later followed by an observation of two AMR bands as signature partners, as well as coexistence of an AMR and an MR band in $^{107}$Cd [31,32], thereby, providing for the first time, confirmed AMR bands and other structure features in odd-A nuclei. Data for 40 AMR bands presented in Table 3 are for 13 odd-A nuclei, twelve even-even nuclei, and five odd-odd nuclei, although, with 29 (out of a total of AMR 40 bands) bands in the Pd (Z=46), Cd (Z=48), and In (Z=49) isotopes. The most well-established examples are for Cd isotopes, followed by those in Pd and In isotopes, all three near Z=50 closed shell, and near N=60 subshell, expected to have low quadrupole deformations. Other ten bands in $^{58}$Fe, $^{61}$Ni, $^{99}$Pd, $^{127}$Xe, $^{130}$Ba, $^{137}$Nd, $^{142}$Eu, $^{143}$Eu and $^{144}$Dy seem much less certain in terms of their AMR structure, as for most of these no level lifetime measurements are available, and for several cases such as $^{61}$Ni, $^{99}$Pd, $^{102}$Pd, $^{110}$Cd, $^{110}$In, $^{112}$In and $^{144}$Dy, reported AMR bands deviate in trend of increasing level energies with ascending spins for a typical rotor.

## 5. Cases excluded due to ambiguous MR character

It should be noted that we have omitted in Table 2, a proposed MR band in $^{131}$La [33], based on a 509γ-469γ-428γ cascade with level spins from $27/2^-$ to $21/2^-$, and suggested configuration=$\pi^3$(11/2[505], 5/2[402], 5/2[413]), as, with only proton configuration, it appears to be a different type of excitation. We have also omitted two MR-like bands, with three $\Delta J=1$ dipole (assumed magnetic) transitions in each band, as reported for $^{29}$Al [34], due to lack of adequate experimental data, and theoretical model calculations for magnetic rotational behaviour in *sd*-shell nuclei. It should be noted that MR bands for $^{77}$Br, $^{131}$La, $^{134,136,137}$Nd, and $^{192,193,195}$Hg in the 2000 compilation by Amita et al. [1] have been omitted in the present work. Although, each of these nuclides has experimental evidence for well-defined band(s) with dominant magnetic dipole (M1) transitions, but dipole bands in these nuclei have been interpreted in literature, through model calculations, due to collective or other mechanism of excitations, different from magnetic-rotational structure.

## 6. Conclusions

In the present work we have presented a compendium of relevant spectroscopic data for all the observed or proposed magnetic-rotational (MR or shears), and antimagnetic-rotational structures in different regions of the nuclear chart, based on experiments and theory, several bands (especially for Z=82, Pb nuclei for MR band, and in Pd, Cd and In regions for AMR bands) with firm MR assignments, while many others with possible or tentative assignments, as indicated in original publications. The magnetic-rotational high-spin structures display a behaviour different from the normal rotational bands and are mostly dominated by magnetic dipole (M1) transitions. However, many of these bands may have a mixed character, containing contribution from collective rotational motion or open to other modes of excitations, especially, when the bands are regular in gamma-ray energies with ascending spins but have fairly significant crossover E2 transitions, and, also when the bands are irregular, where these could be signature partners. The B(M1) and B(E2) values from lifetime measurements, and detailed theoretical model calculations are available for only about 30% of the MR bands listed in this review, which can provide the most stringent tests for the magnetic-rotational character of a band. Level lifetime data are available for 11 MR bands (out of a total of 32) in the A=58-87 region, for 26 MR bands (out of a total of 56 bands) in the A=100-112 region, for 21 MR bands (out of a total of 62 bands) in the A=124-146 region, and for only 14 MR bands (out of a total of 78 bands) in the A=189-209 region. In the case of AMR bands, level lifetime data are available for 15 bands in 12 nuclei (out of a total of 40 bands in 28 nuclei) with the most and high-quality data in the Cd nuclei. As research on high-spin structures in nuclear physics is no longer a high priority in major nuclear laboratories, it is less likely that lifetime data for many of these bands listed here will become available in the immediate future, yet, based on existing experimental data, and theoretical model calculations, a more critical assessment of each band for its magnetic- or antimagnetic rotational character is needed for definite assignments.

## Acknowledgements

SS and SK acknowledge financial support from the International Atomic Energy Agency (IAEA), Vienna, Austria (RC No. 28496 and RC-17642-R0), Department of Atomic Energy (DAE)-Board of Nuclear Science (BRNS), Government of India (grant no.: 36(6)/14/60/2016-BRNS/36145), and from Akal University, Talwandi Sabo, Punjab (India). AKJ acknowledges financial support from the Science and Engineering Research Board (SERB), Government of India (grant no.: CRG/2020/000770), and from the Amity University, UP (India).

## 7. POLICIES

| | |
|---|---|
| Source of Data/Datum | All datum/data are/is taken as presented in the original publications, generally, from the first reference (in bold) listed for a band. If the source of any datum/data is/are not from the bold reference when more than one is listed, the reference is mentioned. |
| Level Energies | The listed level energies are taken from the first reference (in bold) given for a band. If the level-energies are not available in the original articles, then these are either from ENSDF or deduced by the authors and indicated explicitly. |
| Level Lifetimes | The mean lifetimes (in ps) are quoted without reference (as an example in $^{59}$Co) where only one reference is given, or quoted with reference in the comments (as an example in $^{83}$Kr) in case of many references. |
| B(M1) and B(E2) | The values are quoted without reference (as an example in $^{83}$Kr) where only one reference is given, or quoted with reference (as an example in $^{85}$Sr) in case of many references given. |

## 8. EXPLANATION OF TABLE

Table 2: Magnetic Rotational Structure in Nuclei
Table 3: Antimagnetic Rotational Structure in Nuclei

| | |
|---|---|
| $^{A}_{Z}X_{N}$ | Denotes the specific nuclide with<br>X Chemical symbol<br>A Mass number<br>Z Atomic number<br>N Neutron number<br>A single blank row marks the end of entries for each band. The number in the first column denotes band number. |
| $E_{level}$ | Level energy in units of keV. Energies in parentheses denote tentative levels. Labels X, Y, Z, etc., indicate that absolute excitation energies are unknown due to lack of knowledge about linking transitions to the lower levels. |
| $I^{\pi}$ | I denotes the level spin. $\pi$ denotes the parity (+ or -). $I^{\pi}$ given in parentheses denotes uncertain spin and/or parity assignment. |
| $E_\gamma$(M1) | Gamma ray energies in units of keV for the M1($\Delta$I=1) transition I$\rightarrow$I-1. |
| $E_\gamma$(E2) | Gamma ray energies in units of keV for the E2($\Delta$I=2) transition I$\rightarrow$I-2. |
| B(M1)/B(E2) | The ratio of reduced transition probabilities in units of $(\mu_N/eb)^2$ are calculated using Eq. (1). For some bands where E2 transitions are not observed, the lower limits for B(M1)/B(E2) are given. |
| Configurations and Comments | The quasiparticle configuration for a band is given wherever assigned by the original authors. 'π' here is for protons and 'ν' is for neutrons. s, p, d, f, g and h are the orbitals. A positive integer in the superscript of the orbital denotes number of particles while a negative integer denotes number of holes in that orbital.<br>The abbreviations in this item are explained below: DSM: deformed shell model; TAC: tilted axis cranking; CSM: cranked shell model; CNS: cranked Nilsson-Strutinsky; TRS: total Routhian surface; PSM: projected shell model; SPAC: Shears mechanism with the Principal Axis Cranking, IBFM: interacting boson fermion model; FAL: Fermi aligned; HF: Hartree-Fock; BCS: pairing theory of Bardeen, Cooper and Schrieffer; CWS: cranked Woods-Saxon; RMF: relativistic mean-field; CDFT: covariant-density functional theory. SCM: Semiclassical Particle Rotor Model |
| ($\beta_2$, $\gamma$) or ($\varepsilon_2$, $\gamma$) | Deformation parameters. |
| Backbending | In a rotational band, the transition energies increase with increase in spin reflecting the I(I+1) behavior, but in some cases, e.g. in $^{108}$Cd, band 1, the moment of inertia increases drastically after the spin $16^-$ and the transition energy decreases and again starts rising after $18^-$ . This phenomenon is known as backbending and is usually attributed to the crossing of two rotational bands due to the alignment of a pair of either kind of quasiparticles. |
| Regular band | A band where the excitation energy varies more or less smoothly with spin, though not necessarily as I(I+1). |
| Irregular band | A band for which energy variation with spin is quite abrupt. |
| References | The references are listed in chronological order in terms of eight-digit key numbers as assigned in the Nuclear Science References database at NNDC, Brookhaven National Laboratory. Separate reference lists for Tables 2 and Table 3, as well as for theory articles related to MR and AMR bands are supplied. |

## Table 2: Magnetic Rotational Structures in Nuclei

### $^{58}_{26}Fe_{32}$

| | $E_{level}$ keV | $I^{\pi}$ | $E_{\gamma}$(M1) keV | $E_{\gamma}$(E2) keV | B(M1)/B(E2) $(\mu_N/eb)^2$ | References | Configurations and Comments: |
|---|---|---|---|---|---|---|---|
| 1. | 7524.2 | (8) | | | | **2012St06** | 1. On the basis of TAC-RMF calculations, the authors suggest that negative parity is more likely.<br>2. From near degeneracy of band 1 and band 2, the authors suggest that, these bands may also be chiral doublet partners.<br>3. Regular band. |
| | 7915.7 | (9) | 392.1 | | | | |
| | 8428.9 | (10) | 513.3 | | | | |
| | 9174.6 | (11) | 745.6 | | | | |
| | 10130.5 | (12) | 956.0 | | | | |
| 2. | 7657.8 | (8) | | | | **2012St06** | 1. Probable configuration as $\pi(f_{7/2})^{-2} \otimes \nu\{(r)^3(g_{9/2})^1\}$ for lower spin region and $\pi[(f_{7/2})^{-3} (r)^1] \otimes \nu[(r)^3(g_{9/2})^1]$ for higher spin region ( 'r'= $f_{5/2}$, $p_{3/2}$, $p_{1/2}$) is assigned on the basis of TAC-RMF.<br>2. $(\beta_2, \gamma) \approx (0.22, 22°-50°)$ before band crossing ($\hbar\omega$=0.6-1.3 MeV) and (~0.25, ~16°- 4°) after band crossing ($\hbar\omega$ =1.0- 1.3 MeV) from TAC-RMF calculations.<br>3. On the basis of TAC-RMF calculations, the authors suggest that negative parity is more likely.<br>4. From near degeneracy of band 1 and band 2, the authors suggest that, these bands may also be chiral doublet partners.<br>5. Band 1 crosses band 2 in the spin range I=11ħ -13ħ.<br>6. Regular band. |
| | 8041.9 | (9) | 384.1 | | | | |
| | 8540.2 | (10) | 498.1 | | | | |
| | 9163.6 | (11) | 623.3 | | | | |
| | 9924.7 | (12) | 761.2 | | | | |
| | 10952.4 | (13) | 1027.6 | | | | |
| | 12017.0 | (14) | 1064.6 | | | | |
| | 13173.4 | (15) | 1156.7 | | | | |

### $^{59}_{27}Co_{32}$

| | $E_{level}$ keV | $I^{\pi}$ | $E_{\gamma}$(M1) keV | $E_{\gamma}$(E2) keV | B(M1)/B(E2) $(\mu_N/eb)^2$ | References | Configurations and Comments: |
|---|---|---|---|---|---|---|---|
| 1. | 4178.7 | $13/2^+$ | | | | **2024AJ01** | 1. Probable configuration as $\pi(f_{7/2})^{-2}(p_{3/2})^1 \otimes \nu\{(p_{3/2})^2 (g_{9/2})^1\}$ on the basis of Shell Model Calculations<br>2. The mean lifetimes (in ps) of levels from 4717 to 6364 keV are 1.1(4), 0.10(5) and <0.20, respectively.<br>3. Regular band |
| | 4414.6 | $15/2^+$ | 235.9 | | | | |
| | 4717.5 | $17/2^+$ | 302.9 | | | | |
| | 5370.6 | $19/2^+$ | 653.1 | | | | |
| | 6364.6 | $21/2^+$ | 994.0 | | | | |
| | 7842.7 | $23/2^{(+)}$ | 1478.1 | | | | |

### $^{61}_{27}Co_{34}$

| | $E_{level}$ keV | $I^{\pi}$ | $E_{\gamma}$(M1) keV | $E_{\gamma}$(E2) keV | B(M1)/B(E2) $(\mu_N/eb)^2$ | References | Configurations and Comments: |
|---|---|---|---|---|---|---|---|
| 1. | 3472.5 | $13/2^+$ | | | | **2024AJ01** | 1. Configuration assigned as $\pi\{(f_{7/2})^{-2}(p_{3/2})^1\} \otimes \nu\{(p_{3/2})^{-1} (f_{5/2})^{-2}(g_{9/2})^1\}$ from Shell Model calculations. It is tentatively assigned as MR band on the basis of semiclassical calculations by 1998Ma09 and also on the basis of comparison with MRB1observed in $^{58}$Fe (2012St06).<br>2. $(\beta_2, \gamma) = (0.3,0°)$ from CSM calculations.<br>3. Regular band. |
| | 3658.8 | $15/2^+$ | 186.3 | | | 2015Ay02 | |
| | 4094.4 | $17/2^+$ | 435.6 | | | | |
| | 4803.6 | $19/2^+$ | 709.2 | | | | |
| | 5833.1 | $21/2^+$ | 1029.5 | | | | |
| | 6894.9 | $23/2^+$ | 1061.8 | | | | |
| | 8216.7 | $25/2^+$ | 1321.8 | | | | |

## $^{61}_{27}Co_{34}$

| | $E_{level}$ keV | $I^{\pi}$ | $E_{\gamma}$(M1) keV | $E_{\gamma}$(E2) keV | B(M1)/B(E2) $(\mu_N/eb)^2$ | References | **Configurations and Comments:** |
|---|---|---|---|---|---|---|---|
| 2. | 3909.5 | $13/2^-$ | | | | **2015Ay02** | 1. Tentatively assigned as MR band on the basis of semiclassical calculations (1998Ma09). |
| | 4116.5 | $15/2^-$ | 207.4 | | | | 2. Regular band. |
| | 4385.9 | $17/2^-$ | 269.8 | | | | |
| | 5117.6 | $19/2^-$ | 731.7 | | | | |
| | 6065.9 | $21/2^-$ | 947.7 | | | | |
| | 7173.9 | $(23/2^-)$ | (1108.3) | | | | |

## $^{62}_{27}Co_{35}$

| | $E_{level}$ keV | $I^{\pi}$ | $E_{\gamma}$(M1) keV | $E_{\gamma}$(E2) keV | B(M1)/B(E2) $(\mu_N/eb)^2$ | References | **Configurations and Comments:** |
|---|---|---|---|---|---|---|---|
| 1. | 3564.8 | $9^+$ | | | | **2022Se03** | 1. Configuration assigned as $\pi\{(1f_{7/2})^{-2}(2p_{3/2})^{1}\otimes \nu(2p_{3/2})^{-1}\}$ from CDFT calculations. |
| | 4048.8 | $10^+$ | 483.4 | | | | 2. β=0.20. |
| | 4723.2 | $11^+$ | 674.4 | | | | 3. Tentatively assigned as MR band (2022Se03). |
| | 5812.6 | $12^+$ | 1089.4 | | | | 4. Regular band. |
| | 6948 | $(13^+)$ | 1135.2 | | | | |

## $^{59}_{28}Ni_{31}$

| | $E_{level}$ keV | $I^{\pi}$ | $E_{\gamma}$(M1) keV | $E_{\gamma}$(E2) keV | B(M1)/B(E2) $(\mu_N/eb)^2$ | References | **Configurations and Comments:** |
|---|---|---|---|---|---|---|---|
| 1. | 6505.6 | $19/2^-$ | | | | **2024AJ01** | 1. Configuration assigned as $\pi\{(f_{7/2})^{-2}(p_{3/2})^{1}\}\otimes \nu\{(p_{3/2})^{2}(f_{7/2})^{-1}(f_{5/2})^{2}\}$ from Shell Model calculations. |
| | 7167.5 | $21/2^{(-)}$ | 661.9 | | | | 2. Irregular band. |
| | 7954.2 | $23/2^{(-)}$ | 786.7 | | | | |
| | 8632.3 | $25/2^{(-)}$ | 678.1 | | | | |
| | 9309.7 | $27/2^{(-)}$ | 677.4 | | | | |
| | 10089.1 | $29/2^{(-)}$ | 779.4 | | | | |
| | 12007.5 | $31/2^{(-)}$ | 1918.4 | | | | |

## $^{60}_{28}Ni_{32}$

| | $E_{level}$ keV | $I^{\pi}$ | $E_{\gamma}$(M1) keV | $E_{\gamma}$(E2) keV | B(M1)/B(E2) $(\mu_N/eb)^2$ | References | **Configurations and Comments:** |
|---|---|---|---|---|---|---|---|
| 1. | 8044.1 | $9^-$ | | | | **2008TO15** | 1. Configuration assigned as $\pi\{(1f_{7/2})^{-1}(fp)^{1}\}\otimes \nu\{(1g_{9/2})^{1}(fp)^{3}\}$ from CNS calculations. |
| | 8520.5 | $10^-$ | 476.7 | | | 2011Zh57 | 2. Comparison of experimental BM1/BE2 with TAC-RMF calculations (2011Zh57). |
| | 9132.2 | $11^-$ | 611.5 | 1088.2 | 164(24) | | 3. $(\varepsilon_2, \gamma)$ =(0.21,22°) from CNS calculations. |
| | 9989.3 | $12^-$ | 856.9 | 1468.3 | 178(35) | | 4. Regular band. |
| | 11112.8 | $13^-$ | 1123.4 | 1981.1 | 215(56) | | |
| | 12273.8 | $14^-$ | 1160.8 | 2284.6 | 178(48) | | |
| | 13810.0 | $(15^-)$ | 1536.2 | 2697.2 | 27(19) | | |
| 2. | 8485.3 | $9^-$ | | | | **2008TO15** | 1. Configuration assigned as $\pi\{(1f_{7/2})^{-1}(fp)^{1}\}\otimes \nu\{(1g_{9/2})^{1}(fp)^{3}\}$ from CNS calculations. |
| | 9122.8 | $10^-$ | 637.5 | | | 2011Zh57 | 2. $(\varepsilon_2, \gamma) \approx$(0.21,30°) from CNS calculations. |
| | 9960.0 | $11^-$ | 836.4 | | | | 3. Regular band with backbending at 11 ħ. |
| | 10788.6 | $12^-$ | 828.5 | | | | |
| | 11552.8 | $13^-$ | 764.2 | | | | |
| 3. | 11224.8 | $(11^+)$ | | | | **2008TO15** | 1. Probable configuration assignment based on CNS calculations is $\pi\{(1f_{7/2})^{-1}(fp)^{1}\}\otimes\nu\{(1g_{9/2})^{2}(fp)^{2}\}$. |
| | 11785.5 | $(12^+)$ | 560.8 | | | 2011Zh57 | 2. $(\varepsilon_2, \gamma)$ = (0.220,9.1°) and (0.20,58.5°) at I=10ħ and 16ħ, respectively from CNS calculations. |
| | 12486.1 | $(13^+)$ | 700.8 | | | | 3. Regular band with backbending at 14 ħ. |
| | 13353.0 | $(14^+)$ | 866.8 | | | | |
| | 14201.0 | $(15^+)$ | 848.0 | | | | |
| | 15164.6 | $(16^+)$ | 963.8 | | | | |
| | 16241.6 | $(17^+)$ | 1077 | | | | |

### $^{60}_{28}\mathrm{Ni}_{32}$

| | $E_{level}$ keV | $I^\pi$ | $E_\gamma$(M1) keV | $E_\gamma$(E2) keV | B(M1)/B(E2) $(\mu_N/eb)^2$ | References | **Configurations and Comments:** |
|---|---|---|---|---|---|---|---|
| 4. | 11255.0 | $12^+$ | | | | **2008TO15** | 1. Configuration assigned as $\pi\{(1f_{7/2})^{-1}(fp)^1\}\otimes\nu\{(1g_{9/2})^2(fp)^2\}$ from CNS calculations.<br>2. $(\varepsilon_2, \gamma)$= (0.220,9.1°) and (0.20,58.5°) at I=10ħ and 16ħ, respectively from CNS calculations.<br>3. Regular band. |
| | 11851.2 | $13^+$ | 596.0 | | 36(12) | 2011Zh57 | |
| | 12578.4 | $14^+$ | 727.1 | 1323.9 | 114(39) | | |
| | 13662.3 | $15^+$ | 1083.9 | 1811.0 | 95(25) | | |
| | 14803.0 | $16^+$ | 1141.1 | 2224.5 | 88(23) | | |
| | 16097.8 | $(17^+)$ | 1294.8 | | | | |

### $^{61}_{28}\mathrm{Ni}_{33}$

| | $E_{level}$ keV | $I^\pi$ | $E_\gamma$(M1) keV | $E_\gamma$(E2) keV | B(M1)/B(E2) $(\mu_N/eb)^2$ | References | **Configurations and Comments:** |
|---|---|---|---|---|---|---|---|
| 1. | 6972 | $21/2^+$ | | | | **2023Bh02** | 1. Configuration assigned as $\pi[(f_{7/2})^{-1}[p_{3/2}/f_{5/2}]^1\otimes\nu[(g_{9/2})^1(p_{3/2})^2[f_{5/2}/p_{1/2}]^2$ from Shell Model (SM) calculations (2023Bh02) whereas, $\pi[(1f_{7/2})^{-1}(fp)^1]\otimes\nu[(1g_{9/2})^1(fp)^4]$ was suggested from TAC-CDFT calculations (2023Li05).<br>2. The B(M1) ≈1.0 $\mu_N^2$ from TAC- CDFT.<br>3. Regular band. |
| | 7791 | $23/2^+$ | 819 | | | **2023Li05** | |
| | 8663 | $25/2^+$ | 872 | | | | |
| | 9748 | $27/2^+$ | 1085 | | | | |
| | 11043 | $(29/2^+)$ | (1295) | | | | |
| 2. | 7679 | $21/2^+$ | | | | **2023Bh02** | 1. Form the SM calculations, the configuration of band 2 is predicted to be similar as band 1.<br>2. Regular band. |
| | 8028 | $23/2^+$ | 349 | | | | |
| | 8553 | $25/2^+$ | 525 | | | | |
| | 9337 | $27/2^+$ | 784 | | | | |
| | 10200 | $(29/2^+)$ | 863 | | | | |
| | 11138 | $21/2^+$ | 938 | | | | |

### $^{62}_{29}\mathrm{Cu}_{33}$

| | $E_{level}$ keV | $I^\pi$ | $E_\gamma$(M1) keV | $E_\gamma$(E2) keV | B(M1)/B(E2) $(\mu_N/eb)^2$ | References | **Configurations and Comments:** |
|---|---|---|---|---|---|---|---|
| 1. | 2892.7 | $7^-$ | | | | **2022Lu01** | 1. Configuration assigned as $\pi\{(f_{7/2})^{-1}(p_{3/2},f_{5/2})^2\}\otimes\nu\{(g_{9/2})^1(p_{3/2},f_{5/2})^4\}$ from CNS calculations.<br>2. $\beta_2$≈0.26-0.32 from TAC-CDFT calculations.<br>3. Tentatively assigned as MR band. |
| | 3435.0 | $8^-$ | 544.1 | | | 2001Mu14 | |
| | 3979.5 | $9^-$ | 544.5 | | | | |

### $^{75}_{33}\mathrm{As}_{42}$

| | $E_{level}$ keV | $I^\pi$ | $E_\gamma$(M1) keV | $E_\gamma$(E2) keV | B(M1)/B(E2) $(\mu_N/eb)^2$ | References | **Configurations and Comments:** |
|---|---|---|---|---|---|---|---|
| 1. | 2763.3 | $(15/2^-)$ | | | | **2017Li02** | 1. Configuration assigned as $\pi\{(1g_{9/2})^1(1f_{5/2})^{-2}\}\otimes\nu\{(1g_{9/2})^5(fp)^{-3}\}$ from TAC-CDFT calculations.<br>2. $(\beta_2, \gamma)$ =(0.32-0.33, 27°) from TAC-CDFT calculations.<br>3. 2017Li02 interpret this structure as a stapler band.<br>4. Regular band. |
| | 3006.1 | $(17/2^-)$ | 242.8 | | | | |
| | 3271.5 | $(19/2^-)$ | 265.4 | | | | |
| | 3703.3 | $(21/2^-)$ | 431.8 | | | | |

## $^{79}_{35}\mathrm{Br}_{44}$

| | $E_{level}$ keV | $I^{\pi}$ | $E_{\gamma}$(M1) keV | $E_{\gamma}$(E2) keV | B(M1)/B(E2) $(\mu_N/eb)^2$ | References | **Configurations and Comments:** |
|---|---|---|---|---|---|---|---|
| 1. | 2392.8 | $13/2^-$ | | | | **2002Sc13** | 1. $\pi(g_{9/2}) \otimes \nu[g_{9/2}(p_{3/2}/f_{5/2})^1]$ from TAC calculations.<br>2. $(\beta_2, \gamma)$ = (0.18, >20°) from TRS calculations.<br>3. Regular band.<br>4. The mean lifetimes (in ps) of levels from 3088 to 4802 keV are 1.1(3), 0.55(15), 0.20(4) and 0.17(3), respectively.<br>5. The B(M1) values for the transitions from 314.0 to 649.9 keV are 1.55(+62-32), 0.76(+35-23), 0.47(+19-14) and 0.93(+30-21) $\mu_N^2$, respectively.<br>6. The B(E2) values for the transitions from 760.3 to 1268.6 keV are 0.22(+15-9), 0.09(+5-3) and 0.10(+6-4) $(eb)^2$, respectively. |
| | 2580.6 | $15/2^-$ | 187.6 | | | 1999Ra02 | |
| | 2774.2 | $17/2^-$ | 193.8 | | | 1988Sc13 | |
| | 3088.2 | $19/2^-$ | 314.0 | | | 1995Ta21 | |
| | 3534.9 | $21/2^-$ | 446.7 | 760.3 | 3.45(+62-45) | | |
| | 4152.3 | $23/2^-$ | 617.4 | 1064.9 | 5.22(+28-51) | | |
| | 4802.2 | $25/2^-$ | 649.9 | 1268.6 | 9.3(+27-16) | | |
| | 5577.6 | $27/2^-$ | 775.4 | 1426.5 | | | |
| | 6383.0 | $(29/2^-)$ | 805.4 | 1580.9 | | | |

## $^{81}_{35}\mathrm{Br}_{46}$

| | $E_{level}$ keV | $I^{\pi}$ | $E_{\gamma}$(M1) keV | $E_{\gamma}$(E2) keV | B(M1)/B(E2) $(\mu_N/eb)^2$ | References | **Configurations and Comments:** |
|---|---|---|---|---|---|---|---|
| 1. | 2549.4 | $(13/2^-)$ | | | | **1986Fu04** | 1. Tentatively assigned as $\pi(g_{9/2}^2 p_{3/2})$, but this may only be just one component. 1995Ta21 suggest fp-neutron and $g_{9/2}$ neutron orbitals.<br>2. Regular band. |
| | 2668.5 | $(15/2^-)$ | 119.1 | | | 1995Ta21 | |
| | 2942.1 | $(17/2^-)$ | 273.6 | | | | |
| | 3333.5 | $(19/2^-)$ | 391.4 | | | | |
| | 3798.7 | $(21/2^-)$ | 465.2 | | | | |

## $^{79}_{36}\mathrm{Kr}_{43}$

| | $E_{level}$ keV | $I^{\pi}$ | $E_{\gamma}$(M1) keV | $E_{\gamma}$(E2) keV | B(M1)/B(E2) $(\mu_N/eb)^2$ | References | **Configurations and Comments:** |
|---|---|---|---|---|---|---|---|
| 1. | 2857.4 | $(17/2^-)$ | | | | **1994Jo08** | 1. Tentatively assigned as $\pi[g_{9/2} \otimes (p_{3/2}f_{5/2})^1] \otimes \nu(g_{9/2})$ from the TAC calculations (2004Ma09).<br>2. $(\beta_2, \gamma)$ = (0.192, 62°) from TAC calculations.<br>3. Regular band. |
| | 3214.2 | $19/2^-$ | 283.9 | | | **1990Sc07** | |
| | 3585.0 | $21/2^-$ | 371.0 | | | | |
| | 4132.7 | $23/2^-$ | 547.7 | | | | |

## $^{81}_{36}\mathrm{Kr}_{45}$

| | $E_{level}$ keV | $I^{\pi}$ | $E_{\gamma}$(M1) keV | $E_{\gamma}$(E2) keV | B(M1)/B(E2) $(\mu_N/eb)^2$ | References | **Configurations and Comments:** |
|---|---|---|---|---|---|---|---|
| 1. | 2419.7 | $13/2^-$ | | | | **1986Fu03** | 1. Tentatively assigned as $\pi[g_{9/2} \otimes (p_{3/2}f_{5/2})^1] \otimes \nu(g_{9/2})$ from the TAC calculations (2004Ma09).<br>2. $(\beta_2, \gamma)$ = (0.16, 60°) from TAC calculations.<br>3. Regular band.<br>4. B(M1) values for the transitions 113 keV and from 362 to 615 keV are 1.3(+13-4), 0.53(+41-17) , 0.25(+19-9), 0.06(+3-1) and 0.05(+5-2) $\mu_N^2$, respectively.<br>5. B(E2) values for the transitions from 528 to 1223 keV are 0.013(+13-7), 0.010(+8-4), 0.010 (+6-4) and 0.014(+16-7) $(eb)^2$, respectively.<br>6. The mean lifetimes (in ps) of levels from 2533 to 4714 keV are 2(1), <20, 2.1(9), 2.6(11), 2.7(8) and 1.5(8), respectively. |
| | 2533.2 | $15/2^-$ | 113.3 | | | | |
| | 2699.3 | $17/2^-$ | 166.1 | | | | |
| | 3061.3 | $19/2^-$ | 362.0 | 528.1 | 40.8(+192-46) | | |
| | 3490.1 | $21/2^-$ | 428.8 | 791 | 25.0(+17-5) | | |
| | 4098.8 | $23/2^-$ | 608.7 | 1038 | 6.0(+23-4) | | |
| | 4714 | $(25/2^-)$ | 615 | 1223.5 | 3.6(+7-3) | | |

### $^{83}_{36}\mathrm{Kr}_{47}$

| | $E_{level}$ keV | $I^{\pi}$ | $E_\gamma$(M1) keV | $E_\gamma$(E2) keV | B(M1)/B(E2) $(\mu_N/eb)^2$ | References | **Configurations and Comments:** |
|---|---|---|---|---|---|---|---|
| 1. | 2510.0 | $13/2^-$ | | | | **2011Ga44** | 1. Tentatively assigned as $\pi[g_{9/2}\otimes(p_{3/2}f_{5/2})^1]\otimes\nu(g_{9/2})$ from the TAC calculations as given in 2004Ma09 and 2011Ga44.<br>2. $(\beta_2, \gamma)$ = (0.14, 59°) from TAC calculations.<br>3. Regular band.<br>4. B(M1) values for the transitions from 200 to 651 keV are 2.38(+68-40), 1.12(7), 0.54(+14-9), 0.31(+8-6) and >0.065$\mu_N^2$ .<br>5. B(E2) values for the transitions from 761 to 1267 keV are ≈ 0.03, ≈ 0.012 and 0.017(6) $(eb)^2$, respectively.<br>6. The mean Lifetimes (in ps) for levels from 2510 to 2640 as given in 1984Ku23 are 3.0(+30-15), 6(3) and, respectively and that for levels from 2841 to 4868 keV are 2.5 (5), 1.6(1), 1.0(2), 0.60(13) and <1.0 (2011Ga44), respectively. |
| | 2640.5 | $15/2^-$ | 130.5 | | | 1986Ke12 | |
| | 2841.1 | $17/2^-$ | 200.6 | | | 1984Ku23 | |
| | 3157.5 | $19/2^-$ | 316.3 | | | 2004Ma09 | |
| | 3603.1 | $21/2^-$ | 445.6 | 761 | ≈ 13 | | |
| | 4218.4 | $23/2^-$ | 615.3 | 1060.8 | ≈ 21 | | |
| | 4869.9 | $25/2^-$ | 651.4 | 1267.1 | 5.3(7) | | |
| | 5641 | $(27/2^-)$ | 771 | (1423) | | | |

### $^{79}_{37}\mathrm{Rb}_{42}$

| | $E_{level}$ keV | $I^{\pi}$ | $E_\gamma$(M1) keV | $E_\gamma$(E2) keV | B(M1)/B(E2) $(\mu_N/eb)^2$ | References | **Configurations and Comments:** |
|---|---|---|---|---|---|---|---|
| 1. | 3309.4 | $(19/2^-)$ | | | | **1993Ho15** | 1. Tentatively assigned as $\pi(g_{9/2})\otimes\nu[(g_{9/2})\otimes(pf)^1]$ by comparison with the isotone $^{77}$Br.<br>2. Regular band.<br>3. 601 keV M1 transition is from 1996Sm07. |
| | 3687.5 | $(21/2^-)$ | 378.1 | | | 1995Ta21 | |
| | 4152.2 | $(23/2^-)$ | 464.7 | 842.8 | | 1996Sm07 | |
| | 4686.4 | $(25/2^-)$ | 534.2 | | | | |
| | 5287.4 | $(27/2^-)$ | 601 | | | | |

### $^{81}_{37}\mathrm{Rb}_{44}$

| | $E_{level}$ keV | $I^{\pi}$ | $E_\gamma$(M1) keV | $E_\gamma$(E2) keV | B(M1)/B(E2) $(\mu_N/eb)^2$ | References | **Configurations and Comments:** |
|---|---|---|---|---|---|---|---|
| 1. | 2636.0 | $(15/2^-)$ | | | | **1994Do18** | 1. Tentatively assigned as $\pi(g_{9/2})\otimes\nu[(g_{9/2})\otimes(pf)^1]$<br>2. Regular band. |
| | 2697.2 | $17/2^-$ | 61.0 | | | 1995Ta21 | |
| | 2997.7 | $19/2^-$ | 300.5 | | | | |
| | 3427.5 | $21/2^-$ | 429.8 | | | | |
| | 3993.1 | $23/2^-$ | 565.6 | (996) | | | |
| | 4529 | $(25/2^-)$ | 599 | | | | |

### $^{82}_{37}\mathrm{Rb}_{45}$

| | $E_{level}$ keV | $I^{\pi}$ | $E_\gamma$(M1) keV | $E_\gamma$(E2) keV | B(M1)/B(E2) $(\mu_N/eb)^2$ | Reference | **Configurations and Comments:** |
|---|---|---|---|---|---|---|---|
| 1. | 2616.3 | $11^-$ | | | | **2002Sc35** | 1. $\pi[(g_{9/2})^2\otimes(p_{3/2}/f_{5/2})^1]\otimes\nu(g_{9/2})$ from TAC.<br>2. $(\beta_2, \gamma)$ = (0.16, 20°) from TAC calculations.<br>3. Regular band.<br>4. B(M1) values for the transitions from 410.6 keV to 668.4 keV are 1.24(+37-24), 0.77(+20-13), 0.74(+22-14) and >0.11 $\mu_N^2$, respectively.<br>5. B(E2) values for the transitions from 883.5 keV to 1215.8 keV are 0.038(+15-11), 0.051(+19-13) and >0.011 $(eb)^2$, respectively.<br>6. The mean lifetimes (in ps) of states from 3027 to 4716 keV are 0.58(13), 0.59(11), 0.35(7) and <1, respectively.<br>7. g-factors of the states from I=$12^-$ to $15^-$ as given in 2010Yu03 are 1.12 (34), 1.03 (30), 0.87(30) and 0.82(30), respectively. |
| | 3026.9 | $12^-$ | 410.6 | | | 1999Sc14 | |
| | 3499.9 | $13^-$ | 473.0 | 883.5 | 20(+15-8) | 1999Do02 | |
| | 4047.5 | $14^-$ | 547.7 | 1019.2 | 14(+12-6) | 2000Sc17 | |
| | 4715.8 | $15^-$ | 668.4 | 1215.8 | ~10 | 2010Yu03 | |
| | 5484.6 | $(16)^-$ | 768.8 | 1436.2 | | | |

## $^{83}_{37}Rb_{46}$

| | $E_{level}$ keV | $I^\pi$ | $E_\gamma$(M1) keV | $E_\gamma$(E2) keV | B(M1)/B(E2) $(\mu_N/eb)^2$ | References | Configurations and Comments: |
|---|---|---|---|---|---|---|---|
| 1. | 2067.4 | $11/2^-$ | | | | **2009Sc22** | 1. $\pi(g_{9/2}) \otimes \nu[(g_{9/2}) \otimes (pf)^1]$ from TAC |
| | 2313.6 | $13/2^-$ | 246.6 | | | 2006Ga10 | calculations as given in 2001Am08. |
| | 2413.8 | $15/2^-$ | 100.1 | | | 2000Sc17 | 2. $(\beta_2, \gamma)$ = (0.18, 10°) from TAC calculations. |
| | 2595.9 | $17/2^-$ | 181.9 | | | 1980Ga17 | 3. Tentatively assigned as MR band (2006Ga10). |
| | 2958.1 | $19/2^-$ | 362.4 | | | 1995Ta21 | 4. The mean lifetimes (in ps) for the levels from |
| | 3363.2 | $21/2^-$ | 405.1 | | | 2001Am08 | 2313 to 3363 keV are 0.99(30), 6.0(30) ns, |
| | 4134.8 | $23/2^-$ | 771.9 | | | | 101(50), 9.9(+99-50) and 2.7(7), respectively. |
| | 5349.6 | $27/2^-$ | | 1214.8 | | | 5. B(M1) values for the transitions from 182 to 601 |
| | 6557.0 | (31/2-) | | 1207.3 | | | keV are 3.38(100), 0.89(+29-25) and 0.61(+46 |
| | 8032.7 | (35/2-) | | 1476.0 | | | -21) $\mu_N^2$, respectively. |
| | 9633.7 | (39/2-) | | 1601.0 | | | 6. Regular band. |

## $^{84}_{37}Rb_{47}$

| | $E_{level}$ keV | $I^\pi$ | $E_\gamma$(M1) keV | $E_\gamma$(E2) keV | B(M1)/B(E2) $(\mu_N/eb)^2$ | Reference | Configurations and Comments: |
|---|---|---|---|---|---|---|---|
| 1. | 3394.8 | $11^{(-)}$ | | | | **2002Sc35** | 1. $\pi[(g_{9/2})^2 \otimes (p_{3/2}/f_{5/2})^1] \otimes \nu(g_{9/2})$ from TAC |
| | 3721.5 | $12^{(-)}$ | 326.6 | | | **2010Sh12** | calculations. |
| | 4166.7 | $13^{(-)}$ | 445.1 | 771.3 | 20(+8-6) | 1999Sc14 | 2. $(\beta_2, \gamma)$ = (0.14, -15°) from TAC calculations. |
| | 4714.7 | $14^{(-)}$ | 548.0 | 994.8 | 9.4(+32-24) | 2000Sc17 | 3. Regular band. |
| | 5371.9 | $15^{(-)}$ | 656.9 | 1205.4 | 7.8(+41-29) | | 4. B(M1) values for the transitions from 445.1 keV |
| | 6094.8 | $(16^-)$ | 722.6 | 1380.7 | 10.2(+94-49) | | to 766.4 keV are 0.70(+14-10), 0.63(+8-7), |
| | 6861.1 | $(17^-)$ | 766.4 | 1489.3 | ~12 | | 0.44(+11-8), 0.49(+21-12) and >0.13 $\mu_N^2$, |
| | | | | | | | respectively. |
| | | | | | | | 5. B(E2) values for the transitions from 771.3 keV |
| | | | | | | | to 1489.3 keV are 0.036(+7-5), 0.067(+13-11), |
| | | | | | | | 0.058(+16-12), 0.048(+21-13) and >0.011 $(eb)^2$, |
| | | | | | | | respectively. |
| | | | | | | | 6. The mean lifetimes (in ps) of states from 4167 to |
| | | | | | | | 6861 keV are 0.82(12), 0.38(3), 0.25(4), 0.16(4) |
| | | | | | | | and <0.45, respectively. |
| 2. | 3679.8 | $12^{(-)}$ | | | | **2002Sc35** | 1. Configuration assigned as $\pi[(p_{3/2}, (g_{9/2})^2] \otimes \nu(g_{9/2})$ |
| | 4129.9 | $13^{(-)}$ | 450.3 | | | 2010Sh12 | 2. B(M1) values for the transitions from 450.3 keV |
| | 4800.4 | $14^{(-)}$ | 670.6 | | | | to 911.0 keV are 0.78(+21-15), 1.45(+70-38), |
| | 5253.5 | $15^{(-)}$ | 453.1 | | | | 0.96(+16-12), 1.64(+36-25) and >0.69 $\mu_N^2$, |
| | 5932.3 | $16^{(-)}$ | 678.8 | | | | respectively. |
| | 6470.7 | $17^{(-)}$ | 538.4 | | | | 3. Mean Lifetimes (in ps) of states from 4129.9 to |
| | 7381.7 | $(18^-)$ | 911.0 | | | | 6470.7 keV are 0.41(7), 0.63(9), 0.11(2) and |
| | | | | | | | <0.52, respectively. |
| | | | | | | | 4. Irregular band |

## $^{85}_{37}Rb_{48}$

| | $E_{level}$ keV | $I^\pi$ | $E_\gamma$(M1) keV | $E_\gamma$(E2) keV | B(M1)/B(E2) $(\mu_N/eb)^2$ | References | Configurations and Comments: |
|---|---|---|---|---|---|---|---|
| 1. | 3198.2 | $17/2^{(-)}$ | | | | **1995Sc04** | 1. Tentatively assigned as $\pi(g_{9/2}^{-1}) \otimes \nu(g_{9/2}^{-1} f_{5/2}^{-1})$. |
| | 3813.1 | $19/2^{(-)}$ | 614.9 | | | 1995Ta21 | 2. Irregular band. |
| | 4356.1 | $21/2^{(-)}$ | 543.5 | | | | |
| | 4940.0 | $(23/2^-)$ | 583.9 | | | | |

## $^{85}_{38}\mathrm{Sr}_{47}$

| | $E_{level}$ keV | $I^\pi$ | $E_\gamma$(M1) keV | $E_\gamma$(E2) keV | B(M1)/B(E2) $(\mu_N/eb)^2$ | References | **Configurations and Comments:** |
|---|---|---|---|---|---|---|---|
| 1. | 3384.0 | $19/2^+$ | | | | **2014Ku19** | 1. Configuration assigned as $\pi(g_{9/2})^2\otimes\nu(1g_{9/2})^{-1}$ from TAC calculations. |
| | 3511.6 | $21/2^+$ | 127.7 | | | 2017Ku04 | 2. $(\varepsilon_2, \gamma)$ =(0.11,60°) from TAC calculations. |
| | 3965.8 | $23/2^+$ | 454.2 | | | | 3. The B(M1) values for transition 454.2 keV to 658.5 keV as given in 2017Ku04 are 1.08(+25-30), 1.05(+22-16), 0.82(+18-23) and > 0.69 $\mu_N^2$, respectively. |
| | 4491.5 | $25/2^+$ | 525.7 | | | | 4. The mean lifetimes (in ps) of the levels from 3965.8 keV to 5749.7 keV as given in 2017Ku04 are 0.56(+14-16), 0.37(+8-6), 0.33(+7-9) and < 0.29, respectively. |
| | 5091.2 | $27/2^+$ | 599.7 | | | | 5. Regular band with backbending at 15 ħ. |
| | 5749.7 | $29/2^+$ | 658.5 | | | | |
| | 6360.8 | $(31/2^+)$ | 611.1 | | | | |
| 2. | 4779.6 | $(21/2^+)$ | | | | **2014Ku19** | 1. Configuration assigned as $\pi(g_{9/2})^2(f_{5/2})^2\otimes\nu(1g_{9/2})^{-1}$ from TAC calculations. |
| | 4969.0 | $23/2^{(+)}$ | 189.2 | | | 2017Ku04 | 2. $(\beta_2, \gamma)$ = (0.115,60°) from TAC calculations. |
| | 5181.0 | $25/2^+$ | 212.0 | | | | 3. The B(M1) values for transitions 585.0 keV and 458.8 keV as given in 2017Ku04 are 2.17(+50- 76) and >1.34 $\mu_N^2$, respectively. |
| | 5422.9 | $27/2^+$ | 241.9 | | | | 4. The mean lifetimes (in ps) of the levels from 6007.9 keV and 6466.7 keV as given in 2017Ku04 are 0.11(+3-4), <0.44, respectively. |
| | 6007.9 | $29/2^+$ | 585.0 | | | | 5. Regular band with backbending at 15 ħ. |
| | 6466.7 | $31/2^{(+)}$ | 458.8 | | | | |
| 3. | 6878.3 | $12^+$ | | | | **2016Ku17** | 1. Configuration assigned as $\pi(g_{9/2})^2\otimes\nu(g_{9/2})^{-2}$ before I=16ħ and $\pi[(g_{9/2})^2(f_{5/2})^1(p_{3/2\,/}\,p_{1/2})^1\otimes\nu(g_{9/2})^{-2}$ for higher spin region from TAC calculations (2017Ku04, 2016Ku17). |
| | 7336.0 | $13^+$ | 457.6 | | | 2017Ku04 | 2. $(\beta_2, \gamma)$ = (0.095,60°) and (0.010,60°) for 4qp and 6qp configuration, respectively from TAC calculations of 2017Ku04 and 2016Ku17. |
| | 7843.7 | $14^+$ | 507.6 | | | 2015Zh18 | 3. The mean lifetimes (in ps) of the levels from 7336.0 keV to 10004.7 keV as given in 2017Ku04 are 0.248(+91-80), 0.204(+84-75), 0.241(+73-87), 0.921(+164-184), 0.203(+81-72) and <0.193, respectively. |
| | 8337.3 | $15^+$ | 493.5 | | | | 4. The B(M1) values for the transition from 457.6 keV to 574.5 keV as given in 2017Ku04 are 2.39(+86-76), 2.18(+87-76), 1.98(+62-71), 0.57(+10-8), 1.22(+61-54) and > 1.58 $\mu_N^2$, respectively. |
| | 8813.5 | $16^+$ | 476.2 | | | | 5. The band is tentatively assigned as MR band (2017Ku04, 2016Ku17). |
| | 9430.2 | $17^+$ | 616.7 | | | | 6. Irregular band. |
| | 10004.7 | $18^+$ | 574.5 | | | | |
| | 10872.9 | $19^+$ | 868.2 | | | | |

### $^{86}_{39}\mathrm{Y}_{47}$

| | $E_{level}$ keV | $I^{\pi}$ | $E_{\gamma}$(M1) keV | $E_{\gamma}$(E2) keV | B(M1)/B(E2) $(\mu_N/eb)^2$ | References | **Configurations and Comments:** |
|---|---|---|---|---|---|---|---|
| 1. | 2758 | $11^{(-)}$ | | | | **2009Ru03** | 1. Configuration assigned as $\pi(g_{9/2})^2(p_{3/2}/f_{5/2})^{-1} \otimes \nu(g_{9/2})^{-1}$ before and $\pi(g_{9/2})^2 (p_{3/2}/f_{5/2})^{-1} \otimes \nu(g_{9/2})^{-3}$ after alignment from TAC-RMF calculations (2013Li33). |
| | 3090 | $12^{(-)}$ | 332.6 | | | 2013Li33 | |
| | 3454 | $13^{(-)}$ | 363.8 | | | | |
| | 4010 | $14^{(-)}$ | 556.4 | | | | |
| | 4710 | $15^{(-)}$ | 699.4 | | | | 2. The band is tentatively assigned as MR band (2009Ru03, 2013Li33). |
| | 5430 | $16^{(-)}$ | 719.9 | 1419.3 | 11.4(16) | | |
| | 6087 | $17^{(-)}$ | 657.3 | 1377.2 | 21.3(74) | | 3. The B(M1) ≈ 0.4 - 1.4 $\mu_N^2$ using TAC-RMF calculations (2013Li33). |
| | | | | | | | 4. Irregular band with backbending at ≈16 ħ |

### $^{85}_{40}\mathrm{Zr}_{45}$

| | $E_{level}$ keV | $I^{\pi}$ | $E_{\gamma}$(M1) keV | $E_{\gamma}$(E2) keV | B(M1)/B(E2) $(\mu_N/eb)^2$ | References | **Configurations and Comments:** |
|---|---|---|---|---|---|---|---|
| 1. | 2625.3 | $17/2^-$ | | | | **2003Wa36** | 1. Probable Configuration as $\pi(g_{9/2})^2{}_8 \otimes \nu\,(f_{7/2})$ using semi-classical model (2007Yu03). |
| | 2958.3 | $19/2^-$ | 333 | | | 2007Yu03 | |
| | 3387.1 | $21/2^-$ | 429 | | | 2002Ta11 | 2. g-factors of the states from I=$17/2^-$ to $23/2^-$ are 1.3 (4), 1.1 (3), 0.85(25) and 0.43(27), respectively(2007Yu03). |
| | 3838.0 | $23/2^-$ | 451 | 879 | | 1995Ju04 | |
| | 4374.2 | $25/2^-$ | 536 | 987 | | | |
| | 5023.2 | $(27/2^-)$ | 649 | | | | 3. Regular band |

### $^{87}_{40}\mathrm{Zr}_{47}$

| | $E_{level}$ keV | $I^{\pi}$ | $E_{\gamma}$(M1) keV | $E_{\gamma}$(E2) keV | B(M1)/B(E2) $(\mu_N/eb)^2$ | References | **Configurations and Comments:** |
|---|---|---|---|---|---|---|---|
| 1. | 2896.5 | $21/2^+$ | | | | **2018Ba34** | 1. Probable Configuration as $\pi\,[(p_{3/2}\,f_{5/2}\,p_{1/2})^{-2}\,(g_{9/2})^2] \otimes \nu[\,(g_{9/2})^{-1}]$ using semi-classical model. |
| | 3383.2 | $23/2^+$ | 486.7 | | | | |
| | 3945.9 | $25/2^+$ | 562.8 | 1049.2 | 38(15) | | 2. The B(M1) values for the transitions 486.7, 562.8, 588.3 and 525.1 keV are 0.77(19), 0.80(+24-19), 0.53 (+21-14) and 0.29(4-3) $\mu_N^2$, respectively. |
| | 4534.2 | $27/2^+$ | 588.3 | 1151.0 | 17(8) | | |
| | 5059.3 | $29/2^+$ | 525.1 | 1113.3 | 68(12) | | |
| | | | | | | | 3. The B(E2) values for the transition from 1049.2, 1151.0 and 1113.3 keV are 9.2(+27-22), 14.1 (+55-37) and 1.9(2) W.u., respectively. |
| | | | | | | | 4. The mean lifetime of levels from 3383.2 to 5059.3 keV are 0.62(+19-13), 0.35(+11-8), 0.25(+8-7) and 1.19(+15-13) ps, respectively. |
| | | | | | | | 5. Regular band |

### $^{105}_{45}\mathrm{Rh}_{60}$

| | $E_{level}$ keV | $I^{\pi}$ | $E_{\gamma}$(M1) keV | $E_{\gamma}$(E2) keV | B(M1)/B(E2) $(\mu_N/eb)^2$ | Reference | **Configurations and Comments:** |
|---|---|---|---|---|---|---|---|
| 1. | 2019.3 | $13/2^-$ | | | | **2004Al03** | 1. Tentatively assigned as MR band. |
| | 2170.4 | $15/2^-$ | 151.1 | | | | 2. Regular band with small backbending at the top of the band. |
| | 2310.8 | $17/2^-$ | 140.4 | | | | |
| | 2496.1 | $19/2^-$ | 185.3 | | | | |
| | 2718.8 | $21/2^-$ | 222.7 | | | | |
| | 2993.2 | $23/2^-$ | 274.4 | 496.8 | | | |
| | 3308.6 | $25/2^-$ | 315.4 | | | | |
| | 3769.4 | $27/2^-$ | 460.8 | | | | |
| | 4183.7 | $(29/2^-)$ | 414.3 | | | | |

## $^{105}_{45}Rh_{60}$

| | $E_{level}$ keV | $I^\pi$ | $E_\gamma$(M1) keV | $E_\gamma$(E2) keV | B(M1)/B(E2) $(\mu_N/eb)^2$ | Reference | **Configurations and Comments:** |
|---|---|---|---|---|---|---|---|
| 2. | 2417.4 | $15/2^-$ | | | | **2004Al03** | 1. $\pi(g_{9/2}) \otimes \nu(h_{11/2}g_{7/2})$ from TAC calculations.<br>2. Triaxial deformation $(\beta_2, \gamma) = (0.22, 30°)$ from TAC calculations.<br>3. Tentatively assigned as MR band.<br>4. B(M1)/B(E2) ≥ 6 $(\mu_N/eb)^2$ for the 21/2 state.<br>5. Regular band. |
| | 2512.7 | $17/2^-$ | 95.3 | | | | |
| | 2645.7 | $19/2^-$ | 133.0 | | | | |
| | 2825.1 | $21/2^-$ | 179.4 | | | | |
| | 3077.9 | $23/2^-$ | 252.8 | | | | |
| | 3469.9 | $(25/2^-)$ | 392.0 | | | | |
| 3. | 2477.0 | $17/2^-$ | | | | **2004Al03** | 1. Bands 2 and 3 are possibly chiral partners. Thus, band 3 is assigned the same configuration and deformation parameters as those for band 2.<br>2. Tentatively assigned as MR band.<br>3. B(M1)/B(E2) ≥ 6 $(\mu_N/eb)^2$ for the 21/2 state.<br>4. Regular band. |
| | 2668.9 | $19/2^-$ | 191.9 | | | | |
| | 2914.1 | $21/2^-$ | 245.2 | | | | |
| | 3266.9 | $23/2^-$ | 352.8 | | | | |
| | 3667.5 | $(25/2^-)$ | 400.6 | | | | |
| | 4092.3 | $(27/2^-)$ | 424.8 | | | | |
| 4. | 2981.6 | $23/2^+$ | | | | **2004Al03** | 1. Tentatively assigned as MR band.<br>2. Regular band. |
| | 3197.6 | $25/2^+$ | 216.0 | | | | |
| | 3478.0 | $27/2^+$ | 280.4 | 496.5 | | | |
| | 3839.3 | $29/2^+$ | 361.3 | 642.0 | | | |
| | 4215.4 | $31/2^+$ | 376.1 | 736.9 | | | |
| | 4702.2 | $(33/2^+)$ | 486.8 | | | | |

## $^{103}_{47}Ag_{56}$

| | $E_{level}$ keV | $I^\pi$ | $E_\gamma$(M1) keV | $E_\gamma$(E2) keV | B(M1)/B(E2) $(\mu_N/eb)^2$ | Reference | **Configurations and Comments:** |
|---|---|---|---|---|---|---|---|
| 1. | 3104.0 | $19/2^-$ | | | | **2008Ra06** | 1. Configuration assigned as $\pi g_{9/2} \otimes \nu[g_{7/2} \otimes h_{11/2}]$ from TAC calculations (2003Da07).<br>2. $(\beta_2, \gamma) = (0.13, 28°)$ from TAC calculations (2003Da07).<br>3. The band has mixed character of MR and tilted rotational (2008Ra06).<br>4. B(M1) values for the transitions from 515.6 keV to 697.0 keV as given in 2008Ra06 are 1.02(+30-20), 0.986(+10-20) and 0.51(6) $\mu_N^2$, respectively.<br>5. B(E2) values for the transitions from 515.6 keV to 697.0 keV as given in 2008Ra06 are 0.09(+2-3), 0.12(1) and0.05(5) $(eb)^2$, respectively.<br>6. Mean Lifetimes (in ps) of states from 4941.6keV to 6167.7 keV as given in 2008Ra06 are 0.401(+110-60), 0.385(+4-6) and 0.330(+4-3), respectively.<br>7. The B(M1)/B(E2) ratios for levels of spin 29/2ħ to 33/2ħ are deduced using B(M1) and B(E2) values given in 2008Ra06 and for other levels, the B(M1)/B(E2) ratios are calculated from experimental data given in 2008Ra06.<br>8. Regular band with backbending at 25/2 ħ. |
| | 3338.9 | $21/2^-$ | 234.9 | | | 2003Da07 | |
| | 3647.9 | $23/2^-$ | 309.0 | | | | |
| | 4064.1 | $25/2^-$ | 416.2 | 725.2 | 16.9(18) | | |
| | 4426.0 | $27/2^-$ | 361.9 | 778.1 | 11.4(11) | | |
| | 4941.6 | $29/2^-$ | 515.6 | 877.5 | 11.3(41) | | |
| | 5470.7 | $(31/2^-)$ | 529.1 | 1044.7 | 8.2(7) | | |
| | 6167.7 | $(33/2^-)$ | 697.0 | 1226.1 | 10.2(15) | | |

## $^{103}_{47}Ag_{56}$

| | $E_{level}$ keV | $I^\pi$ | $E_\gamma$(M1) keV | $E_\gamma$(E2) keV | B(M1)/B(E2) $(\mu_N/eb)^2$ | Reference | Configurations and Comments: |
|---|---|---|---|---|---|---|---|
| 2. | 3221.4 | (17/2$^-$) | | | | **2008Ra06** | 1. Probable proton and neutron orbital involved in configuration are $\pi g_{9/2}$ and $\nu(g_{7/2}/d_{5/2}, h_{11/2})$.<br>2. $(\beta_2, \gamma) \approx (0.13, 15°)$ from TAC calculations.<br>3. The band has mixed character of MR and tilted rotational (2008Ra06).<br>4. B(M1) values for the transitions from 382.9 keV to 542.8 keV are2.6(+6-2), 2.2(2) and 1.9(+3-2) $\mu_N^2$, respectively. B(E2) value of 382.9<br>5. keV level is 0.09 (+4-1) $(eb)^2$<br>6. Mean Lifetimes (in ps) of states from 5157.0keV to 6133.8 keV are 0.39(+3-7), 0.32(+2-3) and 0.19(+1-3), respectively.<br>7. Irregular band. |
| | 3402.2 | (19/2$^-$) | 180.8 | | | | |
| | 3691.0 | (21/2$^-$) | 289.4 | | | | |
| | 3973.9 | (23/2$^-$) | 282.3 | | | | |
| | 4340.8 | (25/2$^-$) | 368.8 | | | | |
| | 4774.9 | (27/2$^-$) | 432.2 | 801.0 | 9.1(6) | | |
| | 5157.0 | (29/2$^-$) | 382.9 | 815.1 | 28.9(87) | | |
| | 5591.0 | (31/2$^-$) | 433.2 | | | | |
| | 6133.8 | (33/2$^-$) | 542.8 | | | | |
| 3. | 3439 | 21/2$^-$ | | | | **2003Da07** | 1. Configuration assigned as $\pi g_{9/2} \otimes \nu d_{5/2} \otimes \nu h_{11/2}$ from TAC calculations (2003Da07).<br>2. $(\beta_2, \gamma) \approx (0.13, 28°)$ from TAC calculations.<br>3. Small signature splitting.<br>4. B(M1)/B(E2) values are from 2003Da07.<br>5. The mean lifetimes (in ps) of the levels from 27/2 to 35/2 ħ as given in 2006De15 are 0.45(+2 -3), 0.37(2), 0.36(2), 0.29 (1) and 0.27(+1-2), respectively. The uncertainties of ~10-15% from stopping powers are not included.<br>6. The B(M1) values for the transitions from 437 to 585 keV as given in 2006De15 are 1.40(+8-6), 1.72(+9-8), 0.92(4), 1.17(5) and 0.73(+4-2) $\mu_N^2$, respectively.<br>7. The B(E2) values for the transitions from 774 to 1088 keV as given in 2006De15 are 0.052(+3-2) 0.070(+4-3), 0.037(2), 0.057(3) and 0.060(+3-1) $(eb)^2$, respectively. |
| | 3599 | 23/2$^-$ | 160 | | | 2006De15 | |
| | 3936 | 25/2$^-$ | 337 | | | | |
| | 4373 | 27/2$^-$ | 437 | 774 | 31.2(62) | | |
| | 4793 | 29/2$^-$ | 419 | 857 | 24.2(33) | | |
| | 5323 | 31/2$^-$ | 530 | 949 | 21.6(41) | | |
| | 5825 | 33/2$^-$ | 503 | 1033 | 24.3(41) | | |
| | 6411 | 35/2$^-$ | 585 | 1088 | 9.6(25) | | |
| | (6941) | (37/2$^-$) | (531) | (1116) | | | |
| 4. | 5336.5 | (27/2$^-$) | | | | **2008Ra06** | 1.Configuration assigned as $\pi(g_{9/2})^3 \otimes \nu(g_{7/2}$ $h_{11/2})$ from TAC calculations.<br>2. $(\beta_2, \gamma) = (0.10, 19°)$ from TAC calculations.<br>3. B(M1) values for the transitions from 402.4 keV to 503.6 keV are 4.70(50), 3.72(40) and 3.66(35) $\mu_N^2$, respectively.<br>4. The mean lifetimes (in ps) of states from 6166.7 keV to 7157.8 keV are 0.18(2), 0.13(1) and 0.12(1), respectively.<br>5. Irregular band. |
| | 5764.3 | (29/2$^-$) | 427.8 | | | | |
| | 6166.7 | (31/2$^-$) | 402.4 | 830.2 | 78(11) | | |
| | 6654.0 | (33/2$^-$) | 487.5 | | | | |
| | 7157.8 | (35/2$^-$) | 503.6 | | | | |
| | 7670.4 | (37/2$^-$) | 512.6 | | | | |
| | 8240.3 | (39/2$^-$) | 569.9 | | | | |

## $^{104}_{47}\mathrm{Ag}_{57}$

| | $E_{level}$ keV | $I^\pi$ | $E_\gamma$(M1) keV | $E_\gamma$(E2) keV | B(M1)/B(E2) $(\mu_N/eb)^2$ | Reference | Configurations and Comments: |
|---|---|---|---|---|---|---|---|
| 1. | 1077.1 | $8^-$ | | | | **2013Wa20** | 1. $\pi(g_{9/2})^{-1} \otimes \nu(h_{11/2})$ from TAC calculations.<br>2. $(\beta_2, \gamma)$ = (0.17, 29°) from TAC calculations.<br>3. This band is also interpreted as chiral doublet partner of band 2 (2013Wa20).<br>4. The mean lifetimes (in ps) for the levels from 3301 to 3808 keV are 0.395(1) and 0.349(4) (2017Da13), for level 4328 keV the lifetime is 0.35(6) (2004Da14). The uncertainties of ~10% from stopping powers are not included in level lifetime (2004Da14)<br>5. The B(M1) values for the transitions from 481 to 520 keV are 1.25(30), 0.97(21) and 0.86(20) $\mu_N^2$, respectively.<br>6. The B(E2) values for the transitions from 926 to 1028 keV are 0.104(25), 0.079(17) and 0.053(13) $(eb)^2$, respectively.<br>7. Irregular band. |
| | 1252.4 | $9^-$ | 175.4 | | | 2004Da14 | |
| | 1598.7 | $10^-$ | 346.3 | 521.6 | 129.7(28) | 2017Da13 | |
| | 1931.4 | $11^-$ | 332.7 | 679.0 | 150.9(26) | | |
| | 2375.3 | $12^-$ | 443.9 | 776.7 | 90.0(18) | | |
| | 2819.7 | $13^-$ | 444.6 | 888.2 | 58.9(12) | | |
| | 3301.0 | $14^-$ | 481.4 | 925.6 | 70.7(13) | | |
| | 3808.6 | $15^-$ | 507.8 | 988.7 | 86.6(13) | | |
| | 4328.2 | $16^-$ | 519.8 | 1027.0 | 104.2(8) | | |
| | 4900.3 | $17^-$ | 572.3 | 1091.6 | 86.5(18) | | |
| | 5528.1 | $18^-$ | 628.0 | 1199.7 | 61.2(14) | | |
| | 6133.0 | $(19^-)$ | 604.9 | (1232.3) | 56.7(18) | | |
| 2. | 2211.7 | $(10^-)$ | | | | **2013Wa20** | 1. Tentatively assigned as $\pi(g_{9/2})^{-1} \otimes \nu[(g_{7/2}d_{5/2})^2 h_{11/2}]$ from the arguments based on aligned angular momentum and parity.<br>2. $(\beta_2, \gamma)$ = (0.09, 31°) from TAC calculations.<br>3. This band is also interpreted as chiral doublet partner of band 1 (2013Wa20).<br>4. Irregular band. |
| | 2711.4 | $(11^-)$ | 499.4 | | | 2004Da14 | |
| | 3040.1 | $12^-$ | 328.7 | 828.1 | 3.6(13) | | |
| | 3350.4 | $13^-$ | 310.1 | 639.0 | | | |
| | 3647.5 | $14^-$ | 297.0 | | | | |
| | 4096.9 | $15^-$ | 449.5 | 976.9 | | | |
| | 4624.7 | $16^-$ | 527.8 | | | | |
| 3. | 4423.8 | $14^+$ | | | | **2013Wa20** | 1. $\pi(g_{9/2})^{-1} \otimes \nu[(g_{7/2}/d_{5/2})\, h_{11/2}^2]$ from TAC calculations.<br>2. $(\beta_2, \gamma)$ = (0.18, 25°) from TAC calculations.<br>3. Regular band.<br>4. The mean lifetimes (in ps) for the levels from 5571 to 6595 keV are 0.39(7), 0.30(6) and 0.28(4), respectively.<br>5. The B(M1) values for the transitions from 406 to 544 keV are 2.18(43), 1.71(38) and 1.26(21) $\mu_N^2$, respectively. |
| | 4785.2 | $15^+$ | 361.4 | | | 2004Da14 | |
| | 5165.9 | $16^+$ | 380.9 | | | | |
| | 5571.9 | $17^+$ | 406.0 | | | | |
| | 6052.2 | $18^+$ | 480.2 | | | | |
| | 6595.9 | $19^+$ | 543.7 | | | | |
| | 7160.3 | $20^+$ | 564.4 | | | | |

## $^{105}_{47}\mathrm{Ag}_{58}$

| | $E_{level}$ keV | $I^{\pi}$ | $E_{\gamma}$(M1) keV | $E_{\gamma}$(E2) keV | B(M1)/B(E2) $(\mu_N/eb)^2$ | References |
|---|---|---|---|---|---|---|
| 1. | 2595.7 | $17/2^-$ | | | | **1994Je12** |
| | 2751.2 | $19/2^-$ | 155.4 | | | 1995Je05 |
| | 2935.7 | $21/2^-$ | 184.5 | 340.0 | 3.9(8) | 2006De15 |
| | 3176.0 | $23/2^-$ | 240.3 | 424.8 | 4.5(10) | 2007Ti07 |
| | 3510.5 | $25/2^-$ | 334.5 | 574.8 | 11.2(9) | |
| | 3927.8 | $27/2^-$ | 417.3 | 751.5 | 10.7(12) | |
| | 4361.7 | $29/2^-$ | 433.9 | 851.2 | 11.2(14) | |
| | 4931.7 | $31/2^-$ | 570.0 | 1003.9 | 20.0(8) | |
| | 5445.0 | $33/2^-$ | 513.3 | 1083.8 | 40.6(39) | |
| | 6113.0 | $35/2^-$ | 668.0 | 1181.4 | 5.6(19) | |
| | 6715.0 | $37/2^-$ | 602.0 | 1270.0 | 13.4(70) | |
| | 7438.0 | $(39/2^-)$ | 723.0 | | | |

**Configurations and Comments:**

1. Configuration assigned as $\pi g_{9/2}\otimes\nu[(h_{11/2})(g_{7/2}/d_{5/2})^1]$ from TAC calculations (2006De15).
2. Assigned as MR structure (2006De15).
3. $(\varepsilon_2, \gamma) = (0.19, 24°)$ (2006De15).
4. The mean lifetimes (in ps) of the levels from 3510.5 keV to 5445.0 keV are 0.51(3), 0.49 (4), 0.38 (1) and 0.28 (2), respectively (2006De15). 0.46 (4). The uncertainties of ~10-15% from stopping powers are not included.
5. The B(M1) values for the transitions 334.5 keV, to 513.3 keV are 2.70(+16-14), 1.32(10), 1.13(10), 0.68(2) and 1.26(9) $\mu_N^2$, respectively (2006De15).
6. The B(E2) values for the transitions 574.8 keV to 1083.8 keV are 0.242(+14-12), 0.123(10), 0.101(+9-8), 0.034(1) and 0.031(2) $(eb)^2$, respectively (2006De15).
7. The B(M1)/B(E2) ratios for levels of spin $25/2^-$ to $33/2^-$ are deduced using B(M1) and B(E2) values given in 2006DE15 and for other levels, the B(M1)/B(E2) ratios are calculated from experimental data given in 1994Je12.
8. Irregular band with signature splitting.

## $^{106}_{47}\mathrm{Ag}_{59}$

| | $E_{level}$ keV | $I^{\pi}$ | $E_{\gamma}$(M1) keV | $E_{\gamma}$(E2) keV | B(M1)/B(E2) $(\mu_N/eb)^2$ | References |
|---|---|---|---|---|---|---|
| 1. | 3259 | $12^+$ | | | | **2017Da10** |
| | 3490 | $13^+$ | 230 | | | 2010He05 |
| | 3748 | $14^+$ | 258 | 489 | 1.01(2) | 2006De15 |
| | 4043 | $15^+$ | 295 | 553 | | 2006He31 |
| | 4390 | $16^+$ | 347 | 642 | 4.2(3) | 1994Je11 |
| | 4795 | $17^+$ | 405 | 752 | 9.7(20) | |
| | 5261 | $18^+$ | 466 | 871 | 10.4(16) | |
| | 5764 | $19^+$ | 503 | 969 | 10.9(27) | |
| | 6352 | $20^+$ | 588 | 1091 | 6.5(23) | |
| | 6938 | $21^+$ | 585 | 1174 | 8.0(25) | |
| | 7618 | $22^+$ | 680 | 1266 | 10.3(39) | |
| | 8300 | $23^+$ | 682 | 1362 | 8.8(30) | |
| | 9042 | $24^+$ | 742 | 1424 | 7.1 | |
| | 9838 | $25^+$ | 796 | 1538 | | |

**Configurations and Comments:**

1. Configuration assignment as $\pi(g_{9/2})^{-1}\otimes\nu[(h_{11/2})^2(g_{7/2}/d_{5/2})]$ from SPAC calculations (2017Da10) and as $\pi g_{9/2}\otimes\nu[(h_{11/2})^2(g_{7/2}/d_{5/2})]$ (2006De15)
2. Tentatively assigned as MR band.
3. $(\beta_2, \gamma) = (0.18, 6°)$ (2006De15)
4. The mean lifetimes (in ps) of the levels from 4792.7 keV to 9038.7 keV as given in 2017Da10 are 0.33(4), 0.21(2), 0.18(2) 0.15(3), 0.22(3), 0.14(3), 0.12(2) and 0.11, respectively
5. The B(M1) values for the transition from 405 keV to 742 keV as given in 2017Da10 are 2.03(30), 1.97(23), 1.74(28), 0.97(23), 0.64(12), 0.72(18), 0.70(17) and 0.50 $\mu_N^2$, respectively
6. The B(E2) values for the transition from 752 keV to 1424 keV as given in 2017Da10 are 0.21(3), 0.19(2), 0.16(3), 0.15(4), 0.08(2), 0.07(2), 0.08(2) and 0.07 $(eb)^2$, respectively
7. The B(M1)/B(E2) ratios for spin range 17 ħ to 24 ħ are deduced using B(M1) and B(E2) values given in 2017Da10 and for 14 ħ and 16 ħ, are are calculated from experimental data given in 2010He05.
8. Irregular band with signature splitting.

## $^{106}_{47}Ag_{59}$

| | E<sub>level</sub> keV | $I^{\pi}$ | $E_{\gamma}$(M1) keV | $E_{\gamma}$(E2) keV | B(M1)/B(E2) $(\mu_N/eb)^2$ | References | Configurations and Comments: |
|---|---|---|---|---|---|---|---|
| 2. | 3203.2 | $10^-$ | | | | **2016Da03** | 1. Configuration assigned as $\pi(g_{9/2})^{-3}\otimes\nu(h_{11/2})$ and $\pi(g_{9/2})^{-3}\otimes\nu(h_{11/2}(d_{5/2}/g_{7/2})^2)$ before and after band crossing from SPAC calculations.<br>2. The mean lifetimes (in ps) of the levels from 4263.7 keV to 7276.7 keV are 0.41(3), 0.40(3), 0.35(4), 0.34(5), 0.18(4), 0.14(4) and 0.17, respectively<br>3. The B(M1) values for the transitions 322 keV to 586 keV are 3.60(34), 2.18(22), 1.64(24), 0.83(17) 1.55(42), 0.88(31) and 0.83 $\mu_N^2$, respectively.<br>4. The B(E2) values for the transitions 588 keV to 1211 keV are 0.34(3), 0.24(2), 0.19(3), 0.11(2), 0.15(4), 0.14(5) and 0.09 $(eb)^2$, respectively.<br>5. Irregular band with signature splitting.<br>6. The B(M1)/B(E2) ratios for spin range 14 ħ to 20 ħ are deduced using B(M1) and B(E2) values given in 2016Da03. |
| | 3423.2 | $11^-$ | | | | | |
| | 3674.9 | $12^-$ | 252 | | | | |
| | 3941.0 | $13^-$ | 266 | 518 | | | |
| | 4263.0 | $14^-$ | 322 | 588 | 10.6(14) | | |
| | 4636.0 | $15^-$ | 373 | 695 | 9.1(12) | | |
| | 5051.0 | $16^-$ | 415 | 788 | 8.6(18) | | |
| | 5560.0 | $17^-$ | 509 | 924 | 7.5(21) | | |
| | 6065.0 | $18^-$ | 505 | 1014 | 10.3(39) | | |
| | 6690.0 | $19^-$ | 625 | 1130 | 6.3(32) | | |
| | 7276.0 | $20^-$ | 586 | 1211 | 9.2 | | |
| | 7944.0 | $21^-$ | 668 | 1254 | | | |

## $^{107}_{47}Ag_{60}$

| | E<sub>level</sub> keV | $I^{\pi}$ | $E_{\gamma}$(M1) keV | $E_{\gamma}$(E2) keV | B(M1)/B(E2) $(\mu_N/eb)^2$ | References | Configurations and Comments: |
|---|---|---|---|---|---|---|---|
| 1. | 2297.9 | $15/2^-$ | | | | **2014Ya02** | 1. Configuration assigned as $\pi(1g_{9/2})^7\otimes\nu[(2d_{5/2}, 1g_{7/2})^9\ (1h_{11/2})^1]$ from CNS calculations.<br>2. $(\varepsilon_2, \gamma) \approx (0.15,-10°)$ before bending from CNS calculations<br>3. The mean lifetimes (in ps) of the levels from 3054.2 keV to 4395.5 keV are 1.36(39), 0.83 (17), 0.50(9) and <0.83, respectively.<br>4. The B(M1) values for the transition from 307.9 keV to 470.0 keV are 1.25(36), 0.82(18), 0.85(17) and >0.39$\mu_N^2$, respectively<br>5. The B(E2) values for the transition from 512.3 keV to 931.3 keV are 0.18(6), 0.09(2), 0.08(2) and>0.06 $(eb)^2$, respectively.<br>6. Regular band with backbending at 31/2 ħ.<br>7. The B(M1)/B(E2) ratios for spin range 23/2ħ to 29/2ħ are deduced by authors of this work. |
| | 2411.3 | $17/2^-$ | 113.6 | | | | |
| | 2542.0 | $19/2^-$ | 130.7 | | | | |
| | 2746.4 | $21/2^-$ | 204.4 | | | | |
| | 3054.3 | $23/2^-$ | 307.9 | 512.3 | 6.9(30) | | |
| | 3464.3 | $25/2^-$ | 410.0 | 717.9 | 9.1(28) | | |
| | 3925.6 | $27/2^-$ | 461.3 | 871.3 | 10.6(34) | | |
| | 4395.6 | $29/2^-$ | 470.0 | 931.3 | >6.5 | | |
| | 5004.4 | $31/2^-$ | 608.8 | 1078.8 | | | |
| | 5562.9 | $33/2^-$ | 558.5 | 1167.3 | | | |
| 2. | 3460.4 | $23/2^+$ | | | | **2014Ya02** | 1. Configuration assigned as $\pi(1g_{9/2})^7\otimes\nu[(2d_{5/2}, 1g_{7/2})^8\ (1h_{11/2})^2]$ from CNS calculations.<br>2. The mean lifetimes (in ps) of the levels from 3682.2 keV to 5746.0 keV are 0.88(38), 0.75 (15), 0.50(8), 0.39(5), 0.17 (3) and <0.21, respectively<br>3. The B(M1) values for the transition from 221.9 keV to 501.0 keV are 5.2(23), 2.7(6), 1.8(3), 1.9(3), 1.9(4) and >1.25 $\mu_N^2$, respectively.<br>4. The B(E2) values for the transition from 516.5 keV to 995.1 keV are 0.15(7), 0.23(5), 0.15(3), 0.14(3), 0.23(6) and >0.16, $(eb)^2$, respectively<br>5. Regular band |
| | 3682.3 | $25/2^+$ | 221.9 | | 34(22) | | |
| | 3976.9 | $27/2^+$ | 294.6 | 516.5 | 11.7(36) | | |
| | 4355.6 | $29/2^+$ | 378.7 | 673.3 | 12.0(31) | | |
| | 4752.0 | $31/2^+$ | 396.4 | 775.1 | 13.6(36) | | |
| | 5246.1 | $33/2^+$ | 494.1 | 890.5 | 8.3(28) | | |
| | 5747.1 | $35/2^+$ | 501.0 | 995.1 | >7.8 | | |
| | 6318.4 | $37/2^+$ | 571.3 | 1072.3 | | | |

## $^{109}_{47}Ag_{62}$

| | $E_{level}$ keV | $I^\pi$ | $E_\gamma$(M1) keV | $E_\gamma$(E2) keV | B(M1)/B(E2) $(\mu_N/eb)^2$ | References | **Configurations and Comments:** |
|---|---|---|---|---|---|---|---|
| 1. | 2207.1 | $15/2^-$ | | | | **2020Ma41** | 1. Configuration assigned as $\pi(g_{9/2})^{-1}\otimes\nu h_{11/2}(g_{7/2}/d_{5/2})^{-1}$, from TAC-RMF calculations. |
| | 2420.6 | $17/2^-$ | 213.5 | | | 2020Ma47 | |
| | 2661.1 | $19/2^-$ | 240.5 | 454.2 | | 2008Da12 | 2. The calculated B(E2) and B(M1) are of the order of ≈0.06 $(eb)^2$ and ≈ (1.4-1.7) $\mu_N^2$, respectively. |
| | 2941.1 | $(21/2^-)$ | 280.0 | 520.5 | | 2007TiZZ | |
| | 3204.6 | $(23/2^-)$ | 263.5 | 543.5 | | 1996Po07 | 3. Tentatively assigned as MR band |
| | 3559.2 | $(25/2^-)$ | 354.6 | 618.0 | | | 4. The band is also interpreted as stapler band (2020Ma47) |
| | 4015.7 | $(27/2^-)$ | 456.5 | 811.3 | | | |
| | 4475.2 | $(29/2^-)$ | 459.5 | 916.0 | | | 5. Regular band with small backbending at 23/2ħ. |
| | 4951.2 | $(31/2^-)$ | 476.0 | 935.8 | | | |
| 2. | 2568.5 | $19/2^+$ | | | | **2020Ma41** | 1. Configuration assigned as $\pi(g_{9/2})^{-1}\otimes\nu(h_{11/2})^2$ from TAC-RMF calculations. |
| | 2842.0 | $21/2^+$ | 273.5 | | | 2020Ma47 | |
| | 3091.1 | $23/2^+$ | 249.1 | 522.5 | 24.7(+50-45) | 2007TiZZ | 2. The experimental dynamic moment of inertia $J^{(2)}$ is ≈15 $MeV^{-1}$ $ħ^2$ |
| | 3277.6 | $25/2^+$ | 186.5 | 436.5 | 19.3(+62-56) | 1996Po07 | |
| | 3576.4 | $27/2^+$ | 298.8 | 485.2 | 19.0(+62-56) | | 3. Regular band with backbending at 25/2ħ. |
| | 3969.9 | $29/2^+$ | 393.5 | 692.4 | 16.6(+71-65) | | |
| | 4376.9 | $31/2^+$ | 407.0 | 800.3 | | | |
| | 4887.4 | $33/2^+$ | 510.5 | 917.5 | | | |
| | 5416.4 | $(35/2^+)$ | 529.0 | 1039.5 | | | |
| | 6000.6 | $(37/2^+)$ | 584.2 | 1113.0 | | | |
| | 6556.1 | $(39/2^+)$ | 555.5 | 1139.7 | | | |
| | 7182.1 | $(41/2^+)$ | 626.0 | (1181.0) | | | |

## $^{110}_{47}Ag_{63}$

| | $E_{level}$ keV | $I^\pi$ | $E_\gamma$(M1) keV | $E_\gamma$(E2) keV | B(M1)/B(E2) $(\mu_N/eb)^2$ | References | **Configurations and Comments:** |
|---|---|---|---|---|---|---|---|
| 1. | 2638.4 | $12^+$ | | | | **2018Da15** | 1. Configuration assigned as $\pi(1g_{9/2})^{-1}\otimes\nu[(d_{5/2}/g_{7/2})^1 (h_{11/2})^2]$ from SPAC calculations. |
| | 2823.4 | $13^+$ | 185 | | | 2021Ma15 | |
| | 3017.6 | $14^+$ | 194 | | | | 2. The mean lifetimes (in ps) of the levels from 3263.0 keV to 4548.4 keV are 2.30(40), 1.23(25), 0.89(14), and $<0.62$ (effective lifetime, not corrected from side feeding), respectively. |
| | 3263.0 | $15^+$ | 246 | 440 | 6.3(20) | | |
| | 3582.4 | $16^+$ | 320 | 564 | 8.8(40) | | |
| | 4032.8 | $17^+$ | 450 | 770 | 11.6(67) | | |
| | 4548.4 | $18^+$ | 516 | 966 | $>7.4$ | | 3. The B(M1) values for the transition from 246 keV to 516 keV are 1.44(29), 1.23(30), 0.59(23) and $>0.42$ $\mu_N^2$, respectively. |
| | 5082.1 | $19^+$ | 534 | 1049 | | | |
| | | | | | | | 4. The B(E2) values for the transition from 440 keV to 966 keV are 0.23(6), 0.14(5), 0.051(22), and $>0.057$, $(eb)^2$, respectively |
| | | | | | | | 5. Regular band |
| | | | | | | | 6. The B(M1)/B(E2) ratios for spin range 15 to 18 are deduced from given B(M1) and B(E2) values. |

## $^{102}_{48}\mathrm{Cd}_{54}$

| | $E_{level}$ keV | $I^{\pi}$ | $E_{\gamma}$(M1) keV | $E_{\gamma}$(E2) keV | B(M1)/B(E2) $(\mu_N/eb)^2$ | Reference | **Configurations and Comments:** |
|---|---|---|---|---|---|---|---|
| 1. | 3908.5 | $10^{+}$ | | | | **1997Pe25** | 1. Tentatively assigned as $\pi(g_{9/2}^{-2}) \otimes \nu(g_{7/2}^{1}d_{5/2}^{3})$ from the shell model calculations.<br>2. The assignment of 368.5 keV transition to the band is from 2000JeZX.<br>3. The B(M1) values for the transitions from 367 keV to 617 keV as given in 2001Li24 are 0.18(3), 0.87(8), 0.16(4) and >0.06 W.u., respectively.<br>4. The mean lifetimes (in ps) of levels from 4277 to 5926 keV as given in 2001Li24 are 1.5(2), 2.5(2), 0.4(1) and 2.2(2), respectively.<br>5. Irregular band. |
| | 4277.0 | $11^{(+)}$ | 368.5 | | | 2001Li24 | |
| | 4518.2 | $12^{(+)}$ | 241.1 | | | 2000JeZX | |
| | 5308.7 | $13^{(+)}$ | 790.5 | | | | |
| | 5926.1 | $14^{(+)}$ | 617.4 | | | | |
| | 6773.1 | $15^{(+)}$ | 847.1 | | | | |
| | 7331.9 | $16^{(+)}$ | 558.81 | 1405.4 | | | |
| | 8367.3 | $17^{(+)}$ | 1035.4 | | | | |

## $^{104}_{48}\mathrm{Cd}_{56}$

| | $E_{level}$ keV | $I^{\pi}$ | $E_{\gamma}$(M1) keV | $E_{\gamma}$(E2) keV | B(M1)/B(E2) $(\mu_N/eb)^2$ | Reference | **Configurations and Comments:** |
|---|---|---|---|---|---|---|---|
| 1. | 4102.4 | $10^{+}$ | | | | 2000JeZX | 1. Probable configuration might arise due to the coupling of two $g_{9/2}$ protons holes with $d_{5/2}$, $g_{7/2}$ neutrons.<br>2. Tentatively assigned as MR band.<br>3. Band exhibits large signature splitting. |
| | 4738.0 | $11^{+}$ | 635.4 | | | | |
| | 5077.8 | $12^{+}$ | 341.2 | 974.5 | | | |
| | 5794.7 | $13^{+}$ | 716.9 | 1056.6 | | | |
| | 6242.7 | $14^{+}$ | 448.0 | (1165) | | | |
| | 7150.4 | $(15^{+})$ | 907.7 | (1356) | | | |

## $^{107}_{48}\mathrm{Cd}_{59}$

| | $E_{level}$ keV | $I^{\pi}$ | $E_{\gamma}$(M1) keV | $E_{\gamma}$(E2) keV | B(M1)/B(E2) $(\mu_N/eb)^2$ | References | **Configurations and Comments:** |
|---|---|---|---|---|---|---|---|
| 1. | 5031.0 | $29/2^{-}$ | | | | **2015Ch05** | 1. Configuration assigned as $\pi(g_{9/2})^{-2}\otimes\nu(h_{11/2})\,(g_{7/2})^{2}$ before and $\pi(g_{9/2})^{-2}\otimes\nu(h_{11/2})^{3}$ after band crossing from TAC calculations.<br>2. $(\beta_2, \gamma)$ =(-0.118, 48°) and (0.14, 0°) before and after band crossing from TAC calculations<br>3. The mean lifetimes (in ps) of the levels from 5502.2 keV to 7790.7 keV are 0.671(91), 0.378 (44), 0.231(27) ,0.194(29) and 0.135 (20), respectively.<br>4. The B(M1) values for the transition from 285.7 keV to724.3 keV are 3.624(491), 2.638(307), 2.084(244), 0.898(134) and 1.112(164) $\mu_N^2$, respectively.<br>5. Regular band with backbending at 41/2 ħ. |
| | 5216.5 | $31/2^{-}$ | 185.5 | | | | |
| | 5502.2 | $33/2^{-}$ | 285.7 | | | | |
| | 5886.8 | $35/2^{-}$ | 384.6 | | | | |
| | 6377.6 | $37/2^{-}$ | 490.8 | | | | |
| | 7066.4 | $39/2^{-}$ | 688.8 | | | | |
| | 7790.7 | $41/2^{-}$ | 724.3 | | | | |
| | 8493.1 | $43/2^{-}$ | 702.4 | | | | |
| | 9191.6 | $45/2^{(-)}$ | 698.5 | | | | |
| | 9924.0 | $47/2^{(-)}$ | 732.4 | | | | |
| | 10698.3 | $49/2^{(-)}$ | 774.3 | | | | |

## $^{108}_{48}Cd_{60}$

| | $E_{level}$ keV | $I^\pi$ | $E_\gamma$(M1) keV | $E_\gamma$(E2) keV | B(M1)/B(E2) $(\mu_N/eb)^2$ | References | Configurations and Comments: |
|---|---|---|---|---|---|---|---|
| 1. | (5591.4) | $(11^-)$ | | | | **2000Ke01** | 1. $\pi(g_{9/2}^{-3}g_{7/2}) \otimes \nu[h_{11/2}(g_{7/2}d_{5/2})^1]$ before and $\pi(g_{9/2}^{-3}g_{7/2}) \otimes \nu(h_{11/2}^3(g_{7/2}d_{5/2})^1]$ after the band crossing from TAC calculations. |
| | 5642.4 | $12^-$ | (51) | | | 2010Ro15 | 2. Small prolate deformation $(\beta_2,\gamma) \sim (0.14, -125°)$. from TAC calculations. |
| | 5763.4 | $13^-$ | 121.0 | | | 1993Th05 | 3. Lower limits on B(M1)/B(E2) are from 1993Th05 from the unobserved $\Delta I= 2$ (E2) transitions. |
| | 6079.4 | $14^-$ | 316.0 | | >25 | 1994Th01 | 5. The mean lifetimes (in ps) of the levels from 6079.4 keV to 9882.4 keV as given in 2010Ro15 are 0.95(4), 0.39(4), 0.25(7), 0.47(6), 0.65(6), 0.30(6), 0.23(7) and 0.80(7), respectively. The uncertainties of 10% from stopping power are not included. |
| | 6601.1 | $15^-$ | 521.7 | | >118 | | 6. B(M1) values of the transition from 316.0 keV to 705.6 keV as given in 2010Ro15 are 1.86(10), 1.00(12), 0.72(15), 0.64(12), 1.81(25), 1.42(25), 1.01(20) $\mu_N^2$, respectively. |
| | 7277.8 | $16^-$ | 676.7 | | >164 | | 7. Regular band with backbending at 17 ħ. |
| | 7743.3 | $17^-$ | 465.5 | | >218 | | |
| | 8105.0 | $18^-$ | 361.7 | | >91 | | |
| | 8587.3 | $19^-$ | 482.3 | 845.0 | 21(+49-8) | | |
| | 9176.8 | $20^-$ | 589.5 | 1073.9 | 18(+27-5) | | |
| | 9882.4 | $21^-$ | 705.6 | 1293.6 | 20(+16-5) | | |
| | 10680.3 | $(22^-)$ | 797.9 | 1502.2 | | | |
| 2. | 7216.1 | $(15^-)$ | | | | **2000Ke01** | 1. Tentatively assigned as $\pi(g_{9/2}^{-3}g_{7/2}) \otimes \nu[h_{11/2}(g_{7/2}d_{5/2})^3]$. |
| | 7530.1 | $16^-$ | (314.0) | | | | 2. Regular band with backbending at $19^-$. |
| | 7863.1 | $17^-$ | 333.0 | | | | |
| | 8318.5 | $18^-$ | 455.4 | | | | |
| | 8641.9 | $19^-$ | 323.4 | | | | |
| | 9000.7 | $(20^-)$ | 358.8 | 682.0 | | | |
| | 9421.5 | $(21^-)$ | 420.8 | 780.4 | | | |
| | 9898.1 | $(22^-)$ | 476.6 | 897.8 | | | |
| | 10413.7 | $(23^-)$ | 515.6 | 993.6 | | | |
| | 10977.3 | $(24^-)$ | 563.6 | 1079.0 | | | |

## $^{109}_{48}Cd_{61}$

| | $E_{level}$ keV | $I^\pi$ | $E_\gamma$(M1) keV | $E_\gamma$(E2) keV | B(M1)/B(E2) $(\mu_N/eb)^2$ | Reference | Configurations and Comments: |
|---|---|---|---|---|---|---|---|
| 1. | 3353.8 | $21/2^-$ | | | | **1994Ju05** | 1. $\pi(g_{9/2})^{-2} \otimes \nu(h_{11/2})$ and $\pi(g_{9/2})^{-2} \otimes \nu[h_{11/2}(g_{7/2}d_{5/2})^2]$ before and after the backbending respectively from the TAC calculations (2000Ch04). |
| | 3548.8 | $23/2^-$ | 195.0 | | | 2000Ch04 | 2. $(\beta_2, \gamma) \sim (0.106, 0°)$ before and $(0.085, 12°)$ after the backbending from 2000Ch04. |
| | 4030.5 | $25/2^-$ | 481.7 | | | | 3. B(M1)/B(E2) values range from ~ 40 $(\mu_N/eb)^2$ to ~ 150 $(\mu_N/eb)^2$. |
| | 4630.5 | $27/2^-$ | 600.0 | | | | 4. B(M1) values for the transitions from 290 to 759 keV as given in 2000Ch04 are 1.80(15), 2.56(11), 0.83(7) and 0.39(3) $\mu_N^2$, respectively. |
| | 5279.5 | $29/2^-$ | 649.0 | 1249.0 | | | 5. The mean lifetimes (in ps) of levels from 5731 to 7555 keV as given in 2000Ch04 are 1.40(4), 0.272(5), 0.241(9) and 0.329(14), respectively. |
| | 5441.1 | $31/2^-$ | 161.6 | | | | 6. Regular band with a backbending at 31/2. |
| | 5731.0 | $33/2^-$ | 289.9 | | | | |
| | 6164.3 | $35/2^-$ | 433.3 | | | | |
| | 6795.8 | $37/2^-$ | 631.5 | | | | |
| | 7554.8 | $(39/2^-)$ | 759 | | | | |

## $^{109}_{48}\mathrm{Cd}_{61}$

| | $E_{level}$ keV | $I^{\pi}$ | $E_{\gamma}$(M1) keV | $E_{\gamma}$(E2) keV | B(M1)/B(E2) $(\mu_N/eb)^2$ | Reference | Configurations and Comments: |
|---|---|---|---|---|---|---|---|
| 2. | 5811 | $29/2^+$ | | | | **2000Ch04** | 1. $\pi(g_{9/2})^{-2}\otimes\nu[h_{11/2}^{2}\,(d_{5/2}\,g_{7/2})^{1}]$ from TAC |
| | 6002 | $31/2^+$ | 191 | | | 1994Ju05 | calculations. |
| | 6303 | $33/2^+$ | 300.9 | | | | 2. $(\beta_2, \gamma) \sim (0.116, 10°)$. |
| | 6681 | $35/2^+$ | 378.7 | | | | 3. B(M1)/B(E2) values ≥ 20 $(\mu_N/eb)^2$ for the two |
| | 7144 | $37/2^+$ | 462.6 | | | | levels at 33/2 and 35/2 as given in 1994Ju05. |
| | 7684 | $39/2^+$ | 540.1 | | | | 4. B(M1) values for the transitions from 301 to |
| | 8261 | $41/2^+$ | 577.3 | | | | 577 keV are 4.45(29), 4.19(14), 2.76(4), |
| | 8868 | $43/2^+$ | 606 | | | | 3.15(+32-24) and 3.69(31) $\mu_N^2$ respectively. |
| | 9500 | $(45/2^+)$ | 632 | | | | 5. The mean lifetimes (in ps) of levels from 6303 to |
| | 10163 | $(47/2^+)$ | 663 | | | | 8261 keV are 0.367(15), 0.253(5), 0.210(5), |
| | 10895 | $(49/2^+)$ | 732 | | | | 0.115(4) and 0.084(4), respectively. |

## $^{110}_{48}\mathrm{Cd}_{62}$

| | $E_{level}$ keV | $I^{\pi}$ | $E_{\gamma}$(M1) keV | $E_{\gamma}$(E2) keV | B(M1)/B(E2) $(\mu_N/eb)^2$ | References | Configurations and Comments: |
|---|---|---|---|---|---|---|---|
| 1. | 8015.8 | 17 | | | | **1994Ju04** | 1. Configuration assigned as $\pi(g_{9/2}^{-2})\otimes\nu(h_{11/2}^{2}\,g_{7/2}^{-2}$ |
| | 8277.0 | 18 | 261.2 | | | 1999Cl03 | $d_{5/2}^{-2})$ for lower and $\pi(g_{9/2}^{-2})\otimes\nu(h_{11/2}^{2}\,g_{7/2}^{-1}\,d_{5/2}^{-3})$ |
| | 8594.6 | 19 | 317.6 | | | 2001Cl02 | for higher spin from TAC-RMF calculations |
| | 8966.9 | 20 | 372.3 | | | | (2015Pe06). |
| | 9429.4 | 21 | 462.5 | | >48 | | 2. Assigned as MR band by 1999Cl03. |
| | 9990.4 | 22 | 561 | | >63 | | 3. Prolate deformation. |
| | 10664.2 | 23 | 673.8 | | >60 | | 4. Lower limits on B(M1)/B(E2) values are from |
| | 11450.2 | 24 | 786 | | | | the unobserved $\Delta I= 2$ (E2) transitions. |
| | | | | | | | 5. Regular band. |
| | | | | | | | 6. The mean lifetimes (in ps) of levels with spins |
| | | | | | | | from 20 to 23, as given in 1999Cl03 are |
| | | | | | | | 0.184(+18-22), 0.101(+15-18), 0.094(+14-18) and |
| | | | | | | | 0.092(+17-23), respectively. |
| | | | | | | | 7. B(M1) values for transitions from 372.3 to |
| | | | | | | | 673.8 keV, as given in 1999Cl03 are |
| | | | | | | | 5.40(+65-53), 5.13(+90-75), 3.06(+57-45) |
| | | | | | | | and 1.83(+46-34) $\mu_N^2$, respectively. |

## $^{107}_{49}\mathrm{In}_{58}$

| | $E_{level}$ keV | $I^{\pi}$ | $E_{\gamma}$(M1) keV | $E_{\gamma}$(E2) keV | B(M1)/B(E2) $(\mu_N/eb)^2$ | References | Configurations and Comments: |
|---|---|---|---|---|---|---|---|
| 1. | 3282.8 | $19/2^-$ | | | | **2010Ne05** | 1. Configuration assigned as $\pi(g_{9/2})^{-1}\otimes\nu(h_{11/2}$ |
| | 3441.5 | $21/2^-$ | 158.7 | | | 2010Si14 | $(d_{5/2}/g_{7/2}))$ before the alignment and $\pi(g_{9/2})^{-1}\otimes$ |
| | 3645.4 | $23/2^-$ | 203.9 | | | 2010Id01 | $\nu(h_{11/2}(d_{5/2}/g_{7/2})^{3})$ after the alignment. |
| | 4038.6 | $25/2^-$ | 393.2 | | | 1998Ta26 | 2. $(\varepsilon_2, \gamma)$ =(0.12,15°) and (0.13, 10°) before and after |
| | 4650.2 | $27/2^-$ | 611.6 | 1004.8 | 76.7(+133 -180) | | alignment, respectively from TAC calculations |
| | 5182.4 | $29/2^-$ | 532.2 | 1143.8 | 87.9(+118-153) | | 3. The mean lifetimes (in ps) of the levels from |
| | 5565.2 | $31/2^-$ | 382.8 | 915.0 | 147.8(+367-606) | | 3645.4 keV to 6769.8 keV as given in 2010Ne05 |
| | 6069.3 | $33/2^-$ | 504.1 | | | | are 1.56 (+17- 16), 0.64(+1- 2), 0.72(+2-3), 0.65 |
| | 6769.8 | $35/2^-$ | 700.5 | | | | (2), 0.76 (+4-3) 0.53 (2) and 0.43 (3), respectively. |
| | 7610.8 | $(37/2^-)$ | 841.0 | | | | 4. The B(M1) values for the transition from 203.9 |
| | 8199.3 | $(39/2^-)$ | 588.5 | | | | keV to 700.5 keV are 3.24(+37-33), 1.44(7), |
| | | | | | | | 0.33(+2-1), 0.51(+3-4), 1.02(+9-6), 0.84(3) and |
| | | | | | | | 0.38(+3-2) $\mu_N^2$, respectively |
| | | | | | | | 5. The B(E2) values for the transition from 1004.8 |
| | | | | | | | keV to 915.0 keV are 0.0043(+9-7), 0.0058(+10- |
| | | | | | | | 7) and 0.0069(+28-16) $(eb)^2$, respectively. |
| | | | | | | | 6. Regular band with backbending at 29/2 ħ. |

## $^{108}_{49}\mathrm{In}_{59}$

| | $E_{level}$ keV | $I^{\pi}$ | $E_{\gamma}$(M1) keV | $E_{\gamma}$(E2) keV | B(M1)/B(E2) $(\mu_N/eb)^2$ | Reference | **Configurations and Comments:** |
|---|---|---|---|---|---|---|---|
| 1. | 1119.4 | $8^-$ | | | | **2001Ch71** | 1. $\pi(g_{9/2}^{-1}) \otimes \nu(h_{11/2})$ by comparison with a similar band in $^{109}$Cd. |
| | 1332.5 | $9^-$ | 213.1 | | | 1998Ch35 | |
| | 1861.1 | $10^-$ | 528.6 | | | | 2. Small prolate deformation ($\beta_2$, $\gamma$ = 0.116, 10°) from TAC calculations. |
| | 2465.4 | $11^-$ | 604.3 | 1133.8 | | | |
| | 3006.9 | $12^-$ | 541.5 | 1146.7 | | | 3. Regular band with small backbending at $12^-$. |
| | 3642.4 | $13^-$ | 635.5 | | | | |
| 2. | 2515.4 | $10^-$ | | | | **2001Ch71** | 1. $\pi(g_{9/2}^{-1}) \otimes \nu[(g_{7/2}/d_{5/2})^2\, h_{11/2}]$ by comparison with a similar band in $^{109}$Cd . |
| | 2662.4 | $11^-$ | 147.0 | | | 1998Ch35 | |
| | 2815.9 | $12^-$ | 153.5 | | | | 2. Small prolate deformation ($\beta_2$, $\gamma$ = 0.095, 15°) from TAC calculations. |
| | 3046.7 | $13^-$ | 230.8 | | | | |
| | 3382.1 | $14^-$ | 335.4 | | | | 3. B(M1) values for the transitions from 335.4 to 660.9 keV are 0.91(3), 0.60(+13-8) and 0.40(+5-2) $\mu_N^2$, respectively. |
| | 3909.9 | $15^-$ | 527.8 | 864.2 | 35(+17-10) | | |
| | 4570.8 | $16^-$ | 660.9 | 1189.1 | 50(+13-8) | | |
| | 5155.9 | $17^-$ | 585.1 | 1246.9 | | | 4. The mean lifetimes (in ps) of levels from 3382 to 4571 keV are 1.63(6), 0.60(10) and 0.45(+2-5), respectively. |
| | | | | | | | 5. Regular band with small backbending at $17^-$ |
| 3. | 4331.2 | $13^+$ | | | | **2001Ch71** | 1. $\pi(g_{9/2}^{-1}) \otimes \nu[(g_{7/2}/d_{5/2})\, h_{11/2}^2]$ before and $\pi(g_{9/2}^{-1}) \otimes \nu[(g_{7/2}/d_{5/2})^3\, h_{11/2}^2]$ after the backbending by comparison with $^{109}$Cd. |
| | 4517.4 | $14^+$ | 186.2 | | | 1998Ch35 | |
| | 4773.3 | $15^+$ | 255.9 | | | | |
| | 5130.6 | $16^+$ | 357.3 | | | | 2. Small prolate deformation ($\beta_2$, $\gamma$ = 0.126, 10°) for the configuration before and ($\beta_2$, $\gamma$ = 0.063, 15°) for the configuration after backbending. |
| | 5603.6 | $17^+$ | 473.0 | | | | |
| | 6168.0 | $18^+$ | 564.4 | 1038.8 | 29(+6-5) | | |
| | 6710.4 | $19^+$ | 542.4 | 1107.7 | 31(+25-14) | | 3. B(M1) values for the transitions from 357.3 to 542.4 keV are 2.48(+22-20), 2.38(+17-16), 1.71(+13-11) and 5.2(+26-8) $\mu_N^2$, respectively. |
| | 7234.4 | $(20^+)$ | 524.0 | | | | |
| | 7830.4 | $(21^+)$ | 596.0 | | | | |
| | 8570.7 | $(22^+)$ | 740.3 | | | | 4. The mean lifetimes (in ps) of levels from 5130 to 6710 keV are 0.43(4), 0.22(2), 0.158(11) and 0.055(+10-18), respectively. |
| | | | | | | | 5. Regular band with small backbending at $18^+$. |

## $^{109}_{49}\mathrm{In}_{60}$

| | $E_{level}$ keV | $I^{\pi}$ | $E_{\gamma}$(M1) keV | $E_{\gamma}$(E2) keV | B(M1)/B(E2) $(\mu_N/eb)^2$ | Reference | **Configurations and Comments:** |
|---|---|---|---|---|---|---|---|
| 1. | 3092.0 | $19/2^-$ | | | | **2012Ne03** | 1. Configuration assigned as $\pi(g_{9/2}^{-1}) \otimes \nu[h_{11/2}\,(g_{7/2}/d_{5/2})^1]$ before alignment and $\pi(g_{9/2}^{-1}) \otimes \nu[h_{11/2}\,(g_{7/2}/d_{5/2})^3]$ after alignment from TAC calculations. |
| | 3202.4 | $21/2^-$ | 110.4 | | | 1997Ko51 | |
| | 3410.4 | $23/2^-$ | 208.0 | | | 1997VaZS | |
| | 3800.5 | $25/2^-$ | 390.1 | | | | |
| | 4508.6 | $27/2^{(-)}$ | 708.1 | | | | 2. Small prolate deformation ($\varepsilon_2$, $\gamma$ = 0.08, 9°) before alignment from TAC calculations. |
| | 4833.0 | $29/2^{(-)}$ | 324.4 | | | | |
| | 5242.0 | $31/2^{(-)}$ | 409.0 | | | | 3. Regular band with backbending at $27/2^-$. |
| | 5796.7 | $33/2^{(-)}$ | 554.7 | | | | |
| | 6386 | | 589.0 | | | | |

## $^{110}_{49}\mathrm{In}_{61}$

| | $E_{level}$ keV | $I^{\pi}$ | $E_{\gamma}$(M1) keV | $E_{\gamma}$(E2) keV | B(M1)/B(E2) $(\mu_N/eb)^2$ | Reference | **Configurations and Comments:** |
|---|---|---|---|---|---|---|---|
| 1. | 799.7 | $7^-$ | | | | **2001Ch71** | 1. $\pi(g_{9/2}^{-1}) \otimes \nu(h_{11/2})$ by comparison with a similar band in $^{109}$Cd.<br>2. Small prolate deformation ($\varepsilon_2$, $\gamma$ = 0.11, 10°) from TAC calculations.<br>3. Regular band. |
| | 807.7 | $8^-$ | 8.0 | | | | |
| | 1017.2 | $9^-$ | 209.5 | | | | |
| | 1560.9 | $10^-$ | 543.7 | 753.6 | | | |
| | 2173.9 | $11^-$ | 613.0 | 1157.0 | | | |
| 2. | 2596.8 | $12^-$ | | | | **2001Ch71** | 1. $\pi(g_{9/2}^{-1}) \otimes \nu[(g_{7/2}/d_{5/2})^2\ h_{11/2}]$ by comparison $^{109}$Cd.<br>2. Small prolate deformation ($\varepsilon_2$, $\gamma$ = 0.08, 10°) from TAC calculations.<br>3. B(M1) values for the transitions 354.6 and 521.1 keV are 1.01(3) and 0.56(4) $\mu_N^2$, respectively.<br>4. The mean lifetimes (in ps) of levels from 3193 and 3714 keV are 1.25(4) and 0.72(+5-4), respectively.<br>5. Regular band with small backbending at $17^-$ |
| | 2837.9 | $13^-$ | 241.1 | | | | |
| | 3192.5 | $14^-$ | 354.6 | | | | |
| | 3713.6 | $15^-$ | 521.1 | | | | |
| | 4528.6 | $16^-$ | 815.0 | | | | |
| | 5265.4 | $17^-$ | 736.8 | 1552.0 | | | |
| 3. | 3326.9 | $11^+$ | | | | **2001Ch71** | 1. $\pi(g_{9/2}^{-1}) \otimes \nu[(g_{7/2}/d_{5/2})\ h_{11/2}^2]$ before and $\pi(g_{9/2}^{-1}) \otimes \nu[(g_{7/2}/d_{5/2})^3\ h_{11/2}^2]$ after the backbending by comparison in $^{109}$Cd.<br>2. Small prolate deformation ($\varepsilon_2$, $\gamma$ = 0.11, 10°) for the configuration before and ($\varepsilon_2$, $\gamma$ = 0.08, 20°) for the configuration after backbending from TAC calculations.<br>3. B(M1) values for the transitions from 285.4 to 572.9 keV are 3.73(14), 2.61(+10-5), 1.97(+25-26), 1.35(+15-13) and 2.51(+36-20) $\mu_N^2$, respectively.<br>4. The mean lifetimes (in ps) of levels from 4229 to 6224 keV are 0.627(+24-23), 0.428(+8-16), 0.230(+35-26),0.196(+21-19) and 0.105(+9-13), respectively.<br>5. Regular band with small backbending at $19^+$. |
| | 3512.5 | $12^+$ | 185.6 | | | | |
| | 3720.0 | $13^+$ | 207.5 | | | | |
| | 3943.8 | $14^+$ | 223.8 | | | | |
| | 4229.2 | $15^+$ | 285.4 | | | | |
| | 4598.2 | $16^+$ | 369.0 | | | | |
| | 5085.1 | $(17^+)$ | 486.9 | 855.9 | 34(+10-8) | | |
| | 5650.7 | $(18^+)$ | 565.6 | 1052.3 | 47(+11-10) | | |
| | 6223.6 | $(19^+)$ | 572.9 | 1138.6 | 49(+13-10) | | |
| | 6707.5 | $(20^+)$ | 483.9 | | | | |
| | 7272.9 | $(21^+)$ | 565.4 | | | | |
| | 7981.1 | $(22^+)$ | 708.2 | | | | |
| | 8748.0 | $(23^+)$ | 766.9 | | | | |

## $^{111}_{49}\mathrm{In}_{62}$

| | $E_{level}$ keV | $I^{\pi}$ | $E_{\gamma}$(M1) keV | $E_{\gamma}$(E2) keV | B(M1)/B(E2) $(\mu_N/eb)^2$ | Reference | **Configurations and Comments:** |
|---|---|---|---|---|---|---|---|
| 1. | 3461.0 | $19/2^+$ | | | | **1998Va03** | 1. $\pi(g_{9/2}^{-1}) \otimes \nu(h_{11/2}^2)$ by comparison with $^{110}$Cd.<br>2. Small prolate deformation.<br>3. B(M1)/B(E2) > 50-100 $(\mu_N/eb)^2$ from unobserved $\Delta I$= 2 (E2) transitions.<br>4. The mean lifetimes (in ps) of the levels from 4282.6 keV to 5877.1 keV are 0.66 (+21-14), 0.30 (+10-7),0.65(+23-16) and 0.73 (+25-16), respectively (2011Ba05).<br>5. B(M1) values for the transition from 371.3 keV to 546.4 keV are 1.62 (+43-40),1.30(+50-30), 0.50(+16-13) and 0.35(10) $\mu_N^2$, respectively (2011Ba05).<br>6. B(E2) values for the transition from 884.3 keV to 1081.3 keV as given in 2011Ba05 are 6.4 (+19-16), 4.0(+13-10) and 5.2(+15-13) (W.u), respectively.<br>7. Regular band with small backbending at 23/2 ħ. |
| | 3588.4 | $21/2^+$ | 127.4 | | | 2011Ba05 | |
| | 3707.2 | $23/2^+$ | 118.8 | | | | |
| | 3911.3 | $25/2^+$ | 204.1 | | | | |
| | 4282.6 | $27/2^+$ | 371.3 | | | | |
| | 4795.8 | $29/2^+$ | 513.2 | 884.3 | 70(11) | | |
| | 5330.7 | $31/2^+$ | 534.9 | 1048.4 | 34(3) | | |
| | 5877.1 | $(33/2^+)$ | 546.4 | 1081.3 | 20(2) | | |

## $^{111}_{49}In_{62}$

| | $E_{level}$ keV | $I^{\pi}$ | $E_{\gamma}(M1)$ keV | $E_{\gamma}(E2)$ keV | B(M1)/B(E2) $(\mu_N/eb)^2$ | **Reference** | **Configurations and Comments:** |
|---|---|---|---|---|---|---|---|
| 2. | 4932.0 | $27/2^+$ | | | | **1998Va03** | 1. $\pi(g_{9/2}^{-1}) \otimes \nu(h_{11/2}^2 g_{7/2}^2)$ by comparison with a similar band in $^{110}Cd$ . |
| | 5166.8 | $29/2^+$ | 234.8 | | | 2011Ba05 | 2. Small prolate deformation. |
| | 5398.8 | $31/2^+$ | 232.0 | | | | 3. B(M1)/B(E2) > 50-100 $(\mu_N/eb)^2$ from unobserved $\Delta I$= 2 (E2) transitions. |
| | 5678.1 | $33/2^+$ | 279.3 | | | | 4. The mean lifetimes (in ps) of the levels from 5398.8 keV to 7917.1 keV are 1.60(+25-17), 0.54(+7-5),0.38(+9-7), 0.30(+7- 5), 0.28(7) and <0.20, respectively (2011Ba05) |
| | 6051.0 | $35/2^+$ | 372.9 | | | | 5. B(M1) values for the transition from 232.0 keV to 741.9 keV are2.74(+32-37),4.64(+47-53), 2.71(+61-52),1.63(+33-31), 0.79(+26-16) and >0.70 $\mu_N^2$, respectively (2011Ba05). |
| | 6538.1 | $(37/2^+)$ | 487.1 | | | | 6. Regular band with small backbending at $31/2^+$. |
| | 7175.2 | $(39/2^+)$ | 637.1 | | | | |
| | 7917.1 | $(41/2^+)$ | 741.9 | | | | |
| | 8681.1 | $(43/2^+)$ | 764.0 | | | | |
| 3. | X | $(31/2^-)$ | | | | **1998Va03** | 1. Tentatively assigned as $\pi(g_{9/2}^{-1}) \otimes \nu(h_{11/2} g_{7/2} d_{5/2})$ (configuration of a band in $^{110}Cd$) coupled to an aligned $g_{7/2}$ or $h_{11/2}$ neutron pair. |
| | 390.5+X | $(33/2^-)$ | 390.5 | | | | 2. $I^{\pi}$ and level energies are lower limits as estimated from intensity and feeding considerations, X~ 5500 keV. |
| | 794.7+X | $(35/2^-)$ | 404.2 | | | | 3. B(M1)/B(E2) > 50-100 $(\mu_N/eb)^2$ from unobserved $\Delta I$= 2 (E2) transitions. |
| | 1244.3+X | $(37/2^-)$ | 449.6 | | | | 4. Regular band. |
| | 1774.1+X | $(39/2^-)$ | 529.8 | | | | |
| | 2354.6+X | $(41/2^-)$ | 580.5 | | | | |

## $^{112}_{49}In_{63}$

| | $E_{level}$ keV | $I^{\pi}$ | $E_{\gamma}(M1)$ keV | $E_{\gamma}(E2)$ keV | B(M1)/B(E2) $(\mu_N/eb)^2$ | Reference | **Configurations and Comments:** |
|---|---|---|---|---|---|---|---|
| 1. | 3062.7 | $12^+$ | | | | **2012Tr01** | 1. Configuration assigned as $\pi(g_{9/2})^{-1} \otimes \nu[(h_{11/2})^2 (g_{7/2})]$ is based on similar configuration observed in $^{110}In$ (2011He04). But different configuration $\pi(g_{9/2}) \otimes \nu[(h_{11/2})^2 (d_{5/2}/g_{7/2})]$ is proposed on the basis of TAC calculations (2010He09, 2012Tr11). |
| | 3191.0 | $13^+$ | 128.3 | | | 2011He04 | 2. Tentatively assigned as MR band. |
| | 3369.5 | $14^+$ | 178.5 | | | 2010He09 | 3. The mean lifetimes (in ps) of the levels from 3642 to 5297 keV are 0.83(4), 0.49(3), 0.22(+2-3), and <0.25, respectively (2012Tr01). |
| | 3642.2 | $15^+$ | 272.7 | | | 2009Li66 | 4. B(M1) values for the transition from 272.7 keV to 707.6 keV are 3.28(14), 1.89(+11-10), 1.52 (+20-14) and >0.63 $\mu_N^2$, respectively (2012Tr01). |
| | 4035.5 | $16^+$ | 393.3 | | | | 5. Small prolate deformation $(\beta_2, \gamma) = (0.1, 25^o)$ from PRM calculations. |
| | 4589.7 | $17^+$ | 554.2 | | | | 6. Regular band |
| | 5297.3 | $18^+$ | 707.6 | | | | |
| | 6035.3 | $(19^+)$ | 738 | | | | |

## $^{113}_{49}\mathrm{In}_{64}$

| | $E_{level}$ keV | $I^{\pi}$ | $E_{\gamma}$(M1) keV | $E_{\gamma}$(E2) keV | B(M1)/B(E2) $(\mu_N/eb)^2$ | Reference | **Configurations and Comments:** |
|---|---|---|---|---|---|---|---|
| 1. | 2233.2 | $15/2^{(-)}$ | | | | **1997Ch01**<br>2005Na37 | 1. Tentatively assigned as $\pi(g_{9/2}^{-1})\otimes\nu(g_{7/2}^{-1}h_{11/2})$.<br>2. Small prolate deformation ($\beta_2$=0.09).<br>3. The spin and parity assignments are from 2005Na37.<br>4. Parity assignment is based on comparison with neighboring nuclei.<br>5. Irregular band.<br>6. Fully aligned configuration gives rise to $I^{\pi}$= $27/2^-$ ; $I^{\pi}$ beyond this value is attributed to some collectivity. |
| | 2396.4 | $17/2^{(-)}$ | 163.2 | | | | |
| | 2663.9 | $19/2^{(-)}$ | 267.5 | | | | |
| | 2853.6 | $21/2^{(-)}$ | 189.7 | | | | |
| | 3023.1 | $23/2^{(-)}$ | 169.5 | | | | |
| | 3280.0 | $25/2^{(-)}$ | 256.9 | | | | |
| | 3972.6 | $27/2^{(-)}$ | 692.6 | | | | |
| | 4715.0 | $29/2^{(-)}$ | 742.4 | 1434.9 | 8(2) | | |
| | 5392.7 | $31/2^{(-)}$ | 677.7 | 1418.6 | 17(6) | | |
| 2. | 3122.1 | $21/2^{(+)}$ | | | | **1997Ch01**<br>2005Na37 | 1. Tentative Configuration $\pi(g_{9/2}^{-1})\otimes\nu(h_{11/2}^{2})$.<br>2. Small prolate deformation ($\beta_2$=0.09)<br>3. The 558 keV transition is from 2005Na37.<br>4. The spin and parity assignments are from 2005Na37.<br>5. Parity assignment is based on comparison with neighboring nuclei.<br>6. Regular band. |
| | 3213.9 | $23/2^{(+)}$ | 91.8 | | | | |
| | 3397.2 | $25/2^{(+)}$ | 183.3 | | | | |
| | 3788.1 | $27/2^{(+)}$ | 390.9 | | | | |
| | 4377.5 | $29/2^{(+)}$ | 589.4 | 980.2 | 24(6) | | |
| | 5062.1 | $31/2^{(+)}$ | 684.6 | 1274.2 | 71(14) | | |
| | 5790.3 | $33/2^{(+)}$ | 728.2 | | | | |
| | 6348.3 | $35/2^{(+)}$ | 558 | | | | |
| 3. | 4563.0 | $(27/2^+)$ | | | | **2012Ma27** | 1. Configuration assigned as $\pi(g_{9/2}^{-1})\otimes\nu(h_{11/2}^{2}, g_{7/2}^{2})$. On the bases of TAC-RMF calculations.<br>2. ($\beta_2, \gamma \approx$ 0.15, 30°) and with increasing rotational frequency deformation parameter decreases smoothly.<br>3. Regular band |
| | 4809.0 | $(29/2^+)$ | 246.1 | | | | |
| | 5106.9 | $(31/2^+)$ | 298.1 | | | | |
| | 5380.0 | $(33/2^+)$ | 273.1 | | | | |
| | 5781.1 | $(35/2^+)$ | 401.1 | | | | |
| | 6323.2 | $(37/2^+)$ | 542.1 | | | | |
| | 6980.7 | $(39/2^+)$ | 657.5 | | | | |

## $^{114}_{49}\mathrm{In}_{65}$

| | $E_{level}$ keV | $I^{\pi}$ | $E_{\gamma}$(M1) keV | $E_{\gamma}$(E2) keV | B(M1)/B(E2) $(\mu_N/eb)^2$ | References | **Configurations and Comments:** |
|---|---|---|---|---|---|---|---|
| 1. | 2532 | $(11^+)$ | | | | **2012Li38**<br>2011Li43 | 1. Configuration assigned as $\pi(g_{9/2})^{-1}\otimes\nu[(g_{7/2}/d_{5/2})(h_{11/2})^2]$ from TAC-RMF calculations.<br>2. Regular band with backbending at 15 ħ. |
| | 2680 | $(12^+)$ | 148.4 | | | | |
| | 2931 | $(13^+)$ | 250.7 | | | | |
| | 3299 | $(14^+)$ | 368.3 | | | | |
| | 3792 | $(15^+)$ | 492.6 | | | | |
| | 4256 | $(16^+)$ | 464.2 | | | | |

## $^{115}_{49}\mathrm{In}_{66}$

| | $E_{level}$ keV | $I^{\pi}$ | $E_{\gamma}$(M1) keV | $E_{\gamma}$(E2) keV | B(M1)/B(E2) $(\mu_N/eb)^2$ | References | **Configurations and Comments:** |
|---|---|---|---|---|---|---|---|
| 1. | 2876.9 | $(21/2^+)$ | | | | **2015Ch24** | 1. Configuration assigned as $\pi(g_{9/2})^{-1}\otimes\nu(h_{11/2})^{2}$ from TAC-CDFT calculations.<br>2. Tentatively assigned as of mixed character of MR and tilted rotational.<br>3. ($\beta_2$, $\gamma$) =(0.15,50°) from TAC-CDFT calculations.<br>4. Regular band |
| | 2958.9 | $(23/2^+)$ | 82.0 | | | | |
| | 3095.4 | $(25/2^+)$ | 136.5 | | | | |
| | 3470.9 | $(27/2^+)$ | 375.5 | | | | |
| | 4040.3 | $(29/2^+)$ | 569.4 | | | | |
| | 4716.3 | $(31/2^+)$ | 676.0 | | | | |

## $^{105}_{50}\mathrm{Sn}_{55}$

| | $E_{level}$ keV | $I^{\pi}$ | $E_{\gamma}$(M1) keV | $E_{\gamma}$(E2) keV | B(M1)/B(E2) $(\mu_N/eb)^2$ | Reference | Configurations and Comments: |
|---|---|---|---|---|---|---|---|
| 1. | 7043 | $29/2^+$ | | | | **1997Ga01** | 1. Tentatively assigned as $\pi(g_{9/2}^{-1}g_{7/2})\otimes\nu(h_{11/2}^2 (d_{5/2}g_{7/2})^1)$ from TRS calculations.<br>2. Prolate deformation ($\beta_2$=0.137)<br>3. B(M1)/B(E2)>100 $(\mu_N/eb)^2$ from the unobserved $\Delta$I= 2 (E2) transitions.<br>4. Regular band with a backbending at 37/2. |
| | 7343 | $31/2^+$ | 300 | | | 1999De50 | |
| | 7730 | $33/2^{(+)}$ | 388 | | | | |
| | 8196 | $35/2^{(+)}$ | 466 | | | | |
| | 8682 | $37/2^{(+)}$ | 486 | | | | |
| | 9137 | $39/2^{(+)}$ | 456 | | | | |
| | 9692 | $41/2^{(+)}$ | 555 | | | | |
| | 10287 | $43/2^{(+)}$ | 596 | | | | |

## $^{106}_{50}\mathrm{Sn}_{56}$

| | $E_{level}$ keV | $I^{\pi}$ | $E_{\gamma}$(M1) keV | $E_{\gamma}$(E2) keV | B(M1)/B(E2) $(\mu_N/eb)^2$ | References | Configurations and Comments: |
|---|---|---|---|---|---|---|---|
| 1. | 7598.5 | $14^-$ | | | | **1998Je03** | 1. $\pi(g_{7/2}\, g_{9/2}^{-1})\otimes\nu((g_{7/2}d_{5/2})^3\, h_{11/2})$ from TAC calculations.<br>2. ($\beta_2$, $\gamma$) = (0.11, -13°).<br>3. Level energies, spins and parities are from 1999Je07.<br>4 The mean lifetimes (in ps) of the five uppermost levels are 0.30(3), 0.43(5), 0.51(15), 0.22(2) and 0.22(+1-3), respectively.<br>5. B(M1) values for the transitions from 450 to 599 keV are 2.06(+22-26), 1.12(+15-13), 0.54(+20-13), 0.54(+5-7) and 1.17(17) $\mu_N^2$, respectively.<br>6. Regular band with backbending at the top. |
| | 8013.2 | $15^-$ | 413.4 | | | 1999Je07 | |
| | 8560.3 | $16^-$ | 547.0 | | | 1997Ju01 | |
| | 9103.3 | $17^-$ | 542.9 | | | | |
| | 9552.9 | $18^-$ | 449.7 | | >160 | | |
| | 10040.9 | $19^-$ | 488.1 | | >250 | | |
| | 10632.9 | $20^-$ | 591.9 | | >200 | | |
| | 11413.3 | $21^-$ | 780.4 | | >35 | | |
| | 12047.3 | $22^-$ | 634.0 | | | | |
| 2. | 9236.1 | $17^-$ | | | | **1998Je03** | 1. $\pi(g_{7/2}\, g_{9/2}^{-1})\otimes\nu((g_{7/2}d_{5/2})^3\, h_{11/2})$ from TAC calculations.<br>2. ($\beta_2$, $\gamma$) = (0.11, -13°).<br>3. Regular band. |
| | 9637.8 | $18^-$ | 401.7 | | | | |
| | 10117.0 | $19^-$ | 479.2 | | >155 | | |
| | 10672.4 | $20^-$ | 555.4 | | >290 | | |
| | 11292.7 | $21^-$ | 620.3 | | >220 | | |
| | 11971.5 | $22^-$ | 678.8 | | | | |

## $^{108}_{50}\mathrm{Sn}_{58}$

| | $E_{level}$ keV | $I^{\pi}$ | $E_{\gamma}$(M1) keV | $E_{\gamma}$(E2) keV | B(M1)/B(E2) $(\mu_N/eb)^2$ | References | Configurations and Comments: |
|---|---|---|---|---|---|---|---|
| 1. | 6665 | $12^-$ | | | | **1998Je03** | 1. $\pi(g_{7/2}\, g_{9/2}^{-1})\otimes\nu((g_{7/2}d_{5/2})^1\, h_{11/2})$ from TAC calculations.<br>2. Prolate shape ($\beta_2$, $\gamma$) = (0.08, 0°) from 1999Je07.<br>3. The mean lifetimes (in ps) of levels with spins from 15 to 19 as given in 1999Je07 are 0.66(2), 0.23(1), 0.29(1), 0.44(+5-2) and 0.56(2), respectively.<br>4. B(M1) values for the transitions from 424 to 550 keV are 1.05(3), 1.63(8), 1.16(5), 0.64(+4-8) and 0.48(3) $\mu_N^2$, respectively.<br>5. Regular band. |
| | 6885.0 | $13^-$ | 220.0 | | | 1999Je07 | |
| | 7182.7 | $14^-$ | 297.7 | | | | |
| | 7606.4 | $15^-$ | 423.7 | 720 | 30.0(25) | | |
| | 8116.3 | $16^-$ | 509.9 | 934 | 23.5(40) | | |
| | 8634.5 | $17^-$ | 518.2 | 1028 | 26.0(35) | | |
| | 9169.6 | $18^-$ | 535.1 | 1053 | 19.5(40) | | |
| | 9719.8 | $19^-$ | 550.2 | 1085 | 23(4) | | |
| | 10355.3 | $20^-$ | 635.5 | 1184 | 24(4) | | |

## $^{108}_{50}Sn_{58}$

| | $E_{level}$ keV | $I^\pi$ | $E_\gamma$(M1) keV | $E_\gamma$(E2) keV | B(M1)/B(E2) $(\mu_N/eb)^2$ | **References** | **Configurations and Comments:** |
|---|---|---|---|---|---|---|---|
| 2. | 8103 | $16^-$ | | | | **1998Je03** | 1. $\pi(g_{7/2}\, g_{9/2}^{-1}) \otimes \nu(g_{7/2}^{2}\, (g_{7/2}d_{5/2})^{1}\, h_{11/2})$ from TAC calculations. |
| | 8351.2 | $17^-$ | 248.2 | | | 1999Je07 | 2. Prolate shape $(\beta_2, \gamma) = (0.11, 0°)$ from 1999Je07. |
| | 8695.8 | $18^-$ | 344.6 | (592) | | | |
| | 9105.8 | $19^-$ | 410.0 | 753 | 15.4(40) | | |
| | 9579.4 | $20^-$ | 473.6 | 885 | 14.1(40) | | |
| | 10062.8 | $21^-$ | 483.4 | 956 | 20.2(50) | | |
| | 10572.2 | $22^-$ | 509.4 | 992 | 22.7(70) | | |

## $^{108}_{51}Sb_{57}$

| | $E_{level}$ keV | $I^\pi$ | $E_\gamma$(M1) keV | $E_\gamma$(E2) keV | B(M1)/B(E2) $(\mu_N/eb)^2$ | Reference | **Configurations and Comments:** |
|---|---|---|---|---|---|---|---|
| 1. | 2154.6 | $7^-$ | | | | **1998Je09** | 1. $\pi[(g_{7/2}, d_{5/2})^2 g_{9/2}^{-1}] \otimes \nu(h_{11/2})$ from TAC calculations. |
| | 2246.0 | $8^-$ | 91.4 | | | | 2. $(\beta_2, \gamma) = (0.116, 30°)$ from TAC calculations. |
| | 2438.3 | $9^-$ | 192.3 | 283 | | | 3. B(M1)/B(E2) values range from ~5 $(\mu_N/eb)^2$ to ~20 $(\mu_N/eb)^2$. |
| | 2719.9 | $10^-$ | 281.6 | 474 | | | 4. Regular band. |
| | 3032.4 | $11^-$ | 312.5 | 595 | | | |
| | 3376.8 | $12^-$ | 344.4 | 657 | | | |
| | 3764.7 | $13^-$ | 387.9 | 732 | | | |
| | 4173.6 | $14^-$ | 408.9 | 797 | | | |
| | 4613.3 | $15^-$ | 439.7 | 849 | | | |
| | 5101.9 | $16^-$ | 488.6 | 929 | | | |
| | 5611.5 | $17^-$ | 509.6 | 999 | | | |
| | 6150.0 | $18^-$ | 538.5 | 1049 | | | |
| | 6719.6 | $19^-$ | 569.6 | 1109 | | | |
| 2. | 2753.4 | $10^-$ | | | | **1998Je09** | 1. $\pi(h_{11/2}\, g_{7/2}\, g_{9/2}^{-1}) \otimes \nu(g_{7/2}, d_{5/2})^{1}$ from TAC calculations. |
| | 3057.4 | $11^-$ | 304.0 | | | | 2. $(\beta_2, \gamma) = (0.116, 10°)$ from TAC calculations. |
| | 3376.4 | $12^-$ | 319.0 | 623 | | | 3. B(M1)/B(E2) values range from ~5 $(\mu_N/eb)^2$ to ~25 $(\mu_N/eb)^2$. |
| | 3722.5 | $13^-$ | 346.1 | 665 | | | 4. Regular band with backbending at 15. |
| | 4177.9 | $14^-$ | 455.4 | 801 | | | |
| | 4597.3 | $15^-$ | 419.4 | 874 | | | |
| | 5064.4 | $16^-$ | 467.1 | 886 | | | |
| | 5561.8 | $17^-$ | 497.4 | 964 | | | |
| | 6092.3 | $18^-$ | 530.5 | 1028 | | | |
| | 6645.2 | $19^-$ | 552.9 | 1084 | | | |
| | 7216.3 | $20^-$ | 571.1 | 1124 | | | |

## $^{110}_{51}Sb_{59}$

| | $E_{level}$ keV | $I^\pi$ | $E_\gamma$(M1) keV | $E_\gamma$(E2) keV | B(M1)/B(E2) $(\mu_N/eb)^2$ | Reference | **Configurations and Comments:** |
|---|---|---|---|---|---|---|---|
| 1. | 1921 | $8^-$ | | | | **1997La13** | 1. Tentatively assigned as $\pi(h_{11/2}) \otimes \nu(d_{5/2})$ or $\pi(h_{11/2}) \otimes \nu(g_{7/2})$ by comparison with neighboring odd-odd Sb isotopes. |
| | 2122 | $9^-$ | 201 | | | | 2. Tentatively assigned as MR band. |
| | 2435 | $10^-$ | 313 | 514 | | | 3. Regular band. |
| | 2784 | $11^-$ | 349 | 663 | | | 4. The assignment of this band as MR band is based on the comparison with a band in $^{108}$Sb from 1998Je09. |
| | 3158 | $12^-$ | 374 | 724 | | | |
| | 3556 | $13^-$ | 398 | 772 | | | |
| | 3989 | $14^-$ | 433 | 830 | | | |
| | 4464 | $15^-$ | 475 | 909 | | | |
| | (5016) | | (552) | (1027) | | | |

## $^{112}_{51}Sb_{61}$

| | $E_{level}$ keV | $I^{\pi}$ | $E_\gamma$(M1) keV | $E_\gamma$(E2) keV | B(M1)/B(E2) $(\mu_N/eb)^2$ | Reference | **Configurations and Comments:** |
|---|---|---|---|---|---|---|---|
| 1. | 1675.1 | $7^-$ | | | | **1998La14** | 1. $\pi(g_{9/2}^{-1}) \otimes \nu(h_{11/2})$ from TAC calculations.<br>2. Tentatively assigned as MR band.<br>3. $(\beta_2, \gamma)$ = (0.21, 0°) from TAC calculations.<br>4. The mean lifetimes (in ps) for the states from $11^-$ to $13^-$ as given in 2005De02 are 0.56(+25-26), 0.51(+16-17) and0.50(11), respectively.<br>5. The B(M1) values for the transitions from 354 to 392 keV are 2.28(+69-103), 2.04(+70-63) and 1.90(+38-43) $\mu_N^2$, respectively.<br>6. The B(E2) values for the transitions from 679 to 773 keV as given in 2005De02 are 1.00(+30-45), 0.74(+25-23) and 0.59(+10-11) $(eb)^2$, respectively.<br>7. Regular band. |
| | 1747.5 | $8^-$ | 72.4 | | | 2005De02 | |
| | 1949.7 | $9^-$ | 202.2 | | | | |
| | 2275.2 | $10^-$ | 325.5 | 527.7 | 24(2) | | |
| | 2629.1 | $11^-$ | 353.9 | 679.1 | 16.0(8) | | |
| | 3009.7 | $12^-$ | 380.6 | 734.6 | 14.2(7) | | |
| | 3402.1 | $13^-$ | 392.4 | 773.5 | 13.3(7) | | |
| | 3809.0 | $14^-$ | 406.9 | 799.7 | 6.8(4) | | |
| | 4295.3 | $15^-$ | 486.3 | 893.2 | 10.6(8) | | |
| | 4798.3 | $16^-$ | 503.0 | 989.8 | 9.9(9) | | |
| | 5326.2 | $17^-$ | 527.9 | 1030.8 | | | |
| 2. | X | $(10^+)$ | | | | **1998La14** | 1. $\pi(g_{9/2}^{-1}) \otimes \nu((d_{5/2}, g_{7/2})^1 h_{11/2}^2)$ from TRS calculations and by comparison with similar bands in neighboring isotopes.<br>2. Tentatively assigned as MR band<br>3. Regular band with backbending at 12. |
| | 378.2+X | $(11^+)$ | 378.2 | | | | |
| | 750.8+X | $(12^+)$ | 372.6 | 750.6 | 7.5(5) | | |
| | 1077.6+X | $(13^+)$ | 326.8 | 699.7 | 15.3(12) | | |
| | 1372.5+X | $(14^+)$ | 294.9 | 621.7 | 56(7) | | |
| | 1690.3+X | $(15^+)$ | 317.8 | 613.0 | 200(180) | | |
| | 2046.1+X | $(16^+)$ | 355.8 | 673.9 | 30(3) | | |
| | 2437.7+X | $(17^+)$ | 391.6 | 747.6 | 26(3) | | |
| | 2851.9+X | $(18^+)$ | 414.2 | | | | |
| | 3284.4+X | $(19^+)$ | 432.5 | | | | |

## $^{116}_{51}Sb_{65}$

| | $E_{level}$ keV | $I^{\pi}$ | $E_\gamma$(M1) keV | $E_\gamma$(E2) keV | B(M1)/B(E2) $(\mu_N/eb)^2$ | Reference | **Configurations and Comments:** |
|---|---|---|---|---|---|---|---|
| 1. | 3005.3 | $11^+$ | | | | **2022DA05** | 1. Configuration assigned as $\pi(g_{9/2}^{-1}) \otimes \nu(g_{7/2}/d_{5/2})(h_{11/2})^2$ on the basis of comparison with similar band observed in $^{112}$Sb and further supported by SPAC calculations.<br>2. Tentatively assigned as MR band as dominant contribution is not from shears structure (2012Wa34).<br>3. Regular band<br>4. Level energies are deduced from the gamma energies |
| | 3345.3 | $12^+$ | 340.0 | | | 2012Wa34 | |
| | 3734.5 | $13^+$ | 389.2 | 729.3 | 23.3(27) | | |
| | 4164.6 | $14^+$ | 430.1 | 819.2 | 13.7(16) | | |
| | 4627.6 | $15^+$ | 463.0 | 893.2 | 8.2(10) | | |
| | 5112.8 | $16^+$ | 485.2 | 948.4 | 6.3(8) | | |

## $^{135}_{52}Te_{83}$

| | $E_{level}$ keV | $I^{\pi}$ | $E_\gamma$(M1) keV | $E_\gamma$(E2) keV | B(M1)/B(E2) $(\mu_N/eb)^2$ | Reference | **Configurations and Comments:** |
|---|---|---|---|---|---|---|---|
| 1. | 4023.3 | $(19/2^-)$ | | | | **2001Lu16** | 1. Tentatively assigned as $\pi(g_{7/2}^2) \otimes \nu(f_{7/2}^2 h_{11/2}^{-1})$ as given in 2001Fo02 by comparison with the $^{134}$Te isotope.<br>2. Gamma-ray energies are from Erratum to 2001Lu16 published in PRC 104, 069902(E) (2021).<br>3. Tentatively assigned as MR band (2001Fo02).<br>4. Irregular band. |
| | 4393.9 | $(21/2^-)$ | 370.4 | | | 2001Fo02 | |
| | 4799.4 | $(23/2^-)$ | 405.3 | 776.2 | | | |
| | 5170.7 | $(25/2^-)$ | 371.3 | 776.9 | | | |
| | 5525.4 | $(27/2^-)$ | 355.0 | 726.1 | | | |
| | 5790.7 | $(29/2^-)$ | 265.3 | 620.0 | | | |
| | 6109.8 | $(31/2^-)$ | 319.0 | (585.3) | | | |
| | 6454.9 | $(33/2^-)$ | 344.9 | 664.4 | | | |
| | 6669.5 | $(35/2^-)$ | | 559.7 | | | |

### $^{124}_{54}\mathrm{Xe}_{70}$

| | $E_{level}$ keV | $I^{\pi}$ | $E_{\gamma}$(M1) keV | $E_{\gamma}$(E2) keV | B(M1)/B(E2) $(\mu_N/eb)^2$ | Reference | **Configurations and Comments:** |
|---|---|---|---|---|---|---|---|
| 1. | 5051 | (13) | | | | **1999Sc20** | 1. Tentatively assigned as $\pi(h_{11/2}\otimes (d_{5/2}/g_{7/2})^1)\otimes$ |
| | 5292 | (14) | 241 | | | 2008Al12 | $\nu(h_{11/2}g_{7/2})$. |
| | 5554 | (15) | 262 | 502 | 14.15(+38-31) | 1997Lo12 | 2. Tentatively assigned as MR band. |
| | 5830 | (16) | 276 | 537 | 17.41(+62-49) | 2004Sa47 | 3. $(\beta_2, \gamma)$ = (0.20, 30°) from TAC calculations as. |
| | 6156 | (17) | 326 | 602 | 10.59(+83-46) | | given in 2002Ra34. |
| | 6556 | (18) | 400 | 726 | 14.89(+78-63) | | 4. The B(M1)/B(E2) values are from 2004Sa47. |
| | 6987 | (19) | 431 | 831 | | | 5. The mean lifetimes (in ps) of levels from 5554 to |
| | 7436 | (20) | 449 | 880 | | | 6556 keV as given in 2004Sa47 are 0.89(8), |
| | 7932 | (21) | 496 | 944 | | | 1.84(13), 1.75(8) and 0.40(8), respectively. |
| | 8368 | (22) | 436 | 932 | | | 6. The B(M1) values for the transitions from 262 to |
| | 8914 | (23) | 546 | 982 | | | 400 keV as given in 2004Sa47 are 1.44(+28-24), |
| | 9486 | (24) | 572 | 1118 | | | 1.02(+18-16), 0.75(+11-10) and 1.33(+41-30) |
| | 9929 | (25) | 443 | 1016 | | | $\mu_N^2$, respectively. |
| | | | | | | | 7. Irregular band with backbending at 8368 keV |
| | | | | | | | level and at the top of the band. |

### $^{131}_{55}\mathrm{Cs}_{76}$

| | $E_{level}$ keV | $I^{\pi}$ | $E_{\gamma}$(M1) keV | $E_{\gamma}$(E2) keV | B(M1)/B(E2) $(\mu_N/eb)^2$ | References | **Configurations and Comments:** |
|---|---|---|---|---|---|---|---|
| 1. | 2554.9 | $17/2^+$ | | | | **2005Ku10** | 1. $\pi(d_{5/2}/g_{7/2})\otimes\nu(h_{11/2})^2$ from the decay pattern |
| | 2686.9 | $19/2^+$ | 132.0 | | | 2008Si26 | and TAC calculations. |
| | 2835.1 | $21/2^+$ | 148.2 | | | | 2. Tentatively assigned as MR band. |
| | 3058.7 | $23/2^+$ | 223.6 | | | | 3. Triaxial deformation $(\beta_2, \gamma) \sim (0.11, 46^{o\circ})$ |
| | 3415.3 | $25/2^+$ | 356.6 | | | | 4. Small signature splitting. |
| | 3724.1 | $27/2^+$ | 308.8 | 664.9 | ~10 | | 5. Regular band with backbending at $27/2^+$. |
| | 4145.7 | $29/2^+$ | 421.6 | 730.2 | ~10 | | 6. The calculated B(M1) values decrease with |
| | 4655.3 | $31/2^+$ | 509.6 | 931.2 | ~3 | | frequency. |
| | 5076.9 | $33/2^+$ | 424.7 | 934.0 | | | |
| 2. | 3465.3 | $25/2^-$ | | | | **2005Ku10** | 1. $\pi(h_{11/2})\otimes\nu(h_{11/2})^2$ from the decay pattern |
| | 3621.3 | $27/2^-$ | 156.0 | | | | and TAC calculations. |
| | 4012.2 | $29/2^-$ | 390.9 | | | | 2. Prolate deformation $(\beta_2, \gamma) \sim (0.11, 55^{o\circ})$ |
| | 4387.9 | $31/2^-$ | 375.7 | 766.4 | | | 3. Small signature splitting. |
| | 4905.7 | $33/2^-$ | 517.8 | 893.8 | | | 4. Irregular band. |
| | 5265.6 | $(35/2^-)$ | 359.9 | 877.9 | | | 5. The experimental B(M1)/B(E2) have an average |
| | | | | | | | ~12 $(\mu_N/eb)^2$. |
| | | | | | | | 6. The calculated B(M1) values decrease with |
| | | | | | | | frequency. |

### $^{128}_{56}\mathrm{Ba}_{72}$

| | $E_{level}$ keV | $I^\pi$ | $E_\gamma$(M1) keV | $E_\gamma$(E2) keV | B(M1)/B(E2) $(\mu_N/eb)^2$ | References | **Configurations and Comments:** |
|---|---|---|---|---|---|---|---|
| 1. | 4652 | $12^+$ | | | | **1998Wi20** | 1. $\pi[h_{11/2}(d_{5/2}g_{7/2})]\otimes\nu[h_{11/2}(d_{5/2}g_{7/2})]$ from TAC calculations by 2000Di16.<br>2. Tentatively assigned as MR and collective rotational band (2000Di16).<br>3. Prolate deformation $(\beta_2, \gamma) \sim (0.20, 0^\circ)$.<br>4. B(M1) values for the transitions from 362 to 428 keV as given in 1998Pe17 are 1.14(+21-15), 1.22(+25-18) and 1.41(+30-21) $\mu_N^2$, respectively, for transitions from 305 to 324 keV are 0.32(4), 0.44(+10-7), 1.06(+18-13) and 1.08(+55-27) $\mu_N^2$ (2000Pe20).<br>5. The first three B(M1)/B(E2) ratios have been calculated from the values of B(M1) and B(E2) in 2000Pe20 and the last three ratios have been calculated using the data given in 1998Pe17.<br>6. The mean lifetimes (in ps) of 6215 to 7981 keV states as given in 1998Pe17 are 2.48(7), 1.91(7), 1.54(5), 1.36(11) and 1.16(6) ps, respectively, for levels from 4956 to 5853 keV are 1.44(13), 2.25(40), 1.53(22) and 0.98(33) (2000Pe20).<br>7. Regular band. |
| | 4956 | $13^+$ | 305 | | | 1997Vo12 | |
| | 5233 | $14^+$ | 277 | 582 | 6.29(+29-16) | 2000Di16 | |
| | 5530 | $15^+$ | 296 | 574 | 6.63(+18-13) | 2000Pe20 | |
| | 5853 | $16^+$ | 324 | 619 | 6.00(+56-28) | 1998Pe17 | |
| | 6215 | $17^+$ | 362 | 685 | 5.43(+23-15) | | |
| | 6609 | $18^+$ | 394 | 755 | 5.81(+25-19) | | |
| | 7036 | $19^+$ | 428 | 821 | 5.04(+31-22) | | |
| | 7494 | $20^+$ | 457 | 886 | | | |
| | 7981 | $21^+$ | 487 | 945 | | | |
| | 8497 | $22^+$ | 517 | 1003 | | | |
| | 9032 | $23^+$ | 535 | 1052 | | | |
| | 9601 | $24^+$ | 568 | 1104 | | | |
| | 10168 | $25^+$ | 566 | 1136 | | | |
| | 10785 | $26^+$ | | 1184 | | | |

### $^{130}_{56}\mathrm{Ba}_{74}$

| | $E_{level}$ keV | $I^\pi$ | $E_\gamma$(M1) keV | $E_\gamma$(E2) keV | B(M1)/B(E2) $(\mu_N/eb)^2$ | References | **Configurations and Comments:** |
|---|---|---|---|---|---|---|---|
| 1. | 5960.6 | $15^-$ | | | | 2020Gu21 | 1. Tentatively assigned as $\pi h_{11/2}(g_{7/2}, d_{5/2})\otimes\nu(h_{11/2})^2$ configuration on the basis of comparison of similar bands in neighboring $^{132,134}$Ba and $^{134,136}$Ce nuclides.<br>2. Tentatively assigned as MR band.<br>3. Band exhibits small signature splitting. |
| | 6218.1 | $16^-$ | 257.5 | | | | |
| | 6498.3 | $17^-$ | 280.2 | | | | |
| | 6914.5 | $(18^-)$ | 416.2 | | | | |
| | 7335 | $(19^-)$ | 420 | | | | |
| | 7849 | $(20^-)$ | 514 | | | | |

### $^{132}_{56}\mathrm{Ba}_{76}$

| | $E_{level}$ keV | $I^\pi$ | $E_\gamma$(M1) keV | $E_\gamma$(E2) keV | B(M1)/B(E2) $(\mu_N/eb)^2$ | Reference | **Configurations and Comments:** |
|---|---|---|---|---|---|---|---|
| 1. | 4811.8 | $11^+$ | | | | **1995Ju09** | 1. Tentatively assigned as $\pi(h_{11/2}g_{7/2})\otimes\nu(h_{11/2}d_{3/2})$ by considering the available orbits nearest to the Fermi surface.<br>2. Tentatively assigned as MR band.<br>3. Oblate shape ($\gamma\sim -60°$).<br>4. Regular band. |
| | 4997.2 | $12^+$ | 185.4 | | | 1989Pa17 | |
| | 5201.0 | $13^+$ | 203.9 | | | | |
| | 5436.8 | $14^+$ | 235.9 | | | | |
| | 5771.8 | $15^+$ | 335.0 | | | | |
| | 6196.3 | $16^+$ | 424.6 | | | | |
| | 6665.4 | $(17^+)$ | 469 | | | | |
| | 7144.4 | $(18^+)$ | 479 | | | | |
| 2. | 5721.4 | $14^-$ | | | | **1995Ju09** | 1. $\pi(h_{11/2}g_{7/2})\otimes\nu(h_{11/2}^2)$ by considering the available orbits nearest to the Fermi surface.<br>2. Oblate shape ($\gamma\sim -60°$).<br>3. Regular band. |
| | 5891.3 | $15^-$ | 169.9 | | | 1989Pa17 | |
| | 6107.1 | $16^-$ | 215.7 | | | | |
| | 6414.8 | $17^-$ | 307.7 | | | | |
| | 6821.7 | $18^-$ | 406.9 | | | | |
| | 7287.7 | $(19^-)$ | 466 | (873) | | | |
| | (7751.7) | $(20^-)$ | | (930) | | | |

## $^{134}_{56}Ba_{78}$

| | $E_{level}$ keV | $I^{\pi}$ | $E_{\gamma}$(M1) keV | $E_{\gamma}$(E2) keV | B(M1)/B(E2) $(\mu_N/eb)^2$ | References | **Configurations and Comments:** |
|---|---|---|---|---|---|---|---|
| 1. | 5677.9 | $14^-$ | | | | **2020NE01** | 1. Probable Configuration as $\pi[h_{11/2}(g_{7/2}/d_{5/2})]\otimes$ $\nu\,[h_{11/2}]^2$ using TAC model. |
| | 5854.5 | $(15^-)$ | 176.6 | | | | 2. $\varepsilon_2 \approx 0.090$, $\gamma \approx 60^o$ on the bases of TAC minima. |
| | 6025.8 | $(16^-)$ | 171.3 | | | | 3. Regular band |
| | 6304.8 | $(17^-)$ | 278.9 | | | | |
| | 6693.5 | $(18^-)$ | 388.7 | | | | |
| | 7141.5 | $(19^-)$ | 448.0 | | | | |
| | 7712.9 | $(20^-)$ | 571.4 | | | | |

## $^{135}_{56}Ba_{79}$

| | $E_{level}$ keV | $I^{\pi}$ | $E_{\gamma}$(M1) keV | $E_{\gamma}$(E2) keV | B(M1)/B(E2) $(\mu_N/eb)^2$ | References | **Configurations and Comments:** |
|---|---|---|---|---|---|---|---|
| 1. | 3082.9 | $21/2^+$ | | | | **2010Ku15** | 1. Configuration assignment as $\pi(h_{11/2}g_{7/2}) \otimes$ $\nu(h_{11/2})^{-1}$ from TAC calculations. |
| | 3211.3 | $23/2^+$ | 128.4 | | | 2006Ch51 | 2. Tentatively assigned as MR band. |
| | 3415.7 | $25/2^+$ | 204.4 | | | | 3. Deformation parameters $(\varepsilon_2, \varepsilon_4, \gamma)$ = (0.095,-0.013, 26°) for energy minimization and $(\varepsilon_2, \varepsilon_4, \gamma)$ = (0.095,-0.013, 10°) for B(M1)/B(E2) ratios. |
| | 3758.3 | $27/2^+$ | 342.6 | | | | 4. Regular band with backbending at 33/2 ħ. |
| | 4181.0 | $29/2^+$ | 422.7 | 765.3 | 17.3(25) | | |
| | 4695.8 | $31/2^+$ | 514.8 | 937.5 | 52(31) | | |
| | 5310.3 | $33/2^+$ | 614.5 | 1129.3 | 11.1(68) | | |
| | 5850.3 | $35/2^+$ | 540.0 | 1154.5 | 21.8(62) | | |
| | 6376.3 | $(37/2^+)$ | 526.0 | 1066.0 | | | |
| | | | | | | | |
| 2. | 5355.6 | $(31/2^-)$ | | | | **2010Ku15** | 1. Configuration assigned as $(h_{11/2}g_{7/2})\otimes\nu(h_{11/2})^2 s_{1/2}$ from TAC calculations. |
| | 5522.2 | $(33/2^-)$ | 166.6 | | | | 2. $(\beta_2, \gamma)$ =(0.090,58°) from TAC calculations |
| | 5849.8 | $(35/2^-)$ | 327.6 | | | | 3. Regular band. |
| | 6380.8 | $(37/2^-)$ | 531.0 | | | | |

## $^{133}_{57}La_{76}$

| | $E_{level}$ keV | $I^{\pi}$ | $E_{\gamma}$(M1) keV | $E_{\gamma}$(E2) keV | B(M1)/B(E2) $(\mu_N/eb)^2$ | References | **Configurations and Comments:** ≥ |
|---|---|---|---|---|---|---|---|
| 1. | 3947.8 | $29/2^-$ | | | | **2016Pe21** | 1. Configuration assigned as $\pi(h_{11/2})^1\otimes\nu(h_{11/2})^{-2}$ and $(\pi(h_{11/2})^3\otimes\nu(h_{11/2})^{-2}$ for spins I ≤37 ħ and I >37 ħ, respectively from PRM calculations. |
| | 4396.1 | $31/2^-$ | 448.3 | | | | 2. Tentatively assigned as MR band. |
| | 4844.2 | $33/2^-$ | 448.1 | | | | 3. $(\beta_2, \gamma)$ =(0.20,19.2°) and (0.20,20.5°) for spins I ≤ 37 ħ and I>37ħ, respectively from CDFT calculations |
| | 5318.7 | $35/2^-$ | 474.5 | 922.6 | 5.4(6) | | 4. Irregular band with signature splitting. |
| | 5752.4 | $37/2^-$ | 433.7 | | | | |
| | 6141.7 | $39/2^-$ | 389.3 | | | | |
| | 6677.4 | $(41/2^-)$ | 535.7 | | | | |
| | 7027.5 | $(43/2^-)$ | 350.1 | | | | |
| | 7611.8 | $(45/2^-)$ | 584.3 | | | | |

## $^{135}_{57}La_{78}$

| | $E_{level}$ keV | $I^{\pi}$ | $E_{\gamma}$(M1) keV | $E_{\gamma}$(E2) keV | B(M1)/B(E2) $(\mu_N/eb)^2$ | References | Configurations and Comments: |
|---|---|---|---|---|---|---|---|
| 1. | 3501.4 | $25/2^-$ | | | | **2013Ga11** | 1. Configuration assigned as $\pi(h_{11/2})^1 \otimes \nu(h_{11/2})^{-2}$ from TAC calculations. |
| | 3639.2 | $27/2^-$ | 137.8 | | | 2013Le27 | |
| | 3959.7 | $29/2^-$ | 320.5 | 457.6 | 8.4(1) | 1990XuZW | 2. $(\beta_2, \gamma)$ = (0.120,58°) from TAC calculations. |
| | 4319.5 | $31/2^-$ | 359.8 | 680.3 | 52.1(47) | | 3. Regular band with a band crossing at $I \approx 33/2^-$ |
| | 4822.3 | $33/2^-$ | 502.8 | 863.1 | 83(11) | | |
| | 5381.1 | $35/2^-$ | 558.8 | | | | |
| | 6027.6 | $37/2^{(-)}$ | 646.5 | 1203.4 | 1.26(4) | | |
| 2. | 5049.5 | $(31/2^-)$ | | | | **2013Ga11** | 1. Configuration assigned as $\pi(h_{11/2})^1(g_{7/2}/d_{5/2})^2 \otimes \nu(h_{11/2})^{-2}$ from TAC calculations. |
| | 5212.0 | $33/2^-$ | 162.5 | | | 2013Le27 | |
| | 5464.1 | $35/2^-$ | 252.0 | | | 1990XuZW | 2. Tentatively assigned as MR band. |
| | 5775.1 | $37/2^-$ | 311.0 | | | | 3. $(\beta_2, \gamma)$ = (0.116,60°) from TAC calculations. |
| | 6161.4 | $39/2^-$ | 386.3 | | | | 4. Regular band with a band crossing at $I \approx 33/2^-$ |
| | 6635.3 | $(41/2^-)$ | 473.8 | | | | |
| | 7163.0 | $(43/2^-)$ | 527.7 | | | | |

## $^{134}_{58}Ce_{76}$

| | $E_{level}$ keV | $I^{\pi}$ | $E_{\gamma}$(M1) keV | $E_{\gamma}$(E2) keV | B(M1)/B(E2) $(\mu_N/eb)^2$ | Reference | **Configurations and Comments:** |
|---|---|---|---|---|---|---|---|
| 1. | 5593.6 | $14^-$ | | | | **2016Pe09** | 1. Configuration assigned as $\pi(h_{11/2})^2 \otimes \nu(sd)^{-1} (h_{11/2})^{-1}$ (2016Pe09) from CNS (2016Pe09) and $\pi(g_{7/2}h_{11/2}) \otimes \nu(h_{11/2})^{-2}$ from TAC (2004La03) calculations |
| | 5749.2 | $15^-$ | 155.6 | | | 2004La03 | |
| | 5969.0 | $16^-$ | 219.8 | | | | |
| | 6309.3 | $17^-$ | 340.3 | | | | |
| | 6766.9 | $18^-$ | 457.6 | | | | 2. $(\varepsilon_2, \gamma) \approx$ (0.14-0.18, +23°--30°) (2016Pe09) and $(\varepsilon_2, \gamma)$ = (0.149, 43°) (2004La03). |
| | 7286.8 | $19^-$ | 519.9 | 977.5 | | | |
| | 7776.2 | $20^-$ | 489.4 | 1009.3 | | | 3. Regular band. |
| | 8298.4 | $(21^-)$ | 522.2 | 1011.6 | | | 4. The mean lifetimes (in ps) for levels from $17^-$ to $20^-$ are 0.85(7), 0.34(3), 0.28(3) and <0.28(+3-4), respectively. |
| | 8964.6 | $(22^-)$ | 666.2 | | | | 5. B(M1) values for the transitions from 340 to 548 keV are 1.71(+13-14), 1.76(+17-15), 1.47(+16-14) and >1.24(+17-14) $\mu_N^2$, respectively. |

## $^{135}_{58}Ce_{77}$

| | $E_{level}$ keV | $I^{\pi}$ | $E_{\gamma}$(M1) keV | $E_{\gamma}$(E2) keV | B(M1)/B(E2) $(\mu_N/eb)^2$ | Reference | **Configurations and Comments:** |
|---|---|---|---|---|---|---|---|
| 1. | 3229.8 | $23/2^+$ | | | | **1990Ma26** | 1. Tentatively assigned as $\pi(h_{11/2}g_{7/2}) \otimes \nu(h_{11/2})$ by comparison with the N=75 isotones. |
| | 3431.9 | $25/2^+$ | 202.1 | | | 2005JaZZ | |
| | 3699.9 | $27/2^+$ | 268.0 | | | | 2. Near prolate shape ($\gamma \sim 0°$). |
| | 4128.2 | $29/2^+$ | 428.3 | 696 | >8 | | 3. Irregular band with backbending at $I^{\pi}$ =31/2 and 35/2. |
| | 4486.4 | $31/2^+$ | 358.2 | 786.8 | 23(7) | | |
| | 4979.3 | $33/2^+$ | 492.9 | 851 | >9 | | 4. Small signature splitting. |
| | 5428.5 | $35/2^+$ | 449.2 | 942 | >14 | | 5. Lower limits of B(M1)/B(E2) are from the unobserved $\Delta I$ =2 (E2) transitions. |
| | 5942.5 | $(37/2^+)$ | 514 | 963 | | | |
| | 6444.5 | $(39/2^+)$ | 502 | (1016) | | | |

## $^{135}_{58}Ce_{77}$

| | $E_{level}$ keV | $I^{\pi}$ | $E_{\gamma}$(M1) keV | $E_{\gamma}$(E2) keV | B(M1)/B(E2) $(\mu_N/eb)^2$ | Reference | Configurations and Comments: |
|---|---|---|---|---|---|---|---|
| 2. | 4183.8 | $27/2^-$ | | | | **1990Ma26** | 1. Tentatively assigned as $\pi(h_{11/2}^2) \otimes \nu(h_{11/2})$ by |
| | 4460.9 | $29/2^-$ | 277.1 | | | 2005JaZZ | comparison with the N=75 isotones. |
| | 4830.9 | $31/2^-$ | 370.0 | | | | 2. Near prolate shape ($\gamma \sim 0°$). |
| | 5206.5 | $33/2^-$ | 375.6 | 746 | >6 | | 3. The band is also suggested as chiral doublet |
| | 5651.6 | $35/2^-$ | 445.1 | 821 | >19 | | partner of band 3 (2005JaZZ). |
| | 6086.5 | $37/2^-$ | 434.9 | 880 | >24 | | 4. Irregular band with backbending at 37/2. |
| | 6526.5 | $39/2^-$ | 440.0 | (875) | >13 | | 5. Small signature splitting. |
| | 6994.5 | $41/2^-$ | 468.0 | (908) | >6 | | 6. Lower limits of B(M1)/B(E2) are from the |
| | 7494.5 | $(43/2^-)$ | 500 | 968 | | | unobserved $\Delta I = 2$ (E2) transitions. |
| | 8010.5 | $(45/2^-)$ | 516 | | | | |
| 3. | 4498.8 | $27/2^-$ | | | | **1990Ma26** | 1. Tentatively assigned as $\pi(h_{11/2}g_{7/2}) \otimes \nu(h_{11/2}^2 s_{1/2})$. |
| | 4637.9 | $29/2^-$ | 139.1 | | | 2005JaZZ | 2. Collectively rotating oblate structure ($\gamma \sim -60°$). |
| | 4816.4 | $31/2^-$ | 178.5 | | >4 | | 3. The band is also suggested as chiral doublet |
| | 5065.4 | $33/2^-$ | 249.0 | | 6.7(11) | | partner of band 2 (2005JaZZ). |
| | 5362.9 | $35/2^-$ | 297.5 | | 24.3(42) | | 4. The mean lifetimes (in ps) for levels from 5363 to |
| | 5755.1 | $37/2^-$ | 392.2 | | 5.6(13) | | 6843 keV are 0.92(16), 0.65(3), 0.43(4) and |
| | 6259.7 | $39/2^-$ | 504.6 | | 34.0(64) | | 0.22(1), respectively. |
| | 6843.3 | $(41/2^-)$ | 583.6 | | 19.8(48) | | 5. Limits on B(M1)/B(E2) values are from the |
| | 7473.3 | $(43/2^-)$ | 630 | | >22 | | assumption that the unobserved $\Delta I = 2$ (E2) |
| | 8037.3 | $(45/2^-)$ | (564) | | | | transitions are less than 1% intense as compared |
| | | | | | | | to the strongest transition in the level scheme. |
| | | | | | | | 6. Regular band. |

## $^{136}_{58}Ce_{78}$

| | $E_{level}$ keV | $I^{\pi}$ | $E_{\gamma}$(M1) keV | $E_{\gamma}$(E2) keV | B(M1)/B(E2) $(\mu_N/eb)^2$ | Reference | Configurations and Comments: |
|---|---|---|---|---|---|---|---|
| 1. | 5305.5 | $15^+$ | | | | **2005La29** | 1. $\pi(g_{7/2}h_{11/2}) \otimes \nu(g_{7/2}h_{11/2})$ from TAC calculations. |
| | 5594.4 | $16^+$ | 288.9 | | | | 2. $(\beta_2, \gamma) = (0.116, 28°)$ from TAC calculations. |
| | 6099.3 | $17^{(+)}$ | 504.9 | | | | 3. Regular band. |
| | 6642.9 | $18^{(+)}$ | 543.6 | 1049 | | | 4. The mean lifetimes (in ps) of 6099.3 keV. |
| | (7239) | $(19^+)$ | (596.1) | (1140) | | | level is 0.65(+16-19). |
| | | | | | | | 5. The B(M1) value for the transition 504.9 keV is |
| | | | | | | | 0.69(+20-17) $\mu_N^2$. |
| 2. | 5643.8 | $16^+$ | | | | **2005La29** | 1. $\pi(h_{11/2}^2) \otimes \nu(h_{11/2}^{-2})$ from TAC calculations. |
| | 5878.2 | $17^+$ | 234.4 | | | | 2. $(\beta_2, \gamma) = (0.138, 52°)$ from TAC calculations. |
| | 6171.5 | $18^+$ | 293.3 | | | | 3. Regular band. |
| | 6540.4 | $19^+$ | 368.9 | | | | 4. The mean lifetimes (in ps) of levels from 6540 to |
| | 6934.5 | $20^+$ | 394.1 | | | | 7346 keV are 0.58(+21-26), 0.79(26) and |
| | 7345.9 | $(21^+)$ | 411.4 | | | | 0.45(+16-19), respectively. |
| | 7801.9 | $(22^+)$ | 456 | | | | 5. The B(M1) values for the transitions from |
| | (8317) | $(23^+)$ | (515) | | | | 368.9 to 411.4 keV are 1.97(71), 1.18(39) and |
| | | | | | | | 1.84(+74-61) $\mu_N^2$, respectively. |

## $^{136}_{58}Ce_{78}$

| | $E_{level}$ keV | $I^\pi$ | $E_\gamma$(M1) keV | $E_\gamma$(E2) keV | B(M1)/B(E2) $(\mu_N/eb)^2$ | **Reference** | **Configurations and Comments:** |
|---|---|---|---|---|---|---|---|
| 3. | 5646.5 | $14^-$ | | | | **2005La29** | 1. $\pi(g_{7/2}h_{11/2}) \otimes \nu(h_{11/2}^{-2})$ from TAC calculations.<br>2. $(\beta_2, \gamma)$ = (0.116, 52°) from TAC calculations.<br>3. Regular band.<br>4. The mean lifetimes (in ps) of levels from 6664 to 8626 keV are 0.734(+21-22), 0.454(+17-14), 0.38(+4-5),0.365(+26-41) and 0.577(+41-61), respectively.<br>5. The B(M1) values for the transitions from 380.5 to 515 keV are 1.346(+40-39), 1.39(+4-5), 1.097(+130-107), 0.782(+88-56) and 0.474(+50-34) $\mu_N^2$, respectively.<br>6. The B(E2) values for the transitions from 668 to 1040 keV are 0.038(4), 0.040(4), 0.051(+8-7) 0.059(+10-8) and 0.040(+7-6) $(eb)^2$, respectively. |
| | 5809.9 | $15^-$ | 163.4 | | | 2002La26 | |
| | 5995.8 | $16^-$ | 185.9 | 350 | 13.5(63) | 1990Pa05 | |
| | 6283.5 | $17^-$ | 287.7 | 474 | 14.4(42) | | |
| | 6664.0 | $18^-$ | 380.5 | 668 | 35.5(32) | | |
| | 7100.2 | $19^-$ | 436.2 | 816 | 34.6(15) | | |
| | 7586.6 | $20^-$ | 486.4 | 922 | 21.3(11) | | |
| | 8111.4 | $21^-$ | 524.8 | 1011 | 13.4(6) | | |
| | 8626.6 | $22^-$ | 515.2 | 1040 | 11.8(5) | | |
| | 9229.2 | $23^-$ | 602.6 | 1118 | 12.8(20) | | |

## $^{138}_{58}Ce_{80}$

| | $E_{level}$ keV | $I^\pi$ | $E_\gamma$(M1) keV | $E_\gamma$(E2) keV | B(M1)/B(E2) $(\mu_N/eb)^2$ | References | Configurations and Comments: |
|---|---|---|---|---|---|---|---|
| 1. | 6536.3 | $15^-$ | | | | **2009Bh04** | 1. Configuration assigned as $\pi(h_{11/2}\ g_{7/2}) \otimes \nu(h_{11/2})^{-2}$ based on systematic in neighboring nuclei.<br>2. Tentatively assigned as MR band.<br>3. Regular band. |
| | 6685.4 | $16^-$ | 149.1 | | | 1999Zh28 | |
| | 6888.9 | $17^-$ | 203.5 | | | | |
| | 7211.2 | $18^-$ | 322.3 | | | | |
| | 7685.7 | $19^-$ | 474.5 | | | | |
| | 8350.2 | $20^-$ | 664.5 | 1139.0 | 14.1(53) | | |

## $^{139}_{58}Ce_{81}$

| | $E_{level}$ keV | $I^\pi$ | $E_\gamma$(M1) keV | $E_\gamma$(E2) keV | B(M1)/B(E2) $(\mu_N/eb)^2$ | References | **Configurations and Comments:** |
|---|---|---|---|---|---|---|---|
| 1. | 5698.9 | $31/2^-$ | | | | **2015Ka06** | 1. Configuration assigned as $\pi(h_{11/2})^2 \otimes \nu(h_{11/2})^{-1}$ from TAC calculations<br>2. $(\beta_2, \gamma)$ = (0.12,0°) from CNS calculations<br>3. Regular band. |
| | 5916.5 | $33/2^-$ | 218.6 | | | | |
| | 6142.6 | $35/2^-$ | 226.0 | | | | |
| | 6488.5 | $37/2^-$ | 345.8 | | | | |
| | 6845.3 | $39/2^-$ | 356.6 | | | | |
| | 7333.6 | $41/2^-$ | 488.3 | | | | |
| | 7987.6 | $43/2^-$ | 654.3 | | | | |

## $^{133}_{59}Pr_{74}$

| | $E_{level}$ keV | $I^{\pi}$ | $E_{\gamma}$(M1) keV | $E_{\gamma}$(E2) keV | B(M1)/B(E2) $(\mu_N/eb)^2$ | Reference | **Configurations and Comments:** |
|---|---|---|---|---|---|---|---|
| 1. | 2035.1 | $21/2^-$ | | | | **2003Pa38** | 1. $\pi(5/2[413]) \otimes \nu(9/2[514]\otimes 7/2[404])$, related to |
| | 2475.3 | $23/2^-$ | 440.2 | | | 1988Hi04 | the $\nu(h_{11/2}\ g_{7/2})_{K^{\pi}=8^-}$ isomeric state in the $^{132}$Ce |
| | 2955.2 | $(25/2^-)$ | 479.9 | 920.1 | 8.9(9) | | core. |
| | 3465.5 | $(27/2^-)$ | 510.3 | 990.2 | 8.3(9) | | 2. Tentatively assigned as MR band. |
| | 3960.7 | $(29/2^-)$ | 495.2 | 1005.3 | 6.0(6) | | 3. Prolate configuration. |
| | | | | | | | 4. The B(M1)/B(E2) values decrease from about 10 |
| | | | | | | | to 5 $(\mu_N/eb)^2$ with increasing spin. |
| 2. | 2203.9 | $19/2^+$ | | | | **2003Pa38** | 1. $\pi(3/2[413]) \otimes \nu(9/2[514]\otimes 7/2[404])$, related to |
| | 2352.7 | $21/2^+$ | 148.8 | | | | the $\nu(h_{11/2}\ g_{7/2})$ $K^{\pi}=8^-$ isomeric state in the $^{132}$Ce |
| | 2598.3 | $23/2^+$ | 245.6 | | | | core. |
| | 2925.4 | $25/2^+$ | 327.1 | 572.5 | 23.5(18) | | 2. Tentatively assigned MR band. |
| | 3319.8 | $27/2^+$ | 394.4 | 722.0 | 16.9(14) | | 3. Prolate configuration. |
| | 3767.5 | $29/2^+$ | 447.7 | 842.8 | 12.6(10) | | 4. The B(M1)/B(E2) values decrease from about 25 |
| | 4263.9 | $31/2^+$ | 496.4 | 944.4 | 11.2(20) | | to 10 $(\mu_N/eb)^2$ with increasing spin. |
| | 4793.5 | $(33/2^+)$ | 529.6 | 1026.8 | 8.4(16) | | 5. Regular band. |
| | 5354.4 | $(35/2^+)$ | 560.9 | 1090.2 | 7.9(9) | | |
| | 5907.2 | $(37/2^+)$ | 552.8 | 1112.9 | | | |
| 3. | 3253.0 | $21/2^-$ | | | | **2003Pa38** | 1. $\pi(h_{11/2})_{K=11/2}\otimes\nu(h_{11/2}{}^2)$ by comparison with a |
| | 3371.8 | $23/2^-$ | 118.8 | | | 1988Hi04 | similar band in $^{131}$La isotone. |
| | 3536.8 | $25/2^-$ | 165.0 | | | | 2. Tentatively assigned MR band. |
| | 3787.7 | $27/2^-$ | 250.9 | | | | 3. Oblate configuration. |
| | 4124.4 | $29/2^-$ | 336.7 | | | | 4. The B(M1)/B(E2) values lie between 10-20 |
| | 4533.9 | $31/2^-$ | 409.5 | | | | $(\mu_N/eb)^2$. |
| | 5005.7 | $(33/2^-)$ | 471.8 | 882.2 | 17.2(28) | | 5. Regular band. |
| | 5533.3 | $(35/2^-)$ | 527.6 | 1000.6 | 12.8(19) | | |
| | 6107.4 | $(37/2^-)$ | 574.1 | 1101.0 | 8.9(11) | | |
| | 6725.6 | $(39/2^-)$ | 618.2 | 1191.9 | 15.2(21) | | |
| | 7337.6 | $(41/2^-)$ | 612.0 | 1230.0 | 15.6(32) | | |
| 4. | 4108.1 | $(29/2^-)$ | | | | **2003Pa38** | 1. $\pi(g_{9/2}\ h_{11/2}{}^2) \otimes \nu(h_{11/2}g_{7/2})$ because of the |
| | 4252.3 | $(31/2^-)$ | 144.2 | | | 1988Hi04 | backbend from band 4 to band 1 observed in the |
| | 4379.1 | $(33/2^-)$ | 126.8 | | | | alignment plot. |
| | 4575.0 | $(35/2^-)$ | 195.9 | | | | 2. Tentatively assigned MR band. |
| | 4818.8 | $(37/2^-)$ | 243.8 | 439.7 | | | 3. Prolate configuration. |
| | 5115.2 | $(39/2^-)$ | 296.4 | 540.2 | 16.3(13) | | 4. The B(M1)/B(E2) values lie between 10-25 |
| | 5466.0 | $(41/2^-)$ | 350.8 | 647.2 | 20.4(40) | | $(\mu_N/eb)^2$. |
| | 5869.6 | $(43/2^-)$ | 403.6 | 754.4 | 16.8(16) | | 5. Regular band. |
| | 6323.6 | $(45/2^-)$ | 454.0 | 857.6 | 14.1(22) | | |
| | 6824.6 | $(47/2^-)$ | 501.0 | 955.0 | 15.8(37) | | |
| | 7372.8 | $(49/2^-)$ | 548.2 | 1049.3 | 12.5(22) | | |
| | 7970.4 | $(51/2^-)$ | 597.6 | 1145.7 | | | |
| | 8615.1 | $(53/2^-)$ | 644.7 | 1242.2 | | | |

### $^{135}_{59}\mathrm{Pr}_{76}$

| | $E_{level}$ keV | $I^{\pi}$ | $E_{\gamma}(M1)$ keV | $E_{\gamma}(E2)$ keV | B(M1)/B(E2) $(\mu_N/eb)^2$ | References | Configurations and Comments: |
|---|---|---|---|---|---|---|---|
| 1. | 3380.6 | $25/2^{(-)}$ | | | | **2015Ga39** | 1. Configuration assigned as $\pi(h_{11/2})^1 \otimes \nu(h_{11/2})^{-2}$ and |
| | 3530.0 | $27/2^-$ | 149.5 | | | | $\pi(h_{11/2})^1(g_{7/2})^2 \otimes \nu(h_{11/2})^{-2}$, respectively before |
| | 3863.0 | $29/2^-$ | 332.9 | | | | and after band crossing from TAC calculations. |
| | 4292.7 | $31/2^-$ | 429.7 | 762.7 | 15.8(9) | | 2. $(\varepsilon_2, \gamma) = (0.14, 62°)$ and $(0.14, 60°)$, respectively |
| | 4704.1 | $33/2^-$ | 410.8 | 840.6 | 19.7(10) | | before and after band crossing at $I \approx 33/2^-$. |
| | 5028.6 | $35/2^-$ | 325.1 | 735.9 | 94.9(61) | | 3. Regular band with backbending at 33/2 ħ. |
| | 5452.6 | $37/2^-$ | 424.0 | | | | |
| | 5951.1 | $39/2^-$ | 498.5 | | | | |
| | 6506.4 | $41/2^{(-)}$ | 555.3 | | | | |
| | 7110.4 | $(43/2^-)$ | (604.0) | | | | |

### $^{137}_{59}\mathrm{Pr}_{78}$

| | $E_{level}$ keV | $I^{\pi}$ | $E_{\gamma}(M1)$ keV | $E_{\gamma}(E2)$ keV | B(M1)/B(E2) $(\mu_N/eb)^2$ | Reference | **Configurations and Comments:** |
|---|---|---|---|---|---|---|---|
| 1. | 3438.4 | $25/2^-$ | | | | **2007Ag13** | 1. Configuration assigned as$\pi(h_{11/2}) \otimes \nu(h_{11/2}^{-2})$ |
| | 3550.3 | $27/2^-$ | 111.9 | | | 1989Xu01 | at low spins and $\pi(h_{11/2})\ \pi(g_{7/2}^{2}) \otimes \nu(h_{11/2}^{-2})$ |
| | 3871.3 | $29/2^-$ | 321.0 | 432.6 | 10.4(19) | | near spin 37/2ħ. |
| | 4212.8 | $31/2^-$ | 341.5 | 662.5 | 10.5(23) | | 2. Collective oblate shape ($\gamma \sim -60°$) (1989Xu01) |
| | 4696.2 | $33/2^-$ | 483.4 | 824.1 | 17(3) | | 3. Signature splitting with backbending at 35/2 ħ. |
| | 5174.2 | $35/2^-$ | 478.0 | 961.7 | 41(7) | | Band crossing at 37/2 ħ. |
| | 5515.6 | $37/2^-$ | 341.4 | 819.4 | 62(12) | | |
| | 5924.0 | $39/2^-$ | 408.4 | 749.2 | 33(9) | | |
| | 6389.5 | $41/2^-$ | 465.5 | 872.3 | 17(3) | | |
| | 6896.7 | $43/2^-$ | 507.2 | 972.1 | 18(3) | | |
| | 7473.4 | $45/2^-$ | 576.7 | 1083.6 | 13(3) | | |
| | 8130.8 | $47/2^-$ | 657.4 | 1233.7 | 19(5) | | |

### $^{138}_{60}\mathrm{Nd}_{78}$

| | $E_{level}$ keV | $I^{\pi}$ | $E_{\gamma}(M1)$ keV | $E_{\gamma}(E2)$ keV | B(M1)/B(E2) $(\mu_N/eb)^2$ | Reference | **Configurations and Comments:** |
|---|---|---|---|---|---|---|---|
| 1. | 5492.7 | $(13^-)$ | | | | **2012Pe15** | 1. Tentatively assigned as $\pi(h_{11/2})^1(d_{5/2}\ g_{7/2})^1 \otimes \nu h^2$ |
| | 5576.4 | $14^-$ | 83.6 | | | 1994De11 | from CNS calculations (2012Pe15). |
| | 5769.8 | $15^-$ | 193.4 | 277.0 | | | 2. $(\varepsilon_2, \gamma) \approx (0.15, 30°)$ |
| | 6000.4 | $16^-$ | 230.6 | 424.0 | | | 3. Regular band with a small backbending at $19^-$. |
| | 6286.6 | $17^-$ | 286.2 | 516.7 | | | |
| | 6667.4 | $18^-$ | 380.8 | 667.0 | | | |
| | 7046.2 | $19^-$ | 378.8 | 759.6 | | | |
| | 7563.7 | $20^-$ | 517.5 | 896.3 | | | |
| | 8012.4 | $21^-$ | 448.7 | 966.2 | | | |
| 2. | 7831.7 | $19^{(+)}$ | | | | **2012Pe15** | 1. Tentatively assigned as $\pi(h_{11/2})^1\ (d_{5/2}\ g_{7/2})^2$ |
| | 8059.7 | $20^{(+)}$ | 228.0 | | | | $\otimes \nu(h_{11/2})^2$ from CNS calculations (2012Pe15). |
| | 8353.1 | $21^{(+)}$ | 293.4 | | | | 2. Regular band. |
| | 8710.0 | $22^{(+)}$ | 356.7 | | | | |
| | 9134.2 | $23^{(+)}$ | 424.4 | | | | |
| | 9622.1 | $24^{(+)}$ | 487.9 | | | | |
| | 10233.7 | $(25^+)$ | 611.6 | | | | |

## $^{139}_{60}\mathrm{Nd}_{79}$

| | $E_{level}$ keV | $I^{\pi}$ | $E_{\gamma}$(M1) keV | $E_{\gamma}$(E2) keV | B(M1)/B(E2) $(\mu_N/eb)^2$ | References | **Configurations and Comments:** |
|---|---|---|---|---|---|---|---|
| 1. | 3980.3 | $27/2^-$ | | | | **2011Bh07** | 1. Configuration assigned as $\pi(h_{11/2})^2 \otimes \nu h_{11/2}$ and |
| | 4293.0 | $29/2^-$ | 312.7 | | | 2007Ku12 | $\pi(h_{11/2})^2 (g_{7/2})^2 \otimes \nu h_{11/2}$, respectively, before and |
| | 4756.4 | $31/2^-$ | 463.3 | | | 2013Va10 | after band crossing (2007Ku12). |
| | 5392.6 | $33/2^-$ | 636.2 | | | 2008Xu05 | 2. $(\varepsilon_2, \varepsilon_4, \gamma)$ = (0.125,0.0, 5°) and (0.133,0.0, 10°) |
| | 5695.6 | $35/2^-$ | 303.0 | | | | before and after band crossing, respectively. |
| | 6070.9 | $37/2^-$ | 375.3 | | | | 3. Tentatively assigned as MR band. |
| | 6490.9 | $39/2^-$ | 420.1 | | | | 4. Irregular band with backbending at 33/2ħ. |

## $^{141}_{60}\mathrm{Nd}_{81}$

| | $E_{level}$ keV | $I^{\pi}$ | $E_{\gamma}$(M1) keV | $E_{\gamma}$(E2) keV | B(M1)/B(E2) $(\mu_N/eb)^2$ | References | **Configurations and Comments:** |
|---|---|---|---|---|---|---|---|
| 1. | 5648.0 | $27/2^-$ | | | | **2015Ze02** | 1. Configuration assigned as $\pi[(h_{11/2})^2(d_{5/2}g_{7/2})^2] \otimes$ |
| | 5791.2 | $29/2^-$ | 143.2 | | | | $\nu(h_{11/2})^{-1}$, with rearrangement of (dg) protons in |
| | 5962.1 | $31/2^-$ | 170.9 | | | | 5/2[413] and 3/2[411] Nilsson orbital before and |
| | 6212.0 | $33/2^-$ | 249.9 | | | | after the crossing. |
| | 6559.9 | $35/2^-$ | 347.9 | 598.1 | 27(10) | | 2. $(\varepsilon_2, \gamma)$ = (0.102, +10°) changes to (0.07, +5°) with |
| | 7018.2 | $37/2^-$ | 458.4 | 806.4 | 15(5) | | spin from CNS calculations. |
| | 7498.8 | $39/2^-$ | 480.6 | 938.8 | 16(5) | | 3. Irregular band with backbending at 41/2ħ. |
| | 7851.5 | $41/2^-$ | 353.6 | 833.8 | 8(5) | | |
| | 8263.5 | $43/2^-$ | 411.8 | 765.3 | $> 53$ | | |
| | 8707.3 | $45/2^-$ | 443.6 | 856 | $> 47$ | | |
| | 9060.4 | $47/2^-$ | 353.2 | 797.1 | | | |
| | 9550.3 | $49/2^-$ | 490.3 | 842.7 | | | |
| 2. | 7316.9 | $37/2^{(-)}$ | | | | **2015Ze02** | 1. Configuration assigned as $\pi[(h_{11/2})^2(d_{5/2}g_{7/2})^2] \otimes$ |
| | 7547.7 | $39/2^{(-)}$ | 230.7 | | | | $\nu(h_{11/2})^{-1}$ before and after bandcrossing, with (dg) |
| | 7904.6 | $41/2^{(-)}$ | 356.8 | | | | protons excitation from 3/2[411] to 7/2[404] |
| | 8372.8 | $43/2^{(-)}$ | 467.9 | 824.8 | $> 36$ | | Nilsson orbital. |
| | 8768.7 | $45/2^{(-)}$ | 395.5 | 864.3 | 13(5) | | 2. $(\varepsilon_2, \gamma)$ = (0.102, +10°) changes to (0.04, -30°) with |
| | 9085.9 | $47/2^{(-)}$ | 317.3 | 713.4 | 70(10) | | spin from CNS calculations. |
| | 9497.6 | $49/2^{(-)}$ | 412.2 | 728.6 | 38(5) | | 3. Irregular band with backbending at 45/2ħ. |
| | 10006.8 | $51/2^{(-)}$ | 509.2 | 920.6 | 4(2) | | |
| | 10611.5 | $(53/2^-)$ | 604.4 | 1114.2 | | | |
| 3. | 9361.8 | $(47/2^+)$ | | | | **2015Ze02** | 1. Configuration assigned as $\pi[(h_{11/2})^3(d_{5/2}g_{7/2})^1] \otimes$ |
| | 9653.8 | $(49/2^+)$ | 291.8 | | | | $\nu(h_{11/2})^{-1}$. |
| | 10066.7 | $(51/2^+)$ | 412.6 | 705 | $> 38$ | | 2.Regular band |
| | 10591.6 | $(53/2^+)$ | 524.7 | 938 | 15(10) | | |
| | 11209.6 | $(55/2^+)$ | 618.1 | 1143 | 25(10) | | |
| | 11911.7 | $(57/2^+)$ | 702.2 | 1320 | $>21$ | | |

## $^{139}_{61}Pm_{78}$

| | $E_{level}$ keV | $I^\pi$ | $E_\gamma(M1)$ keV | $E_\gamma(E2)$ keV | B(M1)/B(E2) $(\mu_N/eb)^2$ | References | **Configurations and Comments:** |
|---|---|---|---|---|---|---|---|
| | | | | | | **2011Zh47** | 1. Tentatively assigned as $\pi h_{11/2} \otimes \nu(h_{11/2})^{-2}$ based |
| 1. | 3158.0 | $25/2^-$ | | | | 2010Zh12 | on the similar band observed in $^{141}$Eu(2011Zh47). |
| | 3262.4 | $27/2^-$ | 104.2 | | | 2011ZhZU | 2. Tentatively assigned as MR band. |
| | 3592.0 | $29/2^-$ | 329.6 | | | 2009Dh01 | 3. The B(M1)/B(E2) ratios for 31/2 ħ and 35/2 ħ |
| | 3908.7 | $31/2^-$ | 316.7 | 646.2 | 26.8(63) | | levels are calculated using γ-intensities |
| | 4381.7 | $33/2^-$ | 473.0 | 790.0 | | | (2011Zh47) |
| | 4914.6 | $35/2^-$ | 533.0 | 1005.6 | 18.9(25) | | 4. Regular band with backbending at 29/2 ħ. |
| 2. | 5183.2 | $35/2^-$ | | | | **2011ZH47** | 1. Tentatively assigned as $\pi h_{11/2} \otimes \nu(h_{11/2})^{-4}$ based |
| | 5407.5 | $37/2^-$ | 224.4 | | | 2011ZhZU | on the similar band observed in $^{141}$Eu. |
| | 5669.6 | $39/2^-$ | 262.1 | | | 2009Dh01 | 2. Tentatively assigned as MR band. |
| | 6123.9 | $41/2^-$ | 454.3 | | | | 3. Irregular band. |
| | 7433.8 | $(47/2^-)$ | 313.3 | | | | |

## $^{139}_{62}Sm_{77}$

| | $E_{level}$ keV | $I^\pi$ | $E_\gamma(M1)$ keV | $E_\gamma(E2)$ keV | B(M1)/B(E2) $(\mu_N/eb)^2$ | References | **Configurations and Comments:** |
|---|---|---|---|---|---|---|---|
| 1. | 3327.0 | $25/2^-$ | | | | **2008Pa36** | 1. Configuration assigned as $\pi(h_{11/2})^2 \otimes \nu(h_{11/2})^{-1}$ |
| | 3445.4 | $27/2^-$ | 118.4 | | | 1996Br33 | from SPAC calculations (2008Pa36). |
| | 3710.3 | $29/2^-$ | 264.7 | 382.9 | 6.0(+105-8) | 1996Ro04 | 2. Tentatively assigned as MR band. |
| | 4047.5 | $31/2^-$ | 337.0 | 601.7 | 8.2(+23-11) | | 3. $(\varepsilon_2, \gamma)$ = (0.10, 0° to - 25°). |
| | 4457.0 | $33/2^-$ | 409.1 | 746.3 | 7.3(+38-11) | | 4. The mean lifetimes (in ps) of levels from 3708.7 |
| | 4929.7 | $35/2^-$ | 472.4 | 881.8 | 5.6(+20-10) | | keV to 5931.1 keV are 1.0(+6-3), 0.85(+21-17), |
| | 5443.2 | $37/2^-$ | 513.5 | 985.7 | 4.9(+26-12) | | 0.78(+14-10), 0.49(+12-9), 0.9(+6-3) and >1.0, |
| | 5934.6 | $(39/2^-)$ | 490.4 | 1003.2 | 2.7(+16-10) | | respectively |
| | | | | | | | 5. The B(M1) values for the transitions from |
| | | | | | | | 264.7 keV to 490.4 keV are 2.03(+184-37), |
| | | | | | | | 1.36(+57-23), 0.76(+24-12), 0.62(+26-11), |
| | | | | | | | 0.16(+18-4) and <0.15 $\mu_N^2$, respectively |
| | | | | | | | (1996Br33). |
| | | | | | | | 6. The B(E2) values for the transitions from 382.9 |
| | | | | | | | keV to 1003.2 keV are 0.23(+31-10), 0.16 (+8- 3) |
| | | | | | | | ,0.09(+4-2), 0.10(+5-2), 0.03(+3-1) and < 0.06 |
| | | | | | | | $(eb)^2$, respectively (1996Br33). |
| | | | | | | | 7. Regular band with backbending at 39/2 ħ. |

## $^{141}_{62}\mathrm{Sm}_{79}$

| | $E_{level}$ keV | $I^{\pi}$ | $E_{\gamma}$(M1) keV | $E_{\gamma}$(E2) keV | B(M1)/B(E2) $(\mu_N/eb)^2$ | References | **Configurations and Comments:** |
|---|---|---|---|---|---|---|---|
| 1. | 3377 | $25/2^-$ | | | | **2016Ra33** | 1. Configuration assigned as $\pi(h_{11/2})^2 \otimes \nu(h_{11/2})^{-1}$ on the basis of systematics of MR bands observed in $^{139}$Sm and $^{141}$Eu and also supported by SPAC calculations. |
| | 3509.6 | $27/2^-$ | 132.6 | | | 1991Ca24 | 2. $\beta_2 \approx 0.15$ using TRS calculations. |
| | 3819.0 | $29/2^-$ | 309.4 | | | | 3. The mean lifetimes (in ps) of the levels from 3509.6 keV to 5366.2 keV are 2.37 (+45-39), 1.05(+21- 18), 0.72(15-13),1.11(+23-15) and 0.40(8), respectively. |
| | 4265.4 | $31/2^-$ | 446.4 | | | | 4. The B(M1) values for the transition from 132.6 keV to 573.0 keV are 2.92(+55-48), 1.54(+31-26), 0.75(+16-14), 0.29(+6-4) and 0.72(14) $\mu_N^2$·, respectively. |
| | 4793.2 | $33/2^-$ | 527.8 | | | | 5. Regular band. |
| | 5366.2 | $(35/2^-)$ | 573.0 | | | | |
| 2. | 5340.6 | $35/2^-$ | | | | **2016Ra33** | 1. Configuration assigned as $\pi(h_{11/2})^2 \otimes \nu(h_{11/2})^{-3}$ on the basis SPAC calculations. |
| | 5594.3 | $37/2^-$ | 253.7 | | | 1991Ca24 | 2. $\beta_2 \approx 0.15$ using TRS calculations. |
| | 5940.0 | $39/2^-$ | 345.5 | | | | 3. Tentatively assigned as MR band. |
| | 6413.0 | $41/2^{(-)}$ | 473.0 | | | | 4. The mean lifetimes (in ps) of the levels 5594 keV and 5940 keV are 1.88 (+40-35) and < 1.15, respectively. |
| | 6894.4 | $43/2^{(-)}$ | 481.4 | | | | 5. The B(M1) values for the transition from 253.7 keV and 345.5 keV are 1.45(+31-27) and > 0.97$\mu_N^2$·, respectively. |
| | 7384.4 | $45/2^{(-)}$ | 490.0 | | | | 6. Regular band. |

## $^{142}_{62}\mathrm{Sm}_{80}$

| | $E_{level}$ keV | $I^{\pi}$ | $E_{\gamma}$(M1) keV | $E_{\gamma}$(E2) keV | B(M1)/B(E2) $(\mu_N/eb)^2$ | References | **Configurations and Comments:** |
|---|---|---|---|---|---|---|---|
| 1. | 6459.3 | $16^-$ | | | | **2014Ra03** | 1. Configuration assignment as $\pi(h_{11/2})^1 \otimes \nu(h_{11/2})^{-2}\, \pi(g_{7/2})^{-1}$ and $\pi(h_{11/2})^1 \otimes \nu(h_{11/2})^{-2}\, \pi(g_{7/2})^{-3}$ before and after band crossing from SPAC calculations. |
| | 6746.0 | $17^-$ | 286.7 | | | | 2. The mean lifetimes (in ps) of the levels from 6746.0 keV to 8660.2 keV are 1.05 (+27-21), 0.91(+25-22), 0.76 (+37-26),1.08(+42-26) and < 0.57, respectively. Lifetimes do not include 15% systematic uncertainty from stopping powers. |
| | 7131.2 | $18^-$ | 385.2 | | | | 3. The B(M1) values for the transition from 286.7 keV to 408.0 keV are 1.92(+51-39), 1.10(+30-27), 0.60(+30-21), 0.08(+4-2) and >1.50 $\mu_N^2$·, respectively. |
| | 7630.9 | $19^-$ | 499.6 | | | | 4. Regular band with backbending at 20 ħ. |
| | 8252.4 | $20^-$ | 621.4 | | | | |
| | 8660.4 | $21^-$ | 408.0 | | | | |
| | 9167.1 | $22^{(-)}$ | 506.7 | | | | |

## $^{143}_{62}Sm_{81}$

| | $E_{level}$ keV | $I^{\pi}$ | $E_{\gamma}(M1)$ keV | $E_{\gamma}(E2)$ keV | B(M1)/B(E2) $(\mu_N/eb)^2$ | References | **Configurations and Comments:** |
|---|---|---|---|---|---|---|---|
| 1. | 8613.8 | $43/2^-$ | | | | **2018Ra14** | 1. Probable Configuration as $\pi h^4_{11/2} \otimes \nu h^{-1}_{11/2}$ $\pi(g_{7/2}/d_{5/2})^{-2}$ using SPAC model. |
| | 8853.3 | $45/2^-$ | 239.5 | | | 2018Ra26 | 2. The B(M1) values for the transition from 239.5 to 728.0 keV are 3.52(+55-42), 2.16(+34-28), 1.20(+20-16), 0.61(+10-9) and > 0.18 $\mu_N^2$, respectively. |
| | 9192.9 | $47/2^-$ | 339.6 | 579.0 | 12.70(+250-222) | 2006Ra10 | 3. The B(E2) values for the transition from 579.0 to 1023.0 keV are 0.17(2), 0.9(+2-1) and 0.03(1) $(eb)^2$, respectively. |
| | 9637.7 | $49/2^-$ | 444.8 | 784.3 | 13.33(+370-231) | | 4. The mean lifetime of levels from 8853.3 to 10943.9 keV are 1.02(+16-12), 0.58(+9-8), 0.44(+7-6), 0.41(+7-6) and < 0.78 ps, respectively. |
| | 10215.9 | $51/2^-$ | 578.2 | 1023.0 | 20.33(+491-420) | | 5. Regular band |
| | 10943.9 | $53/2^-$ | 728.0 | | > 46.78 | | |
| 2. | 10080 | $49/2^-$ | | | | **2018Ra26** | 1. Probable Configuration as $\pi h^6_{11/2} (g_{7/2}/d_{5/2})^2 \otimes \nu h^{-1}_{11/2}$. |
| | 10268.9 | $51/2^-$ | 188.9 | | | | 2. Tentatively assigned as MR band. |
| | 10621.6 | $53/2^-$ | 352.7 | 542.0 | 5.7(12) | | 3. The B(M1) values for the transition from 188.9 to 462.1 keV are 1.17(+39-34), 0.74(+11-10), 0.80(12), 0.58(9) and > 0.63 $\mu_N^2$, respectively. |
| | 11071 | $55/2^-$ | 449.4 | 802.0 | 7.3(17) | | 4. The B(E2) values for the transition from 542.0 to 977.0 keV are 0.13(2), 0.11(2), 0.07(1) and > 0.05 $(eb)^2$, respectively. |
| | 11586 | $57/2^-$ | 515.0 | 965.0 | 8.3(17) | | 5. The mean lifetime of level from 10268.9 to 12048.1 keV are 1.17(+33-26), 0.86(+13-12), 0.55(8), 0.43(7) and < 0.62 ps, respectively. |
| | 12048.1 | $59/2^{(-)}$ | 462.1 | 977.0 | | | 6. Regular band |

## $^{141}_{63}Eu_{78}$

| | $E_{level}$ keV | $I^{\pi}$ | $E_{\gamma}(M1)$ keV | $E_{\gamma}(E2)$ keV | B(M1)/B(E2) $(\mu_N/eb)^2$ | References | **Configurations and Comments:** |
|---|---|---|---|---|---|---|---|
| 1. | 3075.0 | $27/2^-$ | | | | **2004Po13** | 1. $\pi(h_{11/2}) \otimes \nu(h_{11/2}^{-2})$ from TAC calculations. |
| | 3416.2 | $29/2^-$ | 341.2 | | | 2003Ma95 | 2. $(\varepsilon_2, \gamma)$ = (0.132, 58°) (2012KuZT). |
| | 3682.5 | $31/2^-$ | 266.3 | 607.6 | 17(+3-2) | | 3. The mean lifetimes (in ps) of levels from 29/2 to 33/2 ħ are 1.1(+4-3), 2.5(+10-5) and 1.1(+4-3), respectively. |
| | 4154.2 | $33/2^-$ | 471.7 | 738.0 | 16(+8-3) | | 4. B(M1) values for the transitions from 341.2 to 471.7 keV are 0.79(+51-14), 0.75(+38-15) and 0.37(+24-6) $\mu_N^2$, respectively. |
| | 4844.6 | $35/2^-$ | 690.4 | 1161.9 | 11(+3-1) | | 5. B(E2) values for the transitions 607.6 and 738.0 keV are 0.041(+23-9) and 0.023(+17-5) $(eb)^2$, respectively. |
| | | | | | | | 6. Irregular band. |
| 2. | 5018.9 | $37/2^-$ | | | | **2004Po13** | 1. $\pi(h_{11/2}) \otimes \nu(h_{11/2}^{-4})$ from TAC calculations. |
| | 5189.1 | $39/2^-$ | 170.2 | | | 2003Ma95 | 2. $(\varepsilon_2, \gamma)$ = (0.12, 58°) (2012KuZT). |
| | 5655.4 | $41/2^-$ | 466.3 | | ≥17 | | 3. The mean lifetimes (in ps) of levels from 41/2 to 45/2 are 0.55(+25-20), 0.8(2) and 0.29(+20-15) respectively. |
| | 5991.7 | $43/2^-$ | 336.3 | 802.6 | 17(+15-2) | | 4. B(M1) values for the transitions from 466.3 to 627.7 keV are 0.77(+76-13), 1.32(+69-22) and 0.50(+84-7) $\mu_N^2$, respectively. |
| | 6619.4 | $(45/2^-)$ | 627.7 | | | | 5. B(E2) value for the 802.6 keV transition is 6.4(+41-20) $(eb)^2$. |
| | | | | | | | 6. Irregular band. |

## $^{141}_{63}Eu_{78}$

| | $E_{level}$ keV | $I^\pi$ | $E_\gamma$(M1) keV | $E_\gamma$(E2) keV | B(M1)/B(E2) $(\mu_N/eb)^2$ | **References** | **Configurations and Comments:** |
|---|---|---|---|---|---|---|---|
| 1. | 5392 | $(39/2^+)$ | | | | **2003Ma95** | 1. According to 2001Rz01, TAC model calculations are required to assign configuration. |
| | 5641 | $(41/2^+)$ | 249 | | | **2001Rz01** | 2. Regular band with backbending at the level energy 7197 keV. |
| | 5976 | $(43/2^+)$ | 335 | | | | 3. Spin and parity suggested by 2001Rz01 and for levels 8544 keV and 9036 keV are assumed by authors of this work. |
| | 6325 | $(45/2^+)$ | 349 | 684 | | | |
| | 6728 | $(47/2^+)$ | 403 | 751 | | | |
| | 7197 | $(49/2^+)$ | 469 | 872 | | | |
| | 7640 | $(51/2^+)$ | 443 | 912 | | | |
| | 8113 | $(53/2^+)$ | 473 | 916 | | | |
| | 8544 | $(55/2^+)$ | 431 | 904 | | | |
| | 9036 | $(57/2^+)$ | 492 | 923 | | | |

## $^{142}_{63}Eu_{79}$

| | $E_{level}$ keV | $I^\pi$ | $E_\gamma$(M1) keV | $E_\gamma$(E2) keV | B(M1)/B(E2) $(\mu_N/eb)^2$ | References | **Configurations and Comments:** |
|---|---|---|---|---|---|---|---|
| 1. | 376.3+X | $10^+$ | | | | **2019Al28** | 1. Configuration assigned as $\pi(h_{11/2})^1 \otimes\nu((h_{11/2})^{-1}$ and $\pi(h_{11/2})^1(d_{5/2}/g_{7/2})^{-2} \otimes\nu((h_{11/2})^{-1}$ respectively, before and after backbending on the basis of TAC-CDFT calculations. |
| | 795.9+X | $11^+$ | 419.6 | | | 1996Pi11 | 2. The mean lifetimes (in ps ) of levels with spin $11^+$ to $20^+$ are 0.79(8), 0.75(+10-13), 0.40(+17-16), <0.90, 3.43(+56-49), 1.46(+22-19), 0.46(+6-5), 1.24(+32-28), <0.74 and <0.81, respectively. |
| | 1099.1+X | $12^+$ | 303.2 | 722.8 | | | 3. The B(M1) values of transitions 419.6 keV to 585.0 keV are 0.55(6), 0.80(+11-14), 0.38(+16-15), >0.06, 1.50(+24-21), 1.11(+17-15), 0.45(+6-5), 0.30(+8-7), >0.06 and >0.04 $\mu_N^2$, respectively. |
| | 1668.9+X | $13^+$ | 569.8 | 873.0 | | | 4. The B(E2) values of transitions 722.8 keV to 1368.6 keV are 55.1(+74-96), 9.1(+39-37), >10.76, 1.18(+19-17), 10.6(+16-14), 16.5(+22-18), 9.6(+25-22), >4.51 and >3.77 (eb)$^2$, respectively |
| | 2130.8+X | $14^+$ | 461.7 | 1031.5 | | | 5. Irregular band |
| | 2288.9+X | $15^+$ | 157.8 | 620.3 | | | |
| | 2543.5+X | $16^+$ | 254.6 | 413.0 | | | |
| | 3057.1+X | $17^+$ | 513.7 | 768.5 | | | |
| | 3434.5+X | $18^+$ | 377.2 | 891.1 | | | |
| | 4218.1+X | $19^+$ | 783.4 | 1161.2 | | | |
| | 4803.1+X | $20^+$ | 585.0 | 1368.6 | | | |
| | 5533.0+X | $21^+$ | 730 | 1315 | | | |
| 2. | 2483.9+X | $13^-$ | | | | **2019Al28** | 1. Configuration assigned as $\pi(h_{11/2})^2(d_{5/2}/g_{7/2})^{-1} \otimes\nu((h_{11/2})^{-1}$ on the basis of TAC-CDFT calculations. |
| | 2610.7+X | $14^-$ | 126.7 | | | 1996Pi11 | 2. The mean lifetimes (in ps ) of levels with spin $14^-$ to $21^{(-)}$ are <4.2, <3.1, 1.36(+22-24), 1.18(+17-18), 0.89(+15-13), 1.17(+25-17), 1.10(+17-15), and <0.46, respectively. |
| | 2751.2+X | $15^-$ | 140.4 | | | | 3. The B(M1) values of transitions 126.7 keV to 585.0 keV are >0.88, >1.84, 2.43(+39-43), 1.95(+28-30), 0.53(+9-8), 0.30(+6-4), 0.17(+3-2), and >0.31 $\mu_N^2$, respectively |
| | 2935.6+X | $16^-$ | 184.5 | 325.0 | | | 4. The B(E2) values of transitions 325.0 keV to 1103.0 keV are 37.6(+61-66), 30.0(+43-46), 12.2(+21-18), 2.56(+55-37), 1.59(+25-22) and >4.94 (eb)$^2$, respectively |
| | 3153.5+X | $17^-$ | 218.4 | 401.8 | | | 5. Regular band. |
| | 3547.8+X | $18^-$ | 394.3 | 611.5 | | | |
| | 3973.6+X | $19^-$ | 425.4 | 820.2 | | | |
| | 4514.5+X | $20^-$ | 540.7 | 967.0 | | | |
| | 5076.6+X | $21^{(-)}$ | 562.0 | 1103.0 | | | |

### $^{143}_{63}\mathrm{Eu}_{80}$

| | $E_{level}$ keV | $I^{\pi}$ | $E_{\gamma}$(M1) keV | $E_{\gamma}$(E2) keV | B(M1)/B(E2) $(\mu_N/eb)^2$ | References | **Configurations and Comments:** |
|---|---|---|---|---|---|---|---|
| 1. | 5869.4 | $35/2^{(+)}$ | | | | **2014Ra18** | 1. $\pi(h_{11/2})^2 \otimes \nu((h_{11/2})^{-2}\pi(g_{7/2})^{-1})$ from SPAC calculations.<br>2. Small prolate deformation ($\beta_2 \approx 0.10$) from SPAC calculations.<br>3. The mean lifetimes (in ps) of levels from 37/2 to 43/2 are 1.10(+20-17), 1.23(+15-12), 0.99(+14-10) and < 1.22 (effective lifetime assuming 10% side feeding), respectively.<br>4. B(M1) values for the transitions from 188.2 to 459.6 keV are 2.86(+52-44), 1.99(+24-19) and 1.15(+16-12) and >0.48 $\mu_N^2$, respectively.<br>5. Regular band with backbending at 47/2 ħ |
| | 6057.7 | $37/2^{+}$ | 188.2 | | | | |
| | 6333.2 | $39/2^{+}$ | 275.5 | | | | |
| | 6694.5 | $41/2^{+}$ | 361.3 | | | | |
| | 7154.2 | $43/2^{+}$ | 459.6 | | | | |
| | 7726.4 | $45/2^{+}$ | 572.2 | | | | |
| | 8263.4 | $47/2^{(+)}$ | 537.0 | | | | |
| 2. | 7272.9 | $41/2^{(-)}$ | | | | **2014Ra18** | 1. $\pi(h_{11/2})^3 \otimes \nu((h_{11/2})^{-2}\pi(g_{7/2})^{-2})$ from SPAC calculations.<br>2. Small prolate deformation ($\beta_2 \approx 0.10$) from SPAC calculations.<br>3. The mean lifetimes (in ps) of levels from 43/2 to 49/2 are 0.79(+14-12), 0.52(+10-8), 0.63(+10-8) and < 0.55 (effective lifetime assuming 10% side feeding), respectively.<br>4. B(M1) values for the transitions from 228.3 to 517.1 keV are 2.93(+52-45), 1.77(+34-27) and 1.32(+21-17) and >0.75 $\mu_N^2$, respectively.<br>5. Regular band |
| | 7500.9 | $43/2^{-}$ | 228.3 | | | | |
| | 7804.1 | $45/2^{-}$ | 303.2 | | | | |
| | 8212.9 | $47/2^{-}$ | 408.8 | | | | |
| | 8730.0 | $49/2^{(-)}$ | 517.1 | | | | |
| | 9295.0 | $(51/2^{-})$ | 565.0 | | | | |

### $^{142}_{64}\mathrm{Gd}_{78}$

| | $E_{level}$ keV | $I^{\pi}$ | $E_{\gamma}$(M1) keV | $E_{\gamma}$(E2) keV | B(M1)/B(E2) $(\mu_N/eb)^2$ | Reference | **Configurations and Comments:** |
|---|---|---|---|---|---|---|---|
| 1. | 4768.2 | $12^{-}$ | | | | **2008Li08** | 1. $\pi(h_{11/2}{}^1\, d_{5/2}{}^{-1}) \otimes \nu(h_{11/2}{}^{-2})$ from the TAC calculations and $\pi(h_{11/2}{}^1\, g_{7/2}{}^{-1}) \otimes \nu(h_{11/2}{}^{-2})$ from SPAC model calculations.<br>2. The mean lifetimes (in ps) of levels from 16 to 20 are 0.50(+16-12), 0.70(+25-15), >1.5, >1.5 and >2.0, respectively.<br>3. B(M1) values for the transitions from 367.7 to 384.0keV are 1.65(+95-31), 0.64(+32-12), < 1, < 0.4 and < 0.5 $\mu_N^2$ respectively.<br>4. Irregular band. |
| | 4990.6 | $13^{-}$ | 222.3 | | | 2005Pa07 | |
| | 5184.0 | $14^{-}$ | 193.3 | | | 2002Li22 | |
| | 5445.4 | $15^{-}$ | 261.4 | | | | |
| | 5813.1 | $16^{-}$ | 367.7 | | | | |
| | 6286.8 | $17^{-}$ | 473.7 | | | | |
| | 6620.6 | $18^{-}$ | 333.8 | | | | |
| | 7071.1 | $19^{-}$ | 450.5 | | | | |
| | 7455.1 | $(20^{-})$ | 384.0 | | | | |
| 2. | 5418.5 | $15^{-}$ | | | | **2008Li08** | 1. $\pi(h_{11/2}{}^1\, g_{7/2}{}^{-1}) \otimes \nu(h_{11/2}{}^{-2})$ from the TAC calculations.<br>2. Small oblate deformation ($\beta_2 \approx 0.106$) suggested from the TAC calculations.<br>3. Irregular band.<br>4. The mean lifetimes (in ps) of states from $17^{-}$ to $21^{-}$ are 1.3(5), 1.1(3), 1.6(+5-3), 0.62(13) and 0.71(+17-15), respectively.<br>5. B(M1) values for the transitions from 286 to 466 keV are 1.48(+156-24), 0.79(+47-13), 0.87(+37-17), 0.53(+22-8) and 0.40(+18-7) $\mu_N^2$, respectively.<br>6. B(E2) values for the transitions from 660 to 993 keV are 5.8(+38-11), 8.8(+39-16), 3.5(+18-8) and 4.8(+23-9) $(eb)^2$, respectively. |
| | 5611.1 | $16^{-}$ | 192.6 | | | 2005Pa07 | |
| | 5896.8 | $17^{-}$ | 285.6 | | | 2002Li22 | |
| | 6271.1 | $18^{-}$ | 374.4 | 660.2 | 13.1(+27-14) | 1997Su11 | |
| | 6566.2 | $19^{-}$ | 295.1 | 669.3 | 9.7(+12-8) | | |
| | 7093.2 | $20^{-}$ | 527.0 | 822.3 | 14.0(+52-18) | | |
| | 7559.5 | $21^{-}$ | 466.4 | 993.4 | 8.1(+15-11) | | |

## $^{142}_{64}Gd_{78}$

| | $E_{level}$ keV | $I^{\pi}$ | $E_{\gamma}$(M1) keV | $E_{\gamma}$(E2) keV | B(M1)/B(E2) $(\mu_N/eb)^2$ | Reference | Configurations and Comments: |
|---|---|---|---|---|---|---|---|
| 3. | 5912.6 | $16^{+}$ | | | | **2008Li08** | 1. $\pi(h_{11/2}^{2}) \otimes \nu(h_{11/2}^{-2})$ from the TAC calculations. |
| | 6176.7 | $17^{+}$ | 264.0 | | | 2005Pa07 | 2. Small oblate deformation ($\beta_2 \approx$ 0.063) suggested from the TAC calculations. |
| | 6477.3 | $18^{+}$ | 300.7 | | | 2002Li22 | 3. Regular band with backbending at $20^{+}$. |
| | 6858.4 | $19^{+}$ | 381.1 | 682.1 | 56(+83-8) | | 4. The mean lifetimes (in ps) of states from $17^{+}$ to $22^{+}$ are 2.2(+8-5), 1.3(+4-3), 0.54(+25-15), 0.52(+20-15), 0.94(+21-16) and 0.98(+40-25), respectively. |
| | 7285.1 | $20^{+}$ | 426.9 | 807.6 | 53(+25-8) | | 5. B(M1) values for the transitions from 264 to 372 keV are 0.19(+11-4), 0.71(+37-14), 1.29(+95-25), 1.10(+77-20), 1.04(+38-18) and 0.91(+58-17) $\mu_N^2$, respectively. |
| | 7645.4 | $21^{+}$ | 360.4 | 787.4 | 55(+20-9) | | 6. B(E2) values for the transitions from 682 to 787 keV are 1.7(+19-7), 1.8(+16-5) and 1.7(+9-4) $(eb)^2$, respectively. |
| | 8017.8 | $22^{+}$ | 372.3 | | | | |
| 4. | 7779.3 | $22^{+}$ | | | | **2008Li08** | 1. $\pi(h_{11/2}^{2}) \otimes \nu(h_{11/2}^{-4})$ from the TAC calculations. |
| | 8248.8 | $23^{+}$ | 469.6 | | | 2005Pa07 | 2. Small oblate deformation ($\beta_2$ = 0.16). |
| | 8592.5 | $24^{+}$ | 343.8 | 812.9 | 13.3(+31-21) | 2002Li22 | 3. Irregular band. |
| | 9140.7 | $25^{+}$ | 548.3 | | | | 4. The mean lifetimes (in ps) of states from $23^{+}$ to $26^{+}$ are 0.33(11), 0.64(15), 0.35(15) and 0.27(15), respectively. |
| | 9700.3 | $26^{+}$ | 559.6 | | | | 5. B(M1) values for the transitions from 470 to 560 keV are 1.22(+100-20), 1.38(+69-24), 0.56(+69-10) and 0.86(+155-12) $\mu_N^2$, respectively. |
| | 10311.7 | $27^{+}$ | 611.4 | | | | 6. B(E2) value for the transition 813 keV is 10.1(+54-19) $(eb)^2$. |
| | 10989.3 | $(28^{+})$ | 677.6 | | | | |
| 5. | 7844.0 | $21^{-}$ | | | | **2008Li08** | 1. $\pi(h_{11/2}^{1}\, g_{7/2}^{-1}) \otimes \nu(h_{11/2}^{-4})$ from the TAC calculations. |
| | 8198.5 | $22^{-}$ | 354.5 | | | 2005Pa07 | 2. Small oblate deformation ($\beta_2$ = 0.16) suggested from the TAC calculations. |
| | 8637.0 | $23^{-}$ | 438.6 | 792.3 | 2.2(+23-4) | 2002Li22 | 3. Irregular band. |
| | 8963.5 | $24^{-}$ | 326.5 | 764.6 | 2.6(+14-7) | | 4. The mean lifetimes (in ps) of states from $22^{-}$ to $24^{-}$ are 0.44(+20-18), 0.77(+25-20) and 1.2(+4-3), respectively. |
| | 9475.0 | $25^{-}$ | 511.3 | 838.2 | 2.3(+16-6) | | 5. B(M1) values for the transitions from 355 to 327 keV are 0.46(+55-8), 0.25(+17-5) and 0.40(+30-10) $\mu_N^2$, respectively. |
| | 9858.6 | $26^{-}$ | 384.4 | 895.0 | 2.3(+16-7) | | 6. B(E2) values for the transitions 792 and 765 are 9.1(+64-29) and 14.4(+95-29) $(eb)^2$, respectively. |

## $^{143}_{64}Gd_{79}$

| | $E_{level}$ keV | $I^{\pi}$ | $E_{\gamma}$(M1) keV | $E_{\gamma}$(E2) keV | B(M1)/B(E2) $(\mu_N/eb)^2$ | Reference | Configurations and Comments: |
|---|---|---|---|---|---|---|---|
| 1. | 3087.2 | $23/2^{-}$ | | | | **2001Rz01** | 1. $\pi(h_{11/2}^{2}) \otimes \nu(h_{11/2}^{-1})$ by comparison with similar band in $^{142}$Gd. |
| | 3158.8 | $25/2^{-}$ | 71.7 | | | 1997Ri16 | 2. According to 2001Rz01, TAC model calculations are required to assign configuration. |
| | 3248.9 | $27/2^{-}$ | 90.3 | | | 2000Li14 | 3. Small oblate deformation. |
| | 3583.1 | $29/2^{-}$ | 334.3 | | | 1998Su04 | 4. Irregular band. |
| | 4015.3 | $31/2^{-}$ | 432.2 | | | | |

## $^{143}_{64}\mathrm{Gd}_{79}$

| | $E_{level}$ keV | $I^{\pi}$ | $E_{\gamma}(M1)$ keV | $E_{\gamma}(E2)$ keV | B(M1)/B(E2) $(\mu_N/eb)^2$ | Reference | **Configurations and Comments:** |
|---|---|---|---|---|---|---|---|
| 2. | 4488.8 | $33/2^-$ | | | | **2001Rz01** | 1. $\pi(h_{11/2}^{2}) \otimes \nu(h_{11/2}^{-3})$ by comparison with band 1. |
| | 4798.6 | $35/2^-$ | 309.8 | | | 1997Ri16 | 2. Tentatively assigned as MR band. |
| | 5027.0 | $37/2^-$ | 228.8 | | | 1998Su04 | 3. Small oblate deformation. |
| | 5310.0 | $39/2^-$ | 283.0 | | | | 4. Irregular band. |
| | 5829 | $(41/2^-)$ | 519 | | | | 5. The level energies and the transition energies |
| | 6259 | $(43/2^-)$ | 430 | 949 | | | are from 1997Ri16. |
| | 6979 | $(45/2^-)$ | 720 | | | | 6. The 949 keV E2 transition is from 2001Rz01. |
| | 7729 | $(47/2^-)$ | 750 | | | | |
| 3. | 5226.3 | $33/2^+$ | | | | **2001Rz01** | 1. Irregular band. |
| | 5399.7 | $35/2^+$ | 173.5 | | | 1997Ri16 | 2. Tentatively assigned as MR band. |
| | 5587.4 | $37/2^+$ | 187.7 | | | 2000Li14 | 3. $B(M1)/B(E2) \geq 10\ (\mu_N/eb)^2$ |
| | 5764.0 | $39/2^+$ | 176.6 | | | 1998Su04 | 4. The level energies and the transition energies |
| | 6159.4 | $(41/2^+)$ | 395.4 | | | | are from 1997Ri16. |
| | 6590.5 | $(43/2^+)$ | 431.1 | 948 | | | 5. The 948 keV E2 transition is from 2000Li14. |
| | 7108 | $(45/2^+)$ | 518 | | | | |
| | 7537 | $(47/2^+)$ | 429 | | | | |
| | 8037 | $(49/2^+)$ | 500 | | | | |
| | 8537 | $(51/2^+)$ | 500 | | | | |

## $^{144}_{64}\mathrm{Gd}_{80}$

| | $E_{level}$ keV | $I^{\pi}$ | $E_{\gamma}(M1)$ keV | $E_{\gamma}(E2)$ keV | B(M1)/B(E2) $(\mu_N/eb)^2$ | Reference | **Configurations and Comments:** |
|---|---|---|---|---|---|---|---|
| 1. | 5370.7 | $14^+$ | | | | **1994Rz01** | 1. Tentatively assigned as $\pi(h_{11/2}^{2})_{K=10^+} \otimes \nu(h_{11/2}^{-2})$ |
| | 5723.6 | $15^+$ | 352.9 | | | | by 1994Rz01 based on dipole rotational bands in |
| | 6214.2 | $16^+$ | 490.6 | | | | Pb nuclei |
| | 6619.0 | $17^+$ | 404.8 | | | | 3. Negative E2/M1 mixing ratios ($\delta_{E2/M1}$) imply an |
| | 7014.6 | $18^+$ | 395.6 | | | | oblate shape ($\beta_2 \sim -0.12$). |
| | 7419.1 | $19^+$ | 404.5 | | | | 4. Irregular band. |
| | 7923.5 | $20^+$ | 504.4 | | | | |
| | 8221.7 | $(21^+)$ | 298.2 | | | | |
| | 8540.4 | $(22^+)$ | 318.7 | | | | |
| | 8993.8 | $(23^+)$ | 453.4 | | | | |

## $^{144}_{65}\mathrm{Tb}_{79}$

| | $E_{level}$ keV | $I^{\pi}$ | $E_{\gamma}(M1)$ keV | $E_{\gamma}(E2)$ keV | B(M1)/B(E2) $(\mu_N/eb)^2$ | Reference | **Configurations and Comments:** |
|---|---|---|---|---|---|---|---|
| 1. | 3515.3 | $16^+$ | | | | **2014Ch22** | 1. Configuration assigned as $\pi(h_{11/2})^3 \otimes \nu(h_{11/2})^{-1}$ |
| | 3789.2 | $17^+$ | 273.9 | | | | from TAC-CDFT calculations. |
| | 4011.9 | $18^+$ | 222.7 | | | | 2. $(\beta_2, \gamma) \approx (0.15, 35^{\circ})$ from TAC-CDFT |
| | 4423.8 | $19^+$ | 411.9 | | | | calculations. |
| | | | | | | | 3. Irregular band. |

## $^{146}_{65}\mathrm{Tb}_{81}$

| | $E_{level}$ keV | $I^\pi$ | $E_\gamma$(M1) keV | $E_\gamma$(E2) keV | B(M1)/B(E2) $(\mu_N/eb)^2$ | Reference | Configurations and Comments: |
|---|---|---|---|---|---|---|---|
| 1. | 7737.5 | $(23^+)$ | | | | **2004Kr14** | 1. Tentatively assigned as $\pi(h_{11/2}^3\ d_{5/2}^{-2})$ |
| | 8004.0 | $(24^+)$ | 266.5 | | | 2004Xi01 | $\otimes\nu(h_{11/2}^{-3}\ f_{7/2}^2)$ from the excitation |
| | 8389.2 | $(25^+)$ | 385.2 | | | | energies and by comparison with the lower |
| | 8875.2 | $(26^+)$ | 486.0 | 870.6 | | | lying bands. |
| | 9304.5 | $(27^+)$ | 429.3 | | | | 2. Tentatively assigned as MR band. |
| | 9717.9 | $(28^+)$ | 413.4 | | | | 3. Irregular band. |
| | 10192.5 | $(29^+)$ | 474.6 | | | | |
| | 10655.6 | $(30^+)$ | 463.1 | | | | |

## $^{144}_{66}\mathrm{Dy}_{78}$

| | $E_{level}$ keV | $I^\pi$ | $E_\gamma$(M1) keV | $E_\gamma$(E2) keV | B(M1)/B(E2) $(\mu_N/eb)^2$ | Reference | Configurations and Comments: |
|---|---|---|---|---|---|---|---|
| 1. | 5153.5 | $15^-$ | | | | **2010Pr04** | 1. Tentatively assigned as $\pi(h_{11/2}^1\ g_{7/2}^{-1})\otimes\nu(h_{11/2}^{-2})$ |
| | 5306.3 | $16^-$ | 152.6 | | | **2009Su09** | configuration based on similar structure observed |
| | 5573.7 | $17^-$ | 267.4 | | | | in $^{142}$Gd. |
| | 5781.3 | $18^-$ | 207.6 | | | | 2. Tentatively assigned as MR band (2009Su09). |
| | 5987.8 | $19^-$ | 206.3 | | | | 3. Irregular band. |
| | 6369.2 | $20^-$ | 381.3 | 588.0 | 1.1(1) | | |
| | 6872.0 | $21^-$ | 502.8 | | | | |
| | 7272.5 | $22^-$ | 400.4 | 903.3 | 6.2(6) | | |
| | 7866.0 | $(23^-)$ | 593.1 | 994.0 | 7.7(11) | | |
| | 8249.2 | $(24^-)$ | 383.2 | 976.8 | 12.7(10) | | |
| | 8762.9 | $(25^-)$ | 513.7 | | | | |

## $^{194}_{81}\mathrm{Tl}_{113}$

| | $E_{level}$ keV | $I^\pi$ | $E_\gamma$(M1) keV | $E_\gamma$(E2) keV | B(M1)/B(E2) $(\mu_N/eb)^2$ | References | Configurations and Comments: |
|---|---|---|---|---|---|---|---|
| 1. | 2516.4 | $16^-$ | | | | **2012PA16** | 1. Configuration assignment as $\pi(h_{9/2})^2(s_{1/2})^{-1}\otimes$ |
| | 2679.0 | $17^-$ | 162.5 | | | 2014Ma55 | $\nu\ (i_{13/2})^{-2}\ p_{3/2}$ from semiclassical calculations. |
| | 2886.1 | $18^-$ | 207.1 | | | 2016Ma13 | 2. Tentatively assigned as MR band (2012Pa16) |
| | 3213.8 | $19^-$ | 327.7 | | | | but authors of 2014Ma55 suggest that 'shears' |
| | 3590.7 | $20^-$ | 376.9 | | | | mechanism not likely involved. |
| | 4019.3 | $21^-$ | 428.6 | | | | 3. $(\beta_2, \gamma)\ \approx (0.06,-80°)$ using TRS calculations. |
| | | | | | | | 4. Regular band. |

## $^{197}_{81}\mathrm{Tl}_{116}$

| | $E_{level}$ keV | $I^\pi$ | $E_\gamma$(M1) keV | $E_\gamma$(E2) keV | B(M1)/B(E2) $(\mu_N/eb)^2$ | References | Configurations and Comments: |
|---|---|---|---|---|---|---|---|
| 1. | 3106.2 | $25/2^+$ | | | | **2019Na08** | 1. Probable Configuration as $\pi\ i_{13/2}\otimes\nu(i_{13/2}^{-2})$ |
| | 3310.9 | $27/2^+$ | 204.7 | | | | using TRS and SPAC calculations. |
| | 3584.1 | $29/2^+$ | 273.1 | 478.0 | 10.4(19) | | 2. $\beta_2 \approx 0.08$, $\gamma \approx -72^o$ on the basis of TRS |
| | 3946.8 | $31/2^+$ | 362.8 | 635.9 | 8.4(22) | | calculations. |
| | 4367.3 | $33/2^+$ | 420.7 | 782.3 | 6.3(5) | | 3. The estimated g-factors for the states $29/2^+$, $31/2^+$ |
| | 4563.9 | $35/2^+$ | 197.1 | 616.5 | | | and $33/2^+$are in the range of 0.57(4)-0.60(4). |
| | | | | | | | 4. Regular band |

## $^{197}_{81}\mathrm{Tl}_{116}$

| | $E_{level}$ keV | $I^{\pi}$ | $E_{\gamma}$(M1) keV | $E_{\gamma}$(E2) keV | B(M1)/B(E2) $(\mu_N/eb)^2$ | References | **Configurations and Comments:** |
|---|---|---|---|---|---|---|---|
| 2. | 3566.1 | $25/2^+$ | | | | **2019Na08** | 1. Probable Configuration as $\pi\, h_{9/2} \otimes \nu(i_{13/2}^{-3})\ (pf)^{-1}$ using TRS and SPAC calculations.<br>2. $\beta2 \approx 0.05$, $\gamma \approx -30^o$ on the basis of TRS minima.<br>3. Irregular band with band crossing at $\hbar\omega = 0.22$ MeV. |
| | 3758.1 | $27/2^+$ | 192.1 | | | | |
| | 3871.4 | $29/2^+$ | 113.4 | | | | |
| | 4075.9 | $31/2^+$ | 204.5 | | | | |
| | 4338.1 | $33/2^+$ | 262.2 | | | | |
| | 4705.6 | $35/2^+$ | 367.5 | | | | |
| | 4881.8 | $37/2^+$ | 176.2 | | | | |
| | 5140.1 | $39/2^+$ | 258.3 | | | | |

## $^{189}_{82}\mathrm{Pb}_{107}$

| | $E_{level}$ keV | $I^{\pi}$ | $E_{\gamma}$(M1) keV | $E_{\gamma}$(E2) keV | B(M1)/B(E2) $(\mu_N/eb)^2$ | References | **Configurations and Comments:** |
|---|---|---|---|---|---|---|---|
| 1. | 2435 | $31/2^-$ | | | | **2015Ho14** | 1. Configuration assigned as $\pi\ [(s_{1/2})^{-2}\ (h_{9/2})\ i_{13/2}]_{11-} \otimes \nu[(i_{13/2})^{-1}]_{13/2+}$ from the comparison with neighboring nuclides.<br>2. The mean lifetime of bandhead is 32 μs (+10-2) (2005Ba51).<br>3. Regular band with backbending at 39/2 ħ. |
| | 2641.4 | $33/2^-$ | 206.4 | | | 2005Ba51 | |
| | 3030.2 | $35/2^{(-)}$ | 388.8 | | | | |
| | 3448.9 | $37/2^{(-)}$ | 418.7 | | | | |
| | 3883.9 | $(39/2^-)$ | 435.0 | | | | |
| | 4297.1 | $(41/2^-)$ | 413.2 | | | | |
| | 4632.2 | $(43/2^-)$ | 335.1 | | | | |

## $^{191}_{82}\mathrm{Pb}_{109}$

| | $E_{level}$ keV | $I^{\pi}$ | $E_{\gamma}$(M1) keV | $E_{\gamma}$(E2) keV | B(M1)/B(E2) $(\mu_N/eb)^2$ | Reference | **Configurations and Comments:** |
|---|---|---|---|---|---|---|---|
| 1. | 2577.5+X | $(29/2^-)$ | | | | **1998Fo02** | 1. Tentatively assigned as $\pi(h_{9/2}i_{13/2}s_{1/2}^{-2})_{K=11}^{-} \otimes \nu(i_{13/2}^{-1})$ below, and $\pi(h_{9/2}i_{13/2}s_{1/2}^{-2})_{K=11}^{-} \otimes \nu(i_{13/2}^{-3})_{K=33/2}^{+}$ above the bandcrossing.<br>2. X ~ 72 keV.<br>3. All $E_{level}$ given here are approximate since $E_{level}$ of $13/2^+$ state is ~ 138 keV.<br>4. Regular band with backbending at 39/2. |
| | 2811.5+X | $(31/2^-)$ | 234.0 | | | | |
| | 3195.1+X | $(33/2^-)$ | 383.6 | | | | |
| | 3604.4+X | $(35/2^-)$ | 409.3 | 792.9 | 23(5) | | |
| | 4030.5+X | $(37/2^-)$ | 426.1 | 835.5 | 20(5) | | |
| | 4377.2+X | $(39/2^-)$ | 346.7 | | | | |
| | 4691.3+X | $(41/2^-)$ | 314.1 | | | | |
| | 4929.9+X | $(43/2^-)$ | 238.6 | | | | |
| | 5207.1+X | $(45/2^-)$ | 277.2 | | | | |
| 2. | 2428.7 | $27/2^+$ | | | | **1998Fo02** | 1. Tentatively assigned as $\pi(h_{9/2}^{2}s_{1/2}^{-2})_{K=8}^{+} \otimes \nu(i_{13/2}^{-1})$ or $\pi(i_{13/2}s_{1/2}^{-1})_{K=7}^{+} \otimes \nu(i_{13/2}^{-1})$ .<br>2. Tentatively assigned as MR band.<br>3. All $E_{level}$ given here are approximate since $E_{level}$ of $13/2^+$ state is ~ 138 keV.<br>4. Regular band. |
| | 2765.9 | $(29/2^+)$ | 337.2 | | | | |
| | 3141.4 | $(31/2^+)$ | 375.5 | | | | |
| | 3551.3 | $(33/2^+)$ | 409.9 | | | | |

## $^{192}_{82}Pb_{110}$

| | $E_{level}$ keV | $I^\pi$ | $E_\gamma$(M1) keV | $E_\gamma$(E2) keV | B(M1)/B(E2) $(\mu_N/eb)^2$ | Reference | **Configurations and Comments:** |
|---|---|---|---|---|---|---|---|
| 1. | 4241.2 | $15^-$ | | | | **1993Pl02** | 1. Tentatively assigned as $\pi(9/2[505] \otimes 13/2[606])$ |
| | 4370.1 | $16^-$ | 128.9 | | | | $\otimes\nu(i_{13/2}^2)$ based on CSM-TRS calculations and |
| | 4519.2 | $17^-$ | 149.1 | | >2.38 | | by comparison with $^{191}$Tl. |
| | 4702.3 | $18^-$ | 183.1 | | >7.69 | | 2. Small oblate deformation. |
| | 4989.6 | $19^-$ | 287.3 | | >6.67 | | 3. Tentatively assigned as MR band by 1993Pl02. |
| | 5276.9 | $20^-$ | 287.3 | | >16.67 | | 4. The limits on B(M1)/B(E2) are by assuming |
| | 5559.5 | $21^-$ | 282.6 | | >11.11 | | that the unobserved E2 transitions are at the |
| | 5708.6 | $(22^-)$ | 149.1 | 431.7 | <20 | | most half intense than the 489.5 keV $\gamma$ ray in |
| | | | | | | | band 2. |
| | | | | | | | 5. Regular band with backbending at spin 21. |
| 2. | 4963.0 | $18^-$ | | | | **1993Pl02** | 1. Tentatively assigned as $\pi(7/2[514] \otimes 13/2[606])$ |
| | 5087.1 | $19^-$ | 124.1 | | >50 | | $\otimes\nu(i_{13/2}^2)$ from the CSM-TRS calculations and |
| | 5286.3 | $20^-$ | 199.2 | | >4 | | by comparison with $^{191}$Tl. |
| | 5531.7 | $21^-$ | 245.4 | | >11.11 | | 2. Small oblate deformation. |
| | 5871.0 | $22^-$ | 339.3 | | >12.5 | | 3. Tentatively assigned as MR band by 1993Pl02. |
| | 6232.1 | $23^-$ | 361.1 | | >5.88 | | 4. The limits on B(M1)/B(E2) are by assuming |
| | 6666.0 | $(24^-)$ | 433.9 | | >4 | | that the unobserved E2 transitions are at the |
| | 7155.5 | $(25^-)$ | 489.5 | | >5.88 | | most half-intense than the 489.5 keV $\gamma$ ray. |
| | | | | | | | 5. Regular band. |

## $^{193}_{82}Pb_{111}$

| | $E_{level}$ keV | $I^\pi$ | $E_\gamma$(M1) keV | $E_\gamma$(E2) keV | B(M1)/B(E2) $(\mu_N/eb)^2$ | References | **Configurations and Comments:** |
|---|---|---|---|---|---|---|---|
| 1. | 2584.8+X | $29/2^-$ | | | | **1996Du18** | 1. $\pi(9/2[505] \otimes 13/2[606])_{K=11^-} \otimes\nu(i_{13/2})$ by |
| | 2686.9+X | $31/2^-$ | 102.1 | | | 1996Ba54 | comparison with similar bands in neighboring |
| | 2939.2+X | $33/2^-$ | 252.3 | | | 1997Ch33 | Pb nuclei. |
| | 3320.7+X | $35/2^-$ | 381.5 | 633.8 | 28(5) | 2005Gl09 | 2. Oblate deformation. |
| | 3722.3+X | $37/2^-$ | 401.6 | 783.1 | 22(4) | 2011Ba02 | 3. X ~ 100 keV from systematics. |
| | 4136.1+X | $39/2^-$ | 413.8 | 815.4 | 16(6) | | 4. For bandhead $T_{1/2}$ = 9.4(7) ns and g factor = |
| | 4470.6+X | $41/2^-$ | 334.5 | 748.3 | | | 0.68(3) (1997Ch33). |
| | 4828.3+X | $43/2^-$ | 357.7 | 692.3 | | | 5. The mean lifetimes (in ps) for the transitions 252 |
| | 5218.6+X | $45/2^-$ | 390.3 | | | | and 381 keV as given in 2005Gl09 are 3.2(8) and |
| | | | | | | | $\leq$ 1, respectively. |
| | | | | | | | 6. The B(M1) values as given in 2005Gl09 for the |
| | | | | | | | transitions 252 and 381 keV are 1.1(2) and $\geq$ 1.4 |
| | | | | | | | $\mu_N^2$, respectively. |
| | | | | | | | 7. The B(E2) value for the transition 633 keV as |
| | | | | | | | given in 2005Gl09 is $\geq$ 0.1 $(eb)^2$. |
| | | | | | | | 8. Regular band with backbending at 41/2. |
| | | | | | | | 9. 2011Ba02 propose band head spin as 27/2-, |
| | | | | | | | implying that all the spins should be less by one |
| | | | | | | | unit in this band. |

## $^{193}_{82}\mathrm{Pb}_{111}$

| | $E_{level}$ keV | $I^\pi$ | $E_\gamma$(M1) keV | $E_\gamma$(E2) keV | B(M1)/B(E2) $(\mu_N/eb)^2$ | References | **Configurations and Comments:** |
|---|---|---|---|---|---|---|---|
| 2. | 4297.7+X | $(39/2^+)$ | | | | **1996Du18** | 1. Tentatively assigned as $\pi(9/2[505]\otimes 13/2[606])_{K=11^-}\otimes\nu(i_{13/2}^2\, p_{3/2})$ by comparison with similar bands in neighboring Pb nuclei.<br>2. Oblate deformation.<br>3. X ~ 100 keV from systematics.<br>4. The B(M1) values as given in 1998Cl06 for the transitions from 291 to 416 keV are 5.27(64), 4.32(+56-75), 4.01(+95-76) and 2.83(34) $\mu_N^2$, respectively.<br>5. The mean lifetimes (in ps) of levels having spin values from 45/2 to 51/2 as given in 1998Cl06 are 0.33(4), 0.23(+4-3), 0.21(+4-5) and 0.25(3), respectively.<br>6. Regular band with backbending at spin 59/2. |
| | 4387.7+X | $(41/2^+)$ | 90.0 | | | 1998Cl06 | |
| | 4536.6+X | $(43/2^+)$ | 148.9 | | | 1996Ba54 | |
| | 4768.6+X | $(45/2^+)$ | 232.0 | | | | |
| | 5060.2+X | $(47/2^+)$ | 291.6 | | | | |
| | 5425.4+X | $(49/2^+)$ | 365.2 | 656.8 | 15(3) | | |
| | 5815.0+X | $(51/2^+)$ | 389.6 | 754.7 | 12(3) | | |
| | 6231.1+X | $(53/2^+)$ | 416.1 | 805.6 | | | |
| | 6657.2+X | $(55/2^+)$ | 426.1 | 842.2 | 15(3) | | |
| | 7089.9+X | $(57/2^+)$ | 432.7 | 858.8 | | | |
| | (7516.0+X) | $(59/2^+)$ | (426.1) | | | | |
| | (7932.1+X) | $(61/2^+)$ | (416.1) | | | | |
| 3. | 4944.8+X | $(43/2^+)$ | | | | **1996Du18** | 1. $\pi(9/2[505]\otimes 13/2[606])_{K=11^-}\otimes\nu(i_{13/2}^2\, f_{5/2})$ by comparison with similar bands in neighboring Pb nuclei.<br>2. Oblate deformation.<br>3. X ~ 100 keV from systematics.<br>4. Regular band. |
| | 5169.1+X | $(45/2^+)$ | 224.3 | | | | |
| | 5436.6+X | $(47/2^+)$ | 267.5 | | | | |
| | 5762.8+X | $(49/2^+)$ | 326.2 | | | | |
| | 6145.2+X | $(51/2^+)$ | 382.4 | | | | |
| 4. | 5092.7+X | $(45/2^-)$ | | | | **1996Du18** | 1. Tentatively assigned as $\pi(9/2[505]\otimes 13/2[606])_{K=11^-}\otimes\nu(i_{13/2})^3$ by comparison with similar bands in neighboring Pb nuclei.<br>2. Oblate deformation.<br>3. X ~ 100 keV from systematics.<br>4. Regular band. |
| | 5331.8+X | $(47/2^-)$ | 239.1 | | | 1996Ba54 | |
| | 5597.4+X | $(49/2^-)$ | 265.6 | | | | |
| | 5926.9+X | $(51/2^-)$ | 329.5 | | | | |
| | 6302.5+X | $(53/2^-)$ | 375.6 | | | | |
| | 6715.4+X | $(55/2^-)$ | 412.9 | | | | |
| | 7154.6+X | $(57/2^-)$ | 439.2 | | | | |
| 5. | 5825.3+X | $(49/2^-)$ | | | | **1996Du18** | 1. Tentatively assigned as $\pi(9/2[505]\otimes 13/2[606])_{K=11^-}\otimes\nu(i_{13/2})^3$ or $\pi(9/2[505]\otimes 13/2[606])_{K=11^-}\otimes\nu(i_{13/2}h_{9/2}^2)$ from HF+BCS calculations.<br>2. X ~ 100 keV from systematics.<br>3. Parity assignment is based on three M1 transitions to band 1.<br>4. Regular band. |
| | 6001.6+X | $(51/2^-)$ | 176.3 | | | | |
| | 6285.3+X | $(53/2^-)$ | 283.7 | | | | |
| | 6597.2+X | $(55/2^-)$ | 311.9 | | | | |
| | 6927.6+X | $(57/2^-)$ | 330.4 | | | | |
| | 7312.1+X | $(59/2^-)$ | 384.5 | | | | |
| | 7713.6+X | $(61/2^-)$ | 401.5 | | | | |

## $^{194}_{82}\mathrm{Pb}_{112}$

| | $E_{level}$ keV | $I^\pi$ | $E_\gamma$(M1) keV | $E_\gamma$(E2) keV | B(M1)/B(E2) $(\mu_N/eb)^2$ | References | **Configurations and Comments:** |
|---|---|---|---|---|---|---|---|
| 1. | 6416.0 | $22^+$ | | | | **2009Ku03** | 1. Tentatively assigned as $\nu\{h_{11/2}^2\, i_{13/2}\,(p_{3/2}f_{5/2})^1\}\otimes\pi(9/2[505]\otimes 13/2[606])_{K=11^-}$ from the excitation energy, spin and parity.<br>2. Irregular band. |
| | 6763.9 | $23^+$ | 347.9 | | | 2002Ka01 | |
| | 7114.7 | $24^+$ | 351.5 | | | | |
| 2. | 4264.6 | $14^-$ | | | | **2009Ku03** | 1. The configuration assigned as $\nu\{(i_{13/2})^{-1}(p_{3/2})^1\}\otimes\pi\{(h_{9/2})^2\}_{8+}$<br>2. Regular band. |
| | 4407.8 | $15^-$ | 143.2 | | | | |
| | 4691.7 | $16^-$ | 283.9 | | | | |
| | 5052.7 | $17^-$ | 361.0 | 644.9 | 4.7(19) | | |
| | 5433.2 | $18^-$ | 380.5 | 741.5 | 6.5(30) | | |
| | 5818.0 | $19^-$ | 384.8 | 765.3 | 7.1(45) | | |

## $^{194}_{82}Pb_{112}$

| | E<sub>level</sub> keV | Iπ | Eγ(M1) keV | Eγ(E2) keV | B(M1)/B(E2) $(\mu_N/eb)^2$ | References | **Configurations and Comments:** |
|---|---|---|---|---|---|---|---|
| 3. | 4640.7 | $(15^+)$ | | | | **2009Ku03** | 1. The configuration $\nu\{(i_{13/2})^{-2}\}\otimes\pi\{(h_{9/2})^2\}_{8+}$ and |
| | 4725.7 | $16^+$ | 85 | | | 2002Ka01 | $\nu\{(i_{13/2})^{-4}\}\otimes\pi\{(h_{9/2})^2\}_{8+}$ are assigned before |
| | 4888.3 | $17^+$ | 162.6 | | | 1993Me12 | and after band crossing. |
| | 5121.1 | $18^+$ | 232.8 | | | | 2. Regular band with backbending at I= 22ħ. |
| | 5409.2 | $19^+$ | 288.1 | | | | |
| | 5756.8 | $20^+$ | 347.6 | 635.7 | 48(25) | | |
| | 6131.0 | $21^+$ | 374.2 | 721.8 | 8.9(39) | | |
| | 6527.8 | $22^+$ | 396.8 | 771.0 | 12.9(61) | | |
| | 6905.0 | $(23^+)$ | 377.2 | | | | |
| | 7276.5 | $(24^+)$ | 371.5 | | | | |
| | 7637.5 | $(25^+)$ | 361.0 | | | | |
| 4. | 4642.6 | $15^+$ | | | | **2009Ku03** | 1. The configuration $\nu\{(i_{13/2})^{-1}(p_{3/2})^1\}\otimes$ |
| | 4700.9 | $(16^+)$ | 58.3 | | | 2002Ka01 | $\pi(9/2[505]\otimes13/2[606])_{K=11^-}$ and $\nu\{(i_{13/2})^{-3}$ |
| | 4766.4 | $17^+$ | 65.5 | | | 1998Ka59 | $(p_{3/2})^1\}\otimes\pi(9/2[505]\otimes13/2[606])_{K=11^-}$ |
| | 5629.2 | $20^+$ | 396.8 | 699.8 | 31(10) | 1995Ka19 | are assigned before and after band crossing. |
| | 6005.5 | $21^+$ | 376.3 | 773.1 | 18.6(62) | | 2. Regular band with backbending at I =20ħ. |
| | 6368.9 | $22^+$ | 363.4 | 739.7 | 16.7(58) | | |
| | 6629.6 | $23^+$ | 260.7 | 624.1 | 52(26) | | |
| | 6841.9 | $24^+$ | 212.3 | | | | |
| | 7069.5 | $25^+$ | 227.6 | | | | |
| | 7336.3 | $26^+$ | 266.8 | | | | |
| | 7643.2 | $27^+$ | 306.9 | | | | |
| | 8004.0 | $28^+$ | 360.8 | | | | |
| | 8398.1 | $29^+$ | 394.1 | | | | |
| | 8819.0 | $(30^+)$ | 420.9 | | | | |
| | 9260.6 | $(31^+)$ | 440.9 | | | | |
| | 9722.6 | $(32^+)$ | 462.0 | | | | |
| | 10206.8 | $(33^+)$ | 484.2 | | | | |
| 5. | 4985.7 | $17^-$ | | | | **2009Ku03** | 1. The configuration $\nu\{(i_{13/2})^{-2}\}\otimes$ |
| | 5105.2 | $18^-$ | 119.5 | | | 2002Ka01 | $\pi(9/2[505]\otimes13/2[606])_{K=11^-}$ and $\nu\{(i_{13/2})^{-4}$ |
| | 5250.2 | $19^-$ | 145.0 | | | 1993Me12 | $\otimes\pi(9/2[505]\otimes13/2[606])_{K=11^-}$ |
| | 5447.2 | $20^-$ | 197.0 | | | 1994Po08 | are assigned before and after band crossing. |
| | 5707.4 | $21^-$ | 260.2 | | | 1995Ka19 | 2. Regular band with backbending at I=26ħ. |
| | 6043.4 | $22^-$ | 336.0 | 596.2 | 10.7(39) | 1998Cl06 | 3. The B(M1) values as given in 1998Cl06 for the |
| | 6419.4 | $23^-$ | 376.0 | 712.0 | 14.4(54) | 1998Ka59 | transitions from 260 to 417 keV are 9.79(+255 |
| | 6836.2 | $24^-$ | 416.8 | 792.8 | 13.9(59) | | -170), 5.86(+56-56), 5.13(+114-143) and 3.90 |
| | 7260.2 | $25^-$ | 424.0 | 840.8 | 13.6(70) | | (87) $\mu_N^2$, respectively. |
| | 7702.0 | $26^-$ | 441.8 | 865.8 | 10.8(53) | | 4. The mean lifetimes of levels having spin values |
| | 8130.5 | $27^-$ | 428.5 | | | | from 20 to 23 as given in 1998Cl06 are |
| | 8515.4 | $(28^-)$ | 384.9 | | | | 0.23(+4-6), 0.21(2), 0.18(+5-4) and 0.18(4) ps, |
| | 8882.4 | $(29^-)$ | 367.0 | | | | respectively. |
| | 9254.9 | $(30^-)$ | 372.5 | | | | |
| 6. | 5993.0 | $20^-$ | | | | **2009Ku03** | 1. The configuration $\nu\{(i_{13/2})^{-2}(p_{3/2}f_{5/2})^1\}\otimes$ |
| | 6122.2 | $21^-$ | 129.2 | | | 2002Ka01 | $\pi(9/2[505]\otimes13/2[606])_{K=11^-}$ and $\nu\{(i_{13/2})^{-4}$ |
| | 6318.5 | $22^-$ | 196.3 | | | | $(p_{3/2}f_{5/2})^1\}\otimes\pi(9/2[505]\otimes13/2[606])_{K=11^-}$ |
| | 6527.0 | $23^-$ | 208.5 | | | | are assigned before and after band crossing. |
| | 6797.0 | $24^-$ | 270.0 | | | | 2. Regular band with backbending at I= 27ħ. |
| | 7125.9 | $25^-$ | 328.9 | | | | |
| | 7488.9 | $26^-$ | 363.0 | | | | |
| | 7861.9 | $27^-$ | 373.0 | | | | |
| | 8258.7 | $28^-$ | 396.8 | | | | |
| | 8646.6 | $29^-$ | 387.9 | | | | |
| | 9038.1 | $(30^-)$ | 391.5 | | | | |
| | 9439.1 | $(31^-)$ | 401.0 | | | | |

## $^{194}_{82}Pb_{112}$

| | $E_{level}$ keV | $I^{\pi}$ | $E_{\gamma}$(M1) keV | $E_{\gamma}$(E2) keV | B(M1)/B(E2) $(\mu_N/eb)^2$ | References | **Configurations and Comments:** |
|---|---|---|---|---|---|---|---|
| 7. | 6307.9 | $(21^-)$ | | | | **2009Ku03** | 1. The configuration $\nu\{(i_{13/2})^{-2}(p_{3/2}f_{5/2})^1\}$ |
| | 6510.0 | $(22^-)$ | 202.1 | | | 2002Ka01 | $\otimes\pi\{(h_{9/2})^2\}_{8+}$ |
| | 6758.8 | $(23^-)$ | 248.8 | | | | 2. Regular band |
| | 7034.8 | $(24^-)$ | 276.0 | | | | |
| | 7351.9 | $(25^-)$ | 317.1 | | | | |
| | 7715.1 | $(26^-)$ | 363.2 | | | | |
| | 8100.0 | $(27^-)$ | 384.9 | | | | |
| | 8513 | $(28^-)$ | 413.0 | | | | |
| | | | | | | | |
| 8. | X | I | | | | **2009Ku03** | 1. The configuration $\nu\{(i_{13/2})^{-1}(f_{5/2})^1\}\otimes$ |
| | X+154.6 | I+1 | 154.6 | | | | $\pi(9/2[505]\otimes13/2[606])_{K=11^-}$ and $\nu\{(i_{13/2})^{-3}(f_{5/2})^1\}$ |
| | X+456.4 | I+2 | 301.8 | | | | $\otimes\pi(9/2[505]\otimes13/2[606])_{K=11^-}$ are assigned before |
| | X+857.8 | I+3 | 401.4 | | | | and after band crossing. |
| | X+1245.5 | I+4 | 387.7 | | | | 2. Irregular band. |
| | X+1643.0 | I+5 | 397.5 | | | | 3. Most probable spin and parity of first observed |
| | X+1928.8 | I+6 | 285.8 | | | | level is $16^+$ or $17^+$. |
| | X+2152.1 | I+7 | 223.3 | | | | |
| | X+2395.0 | I+8 | 242.9 | | | | |

## $^{195}_{82}Pb_{113}$

| | $E_{level}$ keV | $I^{\pi}$ | $E_{\gamma}$(M1) keV | $E_{\gamma}$(E2) keV | B(M1)/B(E2) $(\mu_N/eb)^2$ | References | **Configurations and Comments:** |
|---|---|---|---|---|---|---|---|
| 1. | 2968.3 | $27/2^-$ | | | | **1996Ka15** | 1. $\pi(9/2[505]\otimes13/2[606])_{K^{\pi}=11^-}\otimes\nu(i_{13/2})$ by |
| | 3098.0 | $29/2^-$ | 129.7 | | | 1995Fa19 | comparison with the neighboring $^{194}$Pb and |
| | 3362.0 | $31/2^-$ | 264.0 | | | | $^{195}$Tl. |
| | 3734.7 | $33/2^-$ | 372.7 | 637.0 | 16(4) | | 2. Oblate shape |
| | 4119.9 | $35/2^-$ | 385.2 | 757.7 | 18(3) | | 3. Parities are from 1995Fa19. |
| | 4566.2 | $(37/2^-)$ | 446.3 | 832.0 | 13(3) | | 4. Irregular band with backbending at the top of |
| | 4966.8 | $(39/2^-)$ | 400.6 | | | | the band. |
| | | | | | | | |
| 2. | 5123.6 | $(39/2^-)$ | | | | **1996Ka15** | 1. $\pi(9/2[505]\otimes13/2[606])_{K^{\pi}=11^-}\otimes\nu(i_{13/2}{}^3)$ by |
| | 5270.4 | $(41/2^-)$ | 146.8 | | | 1995Fa19 | comparison with the neighboring $^{194}$Pb and |
| | 5467.7 | $(43/2^-)$ | 197.3 | | | 1998Cl06 | $^{195}$Tl. |
| | 5702.5 | $(45/2^-)$ | 234.8 | | | | 3. Regular band with oblate shape. |
| | 6308.1 | $(49/2^-)$ | 329.7 | | | | 4. The B(M1) values as given in 1998Cl06 for the |
| | 6674.2 | $(51/2^-)$ | 366.1 | | | | transitions from 276 to 366 keV are 7.01(+200 |
| | 7090.8 | $(53/2^-)$ | 416.6 | | | | -125), 6.14(88) and 4.48 (+41-61) $\mu_N^2$, |
| | 7536.8 | $(55/2^-)$ | (446.0) | | | | respectively. |
| | | | | | | | 5. The mean lifetimes (in ps) of levels having spin values from 47/2 to 51/2 as given in 1998Cl06 are 0.28(+5-8), 0.21(3) and 0.22(+3-2), respectively. |
| | | | | | | | |
| 3. | 4465.6 | $(33/2^-)$ | | | | **1996Ka15** | 1. $\pi(9/2[505]\otimes13/2[606])_{K=11}\otimes\nu(i_{13/2}{}^2(f_{5/2}/p_{3/2})^1)$ |
| | 4560.4 | $(35/2^-)$ | 94.8 | | | 1995Fa19 | at low spin and $\pi(9/2[505]\otimes13/2[606])_{K=11}\otimes$ |
| | 4693.9 | $(37/2^-)$ | 133.5 | | | | $\nu(i_{13/2}{}^4(f_{5/2}/p_{3/2})^1)$ at high spin, by comparison |
| | 4866.5 | $(39/2^-)$ | 172.6 | | | | with a similar band of $^{194}$Pb. |
| | 5108.1 | $(41/2^-)$ | 241.6 | | | | 2. Oblate shape ($\beta_2\sim$ -0.15). |
| | 5412.9 | $(43/2^-)$ | 304.8 | | | | 3. Regular band with backbending at the top of |
| | 5770.9 | $(45/2^-)$ | 358.0 | 663.0 | 17(4) | | the band. |
| | 6144.7 | $(47/2^-)$ | 373.8 | 732.0 | 23(5) | | 4. Parities are from 1995Fa19. |
| | 6529.5 | $(49/2^-)$ | 384.8 | 759.0 | 13(3) | | |
| | 6907.2 | $(51/2^-)$ | 377.7 | 763.0 | 10(3) | | |
| | 7281.2 | $(53/2^-)$ | 374.0 | | | | |

## $^{196}_{82}\mathrm{Pb}_{114}$

| | $E_{level}$ keV | $I^\pi$ | $E_\gamma$(M1) keV | $E_\gamma$(E2) keV | B(M1)/B(E2) $(\mu_N/eb)^2$ | References | Configurations and Comments: |
|---|---|---|---|---|---|---|---|
| 1. | 4385.0 | $14^-$ | | | | **2002Si20** | 1. $\pi(h_{9/2}^2)_{K=8^+}\otimes\nu[i_{13/2}^{-1}(p_{3/2}f_{5/2})^1]$ before and $\pi(h_{9/2}^2)_{K=8^+}\otimes\nu[i_{13/2}^{-3}(p_{3/2}f_{5/2})^1]$ after the band crossing from the TAC calculations.<br>2. Oblate deformation ($\lvert\beta_2\rvert<0.1$).<br>3. Irregular band with backbending at $19^-$. |
| | 4561.2 | $15^-$ | 176.2 | | | | |
| | 4864.4 | $16^-$ | 303.2 | | | | |
| | 5212.3 | $17^-$ | 347.9 | 651.1 | 12(4) | | |
| | 5558.8 | $18^-$ | 346.5 | 694.4 | 14(4) | | |
| | 5896.2 | $19^-$ | 337.4 | 683.9 | 9(3) | | |
| | 6160.3 | $20^-$ | 264.1 | | | | |
| | 6369.3 | $21^-$ | 209.0 | | | | |
| | 6602.1 | $22^-$ | 232.8 | | | | |
| | 6881.3 | $23^-$ | 279.2 | | | | |
| | 7211.7 | $24^-$ | 330.4 | | | | |
| | 7564.0 | $25^-$ | 352.3 | | | | |
| | 7940.5 | $26^-$ | 376.5 | | | | |
| | 8383.1 | $27^-$ | 442.6 | | | | |
| | 8850.8 | $28^-$ | 467.7 | | | | |
| | 9374.5 | $29^-$ | 523.7 | | | | |
| 2. | 4658.2 | $14^+$ | | | | **2002Si20** | 1. $\pi(h_{9/2}^2)_{K=8^+}\otimes\nu(i_{13/2}^{-2})$ before and $\pi(h_{9/2}^2)_{K=8^+}\otimes\nu(i_{13/2}^{-4})$ after the band crossing from the TAC calculations.<br>2. Oblate deformation ($\lvert\beta_2\rvert<0.1$).<br>3. Regular band that becomes irregular after $20^+$. |
| | 4748.2 | $15^+$ | 90.0 | | | | |
| | 4852.5 | $16^+$ | 104.3 | | | | |
| | 5035.2 | $17^+$ | 182.7 | | | | |
| | 5283.3 | $18^+$ | 248.1 | | | | |
| | 5577.2 | $19^+$ | 293.9 | 542.0 | 15(4) | | |
| | 5934.0 | $20^+$ | 356.8 | 650.7 | 13(3) | | |
| | 6294.1 | $21^+$ | 360.1 | 716.9 | $<16$ | | |
| | 6689.7 | $22^+$ | 395.6 | 755.7 | 17(5) | | |
| | 7074.4 | $23^+$ | 384.7 | 780.3 | 16(5) | | |
| | 7465.0 | $24^+$ | 390.6 | 775.3 | 14(5) | | |
| | 7825.6 | $25^+$ | 360.6 | 751.2 | | | |
| | 8166.3 | $26^+$ | 340.7 | | | | |
| | 8516.7 | $27^+$ | 350.4 | | | | |
| 3. | 4995.4 | $17^+$ | | | | **2002Si20** | 1. $\pi(i_{13/2}h_{9/2})_{K=11^-}\otimes\nu[i_{13/2}^{-1}(p_{3/2}f_{5/2})^1]$ before and $\pi(i_{13/2}h_{9/2})_{K=11^-}\otimes\nu[i_{13/2}^{-3}(p_{3/2}f_{5/2})^1]$ after the band crossing from the TAC calculations.<br>2. Oblate deformation ($\lvert\beta_2\rvert<0.1$).<br>3. Regular band with backbending at $20^+$.<br>4. The mean lifetimes (in ps) of levels from $18^+$ to $20^+$ as given in 2001Ke12 are 1.28(22), 0.96(20) and 1.34(50), respectively, those of levels from $25^+$ to $27^+$as given in 1998Cl06 are 0.19(+3-4), 0.17(+3-4), and 0.15(3), respectively and those of levels from $28^+$ to $31^+$ are 0.13(+3-2), 0.14(2), 0.13(3) and 0.16(3), respectively (2002Si29).<br>5. B(M1) values for the transitions from 193 to 375 keV as given in 2001Ke12 are 2.4(+5-3), 1.4(+4-2) and 0.7(+4-2) $\mu_N^2$, respectively, for transitions 286 and 339 keV as given in 1998Cl06 are 9.57(+201-151) and 7.05(+166 -124) and those for the transitions from 398 to 526 keV as given in 2002Si29 are 5.12(+80-119), 3.24(50), 2.75(66) and 1.80(36) $\mu_N^2$, respectively.<br>6. B(E2) values for the transitions from 736 to 1017 keV as given in 2002Si29 are 0.269(+82-94), 0.191(63), 0.131(54) and 0.078(31) $(eb)^2$. |
| | 5188.1 | $18^+$ | 192.7 | | | 2002Si29 | |
| | 5502.6 | $19^+$ | 314.5 | 507.2 | 16(4) | 2001Ke12 | |
| | 5877.2 | $20^+$ | 374.6 | 689.1 | 19(4) | 1996Ba53 | |
| | 6232.5 | $21^+$ | 355.3 | 729.9 | 21(4) | 1993Hu01 | |
| | 6574.1 | $22^+$ | 341.6 | 696.9 | 19(5) | 1995Mo01 | |
| | 6817.7 | $23^+$ | 243.6 | 585.2 | 47(12) | 1998Cl06 | |
| | 7027.5 | $24^+$ | 209.8 | 453.4 | 42(13) | | |
| | 7266.9 | $25^+$ | 239.4 | 449.2 | | | |
| | 7553.0 | $26^+$ | 286.1 | 525.5 | | | |
| | 7891.7 | $27^+$ | 338.7 | 624.8 | 25(7) | | |
| | 8289.4 | $28^+$ | 397.7 | 736.4 | 19(5) | | |
| | 8738.1 | $29^+$ | 448.7 | 846.4 | 17(5) | | |
| | 9228.3 | $30^+$ | 490.2 | 938.9 | 21(7) | | |
| | 9754.7 | $31^+$ | 526.4 | 1016.6 | 23(8) | | |
| | 10310.6 | $32^+$ | 555.9 | 1082.3 | 16(8) | | |
| | 10883.7 | $33^+$ | 573.1 | 1129.0 | | | |
| | 11456.3 | $34^+$ | 572.6 | 1145.7 | | | |
| | 12023.0 | $35^+$ | 566.7 | | | | |
| | 12585.5 | $36^+$ | 562.5 | | | | |

## $^{196}_{82}Pb_{114}$

| | $E_{level}$ keV | $I^\pi$ | $E_\gamma$(M1) keV | $E_\gamma$(E2) keV | B(M1)/B(E2) $(\mu_N/eb)^2$ | References | Configurations and Comments: |
|---|---|---|---|---|---|---|---|
| 4. | 5155.3 | $16^-$ | | | | **2002Si20** | 1. $\pi(i_{13/2}h_{9/2})_{K=11^-}\otimes\nu(i_{13/2}^{-2})$ before and |
| | 5236.0 | $17^-$ | 80.7 | | | 2002Vy02 | $\pi(i_{13/2}h_{9/2})_{K=11^-}\otimes\nu(i_{13/2}^{-4})$ after the |
| | 5342.9 | $18^-$ | 106.9 | | | 1995Mo01 | band crossing from the TAC calculations. |
| | 5480.9 | $19^-$ | 138.0 | | | 1993Hu01 | 2. Oblate deformation ($\lvert\beta_2\rvert<0.1$). |
| | 5684.8 | $20^-$ | 203.9 | | | 1996Ba53 | 3. Regular band with backbending at $26^-$. |
| | 5952.6 | $21^-$ | 267.8 | 471.7 | 24(6) | | 4. The value of deformation parameter $\beta_2$ = -0.13 |
| | 6284.4 | $22^-$ | 331.8 | 599.6 | 33(9) | | is based on experimental measurement of Q for |
| | 6651.4 | $23^-$ | 367.0 | 698.8 | 20(5) | | $11^-$ level by 2002Vy02. |
| | 7043.6 | $24^-$ | 392.2 | 759.2 | 19(5) | | 5. The mean lifetimes (in ps) of levels having spin |
| | 7441.3 | $25^-$ | 397.7 | 789.9 | 17(4) | | values from $23^-$ to $28^-$ as given in 1995Mo01 are |
| | 7849.4 | $26^-$ | 408.3 | 805.8 | 15(4) | | ≤0.4, 0.21(+15-12), 0.17(+12 -8), 0.39(11), |
| | 8222.8 | $27^-$ | 373.4 | 781.5 | 29(8) | | 0.47(+10-14) and 0.23(9), respectively. |
| | 8556.3 | $28^-$ | 333.5 | 706.9 | 38(9) | | |
| | 8892.8 | $29^-$ | 336.5 | 670.0 | 27(10) | | |
| | 9251.3 | $30^-$ | 358.5 | 695.0 | 16(5) | | |
| | 9646.8 | $31^-$ | 395.5 | 754.0 | 14(5) | | |
| | 10088.9 | $32^-$ | 442.1 | 837.6 | 11(4) | | |
| | 10578.3 | $33^-$ | 489.4 | | | | |
| | 11111.8 | $34^-$ | 533.5 | | | | |
| | 11683.1 | $35^-$ | 571.3 | | | | |
| | 12282.6 | $36^-$ | 599.5 | | | | |
| 5. | 5265.4 | $16^-$ | | | | **2002Si20** | 1. Irregular band showing a different behavior |
| | 5381.1 | $17^-$ | 115.7 | | | | than the other MR bands. Due to its irregularity |
| | 5658.6 | $18^-$ | 277.5 | | | | the MR assignment to this band is uncertain. |
| | 5870.6 | $19^-$ | 212.0 | | | | 2. No configuration assigned by 2002Si20. |
| | 6196.7 | $20^-$ | 326.1 | | | | |
| | 6498.9 | $21^-$ | 302.2 | | | | |
| | 6857.9 | $22^-$ | 359.0 | | | | |
| | 7213.5 | $23^-$ | 355.6 | | | | |
| | 7593.2 | $24^-$ | 379.7 | | | | |
| 6. | 5886.6 | $18^-$ | | | | **2002Si20** | 1. Tentatively assigned as $\pi(i_{13/2}h_{9/2})_{K=11^-}\otimes$ |
| | 6041.6 | $19^-$ | 155.0 | | | 1996Ba53 | $\nu[i_{13/2}^{-2}(p_{3/2}f_{5/2})^2]$ before and $\pi(i_{13/2}h_{9/2})_{K=11^-}\otimes$ |
| | 6185.3 | $20^-$ | 143.7 | | | 1993Hu01 | $\nu[i_{13/2}^{-2}(p_{3/2}f_{5/2})^4]$ after the band crossing by |
| | 6349.4 | $21^-$ | 164.1 | | | 1995Mo01 | comparison with the configuration of band 4 |
| | 6557.7 | $22^-$ | 208.3 | | | | and the alignment properties. |
| | 6807.9 | $23^-$ | 250.2 | | | | 2. Oblate deformation ($\lvert\beta_2\rvert<0.1$). |
| | 7117.0 | $24^-$ | 309.1 | | | | 3. Regular band with backbending at $26^-$. |
| | 7492.1 | $25^-$ | 375.1 | | | | |
| | 7896.5 | $26^-$ | 404.4 | | | | |
| | 8271.3 | $27^-$ | 374.8 | | | | |
| | 8666.3 | $28^-$ | 395.0 | | | | |
| | 9070.4 | $29^-$ | 404.1 | 799.1 | 19(5) | | |
| | 9498.4 | $30^-$ | 428.0 | | | | |
| | 9951.0 | $31^-$ | 452.6 | 880.6 | 14(5) | | |
| | 10438.6 | $32^-$ | 487.6 | | | | |
| | 10956.2 | $33^-$ | 517.6 | | | | |

## $^{196}_{82}\mathrm{Pb}_{114}$

| | $E_{level}$ keV | $I^{\pi}$ | $E_{\gamma}$(M1) keV | $E_{\gamma}$(E2) keV | B(M1)/B(E2) $(\mu_N/eb)^2$ | **References** | **Configurations and Comments:** |
|---|---|---|---|---|---|---|---|
| 7. | 6780.1 | $22^+$ | | | | **2002Si20** | 1. Tentatively assigned as $\pi(i_{13/2}h_{9/2})_{K=11^-}\otimes\nu[i_{13/2}^{-3}(p_{3/2}f_{5/2})^{1}]$ before and $\pi(i_{13/2}h_{9/2})_{K=11^-}\otimes\nu[i_{13/2}^{-2}(p_{3/2}f_{5/2})^{4}]$ after the gain in alignment at a frequency around 0.45 MeV by comparison with the configuration of band 3 and a similar band in $^{194}$Pb.<br>2. Oblate deformation ($\lvert\beta_2\rvert<0.1$).<br>3. The bandhead could have been one of the $21^+$ states at 6534.1 and 6589.7 keV also.<br>4. The mean lifetimes (in ps) of levels from $26^+$ to $31^+$ as given in 2002Si29 are 0.17(3), 0.15(+3-2), 0.13(+3-2), 0.14(+3-2), 0.11(2) and 0.13(+3-2), respectively.<br>5. B(M1) values for the transitions from 343 to 454 keV as given in 2002Si29 are 6.30(111), 5.60(+75-112), 5.24(+81-121), 4.30(+61-92), 5.12(93) and 4.08(+63-94) $\mu_N^2$, respectively. |
| | 7041.8 | $23^+$ | 261.7 | | | 2002Si29 | |
| | 7336.7 | $24^+$ | 294.9 | | | 1995Mo01 | |
| | 7634.6 | $25^+$ | 297.9 | | | 1996Ba53 | |
| | 7977.5 | $26^+$ | 342.9 | | | | |
| | 8356.9 | $27^+$ | 379.4 | | | | |
| | 8769.2 | $28^+$ | 412.3 | | | | |
| | 9201.7 | $29^+$ | 432.5 | | | | |
| | 9645.4 | $30^+$ | 443.7 | | | | |
| | 10099.4 | $31^+$ | 454.0 | | | | |
| | 10568.0 | $32^+$ | 468.6 | | | | |
| | 11059.4 | $33^+$ | 491.4 | | | | |
| | 11585.6 | $34^+$ | 526.2 | | | | |
| 8. | 7912.0 | $26^{(+)}$ | | | | **2002Si20** | 1. Tentatively assigned as $\pi(i_{13/2}h_{9/2})_{K=11^-}\otimes\nu[i_{13/2}^{-3}(p_{3/2}f_{5/2})^{1}]$ by comparison with the configuration of band 3 and with the similar decay pattern between the similar bands in $^{199}$Pb.<br>2. Oblate deformation ($\lvert\beta_2\rvert<0.1$).<br>3. The mean lifetimes (in ps) of levels from $27^{(+)}$ to $31^{(+)}$ as given in 2002Si29 are 0.23(4), 0.15(+3-2), 0.13(2), 0.14(+3-2) and 0.14(3), respectively.<br>4. B(M1) values for the transitions from 289 to 513 keV as given in 2002Si29 are 6.84(119), 7.37(+98-147), 5.68(+87), 3.57(+51-77), and 2.72(58) $\mu_N^2$, respectively. |
| | 8201.0 | $27^{(+)}$ | 289.0 | | | 2002Si29 | |
| | 8540.3 | $28^{(+)}$ | 339.3 | | | | |
| | 8939.8 | $29^{(+)}$ | 399.5 | | | | |
| | 9403.9 | $30^{(+)}$ | 464.1 | | | | |
| | 9917.1 | $31^{(+)}$ | 513.2 | | | | |
| | 10461.7 | $32^{(+)}$ | 544.6 | | | | |
| | 11027.9 | $33^{(+)}$ | 566.2 | | | | |
| | 11625.1 | $34^{(+)}$ | 597.2 | | | | |
| 9. | X | | | | | **2002Si20** | 1. Tentatively assigned as $\pi(i_{13/2}h_{9/2})_{K=11^-}\otimes\nu(i_{13/2}^{-2})$ before and $\pi(i_{13/2}h_{9/2})_{K=11^-}\otimes\nu[i_{13/2}^{-2}(p_{3/2}f_{5/2})^{2}]$ after the band crossing at a frequency around 0.42 MeV by comparison with the configuration of band 4.<br>2. Oblate deformation ($\lvert\beta_2\rvert<0.1$).<br>3. Bandhead spin tentatively assigned as $17^-$ from the decay pattern.<br>4. The mean lifetimes (in ps) of levels having transitions 349, 363, 414 and 424 keV as given in 2002Si29 are 0.16(+4-3), 0.15(3), 0.20(4), and 0.16(3), respectively.<br>5. B(M1) values for the transitions 349, 363, 414 and 424 keV as given in 2002Si29 are 6.45(+119-159), 6.26(125), 3.38(56) and and 3.97(75) $\mu_N^2$, respectively. |
| | 150.3+X | | 150.3 | | | 2002Si29 | |
| | 331.3+X | | 181.0 | | | | |
| | 529.9+X | | 198.6 | | | | |
| | 774.2+X | | 244.3 | | | | |
| | 1071.9+X | | 297.7 | | | | |
| | 1420.7+X | | 348.8 | | | | |
| | 1783.5+X | | 362.8 | | | | |
| | 2177.1+X | | 393.6 | | | | |
| | 2590.8+X | | 413.7 | | | | |
| | 3011.6+X | | 420.8 | | | | |
| | 3435.3+X | | 423.7 | | | | |
| | 3893.1+X | | 457.8 | | | | |
| | 4386.6+X | | 493.5 | | | | |
| | (4906.8+X) | | (520.2) | | | | |

## $^{197}_{82}Pb_{115}$

| | $E_{level}$ KeV | $I^{\pi}$ | $E_{\gamma}$(M1) KeV | $E_{\gamma}$(E2) keV | B(M1)/B(E2) $(\mu_N/eb)^2$ | References | **Configurations and Comments:** |
|---|---|---|---|---|---|---|---|
| 1. | 3283.4 | $27/2^-$ | | | | **2001Go06** | 1. $\pi(h_{9/2}i_{13/2})_{K=11^-}\otimes\nu(i_{13/2}^{-1})$ below crossing, |
| | 3436.0 | $29/2^-$ | 152.6 | | | 2001Co19 | $\pi(h_{9/2}i_{13/2})_{K=11}\otimes\nu(i_{13/2}^{-3})$ above the first band |
| | 3706.5 | $31/2^-$ | 270.5 | | | 1999Po13 | crossing and $\pi(h_{9/2}i_{13/2})_{K=11^-}\otimes\nu(i_{13/2}^{-3}\,(f_{5/2}p_{3/2})^{-2})$, |
| | 4065.6 | $33/2^-$ | 359.1 | 629.8 | 30(11) | 1995Ba35 | above second crossing by comparison with the |
| | 4435.4 | $35/2^-$ | 369.8 | 729.0 | 26(8) | 1992Ku06 | similar band in neighboring Pb isotopes and |
| | 4820.4 | $37/2^-$ | 385.0 | 754.9 | 21(6) | 1994Cl01 | from the TAC model calculations. |
| | 5185.6 | $39/2^-$ | 365.2 | 750.2 | 21(6) | 1998Cl06 | 2. Small oblate deformation. |
| | 5479.4 | $41/2^-$ | 293.8 | 659.2 | 54(21) | | 3. Regular band showing a backbend at 41/2. |
| | 5707.0 | $43/2^-$ | 227.6 | 521.7 | 35(11) | | 4. The mean lifetimes (in ps) of levels from 3436 to |
| | 5952.4 | $45/2^-$ | 245.4 | 473.2 | 57(29) | | 5707 keV as given in 2001Co19 are 1.1(3), |
| | 6237.6 | $47/2^-$ | 285.2 | 531.0 | 33(15) | | 0.71(19), 0.60(16), 0.49(15), 1.0(5), 0.8(4), |
| | 6564.8 | $49/2^-$ | 327.2 | 612.4 | 25(10) | | 0.7(2) and 0.8(3), respectively and that for |
| | 6903.7 | $51/2^-$ | 338.9 | 666.1 | 84(43) | | levels from 6238 to 7257 keV as given in |
| | 7257.0 | $53/2^-$ | 353.3 | 692.1 | 68(34) | | 1998Cl06 are 0.40(2), 0.29(+3-2), 0.17(+2-1) |
| | 7659.8 | $55/2^-$ | 402.8 | 756.0 | 46(23) | | and 0.17(+2-1), respectively. |
| | 8120.1 | $57/2^-$ | 460.3 | 862.8 | 41(20) | | 5. The B(M1) values for the transitions from |
| | 8635.2 | $59/2^-$ | 515.1 | 975.1 | 16(6) | | 153 to 228 keV as given in 2001Co19 are |
| | 9197.8 | $61/2^-$ | 562.6 | 1077.4 | 18(8) | | 3.8(9), 2.5(4), 1.6(3), 1.8(4),0.8(4), 1.1(4), |
| | 9793.8 | $63/2^-$ | 596.0 | 1158.1 | 57(29) | | 2.1(5) and 3.1(12) $\mu_N^2$, respectively and |
| | 10405.5 | $65/2^-$ | 611.7 | 1207.4 | 47(25) | | that for transitions from 285 to 353 keV as given |
| | | | | | | | in 1998Cl06 are 4.59(23), 4.53(+31-47), |
| | | | | | | | 7.05(+41-83) and 6.35 (+37-75) $\mu_N^2$, |
| | | | | | | | respectively. |
| 2. | 4794.0 | $37/2^+$ | | | | **2001Go06** | 1. $\pi(h_{9/2}i_{13/2})_{K=11^-}\otimes\nu(i_{13/2}^{-2}f_{5/2}^{-1})$ below and |
| | 4906.4 | $39/2^+$ | 112.4 | | | 1999Po13 | $\pi(h_{9/2}i_{13/2})_{K=11^-}\otimes\nu(i_{13/2}^{-4}f_{5/2}^{-1})$ above the |
| | 5057.7 | $41/2^+$ | 151.3 | | | 1995Ba35 | bandcrossing from the TAC model calculations. |
| | 5258.3 | $43/2^+$ | 200.6 | | | 1992Ku06 | 2. Small oblate deformation. |
| | 5525.0 | $45/2^+$ | 266.7 | | | 1993Hu08 | 3. The mean lifetimes (in ps) for the transitions from |
| | 5861.7 | $47/2^+$ | 336.7 | | | 1994Cl01 | 151.3 to 266.7 keV as given in 1994Cl01 are |
| | 6265.6 | $49/2^+$ | 403.9 | 740.7 | 85(47) | 1998Cl06 | 1.8(8), 0.9(4) and 1.2(3), and from 337 to |
| | 6711.7 | $51/2^+$ | 446.1 | 849.9 | 28(11) | | 467 as given in 1998Cl06 are 0.17(3), 0.13(+3 |
| | 7178.8 | $53/2^+$ | 467.1 | 913.3 | 38(17) | | -2), 0.16(2) and 0.28(+5-6), respectively. |
| | 7612.5 | $55/2^+$ | 433.7 | 900.6 | 65(35) | | 4. The B(M1) values as given in 1994Cl01 for the |
| | 7983.9 | $57/2^+$ | 371.4 | | | | transitions from 151 to 267 keV are 2.32(+232 |
| | 8371.5 | $59/2^+$ | 387.6 | | | | -77), 3.66(+220-100) and 1.78(+118-51) $\mu_N^2$, |
| | 8794.1 | $61/2^+$ | 422.6 | | | | respectively, and that for transitions from 337 |
| | 9245.8 | $63/2^+$ | 451.7 | | | | to 467 keV, as given in 1998Cl06 are |
| | 9722.9 | $65/2^+$ | 477.1 | | | | 7.18(127), 5.88(+90-136), 3.72 (47) and |
| | | | | | | | 1.90(+41-34) $\mu_N^2$, respectively. |
| 3. | 5232.6 | $39/2^{(+)}$ | | | | **2001Go06** | 1. $\pi(h_{9/2}i_{13/2})_{K=11^-}\otimes\nu(i_{13/2}^{-2}p_{3/2}^{-1})$ below and |
| | 5395.3 | $41/2^{(+)}$ | 162.7 | | | 1995Ba35 | $\pi(h_{9/2}i_{13/2})_{K=11^-}\otimes\nu(i_{13/2}^{-4}f_{5/2}^{-1})$ above the |
| | 5614.1 | $43/2^{(+)}$ | 218.8 | | | 1999Po13 | bandcrossing by comparison with the |
| | 5878.8 | $45/2^{(+)}$ | 264.7 | | | | configuration of band 2. |
| | 6195.4 | $47/2^{(+)}$ | 316.6 | | | | 2. Small oblate deformation. |
| | 6558.7 | $49/2^{(+)}$ | 363.3 | | | | 3. Irregular band. |
| | 6912.5 | $51/2^{(+)}$ | 353.8 | | | | |
| | 7286.4 | $53/2^{(+)}$ | 373.9 | | | | |
| | 7677.5 | $55/2^{(+)}$ | 391.1 | | | | |
| | 8067.7 | $57/2^{(+)}$ | 390.2 | | | | |
| | 8438.7 | $59/2^{(+)}$ | 371.0 | | | | |
| | 8830.5 | $61/2^{(+)}$ | 391.8 | | | | |

## $^{197}_{82}\mathrm{Pb}_{115}$

| | $E_{level}$ KeV | $I^\pi$ | $E_\gamma$(M1) KeV | $E_\gamma$(E2) keV | B(M1)/B(E2) $(\mu_N/eb)^2$ | References | **Configurations and Comments:** |
|---|---|---|---|---|---|---|---|
| 4. | 6014.1 | $43/2^-$ | | | | **2001Go06** | 1. $\pi(h_{9/2}i_{13/2})_{K=11^-}\otimes\nu(i_{13/2}^{-3} f_{5/2}^{-2})$, by comparison |
| | 6202.1 | $45/2^-$ | 188.0 | | | | with band 1 and TAC calculations. |
| | 6407.9 | $47/2^-$ | 205.8 | | | | 2. Small oblate deformation. |
| | 6659.3 | $49/2^-$ | 251.4 | | | | 3. Regular band. |
| | 6993.4 | $51/2^-$ | 334.1 | | | | |
| | 7406.6 | $53/2^-$ | 413.2 | | | | |
| | 7859.5 | $55/2^-$ | 452.9 | 866.1 | 27(10) | | |
| | 8352.7 | $57/2^-$ | 493.2 | 946.1 | 23(9) | | |
| | 8878.1 | $59/2^-$ | 525.4 | | | | |
| | 9441.0 | $61/2^-$ | 562.9 | | | | |
| | 10022.9 | $63/2^-$ | 581.9 | | | | |
| 5. | 6262.6 | $45/2^{(+)}$ | | | | **2001Go06** | 1. $\pi(h_{9/2}i_{13/2})_{K=11^-}\otimes\nu(i_{13/2}^{-2} (f_{5/2}p_{3/2})^{-3}$, by |
| | 6517.9 | $47/2^{(+)}$ | 255.3 | | | | comparison with band 1, 4 and TAC |
| | 6806.5 | $49/2^{(+)}$ | 288.6 | | | | calculations. |
| | 7147.2 | $51/2^{(+)}$ | 340.7 | | | | 2. Small oblate deformation. |
| | 7550.8 | $53/2^{(+)}$ | 403.6 | | | | 3. Regular band. |
| | 8015.4 | $55/2^{(+)}$ | 464.6 | 868.2 | 54(23) | | |
| | 8516.6 | $57/2^{(+)}$ | 504.2 | 968.8 | 17(8) | | |
| | 9041.4 | $59/2^{(+)}$ | 521.8 | | | | |
| | 9581.3 | $61/2^{(+)}$ | 539.9 | | | | |

## $^{198}_{82}\mathrm{Pb}_{116}$

| | $E_{level}$ keV | $I^\pi$ | $E_\gamma$(M1) keV | $E_\gamma$(E2) keV | B(M1)/B(E2) $(\mu_N/eb)^2$ | References | **Configurations and Comments:** |
|---|---|---|---|---|---|---|---|
| 1. | 4882.7 | $(14^+)$ | | | | **2001Go06** | 1. Tentatively assigned as $\pi(h_{9/2}i_{13/2})_{K=11^-}\otimes$ |
| | 4975.7 | $(15^+)$ | 93.0 | | | 1993Cl05 | $\nu(i_{3/2}^{-1}f_{5/2}^{-1})$ before, $\pi(h_{9/2}i_{13/2})_{K=11^-}\otimes\nu(i_{13/2}^{-3}f_{5/2}^{1})$ |
| | 5092.2 | $(16^+)$ | 116.5 | | | 1992Wa20 | above first band crossing and $\pi(h_{9/2}i_{13/2})_{K=11^-}\otimes$ |
| | 5248.9 | $(17^+)$ | 156.7 | | | 1994Cl01 | $\nu(i_{3/2}^{-3}(f_{5/2}p_{3/2})^{-3})$ above the second band |
| | 5476.5 | $(18^+)$ | 227.6 | | | 1997Cl03 | crossing from TAC calculations and by |
| | 5812.5 | $(19^+)$ | 336.0 | | | | comparison with similar bands in neighboring |
| | 6241.0 | $(20^+)$ | 428.5 | | | | Pb isotopes. |
| | 6659.4 | $(21^+)$ | 418.4 | | | | 2. Nearly oblate shape. |
| | 6866.8 | $(22^+)$ | 206.5 | | | | 3. The mean lifetimes (in ps) for the transitions from |
| | 7073.3 | $(23^+)$ | 206.5 | | | | 207 to 506 keV as given in 1994Cl01 are |
| | 7311.0 | $(24^+)$ | 237.7 | | | | 2.1(4), 0.85(30), 1.1(6), 0.58(15), 0.36(10), |
| | 7590.5 | $(25^+)$ | 279.5 | | | | 0.20(4), 0.099(25) and 0.052(11), |
| | 7916.1 | $(26^+)$ | 325.6 | | | | respectively. |
| | 8290.5 | $(27^+)$ | 374.4 | 700.2 | 76(41) | | 4. B(M1) values for the transitions from 207 to |
| | 8712.2 | $(28^+)$ | 421.7 | 796.1 | >91 | | 506 keV as given in 1994Cl01 are 1.32(+32-21), |
| | 9175.8 | $(29^+)$ | 463.6 | 885.3 | >79 | | 2.64(+140-70), 1.41(+176-53), 1.94(+88-53), |
| | 9681.1 | $(30^+)$ | 505.3 | 968.9 | 68(34) | | 2.11(+88-53), 2.82 (+88-53), 4.58(+158-88) and |
| | 10230.5 | $(31^+)$ | 549.4 | 1054.6 | 40(24) | | 6.51(+194-158) $\mu_N^2$, respectively. |
| | 10820.8 | $(32^+)$ | 590.3 | 1139.7 | >35 | | 5. Regular band. |
| | 11438.5 | $(33^+)$ | 617.7 | 1208.0 | 9(5) | | |
| | 12059.5 | $(34^+)$ | 621.0 | 1238.7 | >16 | | |
| | 12699.0 | $(35^+)$ | 639.5 | | | | |

## $^{198}_{82}Pb_{116}$

| | $E_{level}$ keV | $I^{\pi}$ | $E_{\gamma}$(M1) keV | $E_{\gamma}$(E2) keV | B(M1)/B(E2) $(\mu_N/eb)^2$ | References | Configurations and Comments: |
|---|---|---|---|---|---|---|---|
| 2. | 6518.9 | $(20^-)$ | | | | **2001Go06** | 1. Tentatively assigned as $\pi(h_{9/2}i_{13/2})_{K=11}^{-}\otimes\nu(i_{3/2}^{-2}(f_{5/2}p_{3/2})^{-2})$ before and $\pi(h_{9/2}i_{13/2})_{K=11}^{-}\otimes\nu(i_{3/2}^{-4}(f_{5/2}p_{3/2})^{-2})$ after the band crossing from TAC calculations and by comparison with band 3.<br>2. Small oblate deformation.<br>3. Regular band. |
| | 6734.2 | $(21^-)$ | 215.3 | | | 1993Cl05 | |
| | 7016.7 | $(22^-)$ | 282.5 | | | | |
| | 7360.4 | $(23^-)$ | 343.7 | | | | |
| | 7778.9 | $(24^-)$ | 418.5 | | | | |
| | 8255.5 | $(25^-)$ | 476.6 | | | | |
| | 8739.4 | $(26^-)$ | 483.9 | | | | |
| | 9154.4 | $(27^-)$ | 415.0 | | | | |
| 3. | 5379.1 | $16^-$ | | | | **2001Go06** | 1. Tentatively assigned as $\pi(h_{9/2}i_{13/2})_{K=11}^{-}\otimes\nu(i_{3/2}^{-2})$ before and $\pi(h_{9/2}i_{13/2})_{K=11}^{-}\otimes\nu(i_{3/2}^{-4})$ after the band crossing from TAC calculations.<br>2. Small oblate deformation.<br>3. The mean lifetimes (in ps) for the transitions from 156 to 476 keV as given in 1994Cl01 are 2.7(9), 1.8(5), 2.1(5), 1.14(23), 0.72(10), 0.46(10), 0.24(4), 0.22(6), and 0.27(7), respectively and B(M1) values for these transitions are 1.18(+60-30), 1.46(+56-32), 0.79 (+21-16), 0.84(+19-14), 0.97(+19-14), 1.21(+46 -21), 1.88(+67-30), 1.74(+86-33) and 1.30(+60 -25) $\mu_N^2$, respectively.<br>4. The mean lifetimes (in ps) for the transitions from 156 to 342.8 keV as given in 1998Kr20 are 0.63(10), 0.70(+10-20), 0.34(+15-10) and 0.20(+20-10) and the B(M1) values for these transitions are 6.2(+11-9), 3.8(+15-5), 4.9(+20-15) and 4.9(+48-28) $\mu_N^2$ respectively. |
| | 5492.7 | $17^-$ | 113.6 | | | 1993Cl05 | |
| | 5648.4 | $18^-$ | 155.7 | | | 1992Wa20 | |
| | 5863.4 | $19^-$ | 215.0 | | | 1994Cl01 | |
| | 6141.8 | $20^-$ | 278.4 | | | 1997Cl03 | |
| | 6484.0 | $21^-$ | 342.2 | 621.0 | 40(17) | 1998Kr20 | |
| | 6872.8 | $22^-$ | 388.8 | 731.0 | 39(17) | | |
| | 7295.2 | $23^-$ | 422.4 | 811.2 | 38(17) | | |
| | 7739.3 | $24^-$ | 444.1 | 866.5 | 24(7) | | |
| | 8210.8 | $25^-$ | 471.5 | 915.6 | 21(6) | | |
| | 8686.0 | $26^-$ | 475.2 | 946.7 | 26(10) | | |
| | 9112.3 | $27^-$ | 426.3 | 901.5 | 34(14) | | |
| | 9512.3 | $28^-$ | 400.0 | 826.3 | 50(16) | | |
| | 9930.5 | $29^-$ | 418.2 | 818.2 | >40 | | |
| | 10380.3 | $30^-$ | 449.8 | | | | |
| | 10869.3 | $31^-$ | 489.0 | | | | |
| | 11398.7 | $32^-$ | 529.4 | | | | |
| | 11970.8 | $33^-$ | 572.1 | | | | |
| | 12579.8 | $34^-$ | 609.0 | | | | |
| 4. | (6392.6) | $(18^-)$ | | | | **2001Go06** | 1. Tentatively assigned as $\pi(h_{9/2}i_{13/2})_{K=11}^{-}\otimes\nu(i_{13/2}^{-2}(f_{5/2}p_{3/2})^{-2})$ before and $\pi(h_{9/2}i_{13/2})_{K=11}^{-}\otimes\nu(i_{13/2}^{-4}(f_{5/2}p_{3/2})^{-2})$ after the bandcrossing from TAC calculations.<br>2. Small oblate deformation.<br>3. Regular band. |
| | (6515.3) | $(19^-)$ | 122.7 | | | 1993Cl05 | |
| | (6674.4) | $(20^-)$ | 159.1 | | | | |
| | (6878.3) | $(21^-)$ | 203.9 | | | | |
| | (7142.9) | $(22^-)$ | 264.6 | | | | |
| | (7480.1) | $(23^-)$ | 337.2 | | | | |
| | (7835.0) | $(24^-)$ | 354.9 | | | | |
| | (8243.5) | $(25^-)$ | 408.5 | | | | |
| | (8695.0) | $(26^-)$ | 451.5 | | | | |
| | (9146.5) | $(27^-)$ | 451.5 | 903.0 | | | |
| 5. | 7333.4 | $(23^+)$ | | | | **2001Go06** | 1. Tentatively assigned as $\pi(h_{9/2}i_{13/2})_{K=11}^{-}\otimes\nu(i_{13/2}^{-1}P_{3/2}^{-1})$ before and $\pi(h_{9/2}i_{13/2})_{K=11}^{-}\otimes\nu(i_{13/2}^{-3}P_{3/2}^{-1})$ after the bandcrossing from TAC calculations and by comparison with the neighboring Pb isotopes.<br>2. Small oblate deformed structure.<br>3. Regular band. |
| | 7554.4 | $(24^+)$ | 221.0 | | | 1993Cl05 | |
| | 7794.8 | $(25^+)$ | 240.4 | | | | |
| | 8076.1 | $(26^+)$ | 281.3 | | | | |
| | 8408.2 | $(27^+)$ | 332.1 | | | | |
| | 8799.7 | $(28^+)$ | 391.5 | | | | |
| | 9254.9 | $(29^+)$ | 455.2 | | | | |
| | 9770.1 | $(30^+)$ | 515.2 | 969.8 | | | |
| | 10329.1 | $(31^+)$ | 559.0 | | | | |
| | 10921.3 | $(32^+)$ | 592.2 | | | | |

## $^{199}_{82}Pb_{117}$

| | $E_{level}$ keV | $I^\pi$ | $E_\gamma$(M1) keV | $E_\gamma$(E2) keV | B(M1)/B(E2) $(\mu_N/eb)^2$ | References | **Configurations and Comments:** |
|---|---|---|---|---|---|---|---|
| 1. | 3604.2 | $(25/2^-)$ | | | | **1995Ne09** | 1. $\pi(h_{9/2}i_{13/2})_{K=11^-}\otimes\nu(i_{13/2}^{-1})$ below and |
| | 3694.1 | $(27/2^-)$ | 89.9 | | | 1999Po13 | $\pi(h_{9/2}i_{13/2})_{K=11^-}\otimes\nu(i_{13/2}^{-3})$ above the |
| | 3868.0 | $(29/2^-)$ | 173.9 | | | 1994Ba43 | bandcrossing from the TAC model |
| | 4143.4 | $(31/2^-)$ | 275.4 | | | 1997Cl03 | calculations. |
| | 4502.8 | $(33/2^-)$ | 359.4 | 634.8 | | | 2. Small oblate deformation $(\beta_2,\gamma)$ ~ (0.1, -70°) |
| | 4904.1 | $(35/2^-)$ | 401.3 | 760.8 | | | 3. The mean lifetimes (in ps) of states with spins |
| | 5324.8 | $(37/2^-)$ | 420.7 | 822.1 | | | from 43/2 to 49/2 are 0.37(+51-29), 0.31(+31-24), |
| | 5746.3 | $(39/2^-)$ | 421.5 | 842.4 | | | 0.17(+6-4) and 0.13(+4-3), and for the states |
| | 6074.9 | $(41/2^-)$ | 328.6 | 750.1 | | | with spins 51/2 to 57/2 as given in 1997Cl03 |
| | 6309.5 | $(43/2^-)$ | 234.6 | | | | are 0.20(5), 0.16(+5-4), 0.15(+5-4) and |
| | 6549.6 | $(45/2^-)$ | 240.1 | | | | 0.21(+6-5), respectively. |
| | 6823.4 | $(47/2^-)$ | 273.8 | | | | 4. B(M1) values for the transitions 234.6, 240.1 |
| | 7139.7 | $(49/2^-)$ | 316.3 | 590.1 | 35(+22-20) | | and 273.8 keV are 6.6(+25-38), 7.4(+24-38) |
| | 7502.8 | $(51/2^-)$ | 363.1 | 679.5 | 27(+18-14) | | and 10.6(+34-29) $\mu_N^2$ and for the transitions |
| | 7914.1 | $(53/2^-)$ | 411.3 | 774.6 | 27(+15-18) | | from 363.1 to 508.3 keV as given in 1997Cl03 |
| | 8373.4 | $(55/2^-)$ | 459.3 | 870.9 | 38(20) | | are 4.8(13), 4.4(+12-15), 3.0(+7-9) and |
| | 8881.7 | $(57/2^-)$ | 508.3 | 967.7 | | | 1.7(+4-5) $\mu_N^2$, respectively. |
| | 9436.5 | $(59/2^-)$ | 554.8 | (1063) | | | 5. Regular band with backbending at 41/2. |
| 2. | X | $(35/2^+)$ | | | | **1999Po13** | 1. $\pi(h_{9/2}i_{13/2})_{K=11^-}\otimes\nu(i_{13/2}^{-2}f_{5/2}^{-1})$ below and |
| | 98.2+X | $(37/2^+)$ | 98.2 | | | 1995Ne09 | $\pi(h_{9/2}i_{13/2})_{K=11^-}\otimes\nu(i_{13/2}^{-4}f_{5/2}^{-1})$ above the |
| | 223.2+X | $(39/2^+)$ | 125.0 | | | 1992Ba13 | bandcrossing from the TAC model |
| | 388.8+X | $(41/2^+)$ | 165.6 | | | 1994Ba43 | calculations. |
| | 603.4+X | $(43/2^+)$ | 214.6 | | | 1997Cl03 | 2. Small oblate deformation $(\beta_2,\gamma)$ ~ (0.1, -70°) |
| | 871.2+X | $(45/2^+)$ | 267.8 | | | | suggested in 1995Ne09. |
| | 1194.3+X | $(47/2^+)$ | 323.1 | | | | 3. The mean lifetimes (in ps) of states with spins |
| | 1571.4+X | $(49/2^+)$ | 377.1 | 700.1 | 28(9) | | from 47/2 to 55/2 are 0.19(+15-8), 0.14(+6-4), |
| | 2001.7+X | $(51/2^+)$ | 430.3 | 807.1 | 49(28) | | 0.10(+3-2), 0.06(2) and 0.11(2), respectively |
| | 2483.6+X | $(53/2^+)$ | 481.9 | 912.4 | 52(20) | | and that for spin 57/2 as given in 1997Cl03 is |
| | 3015.6+X | $(55/2^+)$ | 532.0 | 1014.2 | 30(9) | | 0.14(+3-2). |
| | 3589.2+X | $(57/2^+)$ | 573.6 | 1105.7 | 34(10) | | 4. B(M1) value for the transition 323.1 keV is |
| | 4207.4+X | $(59/2^+)$ | 618.5 | 1192.1 | 45(11) | | 6.6(+47-29) $\mu_N^2$ from 1995Ne09. |
| | 4546.6+X | $(61/2^+)$ | 339.2 | | | | 5. B(M1)/B(E2) values are from 1992Ba13. |
| | 4932.5+X | $(63/2^+)$ | 385.9 | | | | 6. Regular band with backbending at spin 61/2. |
| | 5353.5+X | $(65/2^+)$ | 421.0 | | | | |
| | 5806.9+X | $(67/2^+)$ | 453.4 | | | | |
| | 6303.4+X | $(69/2^+)$ | 496.5 | | | | |
| | 6845.9+X | $(71/2^+)$ | 542.5 | | | | |
| | 7433.6+X | $(73/2^+)$ | 587.7 | | | | |
| 3. | Y | $(39/2^+)$ | | | | **1994Ba43** | 1. Tentatively assigned as $\pi(h_{9/2}i_{13/2})_{K=11^-}\otimes$ |
| | 137.7+Y | $(41/2^+)$ | 137.7 | | | 1999Po13 | $\nu(i_{13/2}^{-2}\,f_{5/2}^{-1})$ from the TAC model calculation. |
| | 302.3+Y | $(43/2^+)$ | 164.6 | | | 1995Ne09 | 2. Small oblate deformation $(\beta_2,\gamma)$ ~ (0.1, -70°) |
| | 510.6+Y | $(45/2^+)$ | 208.3 | | | | 3. The topmost transition is from 1999Po13. |
| | 781.6+Y | $(47/2^+)$ | 271.0 | | | | 4. Regular band. |
| | 1123.6+Y | $(49/2^+)$ | 342.0 | | | | |
| | 1540.6+Y | $(51/2^+)$ | 417.0 | | | | |
| | 2023.3+Y | $(53/2^+)$ | 482.7 | 900.0 | | | |
| | 2560.1+Y | $(55/2^+)$ | 536.8 | 1019.6 | | | |
| | 3145.3+Y | $(57/2^+)$ | 585.2 | 1122.0 | | | |

## $^{199}_{82}Pb_{117}$

| | $E_{level}$ keV | $I^{\pi}$ | $E_{\gamma}(M1)$ keV | $E_{\gamma}(E2)$ keV | B(M1)/B(E2) $(\mu_N/eb)^2$ | **References** | **Configurations and Comments:** |
|---|---|---|---|---|---|---|---|
| 4. | Z | | | | | **1994Ba43** | 1. Tentatively assigned as $\pi(h_{9/2}^{2})_{K=8^{+}} \otimes \nu(i_{13/2}^{-3})$. |
| | 97.7+Z | | 97.7 | | | 1999Po13 | 2. The estimated bandhead spin is 37/2 since it |
| | 232.9+Z | | 135.2 | | | | populates states with spin around 33/2. |
| | 426.1+Z | | 193.2 | | | | 3. The two topmost transitions are from 1999Po13 |
| | 673.5+Z | | 247.4 | | | | 4. Regular band with signature splitting and |
| | 967.6+Z | | 294.1 | 541.4 | | | backbending at the top of the band. |
| | 1349.7+Z | | 382.1 | 676.2 | | | |
| | 1743.9+Z | | 394.2 | 776.4 | | | |
| | 2227.4+Z | | 483.5 | 877.6 | | | |
| | 2737.9+Z | | 510.5 | 994.2 | | | |
| | 3256.7+Z | | 518.8 | 1029.4 | | | |
| | 3594.9+Z | | 338.2 | | | | |
| 5. | U | | | | | **1994Ba43** | 1. Tentatively assigned as $\pi(h_{9/2}^{2})_{K=8^{+}} \otimes \nu(i_{13/2}^{-4}$ |
| | 242.9+U | | 242.9 | | | | $p_{3/2}^{-1})$. |
| | 550.2+U | | 307.3 | | | | 2. The estimated bandhead spin is 45/2 since it |
| | 863.2+U | | 313.0 | 620.5 | | | populates states with spin around 41/2. |
| | 1247.8+U | | 384.6 | 697.6 | | | 3. Regular band with signature splitting. |
| | 1661.8+U | | 414.0 | 798.7 | | | |
| | 2148.8+U | | 487.0 | 901.4 | | | |

## $^{200}_{82}Pb_{118}$

| | $E_{level}$ keV | $I^{\pi}$ | $E_{\gamma}(M1)$ keV | $E_{\gamma}(E2)$ keV | B(M1)/B(E2) $(\mu_N/eb)^2$ | Reference | **Configurations and Comments:** |
|---|---|---|---|---|---|---|---|
| 1. | X | | | | | **1994Ba43** | 1. $\pi(h_{9/2}i_{13/2})_{K=11^{-}} \otimes \nu(i_{13/2}^{-2})$ from the TAC model |
| | 100.6+X | | 100.6 | | | | calculation. |
| | 223.9+X | | 123.3 | | | | 2. Small oblate deformation. |
| | 384.2+X | | 160.3 | | | | 3. Tentative bandhead spin is around 17. |
| | 592.8+X | | 208.6 | | | | 4. Regular band. |
| | 855.3+X | | 262.5 | | | | |
| | 1174.8+X | | 319.5 | | | | |
| | 1549.5+X | | 374.7 | | | | |
| | 1978.9+X | | 429.4 | | | | |
| | 2459.5+X | | 480.6 | | | | |
| | 2992.5+X | | 533.0 | (1014) | | | |
| | 3574.6+X | | 582.1 | | | | |
| | 4207.0+X | | 632.4 | 1214.3 | | | |
| 2. | Y | | | | | **1994Ba43** | 1. Tentatively assigned as $\pi(h_{9/2}i_{13/2})_{K=11^{-}} \otimes$ |
| | 212.5+Y | | 212.5 | | | | $\nu(i_{13/2}^{-3}\ p_{3/2}^{-1})$ from the TAC model calculation. |
| | 452.8+Y | | 240.3 | | | | 2. Tentative bandhead spin is around 23. |
| | 736.1+Y | | 283.3 | | | | 3. Regular band. |
| | 1065.7+Y | | 329.6 | | | | |
| | 1445.8+Y | | 380.1 | | | | |
| | 1884.6+Y | | 438.8 | | | | |
| 3. | Z | | | | | **1994Ba43** | 1. Tentatively assigned as $\pi(h_{9/2}i_{13/2})_{K=11^{-}} \otimes$ |
| | 237.5+Z | | 237.5 | | | | $\nu(i_{13/2}^{-3}\ f_{5/2}^{-1})$ from the TAC model calculation. |
| | 518.8+Z | | 281.3 | | | | 2. Tentative bandhead spin is ~ 23. |
| | 853.4+Z | | 334.6 | | | | 3. Regular band. |
| | 1234.8+Z | | 381.4 | | | | |
| | 1658.3+Z | | 423.5 | | | | |

## $^{201}_{82}Pb_{119}$

| | $E_{level}$ keV | $I^{\pi}$ | $E_{\gamma}$(M1) keV | $E_{\gamma}$(E2) keV | B(M1)/B(E2) $(\mu_N/eb)^2$ | Reference | **Configurations and Comments:** |
|---|---|---|---|---|---|---|---|
| 1. | X | | | | | **1995Ba70** | 1. $\pi(h_{9/2}i_{13/2})_{K=11^-}\otimes\nu(i_{13/2}^{-1})$ by comparison with a similar band in $^{199}$Pb.<br>2. Regular band. |
| | 109.2+X | | 109.2 | | | | |
| | 290.8+X | | 181.6 | | | | |
| | 554.6+X | | 263.8 | | | | |
| | 895.4+X | | 340.8 | | | | |
| | 1299.4+X | | 404.0 | 744.6 | | | |
| | 1758.4+X | | 459.0 | 862.8 | | | |
| | 2264.1+X | | 505.7 | 964.7 | | | |
| | 2822.6+X | | 558.5 | | | | |
| 2. | 6146.0+Y | 35/2 | | | | **1995Ba70** | 1. $\pi(h_{9/2}i_{13/2})_{K=11^-}\otimes\nu(i_{13/2}^{-2}p_{3/2}^{-1})$ by comparison with $^{199}$Pb and TAC model calculations.<br>2. Small oblate deformation.<br>3. From 47/2 and above, there is a forking of the band with very close lying transitions having energies 333.1, 394.8 and 492.5 keV.<br>4. Regular band. |
| | 6247.7+Y | 37/2 | 101.7 | | | | |
| | 6377.4+Y | 39/2 | 129.7 | | | | |
| | 6549.0+Y | 41/2 | 171.6 | | | | |
| | 6769.5+Y | 43/2 | 220.5 | | | | |
| | 7045.4+Y | 45/2 | 275.9 | | | | |
| | 7380.0+Y | 47/2 | 334.6 | | | | |
| | 7773.3+Y | 49/2 | 393.3 | | | | |
| | 8227.2+Y | 51/2 | 453.9 | | | | |
| 3. | Z | | | | | **1995Ba70** | 1. Tentatively assigned as $\pi(h_{9/2}i_{13/2})_{K=11^-}\otimes\nu(i_{13/2}^{-2}f_{5/2}^{-1})$, because of the similarity in the moment of inertia of the bands 2 and 3.<br>2. Regular band. |
| | 139.6+Z | | 139.6 | | | | |
| | 315.4+Z | | 175.8 | | | | |
| | 537.7+Z | | 222.3 | | | | |
| | 814.1+Z | | 276.4 | | | | |
| | 1146.4+Z | | 332.3 | | | | |
| | 1534.5+Z | | 388.1 | | | | |
| | 1975.8+Z | | 441.3 | 829.4 | | | |
| | 2467.5+Z | | 491.7 | 933.1 | | | |
| | 3007.3+Z | | 539.8 | 1031.4 | | | |
| 4. | U | | | | | **1995Ba70** | 1. Regular band.<br>2. No configuration assigned by 2002Si20. |
| | 176.5+U | | 176.5 | | | | |
| | 402.2+U | | 225.7 | | | | |
| | 680.4+U | | 278.2 | | | | |
| | 1007.1+U | | 326.7 | | | | |
| | 1387.5+U | | 380.4 | | | | |
| | 1817.2+U | | 429.7 | | | | |
| | 2300.3+U | | 483.1 | | | | |
| | 2830.5+U | | 530.2 | | | | |
| 5. | V | | | | | **1995Ba70** | 1. Regular band.<br>2. No configuration assigned by 2002Si20. |
| | 152.9+V | | 152.9 | | | | |
| | 351.5+V | | 198.6 | | | | |
| | 601.5+V | | 250.0 | | | | |
| | 913.5+V | | 312.0 | | | | |
| | 1287.9+V | | 374.4 | | | | |
| | 1723.9+V | | 436.0 | | | | |
| | 2217.3+V | | 493.4 | | | | |

## $^{202}_{82}Pb_{120}$

| | $E_{level}$ keV | $I^{\pi}$ | $E_{\gamma}(M1)$ keV | $E_{\gamma}(E2)$ keV | B(M1)/B(E2) $(\mu_N/eb)^2$ | Reference | **Configurations and Comments:** |
|---|---|---|---|---|---|---|---|
| 1. | X | | | | | **2000Go47** | 1. Tentatively assigned as $\pi(h_{9/2}i_{13/2})_{K=11}^{-}\otimes\nu(i_{13/2}^{-2})$ |
| | 161.3+X | | 161.3 | | | 1995Ba70 | by comparison with the lighter mass Pb |
| | 404.6+X | | 243.3 | | | | isotopes. |
| | 737.5+X | | 332.9 | | | | 2. Small oblate deformation. |
| | 1145.1+X | | 407.6 | | | | 3. X > 5.3 MeV and the bandhead spin > 17. |
| | 1611.6+X | | 466.5 | | | | 4. Regular band. |
| | 2129.3+X | | 517.7 | | | | |
| 2. | Y | | | | | **2000Go47** | 1. Tentatively assigned as $\pi(h_{9/2}i_{13/2})_{K=11}^{-}\otimes\nu(i_{13/2}^{-1}$ |
| | 130.0+Y | | 130.0 | | | | $p_{3/2}^{-1})$ by comparison with the lighter mass Pb |
| | 321.7+Y | | 191.7 | | | | isotopes. |
| | 591.5+Y | | 269.8 | | | | 2. Small oblate deformation. |
| | 940.9+Y | | 349.4 | | | | 3. Y > 5.059 MeV. |
| | 1357.3+Y | | 416.4 | | | | 4. Regular band. |
| | 1835.2+Y | | 477.9 | | | | |
| | 2358.6+Y | | 523.4 | | | | |

## $^{194}_{83}Bi_{111}$

| | $E_{level}$ keV | $I^{\pi}$ | $E_{\gamma}(M1)$ keV | $E_{\gamma}(E2)$ keV | B(M1)/B(E2) $(\mu_N/eb)^2$ | **Reference** | **Configurations and Comments:** |
|---|---|---|---|---|---|---|---|
| 1. | 2427.5 | $(16^{+})$ | | | | **2020He17** | 1. Tentatively assigned as $\pi(h^{2}_{9/2}\,i_{13/2})\otimes\nu^{+}$ |
| | 2612.0 | $(17^{+})$ | 184.5 | | | | or $\pi h_{9/2}\otimes\nu(i_{13/2})^{-2}\otimes\nu^{-}$ by comparison with $^{193,195}$Bi |
| | 2966.5 | $(18^{+})$ | 354.5 | | | | isotopes. |
| | 3410.7 | $(19^{+})$ | 444.2 | | | | 2. Tentatively assigned as MR band. |
| | 3843.9 | | 457.9 | | | | 3. Regular band. |
| | 4301.8 | | | | | | |

## $^{195}_{83}Bi_{112}$

| | $E_{level}$ keV | $I^{\pi}$ | $E_{\gamma}(M1)$ keV | $E_{\gamma}(E2)$ keV | B(M1)/B(E2) $(\mu_N/eb)^2$ | **Reference** | **Configurations and Comments:** |
|---|---|---|---|---|---|---|---|
| 1. | 3595.0 | $(35/2^{-})$ | | | | **2017He12** | 1. Tentatively assigned as $\pi(h^{2}_{9/2}\,i_{13/2})\otimes\nu(i_{13/2}\,p_{3/2})$ |
| | 3724.6 | $(37/2^{-})$ | 129.6 | | | | by comparison with $^{194}$Pb and $^{195}$Bi isotopes. |
| | 3953.5 | $(39/2^{-})$ | 228.9 | | | | 2. Tentatively assigned as MR band. |
| | 4255.5 | $(41/2^{-})$ | 302.0 | | | | 3. Regular band. |
| | 4636.4 | $(43/2^{-})$ | 380.9 | | | | |

## $^{197}_{83}Bi_{114}$

| | $E_{level}$ keV | $I^{\pi}$ | $E_{\gamma}(M1)$ keV | $E_{\gamma}(E2)$ keV | B(M1)/B(E2) $(\mu_N/eb)^2$ | Reference | **Configurations and Comments:** |
|---|---|---|---|---|---|---|---|
| 1. | 4019.2 | (37/2) | | | | **2005Ma51** | 1. Tentatively assigned as $\pi(h_{9/2}^{2}i_{13/2})_{K=14.5}$ coupled |
| | 4237.4 | (39/2) | 218.2 | | | | to $\nu[i_{13/2}\,(p_{3/2}f_{5/2})^{1}]$ by comparison with similar |
| | 4492.2 | (41/2) | 254.8 | | | | bands in neighboring $^{196}$Pb and $^{199}$Bi. |
| | 4784.5 | (43/2) | 292.3 | | | | 2. Oblate deformation ($\beta_2 = 0.17$) for the three |
| | 5111.9 | (45/2) | 327.4 | | | | high-K proton configuration from the TRS |
| | 5376.1 | (47/2) | 264.2 | | | | calculations. |
| | 5678.3 | (49/2) | 302.2 | | | | 3. Regular band with backbending at 5376 keV |
| | 6033.8 | (51/2) | 355.5 | | | | level. |
| | (6429.8) | | (396.0) | | | | 4. The ordering of 264, 302 and 356 keV |
| | | | | | | | transitions is arbitrary. |

## $^{198}_{83}\mathrm{Bi}_{115}$

| | $E_{level}$ keV | $I^{\pi}$ | $E_{\gamma}$(M1) keV | $E_{\gamma}$(E2) keV | B(M1)/B(E2) $(\mu_N/eb)^2$ | Reference | **Configurations and Comments:** |
|---|---|---|---|---|---|---|---|
| 1. | U | | | | | **1994Da17** | 1. Tentatively assigned as $\pi(h_{9/2}\, i_{13/2}\, s_{1/2}^{-1})$ coupled to one or three $i_{13/2}$ neutron holes by comparison with a similar band in neighboring Pb isotopes.<br>2. B(M1)/B(E2) ratios are large due to the Non observation of crossover E2 transitions.<br>3. Regular band. |
| | 165+U | | 165 | | | | |
| | 416+U | | 251 | | | | |
| | 731+U | | 315 | | | | |
| | 1108+U | | 377 | | | | |
| | 1517+U | | 409 | | | | |
| 2. | (2290+X) | $(16^+)$ | | | | **2014Pa53** | 1. Configuration assigned as $\pi i_{13/2}\,(h_{9/2})^2\,(s_{1/2})^{-2} \otimes \nu(i_{13/2})^{-1}$ and $\pi i_{13/2}\,(h_{9/2})^2\,(s_{1/2})^{-2} \otimes \nu(i_{13/2})^{-3}$ before and after band crossing, respectively. This configuration is based on the comparison of similar configuration observed in $^{197}$Pb nuclide.<br>2. Irregular band. |
| | 2724.4+X | $17^+$ | (434) | | | 2000Zw02 | |
| | 3301.2+X | $18^+$ | 576.7 | | | | |
| | 3763.7+X | $19^+$ | 462.5 | | | | |
| | 4065.8+X | $(20^+)$ | 301.8 | | | | |
| | 4385.3+X | $(21^+)$ | 319.8 | | | | |
| | 4646.6+X | $(22^+)$ | 261.2 | | | | |
| 3. | 2596.0+X | 17- | | | | **2014Pa53** | 1. Configuration assigned as $\pi h_{9/2} \otimes \nu(i_{13/2})^{-3}$ and $\pi h_{9/2} \otimes \nu(i_{13/2})^{-3}\,(p_{3/2})^{-2}$ before and after band crossing, respectively from TAC calculations.<br>2. $(\varepsilon_2, \varepsilon_4, \gamma)$= (- 0.085, 0.003, 16.4°) and (- 0.090, 0.003, 4.5°) before and after bandcrossing, respectively from TAC calculations.<br>3. The g-factor extracted using experimental B(M1)/B(E2) values for $20^-$ and $21^-$ state is 0.72(10).<br>4. Irregular band with band crossing above I=24 ħ. |
| | 2838.2+X | $(18^-)$ | 242.2 | | | 2000Zw02 | |
| | 3132.3+X | $(19^-)$ | 294.1 | | | | |
| | 3429.3+X | $(20^-)$ | 296.6 | 591.3 | 6.6(14) | | |
| | 3747.0+X | $(21^-)$ | 317.5 | 615.2 | 9.2(21) | | |
| | 4126.8+X | $(22^-)$ | 379.8 | | | | |
| | 4339.7+X | $(23^-)$ | 212.9 | | | | |
| | 4627.5+X | $(24^-)$ | 287.8 | | | | |
| | 4856.7+X | $(25^-)$ | 229.2 | | | | |
| 4. | 3635.8+X | $19^+$ | | | | **2014Pa53** | 1. Configuration assigned as $\pi h_{9/2} \otimes \nu(i_{13/2})^{-2}\,(p_{3/2}\, f_{5/2})^{-3}$ and $\pi h_{9/2} \otimes \nu(i_{13/2})^{-4}\,(p_{3/2}, f_{5/2})^{-3}$ before and after band crossing respectively from TAC calculations.<br>2. $(\varepsilon_2, \varepsilon_4, \gamma)$= (- 0.083, 0.003, 6.9°) and (- 0.085, 0.002, 1.1°) before and after bandcrossing, respectively from TAC calculations.<br>3. Irregular band with band crossing above I=24 ħ. |
| | 3966.3+X | $20^+$ | 330.6 | | | 2000Zw02 | |
| | 4192.5+X | $(21^+)$ | 226.2 | | | | |
| | 4482.8+X | $(22^+)$ | 290.3 | | | | |
| | 4845.7+X | $(23^+)$ | 362.9 | | | | |
| | 5272.2+X | $(24^+)$ | 426.5 | | | | |
| | 5767.8+X | $(25^+)$ | 495.5 | | | | |
| | 5971.3+X | $(26^+)$ | 203.5 | | | | |
| | 6486.4+X | $(27^+)$ | 515.1 | | | | |

## $^{199}_{83}\mathrm{Bi}_{116}$

| | $E_{level}$ keV | $I^{\pi}$ | $E_{\gamma}$(M1) keV | $E_{\gamma}$(E2) keV | B(M1)/B(E2) $(\mu_N/eb)^2$ | Reference | **Configurations and Comments:** |
|---|---|---|---|---|---|---|---|
| 1. | X | | | | | **1994Da17** | 1. Tentatively assigned as $\pi(h_{9/2}\, i_{13/2}\, s_{1/2}^{-1})$ coupled to two $i_{13/2}$ neutron holes by comparison with a similar band in neighboring Pb isotopes. |
| | 184.4+X | | 184.4 | | | | 2. The band depopulates around 37/2. |
| | 400.2+X | | 215.8 | | | | 3. Tentatively assigned as MR band by 1994Da17. |
| | 642.0+X | | 241.8 | | | | 4. B(M1)/B(E2) ratios are large due to the non observation of crossover E2 transitions. |
| | 923.2+X | | 281.2 | | | | 5. Regular band. |
| | 1236.7+X | | 313.5 | | | | |
| | 1590.3+X | | 353.6 | | | | |
| | 1950.8+X | | 360.5 | | | | |
| | 2316.7+X | | 365.9 | | | | |

## $^{200}_{83}\mathrm{Bi}_{117}$

| | $E_{level}$ keV | $I^{\pi}$ | $E_{\gamma}$(M1) keV | $E_{\gamma}$(E2) keV | B(M1)/B(E2) $(\mu_N/eb)^2$ | Reference | **Configurations and Comments:** |
|---|---|---|---|---|---|---|---|
| 1. | X | | | | | **1994Da17** | 1. Tentatively assigned as $\pi(h_{9/2}\, i_{13/2}\, s_{1/2}^{-1})$ coupled to one or three $i_{13/2}$ neutron holes by comparison with a similar band in neighboring Pb isotopes. |
| | 193+X | | 193 | | | | 2. Tentatively assigned as MR band by 1994Da17. |
| | 431+X | | 238 | | | | 3. B(M1)/B(E2) ratios are large due to the non observation of crossover E2 transitions. |
| | 720+X | | 289 | | | | 4. Regular band. |
| | 1056+X | | 336 | | | | |
| | 1432+X | | 376 | | | | |
| | 1855+X | | 423 | | | | |
| 2. | Y | | | | | **1994Da17** | 1. Tentatively assigned as $\pi(h_{9/2}\, i_{13/2}\, s_{1/2}^{-1})$ coupled to one or three $i_{13/2}$ neutron holes by comparison with a similar band in neighboring Pb isotopes. |
| | 199.0+Y | | 199.0 | | | | 2. Tentatively assigned as MR band by 1994Da17. |
| | 446.2+Y | | 247.2 | | | | 3. B(M1)/B(E2) ≥ 10 $(\mu_N/eb)^2$. |
| | 740.7+Y | | 294.5 | | | | 4. Regular band. |
| | 1083.8+Y | | 343.1 | | | | |
| | 1475.2+Y | | 391.4 | | | | |
| | 1918.8+Y | | 443.6 | | | | |
| | 2417.8+Y | | 499.0 | | | | |
| | 2970.7+Y | | 552.9 | | | | |
| | 3577.7+Y | | 607.0 | | | | |

## $^{202}_{83}\mathrm{Bi}_{119}$

| | $E_{level}$ keV | $I^{\pi}$ | $E_{\gamma}$(M1) keV | $E_{\gamma}$(E2) keV | B(M1)/B(E2) $(\mu_N/eb)^2$ | Reference | **Configurations and Comments:** |
|---|---|---|---|---|---|---|---|
| 1. | X | | | | | **1993Cl02** | 1. Tentatively assigned as $\pi(h_{9/2}\, i_{13/2}\, s_{1/2}^{-1})$ coupled to one or two $i_{13/2}$ neutron holes by comparison with a similar band in neighboring Pb isotopes. |
| | 164+X | | 164 | | | | 2. The estimated bandhead spin is about 10-16. |
| | 423+X | | 259 | | | | 3. Tentatively assigned as MR band by 1993Cl02. |
| | 775+X | | 352 | | | | 4. B(M1)/B(E2) ≥ 12 $(\mu_N/eb)^2$. |
| | 1199+X | | 424 | | | | 5. Regular band. |
| | 1680+X | | 481 | | | | |
| | 2210+X | | 530 | | | | |
| | 2780+X | | 570 | | | | |

## $^{202}_{83}\mathrm{Bi}_{119}$

| | $E_{level}$ keV | $I^{\pi}$ | $E_{\gamma}$(M1) keV | $E_{\gamma}$(E2) keV | B(M1)/B(E2) $(\mu_N/eb)^2$ | **Reference** | **Configurations and Comments:** |
|---|---|---|---|---|---|---|---|
| 2. | Y | | | | | **1993Cl02** | 1. Tentatively assigned as π($h_{9/2}$ $i_{13/2}$ $s_{1/2}^{-1}$) or π($h_{9/2}^2 s_{1/2}^{-1}$) coupled to one or two $i_{13/2}$ neutron holes by comparison with the similar band in neighboring Pb isotopes.<br>2. Tentatively assigned as MR band by 1993Cl02.<br>3. The estimated bandhead spin is about 10-16.<br>4. B(M1)/B(E2) ≥ 6 $(\mu_N/eb)^2$.<br>5. Regular band. |
| | 180+Y | | 180 | | | | |
| | 394+Y | | 214 | | | | |
| | 659+Y | | 265 | | | | |
| | 984+Y | | 325 | | | | |
| | 1374+Y | | 390 | | | | |
| 3. | Z | | | | | **1993Cl02** | 1. Tentatively assigned as π( $h_{9/2} i_{13/2} s_{1/2}^{-1}$) coupled to one or two $i_{13/2}$ neutron holes by comparison with the similar band in neighboring Pb isotopes. The estimated bandhead spin is about 11-19.<br>2. Tentatively assigned as MR band by 1993Cl02.<br>3. B(M1)/B(E2) ≥ 5 $(\mu_N/eb)^2$.<br>4. Regular band. |
| | 250+Z | | 250 | | | | |
| | 550+Z | | 300 | | | | |
| | 907+Z | | 357 | | | | |
| | 1320+Z | | 413 | | | | |
| | 1785+Z | | 465 | | | | |
| | 2302+Z | | 517 | | | | |

## $^{203}_{83}\mathrm{Bi}_{120}$

| | $E_{level}$ keV | $I^{\pi}$ | $E_{\gamma}$(M1) keV | $E_{\gamma}$(E2) keV | B(M1)/B(E2) $(\mu_N/eb)^2$ | Reference | **Configurations and Comments:** |
|---|---|---|---|---|---|---|---|
| 1. | X | | | | | **1994Da17** | 1. Tentatively assigned as π($h_{9/2} i_{13/2} s_{1/2}^{-1}$) coupled to two $i_{13/2}$ neutron holes by comparison with a similar band in neighboring Pb isotopes.<br>3. Tentatively assigned as MR band by 1994Da17.<br>3. B(M1)/B(E2) ratios are large due to the non observation of crossover E2 transitions.<br>4. Regular band. |
| | 175+X | | 175 | | | | |
| | 421+X | | 246 | | | | |
| | 759+X | | 338 | | | | |
| | 1201+X | | 442 | | | | |
| | 1718+X | | 517 | | | | |
| | 2295+X | | 577 | | | | |

## $^{201}_{85}\mathrm{At}_{116}$

| | $E_{level}$ keV | $I^{\pi}$ | $E_{\gamma}$(M1) keV | $E_{\gamma}$(E2) keV | B(M1)/B(E2) $(\mu_N/eb)^2$ | References | **Configurations and Comments:** |
|---|---|---|---|---|---|---|---|
| 1. | (2990) | (23/2$^-$) | | | | **2015Au01** | 1. Configuration assigned as π($i_{13/2}$) ⊗ ν[$(f_{5/2})^{-1}$ $(i_{13/2})^{-1}]_{9-}$before band crossing and π[$(h_{9/2})^2$ $i_{13/2}$)] ⊗ν[$(f_{5/2})^{-1}$( $i_{13/2})^{-1}]_{5-}$ after band crossing. from semi-classical calculations.<br>2. Regular band with backbending at 31/2 ħ. |
| | (3135) | (25/2$^-$) | 145.0 | | | | |
| | (3380) | (27/2$^-$) | 244.4 | | >30 | | |
| | (3667) | (29/2$^-$) | 286.9 | | >25 | | |
| | (3984) | (31/2$^-$) | 317.3 | | >30 | | |
| | (4256) | (33/2$^-$) | 272.3 | | >35 | | |
| | (4454) | (35/2$^-$) | 197.9 | | >8 | | |
| | (4791) | (37/2$^-$) | 335.0 | | >2 | | |

### $^{203}_{85}At_{118}$

| | E$_{level}$ keV | I$^{\pi}$ | E$_{\gamma}$(M1) keV | E$_{\gamma}$(E2) keV | B(M1)/B(E2) ($\mu_N$/eb)$^2$ | References | **Configurations and Comments:** |
|---|---|---|---|---|---|---|---|
| 1. | (3486) | (27/2$^+$) | | | | **2018Au01** | 1. Configuration assigned as $\pi(i_{13/2}) \otimes \nu(i_{13/2}^{-2})$ |
| | (3620) | (29/2$^+$) | 133.8 | | | | and as $\pi(h^2_{9/2}\, i_{13/2}) \otimes \nu(i_{13/2}^{-2}\}$ below and |
| | (3843) | (31/2$^+$) | 178.8 | | >15 | | above the band crossing, respectively using |
| | (4102) | (33/2$^+$) | 193.6 | | >2 | | TAC-CDFT calculations. |
| | (4386) | (35/2$^+$) | 223.4 | | >10 | | 2. Regular band |
| | (4639) | (37/2$^+$) | 253.4 | | >25 | | |
| | (4818) | (39/2$^+$) | 259.4 | | >20 | | |
| | (5011) | (41/2$^+$) | 283.3 | | >30 | | |
| | (5333) | (43/2$^+$) | 321.7 | | >3 | | |
| | (5714) | (45/2$^+$) | 381.0 | | >5 | | |

### $^{204}_{85}At_{119}$

| | E$_{level}$ keV | I$^{\pi}$ | E$_{\gamma}$(M1) keV | E$_{\gamma}$(E2) keV | B(M1)/B(E2) ($\mu_N$/eb)$^2$ | References | **Configurations and Comments:** |
|---|---|---|---|---|---|---|---|
| 1. | 4018 | 16$^+$ | | | | **2022Ka24** | 1. Configuration assigned as $\pi(h^2_{9/2}\, i_{13/2}) \otimes \nu(i_{13/2}^{-2}\}$ |
| | 4149 | 17$^+$ | 131.3 | | | 2008Ha39 | and $\pi(h^4_{9/2}\, i_{13/2}) \otimes \nu(i_{13/2}^{-2}\}$ below and |
| | 4434 | 18$^+$ | 285.1 | 415.6 | 34(6) | | above the band crossing, respectively using |
| | 4733 | 19$^+$ | 298.7 | 583.8 | 20(4) | | SPAC calculations. |
| | 5028 | 20$^+$ | 295.1 | 594.4 | 15(4) | | 2. Oblate deformation ($\beta_2$ = -0.143) from RMF |
| | 5274 | 21$^+$ | 245.9 | 541.2 | 35(8) | | calculations. |
| | 5451 | 22$^+$ | 177.4 | 422.5 | 54(27) | | 3. Irregular band. |
| | 5647 | 23$^+$ | 196.3 | 373.4 | 14(4) | | |

### $^{205}_{86}Rn_{119}$

| | E$_{level}$ keV | I$^{\pi}$ | E$_{\gamma}$(M1) keV | E$_{\gamma}$(E2) keV | B(M1)/B(E2) ($\mu_N$/eb)$^2$ | Reference | **Configurations and Comments:** |
|---|---|---|---|---|---|---|---|
| 1. | 1680+X | (21/2$^+$) | | | | **1999No03** | 1. The most likely configuration is the negative |
| | 1796.7+X | (23/2$^+$) | 116.7 | | | | parity $\pi(h_{9/2} i_{13/2}) \otimes \nu(i_{13/2})$ from the TAC |
| | 1966.9+X | (25/2$^+$) | 170.2 | | | | calculations. Since the observed parities are |
| | 2124.8+X | (27/2$^+$) | 157.9 | | 2.0(2) | | positive, the configuration $\pi(i_{13/2}{}^2) \otimes \nu(i_{13/2})$ is |
| | 2246.0+X | (29/2$^+$) | 121.2 | | >4 | | tentatively assigned. |
| | 2494.0+X | (31/2$^+$) | 248.0 | | >7 | | 2. Tentatively assigned as MR band. |
| | 2861.7+X | (33/2$^+$) | 367.7 | | >10 | | 3. Small oblate deformation, $\beta_2$~ -0.1. |
| | 3164.1+X | (35/2$^+$) | 302.4 | | >33 | | 4. X ~ 600 keV from systematics. |
| | 3452.3+X | (37/2$^+$) | 288.2 | | >18 | | 5. Irregular band. |
| | 3653.6+X | (39/2$^+$) | 201.3 | | | | |
| | 4059.4+X | (41/2$^+$) | 405.8 | | | | |

### $^{209}_{86}Rn_{123}$

| | E$_{level}$ keV | I$^{\pi}$ | E$_{\gamma}$(M1) keV | E$_{\gamma}$(E2) keV | B(M1)/B(E2) ($\mu_N$/eb)$^2$ | Reference | **Configurations and Comments:** |
|---|---|---|---|---|---|---|---|
| 1. | 3553.6 | 27/2$^-$ | | | | **2024DA02** | 1. Configuration assigned as $\pi\, (h_{9/2})^4 \otimes \nu[(f_{5/2})^{-2}$ |
| | 3749.4 | 29/2$^-$ | 195.8 | | | | $(p_{1/2})^{-1}]$ from semiclassical model calculations. |
| | 4064.9 | 31/2$^-$ | 315.5 | | | | 2. Regular band. |
| | 4466.1 | 33/2$^-$ | 401.2 | 716.6 | | | |
| | 5041.3 | 35/2$^-$ | 575.2 | | | | |
| | 5663.5 | 37/2$^{(-)}$ | 622.2 | | | | |

## $^{206}_{87}Fr_{119}$

| | $E_{level}$ keV | $I^{\pi}$ | $E_{\gamma}$(M1) keV | $E_{\gamma}$(E2) keV | B(M1)/B(E2) $(\mu_N/eb)^2$ | References | **Configurations and Comments:** |
|---|---|---|---|---|---|---|---|
| 1. | X | I | | | | **2008Ha39** | 1. Probable configuration may be built over π ($h_{9/2}$ and/or $i_{13/2}$)⊗ν$i_{13/2}$ neutrons based on systematic of neighboring nuclides.<br>2. Tentatively assigned as MR band.<br>3. Irregular band. |
| | 140.4+X | I+1 | 140.4 | | | | |
| | 407.9+X | I+2 | 267.5 | | >5.9 | | |
| | 670.3+X | I+3 | 262.4 | | >17 | | |
| | 964.4+X | I+4 | 294.1 | | >31 | | |
| | 1242.0+X | I+5 | 277.6 | | >16 | | |
| | 1484.6+X | I+6 | 242.6 | | >17 | | |
| | 1683.1+X | I+7 | 198.5 | | >10 | | |
| | 1909.7+X | I+8 | 226.6 | | >8.6 | | |
| | 2214.1+X | I+9 | (304.4) | | >1.5 | | |

Table 3: Antimagnetic Rotational Structures in Nuclei

**$^{58}_{26}Fe_{32}$**

| | $E_{level}$ keV | $I^\pi$ | $E_\gamma$(E2) keV | B(E2) (eb)$^2$ | B(E2) (W.u.) | References | **Configurations and Comments:** |
|---|---|---|---|---|---|---|---|
| 1. | 5086.8 | 6 | | | | **2012St06** | 1. Band 3 in Fig. 2 of 2012St06 was interpreted in 2017Pe15 theory paper as a mixture of collective and possible antimagnetic rotation with proposed configuration of $\pi(f_{7/2})^{-2}$ $\otimes\nu[(g_{9/2})^1(fp)^3]$ from TAC-CDFT calculations with ($\beta_2$, $\gamma$) = (0.28-0.22, 15°-23°), whereas 2012St06 interpreted this band only as a regular rotational band |
| | 6033.9 | 8 | 947.1 | | | | |
| | 7457.1 | 10 | 1423.1 | | | | |
| | 9444.7 | 12 | 1987.5 | | | | |
| | 11852.9 | (14) | 2408.4 | | | | |
| | 14265.2 | (16) | 2412.0 | | | | |
| 2. | 8424.3 | $(10^+)$ | | | | **2012St06** | 1. Band 4 in Fig. 2 of 2012St06 was interpreted in 2017Pe15 theory paper as a mixture of dominantly collective and possible antimagnetic rotation character with proposed configuration of $\pi(f_{7/2})^{-2}\otimes\nu[(g_{9/2})^2(fp)^2]$ from TAC-CDFT calculations, with ($\beta_2$, $\gamma$) = (0.34-0.26, 1.4°- 16°), while 2012St06 interpreted this band only as a regular rotational band. |
| | 9983.2 | $12^{(+)}$ | 1558.9 | | | | |
| | 11906.8 | $14^{(+)}$ | 1924.2 | | | | |
| | 14314.8 | $(16^+)$ | 2407.9 | | | | |

**$^{61}_{28}Ni_{33}$**

| | $E_{level}$ keV | $I^\pi$ | $E_\gamma$(E2) keV | B(E2) (eb)$^2$ | B(E2) (W.u.) | References | **Configurations and Comments:** |
|---|---|---|---|---|---|---|---|
| 1. | 3297.6 | $11/2^+$ | | | | **2023Li05** | 1. The authors of 2023Li05 interpreted band #2 in Fig. 2 as a possible mixture of collective and antimagnetic rotation with proposed configuration of $\pi[(1f_{7/2})^{-2}(fp)^2]\otimes$ $\nu[(1g_{9/2})^1(fp)^4]$ from TAC-CDFT calculations, with predicted B(E2) of < 0.11 (eb)$^2$. 2023Bh02 interpreted this band as a deformed collective bands.<br>2. Quasi-rotational band, as level energies do not seem to follow a monotonically increasing trend with ascending spins as expected for a rotor. |
| | 4476.0 | $15/2^+$ | 1178.4 | | | | |
| | 6065.0 | $19/2^+$ | 1587.7 | | | | |
| | 7603.6 | $23/2^+$ | 1538.6 | | | | |
| | 9104.4 | $27/2^+$ | 1500.8 | | | | |

**$^{82}_{36}Kr_{46}$**

| | $E_{level}$ keV | $I^\pi$ | $E_\gamma$(E2) keV | B(E2) (eb)$^2$ | B(E2) (W.u.) | References | **Configurations and Comments:** |
|---|---|---|---|---|---|---|---|
| 1. | 4016.5 | $8^+$ | | | | **2024Ra20** | 1. Configuration assigned as $\pi(g_{9/2})^2\otimes\nu(g_{9/2})^{-2}$ on the basis of Large Basis Shell Model.<br>2. The mean lifetimes (in ps) of the levels from 4822 keV to 9066 keV are 1.67 (21), 0.39(5), 0.34(5) and <0.52, respectively.<br>3. The $\mathscr{J}^2$/B(E2) ratios of 10 ħ to 16 ħ levels are 74.3(93), 218 (28), 298(60) and <2100 $\hbar^2$ MeV$^{-1}$ (eb)$^{-2}$, respectively. |
| | 4822.4 | $10^+$ | 805.8 | 0.14(2) | 50(8) | | |
| | 6011.3 | $12^+$ | 1188.9 | 0.09(1) | 32(2) | | |
| | 7404.1 | $14^+$ | 1392.8 | 0.05(1) | 17(3) | | |
| | 9066.5 | $16^+$ | 1662.4 | >0.01 | >3 | | |
| | 10919.5 | (18+) | 1853.0 | | | | |

## $^{99}_{46}Pd_{53}$

| | $E_{level}$ keV | $I^{\pi}$ | $E_{\gamma}(E2)$ keV | B(E2) $(eb)^2$ | B(E2) (W.u.) | References | **Configurations and Comments:** |
|---|---|---|---|---|---|---|---|
| 1. | 4014.2 | $23/2^-$ | | | | **2011Si04** | 1. Band B3 in 2011Si04 was interpreted as a possible AMR band in theory paper 2022Si07 with proposed configuration of $\pi(g_{9/2})^{-4}\otimes\nu[h_{11/2}(g_{9/2})^2]$ and predicted B(E2) of ≈ 0.10 $(eb)^2$.<br>2. Quasi-rotational band, as level energies do not seem to follow a monotonically increasing trend with ascending spins as expected for a rotor. |
| | 4772.7 | $27/2^-$ | 758.5 | | | | |
| | 5780.5 | $31/2^-$ | 1007.8 | | | | |
| | 6803.2 | $35/2^-$ | 1022.7 | | | | |
| | 8182.7 | $39/2^-$ | 1379.5 | | | | |
| | 9512.9 | 43/2 | 1330.2 | | | | |
| | 11457 | (47/2) | 1944.1 | | | | |

## $^{100}_{46}Pd_{54}$

| | $E_{level}$ keV | $I^{\pi}$ | $E_{\gamma}(E2)$ keV | B(E2) $(eb)^2$ | B(E2) (W.u.) | References | **Configurations and Comments:** |
|---|---|---|---|---|---|---|---|
| 1 | 6701 | $15^-$ | | | | **2020Si20** | 1. Probable configuration as $\pi(g_{9/2})^{-4}\otimes\nu[(h_{11/2})(g_{7/2})^3]$ is assigned based on semiclassical particle-rotor model calculations.<br>2. The mean lifetimes (in ps) of the levels from 7641 keV to 10100 keV are 0.80 (10), 0.60(7) and <0.21, respectively.<br>3. The $\mathscr{J}^2$/B(E2) ratios of 19 ħ to 21 ħ levels are 169 and 318 $\hbar^2$ $MeV^{-1}$ $(eb)^{-2}$, respectively, 2020Si20. |
| | 7641 | $17^-$ | 940 | | | **2001Zh26** | |
| | 8712 | $19^-$ | 1071 | 0.096(8) | 34.8(29) | | |
| | 10100 | $21^-$ | 1388 | >0.075 | >27 | | |
| | 11682 | $23^{(-)}$ | 1582 | | | | |
| | 13434 | $25^{(-)}$ | 1752 | | | | |

## $^{101}_{46}Pd_{55}$

| 1. | $E_{level}$ keV | $I^{\pi}$ | $E_{\gamma}(E2)$ keV | B(E2) $(eb)^2$ | B(E2) (W.u.) | References | **Configurations and Comments:** |
|---|---|---|---|---|---|---|---|
| | 2640 | $19/2^-$ | 748.2 | 0.21(6) | 72(22) | **2017Si12** | 1. Probable configuration as $\pi(g_{9/2})^{-2}\otimes\nu(h_{11/2})^1(g_{7/2})^2$ is assigned on the basis of CSM and TAC calculations. (2017Si12). However, configuration $\pi(g_{9/2})^{-4}\otimes\nu(h_{11/2})^1(g_{7/2})^2$ is assigned by 2015Su19.<br>2. The mean lifetimes (in ps) of the levels from 2640 keV to 9037 keV are 1.7(5), 0.86(26), 0.89(19), 0.66(14), 0.52(10), 0.49(11) and 0.15(4), respectively. The lifetime values are obtained using weighted averages by the authors of this work after adding in quadrature 20% systematic uncertainty (from stopping powers) in each value from the 2017Si12 and 2015Su09, and in addition, assigning a total uncertainty of 25% in the values of lifetimes for the 2640 keV and 3531 keV levels in 2017Si12, where these are given as approximate. The lifetime value of the 9037 keV level is from 2015Su09.<br>3. The B(E2) values for the transitions from 748.2 keV to 1421.8 keV are deduced by compliers from adopted levels lifetimes.<br>4. The $\mathscr{J}^2$/B(E2) ratios of 19/2 ħ to 39/2 ħ levels are 135(72), 1222(96), 448(33), 278(16), 676(44) and 148(86), $\hbar^2$ $MeV^{-1}$ $(eb)^{-2}$, respectively. The values are deduced by the authors of this work based on averaged level lifetimes. |
| | 3531 | $23/2^-$ | 891.1 | 0.16(5) | 57(18) | **2015Su09** | |
| | 4442 | $27/2^-$ | 911.2 | 0.15(3) | 54(11) | 2012Su18 | |
| | 5413 | $31/2^-$ | 971.3 | 0.14(3) | 50(11) | | |
| | 6487 | $35/2^-$ | 1074.1 | 0.109(21) | 39(8) | | |
| | 7615 | $39/2^-$ | 1127.5 | 0.091(21) | 33(8) | | |
| | 9037 | $43/2^-$ | 1421.8 | 0.09(3) | 32(11) | | |
| | 10628 | $(47/2^-)$ | (1591.0) | | | | |

## $^{102}_{46}Pd_{56}$

| | $E_{level}$ keV | $I^{\pi}$ | $E_{\gamma}(E2)$ keV | B(E2) $(eb)^2$ | B(E2) (W.u.) | References | **Configurations and Comments:** |
|---|---|---|---|---|---|---|---|
| 1. | 8418.2 | $18^+$ | | | | **1997Gi10** | 1. Band #2 in Fig. 1 of 1997Gi10 was interpreted in 2018Ji02 theory paper as a mixture of collective and possible antimagnetic rotation with proposed configuration of $\pi(g_{9/2})^{-4} \otimes \nu[(h_{11/2})^2(g_{7/2},d_{5/2})^4]$ from TAC-CDFT calculations with deformation $\beta > 0.15$, whereas 1997Gi10 interpreted this band only as a regular rotational band.<br>2. Quasi-rotational band, as level energies do not seem to follow a monotonically increasing trend with ascending spins as expected for a rotor. |
| | 9543.2 | $20^+$ | 1125 | | | | |
| | 10792.2 | $22^+$ | 1249 | | | | |
| | 12073.2 | $24^+$ | 1281 | | | | |
| | 13589.2 | $26^+$ | 1516 | | | | |
| | 15252.2 | $28^+$ | 1663 | | | | |
| | 16958.2 | $30^+$ | 1706 | | | | |
| | 18826.2 | $32^+$ | 1868 | | | | |

## $^{103}_{46}Pd_{57}$

| | $E_{level}$ keV | $I^{\pi}$ | $E_{\gamma}(E2)$ keV | B(E2) $(eb)^2$ | B(E2) (W.u.) | References | **Configurations and Comments:** |
|---|---|---|---|---|---|---|---|
| 1. | 1974.9 | $19/2^-$ | 714 | | | **2021Sh09** | 1. The higher spin members of the yrast band in Fig. 1 of 2021Sh09 tentatively assigned as AMR band with proposed configuration of $\pi(g_{9/2})^{-4} \otimes \nu[h_{11/2}(g_{7/2}, d_{5/2})^6]$ in theory paper by 2018Ji02. 1999Ny01 interpret this band as a single-particle collective rotational band built on $h_{11/2}$ neutron orbital.<br>2. The mean lifetimes (in ps) of the levels from 2822.0 keV to 8668 keV are 0.55(5), 0.48(7), 0.42(5), 0.36(5), 0.27(5) and <0.35, respectively.<br>3. The $\mathcal{J}^2$/B(E2) ratios of 23/2 ħ to 43/2 ħ levels are 95(9), 161(+22-23), 495(63), 358(50), 524(+101-96) and < 532 $\hbar^2$ MeV$^{-1}$ (eb)$^{-2}$, respectively (2021Sh09). |
| | 2822.0 | $23/2^-$ | 847 | 0.34(3) | 118(10) | **1999Ny01** | |
| | 3792.1 | $27/2^-$ | 970 | 0.20(3) | 68(10) | 2021De11 | |
| | 4886.4 | $31/2^-$ | 1094 | 0.12(2) | 42(7) | | |
| | 6048.3 | $35/2^-$ | 1162 | 0.11(1) | 38.3(35) | | |
| | 7316 | $39/2^-$ | 1267 | 0.09(2) | 31(7) | | |
| | 8668 | $43/2^-$ | 1350 | > 0.05 | > 17.4 | | |
| | 10119 | $47/2^-$ | 1451 | | | | |
| | 11638 | $51/2^-$ | 1519 | | | | |
| 2. | 4160 | $25/2^+$ | | | | **1999Ny01** | 1. Configuration assigned as $\pi(g_{9/2})^{-4} \otimes \nu[(g_{7/2}), (h_{11/2})^2]$ on the basis of classical particle rotor model calculations. The AMR assignment is based on decreasing tendencies of calculated B(E2) with rotational frequency and calculated $\mathcal{J}^2$/B(E2) ratios (>100 $\hbar^2$ MeV$^{-1}$ (eb)$^{-2}$) (2022Pa27).<br>2. This band was identified as signature partner of band 3. |
| | 5025 | $29/2^+$ | 865 | | | | |
| | 5984 | $33/2^+$ | 959 | | | | |
| | 7056 | $37/2^+$ | 1072 | | | | |
| | 8212 | $41/2^+$ | 1156 | | | | |
| | 9442 | $45/2^+$ | 1230 | | | | |
| | 10741 | $49/2^+$ | 1299 | | | | |
| | 12208 | $53/2^+$ | 1467 | | | | |
| | 13798 | $57/2^+$ | 1590 | | | | |
| | 15487 | $61/2^+$ | 1689 | | | | |
| | 17357 | $65/2^+$ | 1870 | | | | |
| 3. | 4587 | $27/2^+$ | | | | **1999Ny01** | 1. Configuration assigned as $\pi(g_{9/2})^{-4} \otimes \nu[(g_{7/2}), (h_{11/2})^2]$ on the basis of classical particle rotor model calculations. The AMR assignment is based on decreasing tendencies of calculated B(E2) with rotational frequency and calculated $\mathcal{J}^2$/B(E2) ratios (>100 $\hbar^2$ MeV$^{-1}$ (eb)$^{-2}$) (2022Pa27).<br>2. This band was identified as signature partner of band 2. |
| | 5458 | $31/2^+$ | 871 | | | | |
| | 6452 | $35/2^+$ | 994 | | | | |
| | 7593 | $39/2^+$ | 1141 | | | | |
| | 8831 | $43/2^+$ | 1238 | | | | |
| | 10190 | $47/2^+$ | 1359 | | | | |
| | 11643 | $51/2^+$ | 1453 | | | | |
| | 13240 | $55/2^+$ | 1597 | | | | |
| | 14842 | $59/2^+$ | 1602 | | | | |

## $^{104}_{46}\mathrm{Pd}_{58}$

| | $E_{level}$ keV | $I^{\pi}$ | $E_{\gamma}$(E2) keV | B(E2) $(eb)^2$ | B(E2) (W.u.) | References | **Configurations and Comments:** |
|---|---|---|---|---|---|---|---|
| 1. | 4024 | $10^+$ | | | | **2014Ra11** | 1. The probable configuration as $\pi(g_{9/2})^{-2}\otimes \nu(h_{11/2})^2(g_{7/2}/ d_{5/2})^2$ is assigned based on semi-classical particle rotor model.<br>2. The mean lifetimes (in ps) of the levels from 6360 keV to 12707 keV are 0.45(5), 0.27(3), 0.18(3), 0.20(4), 0.24(5) and <0.66, respectively.<br>3. The $\mathcal{J}^2$/B(E2) ratios from 16 ħ to 26 ħ levels are 112(9), 141(13), 334(35), 283(22) and 554(80) $\hbar^2$ MeV$^{-1}$ (eb)$^{-2}$, respectively.<br>4. $\beta_2 \approx 0.19$<br>5. The B(E2) and lifetimes values are listed with statistical and systematically uncertainties. |
| | 4636 | $12^+$ | 612 | | | | |
| | 5433 | $14^+$ | 797 | | | | |
| | 6360 | $16^+$ | 927 | 0.26(3) | 89(10) | | |
| | 7424 | $18^+$ | 1064 | 0.22(3) | 76(10) | | |
| | 8617 | $20^+$ | 1193 | 0.19(3) | 65(10) | | |
| | 9873 | $22^+$ | 1256 | 0.13(3) | 45(10) | | |
| | 11239 | $24^+$ | 1365 | 0.07(1) | 24.0(34) | | |
| | 12707 | $26^+$ | 1468 | >0.02 | >6.9 | | |

## $^{105}_{48}\mathrm{Cd}_{57}$

| | $E_{level}$ keV | $I^{\pi}$ | $E_{\gamma}$(E2) keV | B(E2) $(eb)^2$ | B(E2) (W.u.) | References | **Configurations and Comments:** |
|---|---|---|---|---|---|---|---|
| 1 | 3343.0 | $23/2^-$ | | | | **2010Ch54** | 1.The probable configuration as $\pi(g_{9/2})^2\otimes \nu(h_{11/2})(g_{7/2})^2$ is assigned based on semi-classical particle rotor model.<br>2. The mean lifetimes (in ps) of the levels from 4248.0 keV to 9264.1 keV are 1.066 (+141-102), 0.621(82), 0.399(+45-31), 0.207(+31-29) and 0.237(+46- 49), respectively.<br>3. The $\mathcal{J}^2$/B(E2) ratios of $27/2^-$ to $43/2^-$ levels are 201(+19-27), 249(+38-28), 255(+20-29), 287 (+40-43) and 608(+128-117) $\hbar^2$ MeV$^{-1}$ (eb)$^{-2}$, respectively.<br>4. Theoretical $\beta_2$ =0.12 in theory paper 2013Zh16. |
| | 4248.0 | $27/2^-$ | 905.0 | 0.126(+12-17) | 43(+4-6) | | |
| | 5291.7 | $31/2^-$ | 1043.7 | 0.106(+16-12) | 36(+5-4) | | |
| | 6471.3 | $35/2^-$ | 1179.6 | 0.089(+7-10) | 30.2(+24-34) | | |
| | 7802.3 | $39/2^-$ | 1331.0 | 0.094(+13-14) | 31.9(+44-48) | | |
| | 9264.1 | $43/2^-$ | 1461.8 | 0.052(+11-10) | 17.7(+37-34) | | |
| | 10845.1 | $47/2^-$ | 1581 | | | | |

## $^{106}_{48}\mathrm{Cd}_{58}$

| | $E_{level}$ keV | $I^{\pi}$ | $E_{\gamma}$(E2) keV | B(E2) $(eb)^2$ | B(E2) (W.u.) | References | **Configurations and Comments:** |
|---|---|---|---|---|---|---|---|
| 1. | 4104.9 | $10^-$ | 598.3 | | | **2025DI03** | 1. Configuration assigned as $\pi(g_{9/2})^{-2}\otimes\nu(h_{11/2}d_{5/2})$ from SCM calculations.<br>2. Calculated $\mathcal{J}^2$/B(E2) ratios (>150 $\hbar^2$ MeV$^{-1}$ (eb)$^{-2}$) from SCM calculations. |
| | 4966.2 | $12^-$ | 861.3 | | | | |
| | 5974.8 | $14^-$ | 1008.6 | | | | |
| | 7120.1 | $15^-$ | 1145.3 | | | | |
| | 8410.2 | $16^-$ | 1290.1 | | | | |
| | 9876.2 | $17^-$ | 1466.0 | | | | |
| 2. | 4324.1 | $11^-$ | 645.6 | 0.37(1) | 132(4) | **2025DI03** | 1. Configuration assigned as $\pi(g_{9/2})^{-2}\otimes\nu(h_{11/2}g_{7/2})$ from SCM calculations.<br>2. The mean lifetimes (in ps) of the levels from 4324.1 keV to 7517.9 keV are 1.96(5), 1.05(4), 0.93(4), and 0.62(3), respectively.<br>3. The $\mathcal{J}^2$/B(E2) ratios of 11 ħ to 17 ħ levels are 44(10), 180(5), 291(7) and 933(41) $\hbar^2$ MeV$^{-1}$ (eb)$^{-2}$, respectively. |
| | 5213.9 | $13^-$ | 889.8 | 0.14(1) | 50(4) | | |
| | 6264.3 | $15^-$ | 1050.4 | 0.07(2) | 25(7) | | |
| | 7517.9 | $17^-$ | 1253.6 | 0.04(2) | 14(7) | | |

## $^{106}_{48}Cd_{58}$

| | $E_{level}$ keV | $I^{\pi}$ | $E_{\gamma}$(E2) keV | B(E2) $(eb)^2$ | B(E2) (W.u.) | References | **Configurations and Comments:** |
|---|---|---|---|---|---|---|---|
| 3. | 5418.8 | $12^+$ | 602.4 | 0.42(1) | 150(4) | **2025DI03** | 1. Configuration assigned as $\pi(g_{9/2})^{-2} \otimes \nu(h_{11/2})^2 (g_{7/2})^2$ from SCM calculations.<br>2. The mean lifetimes (in ps) of the levels from 5418.8 keV to 9168.4 keV are 2.44(4), 1.41(6), 0.93(5), 0.64(5) and 0.42(4), respectively.<br>3. The $\mathcal{J}^2$/B(E2) ratios of 12 ħ to 20 ħ levels are 47(3), 278(16), 301(21), 169(12) and 250(25) $\hbar^2$ $MeV^{-1}$ $(eb)^{-2}$, respectively. |
| | 6226.7 | $14^+$ | 807.9 | 0.17(1) | 61(4) | 2005Si23 | |
| | 7119.0 | $16^+$ | 892.3 | 0.15(1) | 54(4) | 2003Si14 | |
| | 8017.8 | $18^+$ | 980.8 | 0.14(1) | 50(4) | | |
| | 9168.4 | $20^+$ | 1150.6 | 0.10(1) | 36(4) | | |

## $^{107}_{48}Cd_{59}$

| | $E_{level}$ keV | $I^{\pi}$ | $E_{\gamma}$(E2) keV | B(E2) $(eb)^2$ | B(E2) (W.u.) | References | **Configurations and Comments:** |
|---|---|---|---|---|---|---|---|
| 1. | 4009 | $23/2^+$ | | | | **2013Ch14** | 1. Configuration assigned as $\pi(g_{9/2})^{-2} \otimes \nu g_{7/2} (h_{11/2})^2$ α = -1/2 from semi-classical particle rotor model calculations, see also 2015Ch05.<br>2. The mean lifetimes (in ps) of the levels from 5231 keV to 10221.8 keV are 1.902(+221-206), 0.631(+28-22), 0.332(+16-13), 0.170(+11-8) and 0.205(32), respectively.<br>3. The $Q_t$ values for the transition from 728.9 2.257(+4-5), 1.994(+4-5), 1.766(+4-6) and 1.131(9) (eb), respectively.<br>4. The $\mathcal{J}^2$/B(E2) ratios of $31/2^+$ to $47/2^+$ levels are 82(+8-9), 108(+4-5), 168(+6-8), 172(+8-11) and 454(72) $\hbar^2$ $MeV^{-1}$ $(eb)^{-2}$, respectively.<br>5. $\beta_2$ =0.20 at $31/2^+$ ħ. |
| | 4502 | $27/2^+$ | 493 | | | | |
| | 5231 | $31/2^+$ | 729 | 0.208(+22-24) | 69(+7-8) | | |
| | 6183.3 | $35/2^+$ | 952.3 | 0.165(+6-7) | 54.7(+19-23) | | |
| | 7316.5 | $39/2^+$ | 1133.2 | 0.131(+5-6) | 43.4(+17-20) | | |
| | 8669.3 | $43/2^+$ | 1352.8 | 0.106(+5-7) | 35.1(+17-23) | | |
| | 10221.8 | $47/2^+$ | 1552.5 | 0.044(7) | 14.6(23) | | |
| | 11852 | $(51/2^+)$ | (1630) | | | | |
| 2. | 4191 | $25/2^+$ | | | | **2013Ch14** | 1. Configuration assigned as $\pi(g_{9/2})^{-2} \otimes \nu g_{7/2} (h_{11/2})^2$ α= +1/2 from semi-classical particle rotor model calculations, see also 2015Ch05.<br>2. The mean lifetimes (in ps) of the levels from 5816.0 keV to 11230.1 keV are 0.609(+30-35), 0.317(12), 0.196(+18-14), 0.142(+12-10), 0.222(44), respectively.<br>3. The $Q_t$ values for the transition from 939.0 keV to 1608.0 keV are 2.399(+7-6), 2.175(4), 1.958(+7-6), 1.668(+6-7) and 0.996(10) (eb), respectively.<br>4. The $\mathcal{J}^2$/B(E2) ratios of $33/2^+$ to $49/2^+$ levels are 86(+5-4), 154(6), 195(+14-18), 252(+19-21) and 671(138) $\hbar^2$ $MeV^{-1}$ $(eb)^{-2}$, respectively.<br>5. $\beta_2$ =0.19 at $33/2^+$ ħ. |
| | 4877 | $29/2^+$ | 686 | | | | |
| | 5816.0 | $33/2^+$ | 939.0 | 0.183(+10-9) | 60.6(+33-30) | | |
| | 6922.6 | $37/2^+$ | 1106.6 | 0.155(6) | 51.3(20) | | |
| | 8188.1 | $41/2^+$ | 1265.5 | 0.128(+9-12) | 42.4(+30-40) | | |
| | 9622.1 | $45/2^+$ | 1434.0 | 0.094(+7-8) | 31.2(+23-27) | | |
| | 11230.1 | $49/2^+$ | 1608.0 | 0.034(7) | 11.3(23) | | |

## $^{108}_{48}Cd_{60}$

| | $E_{level}$ keV | $I^{\pi}$ | $E_{\gamma}$(E2) keV | B(E2) $(eb)^2$ | B(E2) (W.u.) | References | **Configurations and Comments:** |
|---|---|---|---|---|---|---|---|
| 1. | 4708.7 | $12^+$ | | | | **2005Da16** | 1. Configuration assigned as $\pi(g_{9/2})^{-2} \otimes \nu [(g_{7/2})^2 (h_{11/2})^2]$ from TRS calculations.<br>2. The mean lifetimes (in ps) of the levels from 5502.5 keV to 8824.5 keV are 1.32(13), 0.58(7), 0.33(4) and 0.22(4), respectively.<br>3. The $\mathcal{J}^2$/B(E2) ratios of 14 ħ to 20 ħ levels are 129(13), 158(19), 184(20) and 175(40) $\hbar^2$ $MeV^{-1}$ $(eb)^{-2}$, respectively. |
| | 5502.5 | $14^+$ | 794.0 | 0.19(2) | 62(6) | 2005Si23 | |
| | 6458.9 | $16^+$ | 956.3 | 0.17(2) | 56(6) | 1993Th05 | |
| | 7564.2 | $18^+$ | 1105.3 | 0.14(2) | 46(6) | | |
| | 8824.5 | $20^+$ | 1260.3 | 0.11(3) | 36(10) | | |

## $^{109}_{48}Cd_{61}$

| | $E_{level}$ keV | $I^{\pi}$ | $E_{\gamma}(E2)$ keV | B(E2) $(eb)^2$ | B(E2) (W.u.) | References | **Configurations and Comments:** |
|---|---|---|---|---|---|---|---|
| 1. | 3524.7 | $25/2^+$ | | | | **1994Ju05** | 1. Band 4 in Fig. 1 of 1994Ju05 interpreted as possible AMR band in theory paper 2022Pa27 with probable configuration of $\pi(g_{9/2})^{-4}\otimes\nu[(g_{7/2})(h_{11/2})^2]$ from classical rotor model calculations. The authors of 1994Ju05 interpreted this band as a regular collective rotational band, and as signature partner of band 2. |
| | 4246.7 | $29/2^+$ | 722.0 | | | | |
| | 5261.5 | $33/2^+$ | 1014.8 | | | | |
| | 6518.8 | $37/2^+$ | 1257.3 | | | | |
| | 7909.2 | $41/2^+$ | 1390.4 | | | | |
| | 9378.2 | $(45/2^+)$ | 1468.9 | | | | |
| 2. | 3939.9 | $27/2^+$ | | | | **1994Ju05** | 1. Band 3 in Fig. 1 of 1994Ju05 interpreted as possible AMR band in theory paper 2022Pa27 with probable configuration of $\pi(g_{9/2})^{-4}\otimes\nu[(g_{7/2})(h_{11/2})^2]$ from classical rotor model calculations. The authors of 1994Ju05 interpreted this band as a regular collective rotational band, and as signature partner of band 1. |
| | 4725.0 | $31/2^+$ | 785.1 | | | | |
| | 5775.6 | $35/2^+$ | 1050.6 | | | | |
| | 7077.6 | $39/2^+$ | 1302.0 | | | | |
| | 8599.0 | $41/2^+$ | 1521.4 | | | | |
| 3. | 5051.4 | $31/2^-$ | | | | **2000Ch04** | 1. Band 5 in Fig. 1 of 2000Ch04 is interpreted by authors as an AMR band, with deformation parameters $(\varepsilon_2, \gamma)=(0.14, 0^o)$ from semiclassical and TRS calculations, with configuration $\pi(g_{9/2})^{-2}\otimes\nu(h_{11/2})^3$ taken from 1994Ju05. See also the same configuration is proposed by 2014Zh14 from TAC-CDFT calculations, with $\beta_2$ = (0.209-0.167).<br>2. The mean lifetimes (in ps) of the levels from 5971.4 keV to 11131.4 keV are 0.354(31), 0.404(+13-12), 0.184 (9) and 0.179(4), respectively obtained by taking weighted average of mean lifetimes given in 2000Ch04.<br>3. The $\mathcal{J}^2$/B(E2) ratio is about values is 165 $MeV^{-1}\,\hbar^2\,e^{-2}\,b^{-2}$. |
| | 5971.4 | $35/2^-$ | 920.0 | 0.33(3) | 106.7(97) | | |
| | 7009.8 | $39/2^-$ | 1038.4 | 0.166(+4-6), | 53.7(+13-19) | | |
| | 8201.0 | $43/2^-$ | 1191.2 | 0.181(+7-6) | 58.5(+23-19) | | |
| | 9567.4 | $47/2^-$ | 1366.4 | 0.096(2) | 31.0(6) | | |
| | 11131.4 | $(51/2^-)$ | 1564.0 | | | | |

## $^{110}_{48}Cd_{62}$

| | $E_{level}$ keV | $I^{\pi}$ | $E_{\gamma}(E2)$ keV | B(E2) $(eb)^2$ | B(E2) (W.u.) | References | **Configurations and Comments:** |
|---|---|---|---|---|---|---|---|
| 1. | 5026.3 | $14^+$ | | | | **2011Ro01** | 1. Configuration assigned as $\pi(g_{9/2})^{-2}\otimes\nu(h_{11/2})^2$ from semi-classical particle rotor model calculations.<br>2. The mean lifetimes (in ps) of the levels from 6101.3 keV to 12763.3 keV are 0.36(3), 0.23(3), 0.17(3), 0.21(7), 0.27(7) and 0.35, respectively.<br>3. $(\beta_2, \gamma)=(0.20, 13°)$ from TAC-RMF calculations (2015Pe06).<br>4. Quasi-rotational band, as level energies do not seem to follow a monotonically increasing trend with ascending spins as expected for a rotor. |
| | 6101.3 | $16^+$ | 1075 | 0.14(1) | 44.7(32) | | |
| | 7325.3 | $18^+$ | 1224 | 0.12(2) | 38(6) | | |
| | 8648.3 | $20^+$ | 1323 | 0.11(2) | 35(6) | | |
| | 9962.3 | $22^+$ | 1314 | 0.09(3) | 29(10) | | |
| | 11320.3 | $24^+$ | 1358 | 0.06(2) | 19.1(63) | | |
| | 12763.3 | $26^+$ | 1443 | | | | |

## $^{111}_{48}\mathrm{Cd}_{63}$

| | $E_{level}$ keV | $I^{\pi}$ | $E_{\gamma}$(E2) keV | B(E2) $(eb)^2$ | B(E2) (W.u.) | References | **Configurations and Comments:** |
|---|---|---|---|---|---|---|---|
| 1. | 3763.2 | 27/2- | | | | **1994Re06** | 1. Yrast band in Fig. 7 of 1994Re06 interpreted as possible AMR band in theory paper 2018Ma28 with probable configuration of $\pi(g_{9/2})^{-2}\otimes\nu(h_{11/2})^3$ from SCM calculations, and predicted B(E2) of 0.24 to 0.09 with ascending spins. 1994Re06 interpreted this band as a regular collective rotational band. |
| | 4556.1 | 31/2- | 792.9 | | | | |
| | 5501.9 | $35/2^-$ | 945.8 | | | | |
| | 6649.0 | $39/2^-$ | 1147.1 | | | | |
| | 7951.3 | $43/2^-$ | 1302.3 | | | | |
| | 9407.3 | $47/2^-$ | 1456.0 | | | | |

## $^{108}_{49}\mathrm{In}_{59}$

| | $E_{level}$ keV | $I^{\pi}$ | $E_{\gamma}$(E2) keV | B(E2) $(eb)^2$ | B(E2) (W.u.) | References | **Configurations and Comments:** |
|---|---|---|---|---|---|---|---|
| 1. | 1861.5 | $8^{(-)}$ | | | | **2001Ch71** | 1. Configuration assigned as $\pi[(g_{9/2})^{-2}d_{5/2}]\otimes\nu h_{11/2}$ below alignment and $\pi[(g_{9/2})^{-2}d_{5/2}]\otimes\nu[(h_{11/2})^3]$ after alignment with deformation $(\beta_2, \gamma) \approx (0.20, 0°)$ from TAC-RMF calculations, see also 2016Su17 for detailed antimagnetic interpretation. |
| | 2439.4 | $10^{(-)}$ | 577.9 | | | | |
| | 3274.1 | $12^{(-)}$ | 834.7 | | | | |
| | 4265.6 | $14^{(-)}$ | 991.5 | | | | |
| 2. | 3548.1 | 11 | | | | **2001Ch71** | 1. Configuration assigned as $\pi[(g_{9/2})^{-2}g_{7/2}]\otimes\nu h_{11/2}$ below alignment and $\pi[(g_{9/2})^{-2}d_{7/2}]\otimes\nu[(h_{11/2})^3]$ after alignment with deformation $(\beta_2, \gamma) \approx (0.17, 5°)$ and (0.18, 5°), respectively, from TAC-RMF calculations, see also 2016Su17 for detailed antimagnetic interpretation. |
| | 4101.2 | 13 | 553.1 | | | | |
| | 4879.0 | 15 | 777.8 | | | | |
| | 577.3 | 17 | 828.3 | | | | |
| | 6111.5 | 19 | 904.2 | | | | |
| | 7613.9 | 21 | 1002.4 | | | | |
| | 8792.9 | (23) | 1179.0 | | | | |

## $^{109}_{49}\mathrm{In}_{60}$

| | $E_{level}$ keV | $I^{\pi}$ | $E_{\gamma}$(E2) keV | B(E2) $(eb)^2$ | B(E2) (W.u.) | References | **Configurations and Comments:** |
|---|---|---|---|---|---|---|---|
| 1. | 4299.2 | $21/2^{(+)}$ | | | | **2020Wa07** | 1. Configuration assigned as $\pi(d_{5/2})^{-2}(g_{9/2})^{-2}\otimes\nu(h_{11/2})^2$ from TAC-RMF calculations.<br>2. The $\mathscr{J}^2$ values form 29/2ħ to 41/2ħ levels are 19.1, 18.9, 25.0, 26.6 MeV$^{-1}$ $\hbar^2$ |
| | 4742.8 | $25/2^{(+)}$ | 443.6 | | | | |
| | 5396.8 | $29/2^{(+)}$ | 654.0 | | | | |
| | 6261.6 | $33/2^{(+)}$ | 864.8 | | | | |
| | 7286.4 | $(37/2^+)$ | 1024.8 | | | | |
| | 8460.8 | $(41/2^+)$ | 1174.4 | | | | |
| 2. | 4755.7 | $23/2^+$ | | | | **2020Wa07** | 1. Configuration assigned as $\pi(g_{7/2}(g_{9/2})^{-2})\otimes\nu(h_{11/2})^2$ from TAC-RMF calculations.<br>2. The $\mathscr{J}^2$ values form 31/2 ħ to 43/2 ħ levels are 23.8, 21.7, 25.5, 26.5 MeV$^{-1}$ $\hbar^2$ |
| | 5218.7 | $27/2^+$ | 463.0 | | | | |
| | 5849.8 | $31/2^+$ | 631.1 | | | | |
| | 6666.1 | $35/2^+$ | 816.3 | | | | |
| | 7639.1 | $39/2^+$ | 973.0 | | | | |
| | 8782.5 | $43/2^+$ | 1143.4 | | | | |

## $^{110}_{49}\mathrm{In}_{61}$

| | $E_{level}$ keV | $I^{\pi}$ | $E_{\gamma}$(E2) keV | B(E2) $(eb)^2$ | B(E2) (W.u.) | References | Configurations and Comments: |
|---|---|---|---|---|---|---|---|
| 1 | 1617.0 | $8^-$ | | | | **2001Ch71** | 1. Configuration assigned as $\pi[(g_{9/2})^{-2}\, g_{5/2}]\otimes\nu h_{11/2}$ below alignment and $\pi[(g_{9/2})^{-2}\, d_{5/2}]\otimes\nu[(h_{11/2})^3]$ after alignment with deformation ($\beta_2$, $\gamma$) ≈ (0.18, -10°) and (0.17, -5°), respectively, from TAC-RMF calculations, see also 2016Su17 for detailed antimagnetic interpretation. |
| | 2201.4 | $10^-$ | 584.4 | | | | 2. Quasi-rotational band, as level energies do not seem to follow a monotonically increasing trend with ascending spins as expected for a rotor. |
| | 3079.4 | $12^-$ | 878.0 | | | | |
| | 4157.0 | $14^-$ | 1077.6 | | | | |
| | 5181.3 | $16^-$ | 1024.3 | | | | |
| | 6062.6 | $18^-$ | 881.3 | | | | |
| | 6999.7 | $(20^-)$ | 937.1 | | | | |
| | 8088.4 | $(22^-)$ | 1088.7 | | | | |
| | 9398.8 | $(24^-)$ | 1310.4 | | | | |
| 2. | 2220.2 | $9^{(-)}$ | | | | **2001Ch71** | 1. Configuration assigned as $\pi[(g_{9/2})^{-2}\, g_{7/2}]\otimes\nu h_{11/2}$ below alignment and $\pi[(g_{9/2})^{-2}\, d_{7/2}]\otimes\nu[(h_{11/2})^3]$ after alignment with deformation ($\beta_2$, $\gamma$) ≈ (0.20, 0°) and (0.18, 5°), respectively, from TAC-RMF calculations, see also 2016Su17 for detailed antimagnetic interpretation. |
| | 2798.2 | $11^{(-)}$ | 578.0 | | | | 2. The mean lifetimes (in ps) of the levels from 5556.2 keV to 8463.5 keV are 0.774(+105-83), 0.554(+63-41), 0.403(+23-18) and 0.205(+18-14), respectively. |
| | 3628.8 | $13^{(-)}$ | 830.6 | | | | 3. Quasi-rotational band, as level energies do not seem to follow a monotonically increasing trend with ascending spins as expected for a rotor. |
| | 4605.8 | $15^{(-)}$ | 977.0 | | | | |
| | 5556.2 | $17^{(-)}$ | 950.4 | 0.136(16) | 43.4(51) | | |
| | 6445.3 | $19^{(-)}$ | 889.1 | 0.265(+21-27) | 85(+7-9) | | |
| | 7391.5 | $21^{(-)}$ | 946.2 | 0.266(+13-15) | 84.9(+41-48) | | |
| | 8463.5 | $(23^-)$ | 1072.0 | 0.281(+21-23) | 90(7) | | |
| | 9698.4 | $(25^-)$ | 1234.9 | | | | |
| | 11117.5 | $(27^-)$ | 1419.1 | | | | |
| | 12744.3 | $(29^-)$ | 1626.8 | | | | |

## $^{112}_{49}\mathrm{In}_{63}$

| | $E_{level}$ keV | $I^{\pi}$ | $E_{\gamma}$(E2) keV | B(E2) $(eb)^2$ | B(E2) (W.u.) | References | Configurations and Comments: |
|---|---|---|---|---|---|---|---|
| 1. | 2071 | $(11^-)$ | | | | **2012Li51** | 1. Configuration assigned as $\pi(g_{9/2})^{-2}\, g_{7/2}\otimes \nu(h_{11/2})^3$ from TAC-RMF calculations. |
| | 2653 | $(13^-)$ | 582 | | | | 2. $\gamma$ ≈ 28° and 14° with small decrease in deformation for both the configurations. alignment from TAC-RMF calculations. |
| | 3458 | $(15^-)$ | 805 | | | | 3. The dynamic moment of inertia ($\mathcal{J}^2$) from 15 ħ to 23 ħ are 17.9, 31.2, 45.4, 53.3 and 55.6 $MeV^{-1}\,\hbar^2$, respectively. |
| | 4390 | $(17^-)$ | 933 | | | | 4. Quasi-rotational band, as level energies do not seem to follow a monotonically increasing trend with ascending spins as expected for a rotor. |
| | 5235 | $(19^-)$ | 845 | | | | |
| | 6155 | $(21^-)$ | 920 | | | | |
| | 7147 | $(23^-)$ | 992 | | | | |
| | 8327 | $(25^-)$ | 1180 | | | | |

## $^{113}_{49}\mathrm{In}_{64}$

| | $E_{level}$ keV | $I^{\pi}$ | $E_{\gamma}$(E2) keV | B(E2) $(eb)^2$ | B(E2) (W.u.) | References | Configurations and Comments: |
|---|---|---|---|---|---|---|---|
| 1. | 3966 | $23/2^+$ | | | | **2019Ma48** | 1. Configuration assigned as $\pi(g_{9/2})^{-2}\, g_{7/2}\otimes \nu(h_{11/2})^2$ and a gentle up-bend is attributed to the $\pi(g_{9/2})^{-2}\, g_{7/2}\otimes\, \nu(h_{11/2})^2\, (g_{7/2}/d_{5/2})^2$ from TAC-RMF calculations. |
| | 4605 | $27/2^+$ | 638.1 | | | | 2. The dynamic moment of inertia ($\mathcal{J}^2$) from 31/2ħ to 47/2ħ are 56.7, 19.2, 27.7, 46.2 and 49.9 $MeV^{-1}\,\hbar^2$, respectively. |
| | 5314 | $(31/2^+)$ | 708.7 | | | | |
| | 6230 | $(35/2^+)$ | 916.7 | | | | |
| | 7291 | $(39/2^+)$ | 1061.1 | | | | |
| | 8439 | $(43/2^+)$ | 1147.6 | | | | |
| | 9667 | $(47/2^+)$ | 1227.8 | | | | |
| | 10938 | $(51/2^+)$ | 1271.0 | | | | |

## $^{114}_{49}\mathrm{In}_{65}$

| | $E_{level}$ keV | $I^{\pi}$ | $E_{\gamma}$(E2) keV | B(E2) $(eb)^2$ | B(E2) (W.u.) | References | **Configurations and Comments:** |
|---|---|---|---|---|---|---|---|
| 1. | 4007.1 | $(16^-)$ | | | | **2023Zh14** | 1. Configuration assigned as $\pi[(g_{9/2})^{-2} d_{5/2}]\otimes \nu(h_{11/2})^3$ from classical rotor model calculations and the AMR band assignment is tentative. |
| | 4851.6 | $(18^-)$ | 844.5 | | | | |
| | 5746.7 | $(20^-)$ | 895.1 | | | | |
| | (6758.9) | $(22^-)$ | (1012.2) | | | | |

## $^{127}_{54}\mathrm{Xe}_{73}$

| | $E_{level}$ keV | $I^{\pi}$ | $E_{\gamma}$(E2) keV | B(E2) $(eb)^2$ | B(E2) (W.u.) | References | **Configurations and Comments:** |
|---|---|---|---|---|---|---|---|
| 1. | 2306.7 | $19/2^+$ | | | | **2020Ch48** | 1. Configuration assigned as $\nu((g_{7/2})^{-1}(h_{11/2})^{-2})_{j\nu=19/2} \otimes\pi((h_{11/2})^2)$ from semi-classical particle plus rotor model calculations |
| | 2778.8 | $23/2^+$ | 472.1 | | | | 2. The dynamic moment of inertia ($\mathscr{J}^2$) of levels from $27/2^+$ to $39/2^+$ are 22.0, 22.7, 26.3 and 32.7 $\hbar^2$ MeV$^{-1}$. |
| | 3425.4 | $27/2^+$ | 646.6 | | | | |
| | 4237.9 | $31/2^+$ | 812.5 | | | | |
| | 5197.4 | $35/2^+$ | 959.5 | | | | |
| | 6276.4 | $39/2^+$ | 1079.0 | | | | |

## $^{130}_{56}\mathrm{Ba}_{74}$

| | $E_{level}$ keV | $I^{\pi}$ | $E_{\gamma}$(E2) keV | B(E2) $(eb)^2$ | B(E2) (W.u.) | References | **Configurations and Comments:** |
|---|---|---|---|---|---|---|---|
| 1. | 5713.8 | $15^-$ | | | | **2020GU21** | 1. Band N6 in 2020Gu21 is assigned as AMR band simply based on a similar structure observed in $^{137}$Nd in 2019Pe08. The configurations assigned as $\nu(h_{11/2})^{-2}\otimes \nu(g_{7/2})^{-1}\otimes\pi((h_{11/2})^2$ (2019Pe12). The authors of present work consider this assignment as very tentative, as no defining criteria are available. |
| | 6587.5 | $17^-$ | 873.7 | | | | |
| | 7561.1 | $(19^-)$ | 973.6 | | | | |

## $^{137}_{60}\mathrm{Nd}_{77}$

| | $E_{level}$ keV | $I^{\pi}$ | $E_{\gamma}$(E2) keV | B(E2) $(eb)^2$ | B(E2) (W.u.) | References | **Configurations and Comments:** |
|---|---|---|---|---|---|---|---|
| 1. | 6072 | $(35/2^-)$ | | | | **2019Pe08** | 1. Configuration assigned as $\pi(h_{11/2})^2 \otimes \nu(h_{11/2})^{-1}$ $(\pi,\alpha)$= (-,-1/2) from CNS calculations, however, 2023Ka10 do not invoke antimagnetic rotational for this band. |
| | 7036 | $(39/2^-)$ | 964 | | | | 2. $(\varepsilon_2, \gamma) \approx (0.15, -60^o)$ from CNS calculations. |
| | 8094 | $(43/2^-)$ | 1058 | | | | 3. Descending theoretical B(E2) values are predicted from CNS model. |
| | 9254 | $(47/2^-)$ | 1160 | | | | 4. The average moment of inertia is $\approx 39\ \hbar^2$ MeV$^{-1}$. |
| | 10514 | $(51/2^-)$ | 1260 | | | | |
| | 11878 | $(55/2^-)$ | 1364 | | | | |
| | 13351 | $(59/2^-)$ | 1473 | | | | |
| | 14936 | $(63/2^-)$ | 1585 | | | | |
| | 16631 | $(67/2^-)$ | 1695 | | | | |
| | 18435 | $(71/2^-)$ | 1804 | | | | |
| | 20338 | $(75/2^-)$ | (1913) | | | | |

## $^{142}_{63}\mathrm{Eu}_{79}$

| | $E_{level}$ keV | $I^\pi$ | $E_\gamma$(E2) keV | B(E2) (eb)$^2$ | B(E2) (W.u.) | References | **Configurations and Comments:** |
|---|---|---|---|---|---|---|---|
| 1. | 3578 | 17$^-$ | | | | **2017Al30** | 1. Probable configuration: [$\pi$ ($g_{7/2}^{-1}$)($h_{11/2}^{-3}$)$\otimes$ ($h_{11/2}^{2}$)] using SPRM model calculations.<br>2. The mean lifetimes (in ps) of level from 4930.1 to 7286.9 keV are 1.23(+21-16), 0.46(+9-7) and < 0.45, respectively.<br>3. The $\mathcal{J}^2$/B(E2) ratios of 19 ħ to 25 ħ levels are 66.9(87), 92.9(+15-10), 158.2(26) and < 414.7 $\hbar^2$ MeV$^{-1}$ (eb)$^{-2}$, respectively. |
| | 4110 | 19$^-$ | 532.0 | 0.23(3) | 52(7) | | |
| | 4930.1 | 21$^-$ | 820.1 | 0.18(+3-2) | 41(+7-4) | | |
| | 6005.9 | 23$^-$ | 1075.8 | 0.12(2) | 27.2(45) | | |
| | 7286.9 | 25$^-$ | 1281.0 | > 0.05 | > 11.4 | | |
| | 8750.9 | 27$^{(-)}$ | 1464 | | | | |

## $^{143}_{63}\mathrm{Eu}_{80}$

| | $E_{level}$ keV | $I^\pi$ | $E_\gamma$(E2) keV | B(E2) (eb)$^2$ | B(E2) (W.u.) | References | **Configurations and Comments:** |
|---|---|---|---|---|---|---|---|
| 1. | 7389.0 | 43/2$^+$ | | | | **2015Ra12** | 1. Configuration assigned as $\nu(h_{11/2})^{-2}$ $\pi(d_{5/2}/g_{7/2})^{-3} \otimes (\pi(h_{11/2})^{2})$ from semi-classical particle rotor model calculations<br>2. $\beta_2 \approx 0.08$ from particle rotor model calculations.<br>3. The mean lifetimes (in ps) of the levels from 8003.8 keV to 11227.5 keV are 2.76(+42-35), 0.66(+14- 10),0.37(+8-6) and 0.44(+6-5), respectively.<br>4. The $\mathcal{J}^2$/B(E2) ratios of 47/2 ħ to 59/2 ħ levels are 47(+8-7), 66(+13-11), 216(+50-33) and 444(74) $\hbar^2$ MeV$^{-1}$ (eb)$^{-2}$, respectively. |
| | 8003.8 | 47/2$^+$ | 614.8 | 0.34(+5-4) | 76(+11-9) | | |
| | 8870.3 | 51/2$^+$ | 866.5 | 0.25(+5-4) | 56(+11-9) | | |
| | 9977.8 | 55/2$^+$ | 1107.5 | 0.13(+3-2) | 29(+7-4) | | |
| | 11227.5 | 59/2$^+$ | 1249.7 | 0.06(1) | 13.5(22) | | |
| | 12627.5 | (63/2$^+$) | 1400 | | | | |
| 2. | 12018.4 | 63/2$^+$ | 790.9 | 0.35(+6-5) | 79(+14-11) | **2015Ra12** | 1. Tentative configuration assigned as [$\nu(h_{11/2})^{-2}$ $\pi(d_{5/2}/g_{7/2})^{-3}$ $\pi(h_{11/2})^{2}$+core($3\hbar$)] $\otimes(\pi(h_{11/2})^{2}$ from semi-classical particle rotor model calculations. According to 2015Ra12, smooth band termination for this band cannot be ruled out.<br>2. $\beta_2 \approx 0.09$ from particle rotor model calculations.<br>3. The mean lifetimes (in ps) of the levels from 12018.4 keV to 14159.7 keV are 0.48(+8-7), 0.44(+7-5) and <0.38, respectively.<br>4. The $\mathcal{J}^2$/B(E2) ratios of 63/2ħ to 71/2ħ levels are 69(+12-10), 76(+13-10) and > 216 $\hbar^2$ MeV$^{-1}$ (eb)$^{-2}$, respectively. |
| | 12974.0 | 67/2$^+$ | 955.6 | 0.23(+4-3) | 52(+9-7) | | |
| | 14159.7 | 71/2$^+$ | 1185.7 | > 0.09 | > 20.3 | | |

## $^{144}_{66}\mathrm{Dy}_{78}$

| | $E_{level}$ keV | $I^\pi$ | $E_\gamma$(E2) keV | B(E2) (eb)$^2$ | B(E2) (W.u.) | Reference | **Configurations and Comments:** |
|---|---|---|---|---|---|---|---|
| 1. | 3172.4 | 10$^+$ | | | | **2009Su09** | 1. Possible assignment as AMR with configuration of $\pi h^2_{11/2} \otimes \nu h^{-2}_{11/2}$, in comparison with configuration for an AMR band in $^{110}$Cd in 2011Ro01.<br>2. Quasi-rotational band, as level energies do not seem to follow a monotonically increasing trend with ascending spins as expected for a rotor. |
| | 3818.5 | 12$^+$ | 664.1 | | | | |
| | 4542.3 | 14$^+$ | 723.8 | | | | |
| | 5302.7 | 16$^+$ | 760.4 | | | | |
| | 6074.1 | 18$^+$ | 771.4 | | | | |

References for Table 2

References for Table 2 (continued)

References for Table 2 (continued)

References for Table 2 (continued)

References for Table 2 (continued)

References for Table 2 (continued)

References for Table 2 (continued)

References for Table 2 (continued)

## References for Table 2 (continued)

## References for Table 2 (continued)

References for Table 3

## References for Table 3(Continued)

References for Table 3(Continued)

## Theory references for magnetic- and antimagnetic-rotational structures

Theory references for magnetic- and antimagnetic-rotational structures (Continued)